# Topological Metamaterials

Xiang Ni[1,#], Simon Yves[1,#], Alex Krasnok[2,#,*], and Andrea Alu[1,3,4,*]

[1]*Photonics Initiative, Advanced Science Research Center, City University of New York, New York, NY, 10031 USA*

[2]*Department of Electrical and Computer Engineering, Florida International University, Miami, FL 33174, USA*

[3]*Department of Electrical Engineering, City College, The City University of New York, 160 Convent Avenue, New York, 10031, USA*

[4]*Physics Program, The Graduate Center, The City University of New York, 365 Fifth Avenue, New York, 10016, USA*

[#]*These authors contributed equally*

\* *Correspondence to: akrasnok@fiu.edu, aalu@gc.cuny.edu*

## Abstract

Unlike geometrical features, such as shape or volume, the topological properties of an object do not change continuously, but only due to abrupt transitions associated with a change in the integer quantities that describe these properties - known as topological invariants. In turn, this resilience of topological features is associated with intrinsic robustness. For example, the topological invariant known as the Chern number, associated with the quantization of the Hall conductance in a two-dimensional electron gas, supports disorder-immune electron transport at the surface of topological insulators (TIs). One of the most significant breakthroughs in physics of the last decade has been the discovery that materials with non-trivial topological properties for electronic, electromagnetic, acoustic and mechanical responses can be designed and manufactured at our will through engineered metamaterials (MMs). Here, we review the foundation and the state-of-the-art advances of topological photonics, acoustics and mechanical MMs. We discuss how topological MMs enable nontrivial wave phenomena in physics, engineering, of great interest for a broad range of interdisciplinary science disciplines such as classical and quantum chemistry. We first introduce the foundations of topological materials and the main concepts behind their peculiar features, including the concepts of topological charge and geometric phase. We then discuss the topology of electronic band structures in natural topological materials, like topological insulators and gapless Dirac and Weyl semimetals. Based on these concepts, we review the concept, design and response of topologically nontrivial MMs in photonics and phononics, including topological phases in 2D MMs with and without time-reversal symmetry, Floquet TIs based on spatial and temporal modulation, topological phases in 3D MMs, higher-order topological phases in MMs, non-Hermitian and nonlinear topological MMs and the topological features of scattering anomalies. We also discuss the topological properties emerging in other related contexts, such as the topological aspects of chemical reactions and polaritons. This survey aims at connecting the recent advances in a broad range of scientific areas associated with topological concepts, and highlights opportunities offered by topological MMs for the chemistry community at large.

# Content

# 1. Introduction

One of the central concepts in modern physics has been the understanding that the geometry and features of the background in which physical processes unfold is as important as the properties of the system itself. Relativity is a clear example of this important statement, since in this context real space geometry and topology play a decisive role, inseparable from the physical system itself. In general relativity, the geometry of spacetime manifests itself in the fact that a body moves along curved geodesic lines following its curvature, as if some forces were acting on this body. In a different context, Y. Aharonov and D. Bohm demonstrated[1] that the knowledge of the local electromagnetic field may be insufficient to predict the time evolution of a particle wavefunction in the quantum mechanics of systems with nontrivial topology. Unusual topological properties were since then discovered in a wide range of optical and electromagnetic phenomena. Since electromagnetism and quantum mechanics govern complex substances, it is not surprising that topological concepts have also made their way into solid-state physics, complex atomic systems, mechanics, acoustics, and quantum chemistry[2–6]. Topologically nontrivial systems are usually quantified with an invariant – an integer-valued quantity, like the Chern number[2], which does not change upon continuous deformations that preserve the topological nature[7,8]. Topological phases of matter are associated with exciting features of wave propagation, holding the promise for revolutionary technologies from quantum electronics[2,9–14] to photonics[15,16,25–34,17,35,36,18–24], mechanics and acoustics[4,37,46,47,38–45], and quantum computing[48–56].

One seminal nontrivial topological property in condensed matter physics is the robust wave propagation features at the edge of a topologically nontrivial material–topological insulator (TI). Like electronic topological insulators (ETIs), topological photonic/phononic states are distinguished by directional and robust propagation along the interface between two topologically different regions[8,24,26,49]. This robustness has been demonstrated in various electromagnetic systems, ranging from microwaves[17,31] to optics[23,25] but also in acoustic[38,57] and mechanical lattices[58,59]. In turn, this feature holds great promise to revolutionize many applications, including slow-light optical buffers[19] and laser technologies[33–35,60–65]. Topological phenomena also find applications in quantum optics, since integrating quantum emitters into topological photonic structures can lead to stable strong interaction[66] and hybrid light-matter states[67].

In developing topologically nontrivial materials, the concept of metamaterials (MMs) and metasurfaces (MSs) has become essential. MMs, artificial engineered materials composed of subwavelength resonators (meta-atoms), enable electromagnetic properties such as a negative index of refraction, etc., which are not found in nature[68]. Metasurfaces, the two-dimensional (2D) analog of MMs, are artificial structures tailored to achieve a certain functionality like control over wave propagation, reflection, and refraction[69,70,79–83,71–78]. The concepts of MMs[84], MSs[75] and photonic/phononic crystals[85] have brought forward a previously unattainable level of control of visible, infrared, and microwave electromagnetic waves and acoustic and mechanical waves at macroscopic and nanoscopic scales. Here, we do not distinguish between photonic/phononic crystals and MMs, because generally they fulfill a similar purpose: engineering artificial materials for extreme wave manipulation. Recently, the pursuit of wave control in these artificial media has been expanded with tools borrowed from the field of topology[26], enabling exotic phenomena such as unidirectional transport, robustness to imperfections due to fabrication, and immunity to back-scattering from sharp corners[17,22,23]. Although topological aspects of photonic/phononic systems have mostly been driven through the condensed matter physics perspective, recently there has been an expansion of topological photonics/phononics beyond these boundaries, leveraging the distinct features of photons and other classical waves[4,8,44]. Furthermore, the design of artificial media has considerably taken advantage of the fast development of modern additive manufacturing techniques, allowing a straightforward implementation of abstract topological concepts into actual samples[86,87]. The progress made in the past decades in the nano and micro fabrication domain also permits envisioning on-chip topological systems with promising applications in communication, computing and sensing.

In this work, we discuss how the concepts of topologically nontrivial interactions at the level of atoms or meta-atoms enable nontrivial phenomena in physics, engineering and interdisciplinary science such as quantum chemistry. First, we discuss the basics of topological materials and the main concepts behind their peculiar features, including topological charge and geometric phase. We then discuss the topology of electronic band structures in natural topological materials, like TIs and gapless Dirac and Weyl semimetals. We follow by reviewing the topologically nontrivial MMs in photonics and phononics: we start with topological phases in one-dimensional (1D) structures and then consider the topological MMs with broken time-reversal symmetry (TRS) and time-reversible topological MMs in 2D. We then discuss Floquet TIs based on spatial and temporal modulation. Next, we review Dirac, Weyl and gapped topological MMs in three dimensions (3D). We also summarize the recent progress on the MM realization of higher-order topological semimetals and higher-order topological insulators (HOTIs). Last, we briefly discuss the research efforts on hybridized topics of topological phenomena implemented in MMs, like non-Hermitian physics in topological MMs, synthetic TIs, and nonlinearity-induced topological phases. We show that the considered topological properties are ubiquitous in wave phenomena, and we supplement our presentation with a description of topological photonic, acoustic and mechanical MMs.

There are already several reviews on topological photonics[8,24,26,88–93]. The comprehensive work in Ref.[8] reviews the progress in topological photonics in a broad aspect. Other works focus on specific topics, such as 2D topological photonics[24,91], topological nanophotonics[89], active topological photonics[94], nonlinear topological photonics[95], topological photonics using synthetic dimensions[90,96], topological valley photonics[97] and higher order topology[98]. Many reviews also exist in the context of topological phononics[4,44–

[47]. In contrast to other works, our review aims at highlighting the role of geometric (Berry) phase concepts in various systems, in the topological properties of scattering anomalies, and of polarization charges in the bounded states and exceptional points (EPs). It also includes a discussion of topological properties of other nature, e.g., topological aspects of chemical reactions, particularly relevant to the broad readership of this journal. Our review hence focuses on a broad range of topics associated with topological MMs implemented in classical systems, such as photonics, acoustics and elastics, chemistry, and other topological systems for interdisciplinary science.

## 2. Topological concepts: from geometric phase to topological band theory

### 2.1. The ubiquity of topology, topological charge and geometric phase

Examples of the topological properties of space and of objects have been known for a long time. A textbook example is the so-called Euler characteristic $\chi$, defined for an object of any shape as $\chi = \text{Faces+Cornes-Edges}$ (Euler's polyhedron formula). The Euler characteristic of a sphere equals 2, **Fig. 1(a)**. Remarkably, if one deforms a sphere into any other shape with no cuts or holes, the Euler characteristic remains unchanged, **Fig. 1(b)**. In contrast to the sphere, a torus has one hole, and its Euler characteristic is 0, regardless of continuous deformations. The Euler characteristic is an example of *topological charge* or *topological invariant*, a parameter that does not change under *continuous deformations*, such as stretching, twisting and bending. *Discontinuous deformations* that involve cutting, tearing or attachments do change topological charges. Any two objects are topologically equivalent if they can be transformed into each other by continuous deformations. The stability of topological charges to continuous deformations determines the topological protection of such systems: in order to change the topological charge, it is necessary to produce discontinuous deformations.

Other common topologically nontrivial objects in real space are vortices, which can occur in field distributions, media or particles. For example, fields with orbital angular momentum carry a topological charge that may enhance wireless communication concepts, **Fig. 1(c)** [100,101]. Topological charges hold great promise for many other applications, including boosting light trapping, lasing, light-matter interaction enhancement, nonlinear optics, wave-front control and polarization conversion. In some instances, topological charges and their robustness may be undesirable or even harmful. We bring an example from medicine: today's most common cause of human death is cardiac arrhythmia, accompanied by the appearance of an electrochemical topological charge (vortex) in the heart[102], **Fig. 1(d)**. Electric defibrillation can destroy this topological charge by dramatically altering the electrochemical potentials and save the patient's life[102]. Tornados and whirlpools are also systems with a topological charge, whose stability can be a cause of great concern, **Fig. 1(e)**.

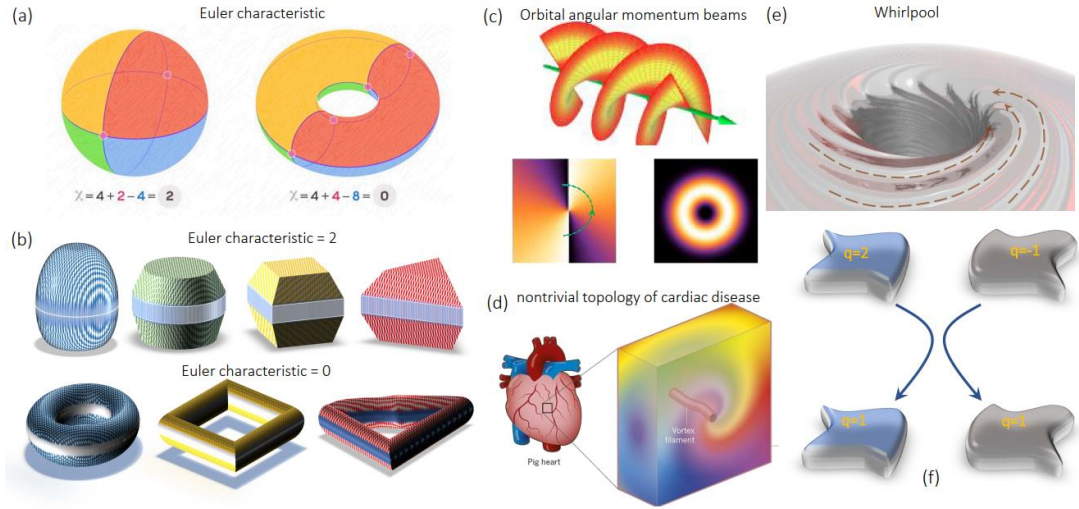

**Figure 1. The ubiquity of topology and topological charges**. (a) Euler characteristic of a sphere and a torus. (b) Continuous deformations leave the Euler characteristic unchanged. (c) Beams with angular momentum carry a topological charge observable as a phase vortex. (d) 3D waves of the electrical and mechanical activity responsible for many sudden cardiac death cases have nontrivial topology (vortexes). Reproduced with permission from[99]. Copyright 2018 Springer Nature. (e) Whirlpools and tornados are other examples of topological charges. (f) Like electric charges, topological charges can change when they compensate each other.

Like electric charges, topological charges can change when they interact with each other. For example, consider a system of two bodies with some electric charges. Changing the shape of the bodies will certainly not lead to a change of charge. However, charges brought into contact can compensate each other, leading the bodies to acquire the same charge, **Fig. 1(f)**. As we will see below, topological charges behave similarly, which is the essence of typological engineering of band diagrams in periodic structures.

The impact of topology on quantum physics has emerged thanks to the introduction of the concept of *Berry phase*, also known as *geometrical phase*[103] (GP). This tool has made it possible to rethink many classical and quantum phenomena in terms of topology. GP can be obtained by integrating or measuring the system response on closed paths in real or reciprocal space. Shortly after its introduction, Simon pointed out a corresponding concept in modern geometry – *holonomy*, i.e., the rotation of a tangent vector during its parallel translation along a closed curve on a curved surface, for example a sphere[104]. Even though the vector never rotates around its radius vector, it acquires a certain rotation after the closed-looped parallel transport, **Fig. 2(a)**. In other words, GP is an example of a failure of parallel transport around closed cycles, the final state not being parallel to the initial one. GP is not attributed to any force applied to the system but to the curvature of space itself.

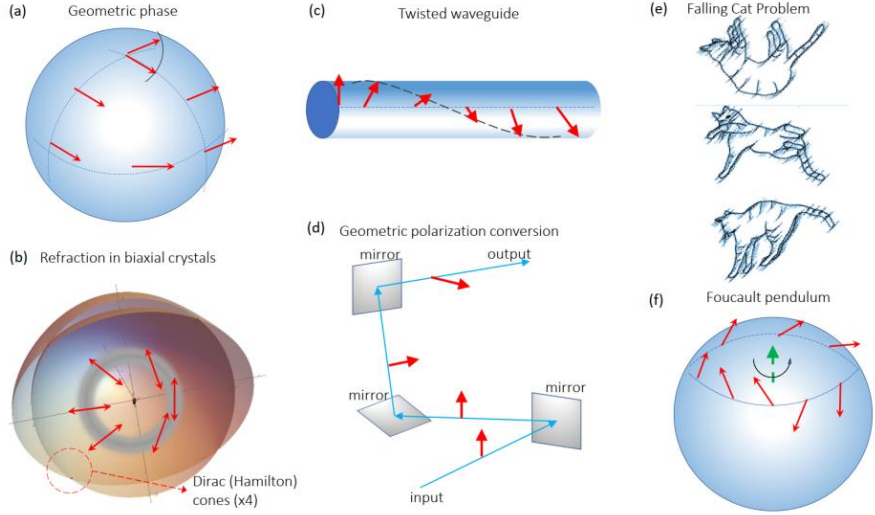

**Figure 2. Geometric phase (GP).** (a) GP is associated with the rotation of a tangent vector during its parallel translation along a closed curve on a curved surface (anholonomy). (b) Normal surface (k-surface) construction for indices of refraction $n_1$ and $n_2$. The insert shows how the liner polarization changes by 180º ($\pi$) when it turns around the conical intersection (Dirac or Hamilton cone). (c) Slow change in polarization of light propagating through an anisotropic medium. (d) 'Spin redirection' when light with a fixed polarization state changes direction. (e) Falling cat succeeds in landing on its feet even when starting in an upside-down position (cat anholonomy). (f) Foucault pendulum: Orientation of the plane of oscillation slowly rotates during the day, not returning to its original orientation.

The first observation of GP in optics was associated with revealing the diabolic point at the conical intersection of refractive index surfaces in biaxial crystals (Hamilton cone) in 1830 by Sir Hamilton[105]. In 1831, Lloyd showed that the linear polarization changes by 180º ($\pi$) when it turns around this conical intersection, **Fig. 2(b)**. Other optical effects of GP are associated with a slow change of the light polarization upon propagating through a twisted medium (**Fig. 2(c)**), giving rise to the *Pancharatnam-Berry phase*[106] and *spin redirection* when light changes polarization after several reflections, **Fig. 2(d)**. In this latter example, while the light beam has completed its trajectory and returned to its starting point, the polarization vector acquires a 90° rotation compared to its original direction. This rotation is due to a global change in the beam trajectory. The value of the GP depends on the path taken, which is associated with topological features. GP gave rise to other early studies concerning conical refraction[107,108], works discussing parallel polarization transport along curved rays[109–111], and a study addressing wave propagation in stratified inhomogeneous media[112]. This phenomenon is also valid for phononic waves. For instance, the transverse mechanical vibrations along a helicoidal spring can be represented as orthogonal circular polarizations[113], similar to sound waves with vortices carrying orbital angular momentum[114].

Due to its fundamental and comprehensive applications[115–122], GP has also been explored in other areas of physics[123,124]. A curious example of GP effects in mechanics is the falling cat problem, **Fig. 2(e)**. Here angular momentum conservation should prevent a cat from changing orientation when falling, but nevertheless the cat lands on its feet, even if it begins its fall upside down. It turns out that the cat can re-orient itself without any particular problem, which has been identified as a geometric effect: the deformation of the shape of a body has immediate consequences on its orientation[125]. Another example is the phenomenon described by Foucault about the orientation of the pendulum's oscillation plane: an observer on Earth witnesses that the orientation of the pendulum slowly rotates during the day and, in general, does not return to its original orientation[126], **Fig. 2(f)**. GP is also shown to underline the dynamics of a broad spectrum of quantum systems[127–129]. Associated concepts known in the literature are the Hannay angle in classical mechanics[130], the Aharonov-Anandan phase in quantum mechanics[131], and the Zak phase in condensed matter physics[132]. As we discuss in the following, GP is known to lead to observable and often unexpected phenomena in chemical reactions, e.g., in the context of molecular electronic degeneracies[133,134]. GP has also influenced high-energy and particle physics[135,136], gravity and cosmology[137–139], fluid mechanics[140,141], chemical physics[142–146], and many other active research areas. The impact of GP in non-Hermitian and open systems has also been discussed[147,148,157–159,149–156]. Several reviews on GPs are available[118], including historical reconstructions[124], applications to molecular physics[143], and a recent overview of its impact on condensed matter[160] and optics[161].

## 2.2. Geometric phase in quantum mechanics and gauge fields

The physics of topologically nontrivial materials and systems is rooted in the concept of gauge fields. As shown in the pioneering works of Pancharatnam[106], Aharonov-Bohm[1], Wu-Yang[162], Mead-Truhlar[134], and Berry[103], there is a deep connection between gauge fields and geometry. Although the gauge potentials were initially introduced in classical electrodynamics, their essence and importance was fully unveiled in quantum mechanics and quantum field theory[1,162]. A study of the physical consequences of gauge fields, as evidenced for the first time by the fundamental Aharonov-Bohm effect[1], led to remarkable theoretical and conceptual progress, culminating in the modern formulation of gauge theories[1,104,163]. Although gauge potentials naturally arise from "real" electromagnetic fields, a large family of gauge fields can also be found in parametric and reciprocal dimensions. The generation of artificial gauge fields has recently been investigated in a great variety of physical platforms[164], such as ultra-cold atomic gases[165,166], photonic crystals[8,24,26,167], graphene and graphene-like materials[168,169], mechanical systems[4], acoustics[170–175] and exciton-polariton systems[176–180].

Gauge fields are described by a Hamiltonian $H(\mathbf{p} - \mathbf{A})$, where $\mathbf{A} = \mathbf{A}(\mathbf{r}, t, \sigma)$ is the gauge potential, related to the so-called *Berry connection*. It is a function of the position operator $\mathbf{r}$, time $t$, and a spin (or pseudospin) degree of freedom $\sigma$. For example, in the case of an electric charge in an external magnetic field, the Hamiltonian reads $H(\mathbf{r}) = [\mathbf{p} - q\mathbf{A}(\mathbf{r})]^2 / 2M$, with $\mathbf{A}(\mathbf{r})$ being the vector potential, $\mathbf{B}(\mathbf{r}) = \nabla \times \mathbf{A}(\mathbf{r})$ [181], and $q$ and $M$ denoting the particle's charge and mass. Another type of model Hamiltonian, which plays an important role in condensed matter physics, is the Hamiltonian describing the motion of a spin 1/2 particle subject to spin-orbit coupling[13,182,183], $H = (\mathbf{p} - \mathbf{A}_{\text{SOC}})^2 / 2M$. Here the Berry connection $\mathbf{A}_{\text{SOC}}$ is a vector $\mathbf{A}_{\text{SOC}} = (\alpha_x \sigma_x, \alpha_y \sigma_y, \alpha_z \sigma_z)$, with $\sigma_{x,y,z}$ being the Pauli matrices. The gauge potential plays a central role in the topological properties of the QHEs[2,13,184], spintronics[182], the anomalous Hall effect[183], TIs, and superconductors[13]. Systems with explicit time dependence $\mathbf{A} = \mathbf{A}(t)$ play an essential role in *Floquet engineering*, e.g., Floquet TIs[23,185–188].

It is instructive to start the discussion of GP, Berry curvature and Berry connection with a recap of vector potentials in electromagnetic theory. To introduce the vector potential $\mathbf{A}(\mathbf{r}, t)$, we consider the Maxwell's divergence equation $\nabla \cdot \mathbf{B}(\mathbf{r}, t) = 0$ in the form $\mathbf{B}(\mathbf{r}, t) = \nabla \times \mathbf{A}(\mathbf{r}, t)$. Substituting this equation in $\nabla \times \mathbf{E}(\mathbf{r}, t) = -\partial \mathbf{B}(\mathbf{r}, t) / \partial t$ yields $\mathbf{E}(\mathbf{r}, t) = -\partial \mathbf{A}(\mathbf{r}, t) / \partial t - \nabla \varphi(\mathbf{r}, t)$ for the electric field through the vector $\mathbf{A}(\mathbf{r}, t)$ and scalar $\varphi(\mathbf{r}, t)$ potentials. The potentials are not unique and are defined up to $\mathbf{A}' = \mathbf{A} + \nabla f(\mathbf{r}, t)$ and $\varphi' = \varphi - \partial f(\mathbf{r}, t) / \partial t$, where $f(\mathbf{r}, t)$ is an arbitrary function of time and space. This is known as *gauge freedom*[164,189–191]. A choice of $f(\mathbf{r}, t)$ selects a particular gauge.

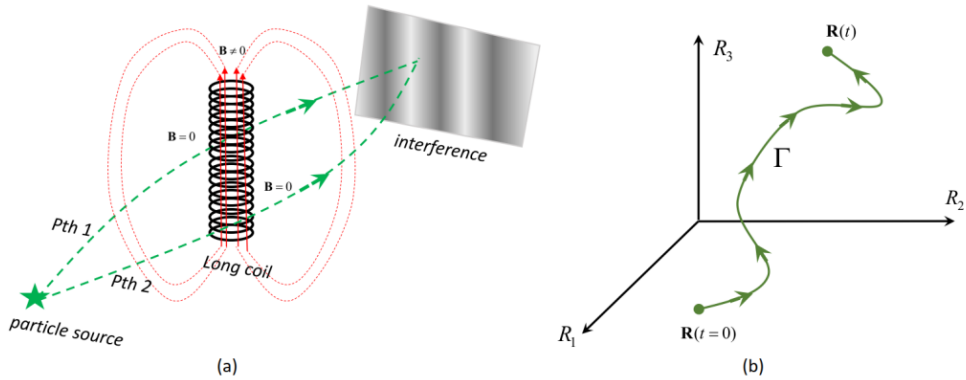

**Figure 3. Aharonov-Bohm effect and parametric space.** (a) Illustration of the Aharonov-Bohm effect. The interference fringes do not depend on the shape of the possible paths 1 and 2 in general, but only on their topological invariants, provided that the particle moves in a field-free region. (b) Example of the configuration space of parameters $\{\mathbf{R}\}$ and a trajectory in this space.

We define the magnetic field flux $\Phi_B$ through an arbitrary surface, and represent it using Stokes' theorem

$$\Phi_B = \int_S \mathbf{B} \cdot d\mathbf{S} = \int_S \nabla \times \mathbf{A} \cdot d\mathbf{S} = \oint_L \mathbf{A} \cdot d\mathbf{l} \ . \tag{1}$$

The magnetic flux through the arbitrary surface area $S$ is equal to the circulation of $\mathbf{A}(\mathbf{r}, t)$ over the closed path along the border of this area.

Let us apply this result to the geometry illustrated in **Fig. 3(a).** Here, a long coil of radius $R_c$ assures that the inner magnetic field $\mathbf{B}$ is homogeneous, and the outer field in the vicinity of the coil vanishes. The magnetic field lines close at infinity for the hypothetical infinitely long coil. For this case, Eq. (1) gives $A(r) = BR_c^2 / (2r)$ for the amplitude of vector potential. Thus, although the outer field vanishes, the vector potential has a finite value. This result is not affected by the gauge transformation since $\oint_L \nabla f(\mathbf{r}, t) \cdot d\mathbf{l} = 0$ for any closed path. Consequently, a particle with a charge $q$ capable of flying around the coil by either part 1 or path 2 from a source to a screen acquires the wave function $\psi_{\text{path1(2)}} \equiv \left\langle \text{sorce} | \text{screen} \right\rangle_{\text{path1(2)}} = \psi_{\mathbf{A}=0} \exp\left( \frac{iq}{\hbar c} \int_{\text{sotce}}^{\text{screen}} \mathbf{A} \cdot d\mathbf{l} \right)$, where $\psi_{\mathbf{A}=0}$ is the solution to Schrodinger's equation for the system with $\mathbf{A} = 0$. The interference of both possible paths gives interference fringes on the screen $\left| \left\langle \text{sorce} | \text{screen} \right\rangle \right|^2 \sim \cos\left( \frac{q}{\hbar c} \Phi_B \right)$. When two electronic wavepackets encircle a solenoid so that the magnetic field is zero along their paths, they will acquire a relative phase - the *Aharonov–Bohm phase* - due to the nonzero vector potential. This phase can be written as an integral over a closed path around the coil

$$\gamma = q \oint_L \mathbf{A} \cdot d\mathbf{l} \ . \tag{2}$$

It is important that this result holds over a regione in which $\nabla \times \mathbf{A} = 0$, because it makes the integral path in **Eq. (2)** independent of the specific choice of path shape. The Aharonov-Bohm phase is topological because it does not depend on the shape or geometric properties of the trajectory. The magnetic Aharonov-Bohm effect reveals the nonlocal aspect of quantum mechanics, allowing electromagnetic fields to influence charge even in areas from which they are excluded[1,162,192–195]. The effect was demonstrated in superconducting films with the completely excluded magnetic field from the electron's path due to the Meissner effect[196]. The Aharonov-Bohm effect has been demonstrated in many systems, including iron whiskers[197], ion traps[198], metallic rings[199], quantum dots[200], carbon nanotubes[201], electronic interferometers[202], optical lattices[203,204], TIs[205–207], and even a whirlpool in a water tank[208]. The dual *Aharonov-Casher effect* was also discovered, predicting that the dipole moments diffracted around charged tubes will acquire a similar phase[209–212]. Nevertheless, discussions about the fundamental aspects of the Aharonov-Bohm effect continue[213–218].

This concept can be generalized to other systems described by a Hamiltonian $H(t)$ with parameters adiabatically depending on time. In the example above, this adiabaticity was implied, assuring that the charge cycling around the coil does not radiate and stays in the same quantum state. The *adiabatic theorem*[104,219] implies that a system remains in the stationary (e.g., ground) state $|\psi_n(t)\rangle$ if the interaction is slow enough. In this case, such a state reads $|\psi(t)\rangle \simeq \exp[i\theta_n(t)]\exp[i\gamma_n(t)]|\psi_n(t)\rangle$, with the standard dynamic phase $\theta_n(t) = -\frac{1}{\hbar}\int_0^t E_n(t')dt'$, where $E_n(t')$ is the energy eigenvalue, and the generalized geometric (Berry) phase $\gamma_n(t) = \int_0^t \nu_n(t')dt'$, where $\nu_n(t) = i\langle\psi_n|\partial_t\psi_n(t)\rangle$. When the parameters of the Hamiltonian complete a cycle, the final state can return to its original value or acquire an additional phase factor. In the latter case, we say that the path encircles a *topological charge*.

To show that this GP does not depend on time but on the system's evolution in the parametric configuration space, we assume that the Hamiltonian depends on a number of slowly varying parameters $\{\mathbf{R}(t)\}$: $H(\mathbf{R}(t))$. In this case, according to the adiabatic theorem, the stationary Schrodinger equation reads $H(\mathbf{R}(t))|\psi_n(\mathbf{R}(t))\rangle = E_n(\mathbf{R}(t))|\psi_n(\mathbf{R}(t))\rangle$. For example, **Fig. 3(b)** shows an example of a configuration space of three parameters where the path $\Gamma$ represents the system's evolution in the configuration space. Since $|\partial_t\psi_n(\mathbf{R}(t))\rangle \equiv \frac{d}{dt}|\psi_n(\mathbf{R}(t))\rangle = \nabla_{\mathbf{R}}|\psi_n(\mathbf{R}(t))\rangle \cdot \frac{d\mathbf{R}}{dt}$, the quantity $\nu_n(t)$ can be rewritten as $\nu_n(t) = i\langle\psi_n(t)|\nabla_{\mathbf{R}}|\psi_n(t)\rangle \cdot \frac{d\mathbf{R}}{dt}$, so that the GP can be represented in the time-independent form[122,143]. For a closed trajectory, it yields

$$\gamma_n(\Gamma) = \oint_{\Gamma} i\langle\psi_n(\mathbf{R})|\nabla_{\mathbf{R}}|\psi_n(\mathbf{R})\rangle \cdot d\mathbf{R}. \qquad (3)$$

Here, the quantity

$$\mathbf{A}_n(\mathbf{R}) = i\langle\psi_n(\mathbf{R})|\nabla_{\mathbf{R}}|\psi_n(\mathbf{R})\rangle \qquad (4)$$

is called *Berry connection*[103,122,220]. It is a vector in a generalized parametric space with the number of components equal to the space dimension. Different states $|\psi_n(\mathbf{R})\rangle$ have different Berry connections. For the considered case of the magnetic Aharonov-Bohm effect, the Berry connection corresponds to the vector potential ($\mathbf{A}$), and the parameter space is simply the coordinate space so that **Eq. (3)** turns into **Eq. (2).** Thus, unlike the time-dependent dynamic phase, the GP is time-independent and arises as a consequence of topological properties of the parametric space.

In electromagnetics, the vector potential obeys the gauge transformation invariance. The Berry connection has a similar property: the eigenstates of the system are defined up to a phase factor, $\left|\psi'_n(\mathbf{R})\right\rangle = e^{-i\beta(\mathbf{R})}\left|\psi_n(\mathbf{R})\right\rangle$, i.e., the quantum system possesses gauge U(1) degree of freedom. In this state, the Berry connection reads $\mathbf{A}'_n(\mathbf{R}) = \mathbf{A}_n(\mathbf{R}) + \nabla_\mathbf{R}\beta(\mathbf{R})$. Thus, under the gauge transformation, the Berry connection is defined up to $\nabla_\mathbf{R}\beta(\mathbf{R})$. The integral of this gradient over a closed path turns to zero, making the GP in **Eq. (3)** invariant.

This analogy between the Berry phase and magnetic field can be continued by applying the Stokes theorem to the Berry phase **Eq. (3)** for a closed path,

$$\gamma_n = \oint_\Gamma \mathbf{A}_n(\mathbf{R}) \cdot d\mathbf{R} = \iint_S \nabla_\mathbf{R} \times \mathbf{A}_n(\mathbf{R}) \cdot d\mathbf{S}, \tag{5}$$

analogous to **Eq. (1)**. The GP phase in **Eq. (5)** generalizes the Aharonov-Bohm phase[1]. The curl of the Berry connection is *Berry curvature*

$$\mathbf{\Omega}_n(\mathbf{R}) = \nabla_\mathbf{R} \times \mathbf{A}_n(\mathbf{R}). \tag{6}$$

In the 3D case, Berry curvature is a (pseudo)vector. In an important 2D case, the Berry curvature has only one component:

$$\Omega_n(\mathbf{R}) = i\left(\left\langle \partial_{R_x}\psi_n(\mathbf{R}) \middle| \partial_{R_y}\psi_n(\mathbf{R})\right\rangle - \left\langle \partial_{R_y}\psi_n(\mathbf{R}) \middle| \partial_{R_x}\psi_n(\mathbf{R})\right\rangle\right). \tag{7}$$

As apparent from **Eq. (5)**, the Berry curvature is also gauge-invariant and plays the role of a fictitious "magnetic field" in parameter space[9]. The Berry curvature $\mathbf{\Omega}_n$ is a pseudovector that is odd under temporal (T-) symmetry but even under parity (P-) symmetry operation. In the presence of both P and T symmetry, the Berry curvature must vanish $\mathbf{\Omega}_n = 0$.

According to Eq. (5), one can write the GP as an integral over the manifold of the Berry curvature. If the manifold is closed, the result is topological and quantized by $2\pi C$, where $C$ is the *Chern number*[2,8,13]. In the 3D case, the topological invariant becomes a triad of Chern numbers, $\{C_x, C_y, C_z\}$, where $C_i$ is the Chern number of the 2D momentum plane normal to the i-axis. The generalization of the Chern number to four dimensions is also possible[163,221].

## 2.3. Geometric phase in optics and acoustics

The first GP effects in optics were discovered in biaxial crystals, **Fig. 2(b)**. In 1956, S. Pancharatnam noticed that a GP separate from the dynamical phase arises when the polarization of a light beam cyclically changes[106]. After discovering Pancharatnam's work[222,223], the equivalence between Berry's GP and the phase introduced by Pancharatnam has been revealed, and hereafter this phase bears both names: *Pancharatnam-Berry phase* (PBP)[161].

To recognize GP in optical systems, one uses the Poincaré or Bloch unit sphere representation to visualize the pure polarization states: linear polarization states are located along the equator, the circular polarisations are on the poles, and other elliptical polarization states are located elsewhere on the sphere, **Fig. 4(a)**. If one starts from the pole of the polarisation Poincaré sphere and goes to the equator, then along the equator, and finally returns to the pole, the polarisation state acquires a GP of $\gamma = \Omega / 2$, where $\Omega$ is the solid angle of the spherical triangle (spherical excess) enclosed by the successive polarisation state points (Pancharatnam's theorem)[106].

Birefringent materials are important instances in which we observe polarization GPs[224–226]. One example is a half-wave plate (HWP). In the case of an impinging circularly polarized beam, the HWP has the effect of flipping the handedness of the circular polarization state from the left- to the right-circular polarization and vice versa. Although the flip of handedness is independent of the orientation of the HWP, the outgoing beam acquires an additional phase of $2\theta$, where $\theta$ is the angle of rotation of the HWP. This extra phase factor is exactly the GP observed by Pancharatnam when the polarization state evolves from one circular state to another. The PBP has also been observed in a Mach-Zehnder[227] and Michelson[228] interferometers.

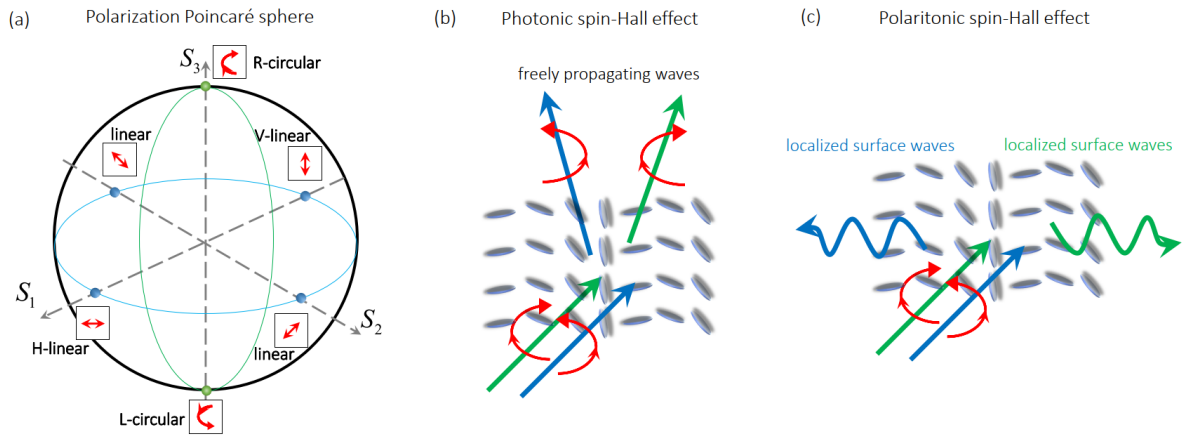

**Figure 4. Berry phase in optics**. (a) Poincaré sphere. When polarization moves along the closed path, the accumulated GP equals half the corresponding solid angle $\Omega$ subtending the associated area. (b) Photonic spin-Hall effect. (c) Polaritonic spin-Hall effect.

In 1997, Bhandari noticed the possibility of changing the local phase of an electromagnetic wave using a twisted uniaxial material, an anisotropic medium whose main axes rotate in a plane perpendicular to the wave vector of the beam[225]. This may result in a so-called *spin-orbit interaction* where the direction of light is locked by its polarization. In other words, the GP, dependent on the beam polarization, drives light propagation. Similar to solids, where the spin-orbit coupling links the spin of a charge carrier with its propagation, the rotation of the electric field or polarization blocks the direction of light propagation, causing the *photonic spin-Hall effect*. This effect exists for freely propagating waves[229,230] (**Fig. 4(b)**), and localized surface waves (polaritons)[231–235] (*polaritonic spin-Hall effect or quantum spin-Hall effect of light*[236]), **Fig. 4(c).** In the case of free-space modes, this effect enables various polarisation selective devices, including polarization-dependent focusing lenses and deflectors[161,225,237–240].

A light beam, as well as light quanta (photons), carry not only spin angular momentum (SAM)[236] but also orbital angular momentum (OAM)[241]. SAM is tightly related to light polarization, and hence the systems discussed above possessing GP in the polarization dimension enable topological manipulation of SAM states. The OAM light states have been widely studied recently due to their ability to carry a topological vortex charge defined by a phase term around azimuthal coordinate ($\varphi$) of $\exp(il\varphi)$, where $l$ is an integer number denoting the OAM value[242–246]. OAM-carrying beams are generated experimentally using refractive elements, spiral phase plates[247], and pitch-fork holograms[248]. In 2006, a device called the q-plate was introduced and demonstrated to generate OAM in the visible domain using a patterned liquid crystal cell[249]. Q-plates are spatially-variable retardation plates whose optical axis rotates with the azimuthal angle. They consist of a birefringent medium with tunable optical retardation, typically set to half a wavelength, and act as polarization converters. The generation of OAM-carrying beams using the q-plate has also been recognized as a spin-orbit coupling in an inhomogeneous medium[243,249–251]. The momentum manipulation in these devices occurs with the conservation of the total angular momentum.

The generation of exotic polarization structures, e.g., vector vortex beams[252–254], Poincaré beams[255], and 3D polarization structures[256], has been reported utilizing the optical GP. MSs with a birefringence resulting from plasmonic resonances have been experimentally demonstrated to generate OAM at the integrated level using devices with thicknesses that can be as small as 1/30 of a wavelength[240,257,258]. Finally, the Sagnac effect[259] and Sagnac interferometry are other manifestations of the GP in atom[260,261] and optical[262–264] interferometers.

Acoustic and mechanical systems are other versatile platforms for studying GP and a wide range of topological effects. Remarkably, even if acoustic pressure is inherently scalar, it is possible to induce extra DoF to their propagation, such as OAM, through GP-based MMs[265–269] to implement enhanced communication schemes. Such vortices can also be used in particle manipulation[270–272]. Moreover, the vector related to the particle velocity in the acoustic field also provides an intrinsic "acoustic spin"[273,274] whose features are related to torques[275,276], skyrmions[277] and spin Hall effect of acoustic surface waves[278–281]. Such features extend to mechanical[282–284] and water waves[285,286].

## 2.4. Geometric phase in chemistry

In chemistry, the impact of Berry phase and Berry connection is associated mainly with so-called conical intersections (CIs) of energy states. These CIs are analogs of Hamilton cone intersections in the optics of birefringent crystals discussed above. Chemical CIs are important in various chemical and biological systems[145,287,296–298,288–295], ranging from elementary reactions involving three to five atoms[295,298] to biomolecules such as DNA[297] and proteins[296]. The Born-Oppenheimer approximation, i.e., the separability of electronic and nuclear motions, is often used to calculate the states of molecular systems and their reactions. Near a CI, two electronic states touch so that the Born-Oppenheimer approximation breaks down. A system with a CI can convert rapidly between electronic states by passing through the intersection. Such rapid transitions are exploited in many light-harvesting and charge-transfer processes. Another consequence of the Born-Oppenheimer breakdown is GP, which occurs even if the system is confined to the lower electronic surface and avoids the neighborhood of the CI[134,143,299,300].

The GP appears in both nuclear $\chi_n(\mathbf{R},t)$ and electronic $\Phi_{\mathbf{R}_n}(\mathbf{r})$ wave functions within the adiabatic representation of the total electron-nuclear wave function $\Psi(\mathbf{r},\mathbf{R},t) = \sum_n \Phi_{\mathbf{R}_n}(\mathbf{r})\chi_n(\mathbf{R},t)$. Here, $\mathbf{R}$ and $\mathbf{r}$ are radius-vectors of the nuclear and electron, respectively. The GP manifests itself in the sign change ($\pi$ phase shift) acquired by the electronic wave function $\Phi_{\mathbf{R}_n}(\mathbf{r})$, when the nuclei complete an odd number of loops around the CI. The GP produces a corresponding sign change in the nuclear wave function $\chi_n(\mathbf{R},t)$ in order to make the total wave function $\Psi(\mathbf{r},\mathbf{R},t)$ single-valued (i.e., gauge-invariant).

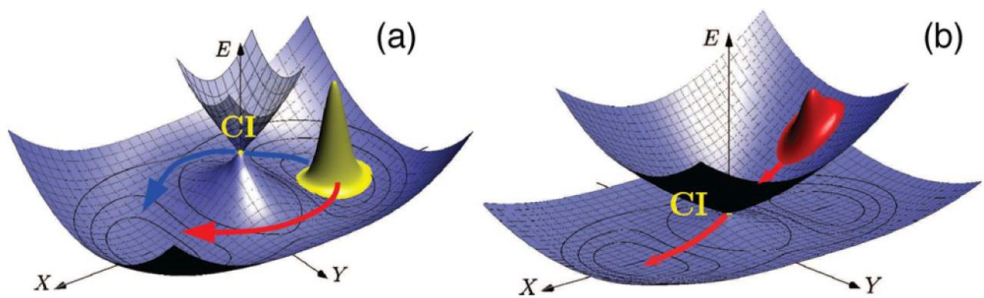

**Figure 5**. **Berry phase in chemistry**. Low energy (a) and excited state (b) nuclear dynamics near CI. In the case of ground-state dynamics (a), the GP leads to destructive interference between paths surrounding the conical intersection. The excited-state dynamics (b) can enhance the rate of nonradiative transitions and the amount of excited-to-ground state population transfer. Reproduced with permission from[310]. Copyright 2015 American Chemical Society.

The effect of the GP on chemical reactions is now well understood[143,301–308]: including the effect of GP is equivalent to the role of the vector potential $\mathbf{A}_\alpha$ (Berry connection) in the nuclear Hamiltonian[134,143,166,309]

$$H_{eff} = \sum_\alpha \frac{(\mathbf{P} - \mathbf{A}_\alpha)^2}{2M_\alpha} + U(\mathbf{R}),$$

(8)

with effective interaction potential energy $U(\mathbf{R}) = V_{NN}(\mathbf{R}) + E_e(\mathbf{R}) + \sum_\alpha \dfrac{\hbar^2}{2M_\alpha} \int d\mathbf{r} \cdot (\nabla_{\mathbf{R}_\alpha} \Phi_{\mathbf{R}}(\mathbf{r}))^2 - \sum_\alpha \dfrac{\mathbf{A}_\alpha^{\ 2}}{2M_\alpha}$.

In this Hamiltonian, the Berry connection $\mathbf{A}_\alpha(\mathbf{R}) = i\hbar \int d\mathbf{r} \cdot \Phi_{\mathbf{R}}^*(\mathbf{r}) \nabla_{\mathbf{R}_\alpha} \Phi_{\mathbf{R}}(\mathbf{r})$ explicitly appears as a gauge potential. Here $\nabla_{\mathbf{R}_\alpha}$ is the spatial derivative with respect to the coordinates of nuclei. The lowest order approximation (Born-Oppenheimer) ignores the Berry phase and Berry connection: $H_{eff} = \sum_\alpha -\dfrac{\hbar^2}{2M_\alpha} \nabla_{\mathbf{R}_\alpha}^2 + V_{NN}(\mathbf{R}) + E_e(\mathbf{R})$, where $M_\alpha$ is the nuclear mass, $V_{NN}$ is the potential energy of nuclei interaction, and $E_e$ is the energy of the electronic subsystem.

Dynamical features associated with the GP are very different for low energy dynamics (**Fig. 5(a)**) and excited-state dynamics (**Fig. 5(b)).** The primary manifestation of GP effects in low-energy nuclear dynamics is destructive interference between two paths around the CI seam (**Fig. 5(a)**). In contrast, in the excited state dynamics (**Fig. 5(b)),** it is the enhancement of population transfer for a fully cylindrical component of a nuclear wave packet and compensation of a repulsive diagonal Born-Oppenheimer correction[310]. In excited state dynamics, it can enhance the rate of nonradiative transitions and the amount of excited to ground state population transfer several times.

The GP effect in chemistry was crucial for modeling the vibrational spectra of Jahn-Teller distorted compounds[311,312] and cross-sections in low-energy atom-molecule reactive scattering[302,303,313–315]. Neglecting the proper GP effect during nonadiabatic dynamics can produce erroneous results[316,317].

## 2.5. Geometric phase in polaritonic systems

Recently, the frontiers of this field have been extended to exciton-polariton systems[176–180,235], resulting from the strong coupling of material excitations (e.g., excitons in solids and 2D materials) and photons. The importance of the exciton-polaritons lies in the possibility of having a high DoF in engineering the particles' Hamiltonian due to the photon energy dispersion in optical resonators[318]. Several strategies to realize topological systems with polaritons have been suggested[319–321], including the use of artificial lattices with honeycomb geometries[322,323], orbital edge modes[324], quantum wells between two distributed Bragg reflectors (DBRs)[178], topological photonic modes strongly coupling with excitons in 2D transition metal dichalcogenides (TMDs)[325,326], and MSs[327–329]. The resonant photonic structures allow for new light-matter interactions, enabling tailoring of effective gauge fields applied to the quantum subsystems (e.g., excitons in various quantum materials) and tuning or adjusting the strength of the gauge field. This nontrivial light-matter coupling enables emulation of various effects hardly achievable otherwise. For example, effective gauge fields have allowed the emulation of matter under strong magnetic fields leading to the realization of the Harper-Hofstadter and Haldane models, the effects otherwise demanding for unfeasibly strong magnetic fields[178]. In other words, neutral excitons can behave as charged particles in strong magnetic fields being subject to artificial gauge potentials. The stability of nontrivial topological interactions could also be advantageous for advanced quantum optical materials. For instance, lattices in organic[330] and perovskite materials[331] with intrinsic robustness of the excitons in these materials have been recently realized.

## 2.6. Geometric phase in lattices

The previous discussion captures the effects of an external gauge field $\mathbf{A}$ on a particle that moves in the continuum, where the coupling is described by the substitution $H(\mathbf{p}) \to H(\mathbf{p} - \mathbf{A})$. However, this substitution also appears in lattice systems, which implies that the Berry curvature plays a crucial role in solid-state physics. **Eq. (5)** indicates how synthetic magnetic fields can be designed in quantum systems by developing nontrivial Berry curvatures in parameter space. These ideas date back to Thouless and his collaborators, who explained the underlying origin of the quantization of Hall conductance of the 2D electron gas (such phenomenon is referred to as integer QHE[184,332]) and explicitly expressed Hall conductance in terms of *Chern number* or Thouless-Kohmoto-Nightingale-den Nijs (TKNN) integer of the U(1) bundle over the magnetic Brillouin zone[333,334]. Haldane[122,220] also pointed out the consequence of Berry curvature on the Fermi surface for Fermi-liquid transport properties, reinterpreting the Karplus-Luttinger anomalous velocity[335].

The topological phenomena in lattices play a vital role in solid-state physics[2,13,122,182,336,337], cold atoms[50,338], photonics[26,339,340] and acoustics[57], where the synthetic gauge fields can be tailored by the symmetry breaking perturbations on the lattice system. Bloch's theorem dictates that the eigenstates of a periodic Hamiltonian are given by $\left| \psi_{n,\mathbf{k}}(\mathbf{r}) \right\rangle = e^{i\mathbf{k}\cdot\mathbf{r}} \left| u_{n,\mathbf{k}}(\mathbf{r}) \right\rangle$, where $n$ is the energy band index, $\mathbf{k}$ is the electronic momentum, and $\left| u_{n,\mathbf{k}}(\mathbf{r}) \right\rangle$ is a periodic function with the periodicity of the underlying Bravais lattice vector. The general model Eqs. (3)-(7) can be used in this case after substituting the parameter $\mathbf{R}(t)$ by the electronic momentum $\mathbf{k}(t)$. If an electron adiabatically traverses a closed path in the reciprocal momentum parameter space from $\mathbf{k}(t_0)$ to $\mathbf{k}(t)$, it obtains the state

$$\left| \psi_{n,\mathbf{k}(t)}(\mathbf{r},t) \right\rangle = \exp[i\gamma_n]\exp[-\frac{i}{\hbar}\int_{t_0}^{t} E_n(\mathbf{k}(t'))dt']\left| \psi_{n,\mathbf{k}(t)}(\mathbf{r},t_0) \right\rangle . \tag{9}$$

Here $\exp[-i/\hbar \int_{t_0}^{t} E_n(\mathbf{k}(t'))dt']$ is the ordinary dynamical phase, with $E_n(\mathbf{k})$ being the energy eigenvalue. The GP for a closed path is $\gamma_n = \oint_\Gamma \mathbf{A}_n(\mathbf{k}) \cdot d\mathbf{k}$, with $\mathbf{A}_n(\mathbf{k})$ being the Berry connection $\mathbf{A}_n(\mathbf{k}) = -i\left\langle u_{n,\mathbf{k}} \left| \nabla_{\mathbf{k}} \right| u_{n,\mathbf{k}} \right\rangle \equiv -i\int u_n^*(r,\mathbf{k})\nabla_{\mathbf{k}} u_n(r,\mathbf{k})d^3r$. The corresponding Berry curvature is $\mathbf{\Omega}_n(\mathbf{k}) = \nabla_{\mathbf{k}} \times \mathbf{A}_n(\mathbf{k})$. Thus, the GP characterizes the topological properties of the energy bands in solids.

The Berry curvature measures the local rotation of the electronic wave packet while it traverses the Brillouin zone. Using Stokes' theorem, one can write the GP as an integral over the manifold of the Berry curvature. If the manifold is closed, the result is quantized by $2\pi$ multiples of the *Chern number*[2,13].

Just like in any parametric space, in $k$-space the Berry curvature acts as an effective magnetic field and it manifests itself through the anomalous transverse velocity, $\dot{\mathbf{k}} \times \mathbf{\Omega}_n(\mathbf{k})$ [122,341–346]:

$$\mathbf{v}_n \equiv \dot{\mathbf{r}} = \frac{1}{\hbar} \frac{\partial E_n(\mathbf{k})}{\partial \mathbf{k}} - \dot{\mathbf{k}} \times \mathbf{\Omega}_n(\mathbf{k}) , \tag{10}$$

$$\hbar \dot{\mathbf{k}} = -e\mathbf{E} - e\dot{\mathbf{r}} \times \mathbf{B} , \tag{11}$$

where $\mathbf{E}$ and $\mathbf{B}$ are the external electric and magnetic fields, $\mathbf{v}_n$ is the velocity of the particles in the $n$-th band. This transverse transport was revealed in the QHE, where the plateau depicted by the Hall conductivity was attributed to the Chern number.

Consider the motion of electrons in a 2D structure and assume no external magnetic field is applied, $\mathbf{B} = 0$. The current density $\mathbf{j} = -1/V \cdot \sum_{n,k} e\mathbf{v}_n(\mathbf{k})$ can be expressed through the Berry curvature (omitting the trivial group velocity term in **Eq. (10)**): $\mathbf{j} = e^2 / (2\pi)^2 \hbar \cdot \int d^2\mathbf{k} \sum_n \Omega_n(\mathbf{k}) \times \mathbf{E}$. Due to the vector product, this current density is defined by the electric conductance tensor $\hat{\sigma}$, $\mathbf{j} = \hat{\sigma}\mathbf{E}$. Assume $\mathbf{E}$ is in-plane and has only one component $E_y$. The field causes the perpendicular component of the electric current density, $j_x = e^2 / 2\pi\hbar \cdot \int dk \sum_n \Omega_{n,z} E_y$ with the transverse Hall conductance $\sigma_{xy} = e^2 / 2\pi\hbar \cdot \sum_n \int dk \Omega_{n,z}$. Here $\Omega_{n,z}$ is the $z$ component of Berry curvature. The quantity $1/2\pi \cdot \sum_n \int dk \Omega_{n,z}$ represents the integer Chern number. Thus the Hall conductance is an integral multiple of $e^2 / \hbar$ (the von Klitzing constant, also called conductance quantum) and is a topological property of the system. This result is known as the *quantum anomalous Hall effect*[9,169,353–355,183,335,347–352]. The term *anomalous* here means that the Hall effect is achieved without an external magnetic field, but due to the breaking of T-symmetry of the material. For example, the seminal model of the anomalous Hall effect is realized in a honeycomb lattice that contains complex next nearest-neighbor hoppings while the overall magnetic flux of the lattice is zero[9].

This approach also describes other intriguing in phenomena in 2D systems, including, e.g., the *valley-Hall effect* (VHE) in 2D TMDCs with non-degenerate valleys K and K' possessing Berry curvature of opposite sign[329,356–362]. Eqs. (10) and (11) show that in an in-plane electric field ($\mathbf{E}$), the single-layer TMDC with nonzero Berry curvature at the valley points acquires anomalous transverse velocity, forcing the K and K' valley electrons to move in different directions[359,363–365]. This property gives rise to various intrinsic topological effects, including valley-selective photoexcitation[329,357,359,364,366] and optical Stark effect[367,368].

Finally, consider a model of a particle in a 2D space-periodic potential $V_l(\mathbf{r}) = V_l(\mathbf{r} + a_x \hat{\mathbf{x}}) = V_l(\mathbf{r} + a_y \hat{\mathbf{y}})$, $H = \mathbf{p}^2/2M + V_l(\mathbf{r})$, where $a_{x,y}$ denote the lattice spacings and $\hat{\mathbf{x}}$ and $\hat{\mathbf{y}}$ are unit vectors along the x- and y-axis in the tight-binding approximation[30,325,369,370]. In this model, the Schrödinger equation $H|\Phi\rangle = E|\Phi\rangle$ yields the state $|\Phi\rangle = \sum_{n,m} \eta_{n,m} |n,m\rangle$, with eigenstates $|n,m\rangle$ associated with the potential wells $(na_x, ma_y)$, and $(n,m)$ are integers. For a square lattice $(a_x = a_y = a)$, the coefficients $\eta_{n,m}$ satisfy an effective Schrödinger equation $E\eta_{n,m} = -J(\eta_{n+1,m} + \eta_{n-1,m} + \eta_{n,m+1} + \eta_{n,m-1}) + \varepsilon\eta_{n,m}$, where $\varepsilon$ is the onsite energy and $J$ describes hopping between neighboring sites. Using the translation operators $T_{x,y} = e^{-iap_{x,y}}$, we rewrite the latter as $H_{eff}\eta_{n,m} = (E - \varepsilon)\eta_{n,m}$, introducing the effective Hamiltonian, $H_{eff} = -J(T_x + T_x^\dagger + T_y + T_y^\dagger)$. The effects of a gauge field can be directly incorporated via the substitution $H_{eff}(\mathbf{p}) \to H_{eff}(\mathbf{p} - q\mathbf{A})$ (Peierls)[371], $H_{eff} = -J(T_x e^{iaqA_x} + T_x^\dagger e^{-iaqA_x} + T_y e^{iaqA_y} + T_y^\dagger e^{-iaqA_y})$. Here the variable $q$ is a coupling charge, and the quantities acting on each link $e^{iaqA_{x,y}}$ are link variables[372]. A simple example of this Peierls substitution can be found in the study of a charged particle moving in a 2D square lattice subjected to a uniform magnetic field perpendicular to the plane[336,337]. In the Landau gauge $A_x = 0$ and $A_y = Bx = \Phi_B n/a$, where we introduced the magnetic flux $\Phi_B = Ba^2$ per cell, the effective Hamiltonian reads

$$H_{eff} = -J(T_x + T_x^\dagger + T_y e^{i\alpha 2\pi n} + T_y^\dagger e^{-i\alpha 2\pi n}),\tag{12}$$

where $\alpha = \Phi_B/\Phi_B^0$ is the normalized magnetic flux, $\Phi_B^0 = 2\pi$ is the flux quantum in the units $\hbar = q = e = 1$. This is the so-called Harper-Hofstadter Hamiltonian, whose spectrum displays a fractal structure known as the *Hofstadter butterfly*[333,337]. This model plays an important role in the physics of topological states of matter[2,13] because it exhibits a wide family of quantum Hall states[333] and offers a natural platform for realizing *fractional Chern insulators*[373]. While the Peierls phase factors $e^{i\alpha 2\pi n}$ are gauge dependent, their product around a closed loop is gauge-invariant. In this model, the Peierls factors around a lattice unit cell $e^{i\Phi_B}$ correspond to the Aharonov-Bohm phase[1] acquired by a particle encircling a region penetrated by the flux $\Phi_B$.

## 2.7. Topological electronic materials

We now discuss the general scenarios enabling topologically nontrivial materials, with a focus on electronic systems. **Fig. 6** exemplifies different types of such systems: as a first example, Hermitian (lossless) 2D electronic systems can be engineered to possess a conical intersection or a *Dirac point*, a conical singularity with linear dispersion in the Brillouin zone with touching conduction and valence bands. Such points have been discovered in the solid-state theory of periodic structures[2,374]. It was shown that the Hamiltonian describing low-energy excitations near Dirac points in momentum space could be written as the Dirac Hamiltonian for massless Dirac fermions in 2D, $H(\mathbf{k}) = v_x k_x \sigma_x + v_y k_y \sigma_z$, where $v_i$ and $\sigma_i$ are group velocities and the Pauli matrices, respectively. Therefore, energy is distributed linearly with momentum at these Dirac points, $\omega(\mathbf{k}) = \pm\sqrt{v_x^2 k_x^2 + v_y^2 k_y^2}$, and spin and momentum are locked. The symmetries of the Dirac Hamiltonian reveal that the Dirac cones are protected by the simultaneous existence of P- and T-symmetries. The Dirac cones appear in pairs[375–380] and disappear when the symmetries are lifted. Every Dirac cone has a quantized Berry phase of $\pi$ looped around it[381,382]. In solid-state physics, graphene, a single layer of carbon atoms arranged on a honeycomb lattice, is famous for hosting Dirac cones at the corners of its Brillouin zone[374]. In the cones' vicinity, the electrons behave as relativistic Dirac fermions, allowing the observation of exotic relativistic phenomena like *Zitterbewegung* motion (trembling motion of electrons)[383] and quantum Klein tunneling on a graphene device[384].

Dirac cones are not tolerant to operations that break P- or T-symmetry. For example, a magnetic bias breaks the T-symmetry of the system, opening up bandgaps. The opened bandgaps are topologically nontrivial because the bulk bands below the bandgap carry integer quantized Chern numbers[9,16,385]. According to the bulk-boundary correspondence[7,386], such a system in a magnetic field possesses gapless and unidirectional edge states, **Fig. 6(a)**, and no second edge state with an opposite group velocity appears in the bandgap, enabling a scattering-less topologically protected mode (Chern isolator). The direction of the magnetic field defines the direction of motion in these modes. The resulting edge states are stable to impurities and inhomogeneities of the surface since these states are determined by the band structure topology of a bulk crystal and not by the specific morphology of the surface[2,13].

The first effect of this kind is the integer QHE, whose observation goes back to 1980 when Klitzing, Dorda and Pepper observed that the Hall conductance of a 2D electron gas under a perpendicular magnetic field is quantized[332]. This effect corresponds to the quantum version of the conventional Hall effect[2,10,332,333,387–391], in which an applied magnetic field deflects electrons in a metal plate by the Lorentz force, which induces a cyclotron motion of electrons in bulk, making it unable to conduct current. However, the electron cyclotron orbits are bounded off at the edge, yielding one-way propagation at the boundaries. The QHE can exist without magnetic bias due to other mechanisms, such as magnetization and spin-orbit coupling, that also break $\mathcal{T}$, giving rise to *anomalous QHE*[9,392].

An example of 2D TIs is quantum wells of a specific width based on HgTe/CdHgTe[387,393], InAs/GaSb[394] compounds, some heterovalent structures such as InSb/CdTe[395], as well as silicene and germanene. The electron spectrum of these materials contains two branches located in the bandgap of the bulk states describing the helical edge states. The edge states spectrum is linear in the vicinity of the branch intersection. These helical edge states are distinguished for their propagation direction locking to the spin of the electron, named the spin-momentum locking effect. Because electrons are fermions, their spin is a half-integer, meaning that the associated time reversal operator squares meet the condition $\mathcal{T}^2 = -1$. This ensures the double degeneracy of edge states, referred to as Kramer degeneracy, at the TRS points of the Brillouin zone, generating topologically protected edge transport which is robust against spin-$S_z$ conserved impurities and sharp bends, in some cases they even survive in the interactions that mix spin-up and spin-down electrons[11]. The most impressive experimental results in the physics of 2D TIs were obtained in structures with HgTe/CdHgTe quantum wells[396–404]. It was shown that the ballistic character of the charge carrier motion along the edge channels is observed only at micrometer scales. In larger samples, the transport demonstrates a diffusive character, which can be associated with scattering on accidental magnetic impurities[405–408] and spin polarization fluctuations of the atomic nuclei[409–411].

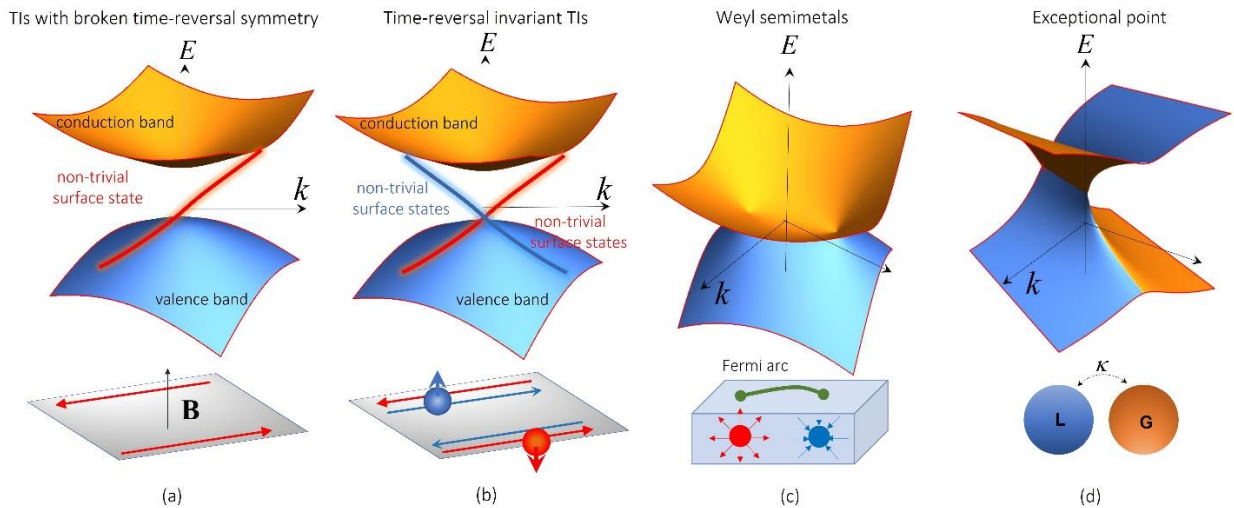

**Figure 6. Various types of topologically nontrivial systems.** (a) TIs with broken T-symmetry through a magnetic field or time modulation. The broken T-symmetry gives rise to the QHEs or anomalous quantum Hall effect. (b) TIs obeying TRS enable quantum spin Hall effect or quantum valley Hall effect. (c) Weyl semimetals and Dirac semimetals. (d) Exceptional points in non-Hermitian structures and materials characterized by the appearance of Riemann surfaces.

A large magnetic bias can become problematic when Chern insulators, having a QHEs phase, are integrated into actual devices. Indeed, for large-scale production, designs with magnetic bias on a chip significantly complicate the fabrication and integration, and magnetic biases can perturb the operation of the surrounding components. The discovery of the quantum spin-Hall effect (QSHE) provides a promising alternative approach for implementing topological electronics[10,11,387] since it does not require breaking $\mathcal{T}$. The necessary conditions for realizing QSHE in solids include the conservation of $\mathcal{T}$ and the incorporation of spin-orbit couplings (SOCs). Kane and Mele proposed that graphene with sufficient large SOCs exhibits a broken Dirac cone with a bandgap[10], **Fig. 6(b)**. This bandgap is characterized by the $Z_2$ invariant[11,412] possessing only two possible integer values: 0 (trivial) and 1 (nontrivial). Although the total Chern number of the two spins of the bandgap is zero ($C = C\uparrow + C\downarrow = 0$), the Chern number calculated for a specific spin has a nonzero value under $\mathcal{T}$. Here, $C\uparrow$ ($C\downarrow$) is the Chern number of spin-up (down) particles. Sometimes the *spin Chern number* is introduced to describe the global topological phase of QSHE $C_{spin} = (C\uparrow - C\downarrow)/2$[413].

Until now, we focused on solids whose topology is defined globally. However, a trivial global band can also possess local topological features near high symmetry points of the Brillouin zone. For instance, if two atoms in the hexagonal lattice are not equivalent, the inversion symmetry of the lattice is broken, which causes the lifting of the point degeneracy of the Dirac cone. Although the bandgap is topologically trivial for the whole BZ, it shows the local topological effect at high symmetry points K/K′ in BZ. At these points, the local Chern number, termed valley Chern number $C_v$, is evaluated around and integrated into half BZ, and thus is half-integer quantized, with its sign depending on the valley DoF. According to the bulk-edge correspondence theorem, an interface between topologically distinct domains with opposite valley Chern numbers supports edge modes whose propagation direction is locked to their valley polarization[12,414].

3D topological phases without gaps, also known as topological semi-metallic phases, form a new topological phase of matter, distinct from the topological insulating phase in 2D[415–420], **Fig. 6(c)**. The band structure of topological semimetals in 3D may possess point degeneracy with linear dispersion (Weyl and Dirac points) or line degeneracy (nodal line), whose topological protection is accompanied by robust surface states[415,416,421–424]. The main signature of 3D gapless topological phases is a Weyl point, with their Hamiltonian $H(\boldsymbol{k}) = v_x k_x \sigma_x + v_y k_y \sigma_y + v_z k_z \sigma_z$, where $v_i$s are the group velocities near Weyl points[415,417,425–427]. Weyl points are distinct from 2D Dirac points, nodal points in 2D momentum space. 2D Dirac points can be easily destroyed if either T-symmetry or P-symmetry is lifted. In contrast, Weyl points exist in systems lacking T, P or both, making them highly robust, as they can only be eliminated when two Weyl points of opposite chiralities meet and annihilate each other. This can be understood through the fact that they are Berry curvature monopoles in 3D momentum space: a Weyl point carries a quantized topological charge and serves as a source or drain of Berry curvature depending on the charge sign. Weyl points exist in two distinct forms: type-I Weyl points with an isofrequency surface that is point-like and type-II Weyl points with an isofrequency surface that is conical[420]. According to the bulk-surface correspondence, the non-zero topological charges of Weyl points result in a surface Fermi arc linking Weyl points with opposite charges[428 415,417,429]. Weyl semimetals exhibit other interesting phenomena, such as the quantum anomalous Hall effect[419] and chiral anomalies[430]. Dirac semimetals, however, are not robust in terms of degeneracy. They will transit into Weyl semimetals if either P or T is broken unless they are protected by extra space-group symmetries[424,425,431]. For example, the Dirac cone protected by non-symmorphic symmetry was explicitly measured in an earth-abundant material, ZrSiS [432]. When the conduction and the valence bands of semimetals are degenerate along a 1D curve in the 3D BZ, it forms a nodal line, or nodal chain, a chain of connected loops in momentum space[433], or even nodal knots, whose nodal lines braid with each other and form topologically nontrivial knots[434]. These 1D lines are robust against the perturbations that preserve a certain symmetry group protecting the nodal line degeneracy and might exhibit anomalous magnetotransport properties.

The 3D gapless phase is not the only phase possible among 3D topological phases. Similar to 2D topological phases, it is possible to break TRS to obtain 3D quantum Hall phases[435,436] or introduce strong SOC for 2D layers and stack them into 3D quantum spin-Hall insulator[437,438]. As predicted by examining 3D electron gases under magnetic fields[435,436,439], the 3D Hall insulator might support the topological surface modes at their boundaries, which has been experimentally demonstrated using semiconductor superlattices constructed by stacking 2D quantum Hall insulators with appropriate interlayer coupling[440]. In the latter case, when TRS is preserved the topological properties of the 3D insulator are characterized by four $Z_2$ invariants $(\nu_0; \nu_1, \nu_2, \nu_3)$. If $\nu_0 = 1$, it has strong topological phase, the surface states appear on arbitrarily terminated surface of the 3D media. While if $\nu_0 = 0$, and other quantities are nonzero, the 3D insulator belongs to the weak topological phase, and the gapless surface states only exist on the boundaries of some specific directions[438]. The known electronic 3D TIs are Bi, Sb, Se, Te compounds[441,442], and strained layers of HgTe [443–446] and a-Sn[447].

Recently, *higher-order TIs* (HOTIs) have been introduced as a new topological phase of matter[369,448,457–466,449,467,468,450–456]. The HOTIs are gapped both in the bulk and boundary states, which are one-dimensional lower than the bulk and support gapless boundary states, which localize in more than one dimension and whose existence is determined by bulk topological invariants related to the medium's symmetries. This means they cannot be destroyed by perturbation of the boundaries that preserve these specific symmetries and offer topological protection.

For completeness, we present one more scenario of nontrivial topological response, dealing with points possessing a topological charge, which have been explored in photonic and phononic systems but are rare in solids. These emerge at so-called exceptional points (EPs), arising in non-Hermitian systems. EPs occur when the system's parameters are tuned to the critical point at which two (second-order EP) or more (higher-order EP) resonant frequencies and their corresponding eigenmodes coalesce[469,470]. If the system obeys parity-time (PT-) symmetry[471], i.e., if the combination of inversion/parity symmetry and time-reversal symmetry is preserved, even though both of them may be separately violated, EPs can emerge at real frequencies and observable in various settings[472–476]. An EP in a PT-symmetric structure coincides with the spontaneous symmetry-breaking threshold, at which the unbroken PT-symmetric phase abruptly transitions to a PT-broken phase. In parameter space, such points are characterized by the appearance of stratified Riemann surfaces, **Fig. 6(d).** Hence, a single adiabatic encircling of a second-order EP results in swapping of the eigenstates, in which only one of them acquires a π-Berry phase[477,478]. This π-phase shift, therefore, has a topological fractional charge, possible only in non-Hermitian systems.

In the two previous sub-sections, we introduced topological band theory concepts and their tantalizing consequences in terms of electronic propagation in natural materials. In the following, we focus on transposing these topological concepts to artificial MMs, implemented in various classical platforms, including photonics, acoustic and elastic systems. Throughout this part of the review, we will first elaborate on the photonic implementation of topological phases and then discuss their acoustic and mechanical counterparts. We will not go too far into the mathematics of their underlying topological origin, and encourage the reader to consult the corresponding referenced papers for more details.

The following context is organized as follows. The first section presents 1D examples of topologically protected modes in classical systems. In the second section, we focus on 2D topological phases realized in MMs, which are split into two categories depending on T-symmetry conservation. The third section reviews 3D topological semimetals and insulators for classical waves. The fourth section is dedicated to unconventional higher-order topological phases, which have recently been explored in both theory and experiments. The next section presents hybridized studies related to topological phenomena and applications in MMs, e.g., topological lasers and topological zero frequency modes in mechanical MMs. We end our review by summarizing the work and envision potential directions of topological MMs for various applications.

## 3. 1D topological phases in MMs

This section reviews various implementations of 1D topological phases in MMs. Notably, we present how the straightforward measurement of band structures and the corresponding classical wave fields in these devices has demonstrated topologically protected defect modes in 1D MMs. In the first part we describe the MM version of the Su-Schrieffer–Heeger (SSH) model in periodic 1D systems. In the second part, we explain how introducing quasi-periodicity in these media enables adiabatic topological pumping, mimicking the 2D integer quantum-Hall effect in lower dimensions.

### 3.1. SSH model

The SSH model consists of a 1D chain of atoms described by a unit cell containing two atoms whose inter-cell couplings $\lambda$ and intra-cell couplings $\gamma$ are tunable[479]. The corresponding Hamiltonian is expressed in the Pauli matrix $\hat{\sigma}_i$ basis

$$\hat{H}(k) = \sum_i d_i \hat{\sigma}_i, i = 1,2, \tag{13}$$

where $d_1 = (\gamma + \lambda \cos(k))$, $d_2 = \lambda \sin(k)$, $k$ is the momentum variable in the 1D Brillouin zone (BZ), its range is $k \in [0, \frac{2\pi}{a_0}]$, where $a_0$ is the lattice constant, and $\hat{\sigma}_i$s are Pauli matrices, $\hat{\sigma}_1 = \begin{pmatrix} 0 & 1 \\ 1 & 0 \end{pmatrix}$, $\hat{\sigma}_2 = \begin{pmatrix} 0 & -i \\ i & 0 \end{pmatrix}$, and $\hat{\sigma}_3 = \begin{pmatrix} 1 & 0 \\ 0 & -1 \end{pmatrix}$. If these two families of couplings are equal, the corresponding band structure presents two bands touching at the edge of the Brillouin zone because of band folding. If they are not equal, these bands are separated by a bandgap, accompanied by a band inversion between the two cases when the coupling ratio satisfies $\frac{\lambda}{\gamma} > 1$ and $\frac{\lambda}{\gamma} < 1$, as present in **Fig. 7(a,c).** Their topological properties can be directly read from the winding number of the vector $\boldsymbol{d}(k) = (d_1, d_2)$ in momentum space. When the wavenumber $k$ runs from 0 to $2\pi$, the endpoint of the vector $\boldsymbol{d}$ traces out a closed circle with radius $\lambda$ in the plane of $d_1$ and $d_2$, as shown in **Fig. 7 (b,d)**. If $\lambda > \gamma$, the circle encloses the origin of the coordinate frame, verifying that the winding number is an integer. If $\lambda < \gamma$, the circle excludes the origin leading to zero winding number. Alternatively, the topological property of the SSH model can be evaluated from the Zak phase (or Berry phase) of the bandgap, which reads[132]

$$\nu = \frac{i}{\pi} \int_{BZ} \langle \psi(k) | \partial_k \psi(k) \rangle dk, \tag{14}$$

where $\psi(k)$ is the eigenstate of the lowest bulk band in the SSH model. The integer winding number $\nu$, or equivalent Zak phase, is responsible for the existence of boundary modes at the edges of the topologically nontrivial chain ($\lambda > \gamma$). When $\nu = 0$, which corresponds to the topologically trivial chain ($\lambda < \gamma$), no edge states are observed at the boundaries of the chain. Note that chiral symmetry is crucial for the preservation of the topological invariant in the SSH model, and its chiral operator $\hat{\Gamma}$ acts on the Hamiltonian as $\hat{\Gamma}\hat{H}(k)\hat{\Gamma}^{-1} = -\hat{H}(k)$, which guarantees the edge states pining to zero energy.

## 3.2. Photonic SSH model

The first 1D topological example in the photonic community comes from the super-lattices supporting Shockley-like surface waves, which are edge modes in the SSH model[480]. This model has been extensively studied in many platforms like photonic crystals[481,482], coupled waveguide arrays[483–485], dielectric resonator chain[486–489], electrical circuits[490,491] and plasmonic nanoparticles[492–495], and in a complementary metal-oxide-semiconductor compatible platform[496]. Moreover, the non-Hermitian version of the SSH model has been investigated, and the role of PT-symmetry in the protection of bound states was explored[497–499]. The topological edge states have also been studied on the active platform, and the topological lasing effect based on these states was observed and showed some immunity to distortion of the structure[34,65,500,501]. Furthermore, a polariton lattice implementing a driven-dissipative SSH model was recently shown to support the gap solitons with topological protection[502].

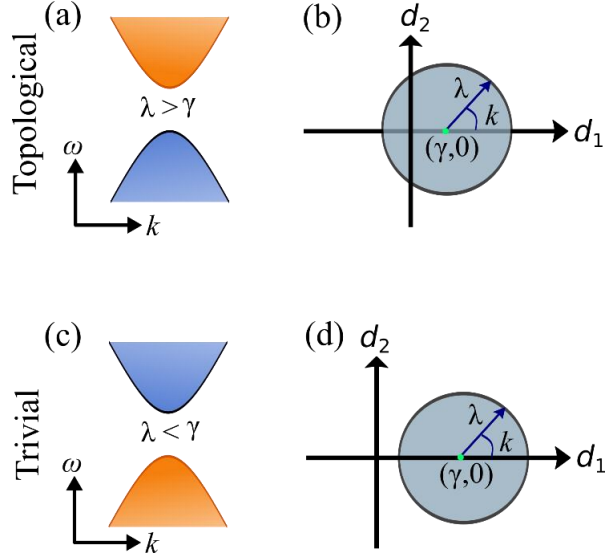

**Figure 7. SSH model and its topological properties.** (a,c) Band diagrams and (b,d) winding behaviors of vector $\boldsymbol{d}(k) = (d_1, d_2)$ in the momentum space for the (a,b) topological lattice when inter-cell coupling strength is larger than the intra-cell coupling strength: $\lambda > \gamma$, and for (c,d) trivial lattice when inter-cell coupling strength is less than the intra-cell coupling strength: $\lambda < \gamma$.

### 3.3. Phononic SSH model

In phononic systems, the 1D SSH model has been demonstrated in a sonic crystal consisting of a waveguide with varying cross-sections[40,503], **Fig. 8(a)**. Using in-situ pressure field measurements within the MM, Xiao et al. evidenced the topological band inversion and the corresponding topological defects occurring at the interface between trivial and non-trivial topological bandgaps[40]. It can also be directly implemented in a tight-binding system of cavities coupled with small tubes whose cross-section controls the coupling amplitude[504] and, more generally, in sonic MM with negative effective properties[505]. A similar phenomenon has also been observed in mechanical media using a cylindrical granular chain[506]. In that case, band inversion is achieved by changing the contact angles between the cylinders resulting in topological boundary modes measured using a laser vibrometer, **Fig. 8(b)**. Many related studies have been performed in other phononic lattices[351,507–512].

Interestingly, the topological nature of these boundary modes gives them strong robustness to perturbations. At the macroscopic scale, such protection against spatial and resonant frequency disorder has been harnessed in the context of topological Fano resonances[515,516] and robust analog computation[517]. This peculiar property also hinders potential fabrication inaccuracies and allows their nanoscale implementation[513,518–521], **Fig. 8(c).** Furthermore, these concepts have also been transposed to microfluidics in the context of micro-particle manipulation[514,522], where the robustness of these topological modes is exploited to design on-chip acoustic tweezers, **Fig. 8(d).** Such advances are very promising for applications in chemistry and biochemistry in on-chip sensing and particle/cell sorting. Finally, the SSH model has been extended to higher dimensions to obtain guided modes along interfaces whose symmetries dictate the Zak phase[523]. Hence, these edge modes have been evidenced at the interface between two soft elastic phononic crystals, whose properties are dynamically tuned through smooth deformations[524].

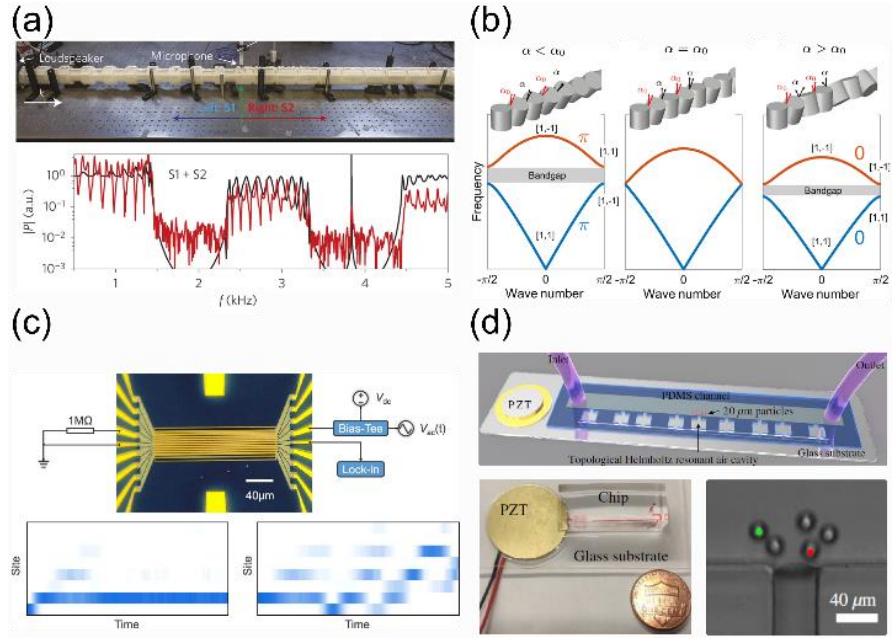

**Figure 8**. **Phononic SSH.** (a) SSH model and corresponding spectrum with a topological defect mode in the bandgap of the tailored acoustic waveguide. Reproduced with permission from[40]. Copyright 2015 Springer Nature. (b) Topological transition in a granular medium. Reproduced with permission from[506]. Copyright 2017 American Physical Society. (c) Reconfigurable on-chip SSH model. Reproduced with permission from[513]. Copyright 2021 American Chemical Society. (d) Topological defect mode on a microfluidic chip for particle manipulation. Reproduced with permission from[514]. Copyright 2021 American Physical Society.

### 3.4. Adiabatic topological pumping in photonics

The previous section reviewed studies related to topological Zak phases in 1D periodic systems. However, topological phenomena are not limited to these commensurate media but also can occur in quasiperiodic chains. Implementing the quasi-periodic 1D tight-binding model (TBM) with tunable MMs opens up new ways to study and observe the topological phenomena attributed to the dimension higher than 1D quasicrystals, such as the quantization of transport, Hofstadter butterfly spectra etc. The strong interest in quasi-periodicity systems has induced many theoretical and experimental works which explored the adiabatic topological pumping and demonstrated edge states transition. Two models demonstrate the adiabatic topological pumpings: Aubry−André−Harper (AAA) model[525] and the Rice-Mele model[526]. To emulate the AAA model, the onsite potentials or the next nearest couplings are spatially or temporally modulated in the 1D system[527]. For example, TBM Hamiltonian simulating the Harper equations with onsite potential modulation may read

$$H(\theta, \phi)\psi_n = \gamma(\psi_{n-1} + \psi_{n+1}) + \lambda \sin(n\theta + \phi)\psi_n, \tag{15}$$

where $\theta$ controls the modulation period, which can be quasi-periodic (periodic) if $\theta$ is incommensurate (commensurate) with $2\pi$. This modulated scheme was studied and experimentally realized in photonic coupled waveguide arrays[527–529], as shown in **Fig. 9(a)**. The parameter phason $\phi$ is tuned by the slowly modified spacing between the waveguides along the propagation direction. As a result, the injected light at the rightmost edge slowly propagates across the waveguide arrays and finally comes out from the leftmost edge (**Fig. 9(b)**). The topological pumping over a Fibonacci quasicrystal, which turned out to be mathematically equivalent to AAA model, was demonstrated[530,531]. Moreover, nonlinear Thouless pumping of photons was experimentally realized recently in which nonlinearity acts to quantize transport via soliton formation and spontaneous symmetry-breaking bifurcations[532].

On the other hand, Thouless pumping can also be realized in the Rice-Mele model, which is modified from the SSH model by inserting the extra pumping parameter $\phi$ in the hoppings and onsite potentials as follows

$$\tilde{H}(k, \phi) = (\lambda + \cos(\phi) + \gamma \cos(k))\hat{\sigma}_1 + \gamma \sin(k)\,\hat{\sigma}_2 + \sin(\phi)\hat{\sigma}_3 \,. \tag{16}$$

This Hamiltonian has been simulated in various systems, including optical superlattices[533,534], disordered optical waveguide arrays[535] and even nonparaxial waveguides, as shown in **Fig. 9(c)**[536]. Interestingly, asymmetric topological pumping occurs when the injected light can transfer from the left boundary to the right boundary, but the reversal is prohibited due to positive far-neighbor interactions among the waveguides, as shown in **Fig. 9(d)**. The non-Hermitian version of the Rice-Mele model was demonstrated recently in fast-modulated plasmonic arrays[537]. Topological invariants were also measurable using 1D quasicrystals based on cavity polaritons or microwave network pumping[500,538] or circuit quantum electrodynamics lattice[539] or fiber loops[540].

### 3.5. Adiabatic topological pumping in phononics

The properties of quasi-periodic chains have also been studied in phononics[541–543]. Remarkably, using a lattice made of 3D printed reconfigurable acoustic cavities, the fractal features in the Hofstadter butterfly spectrum have been revealed in the quasicrystal made of coupled acoustic resonators and edge states spectrum also mapped using the adiabatic pumping of the phason $\varphi$ [543], **Fig. 10(a)**. These topological effects have also been studied in mechanical systems such as chains of spinners[544], quasi-periodically patterned beams[545–547] and plates[548].

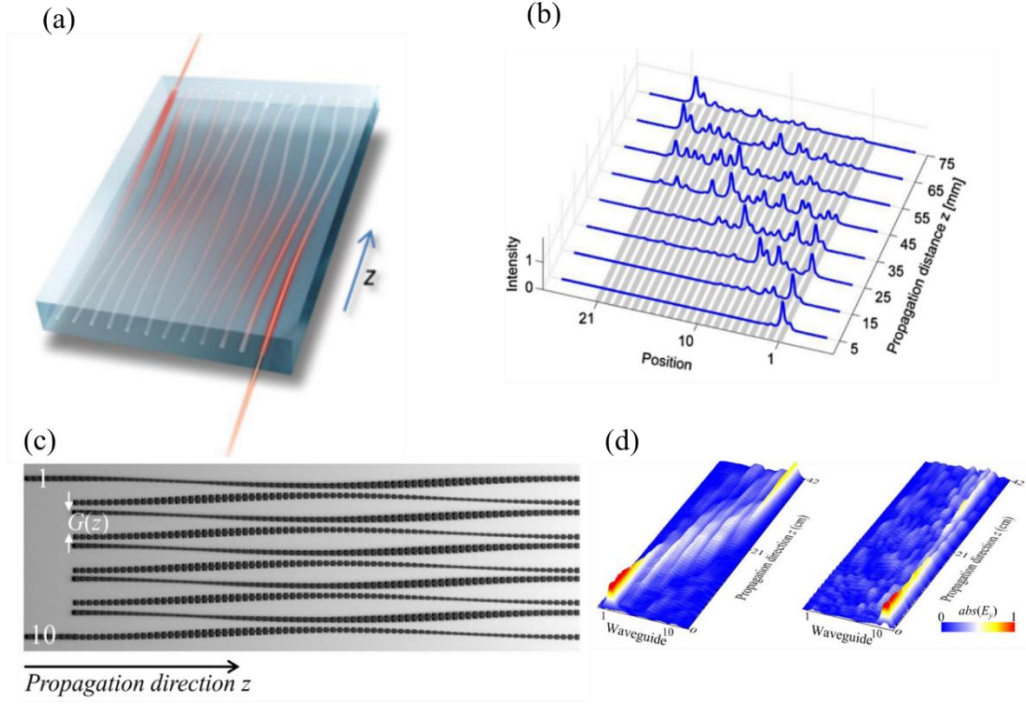

**Figure 9. Adiabatic topological pumping**. (a) Schematic of optical pumping of a coupled waveguide array. (b) Experimental measurement of edge states spatial profiles evolution from one boundary to the other boundary. The "magnetic flux" $\theta$ in Eq. (15) is fixed throughout the pumping process. The figures in (a,b) are reproduced from reference[527]. Copyright 2012 American Physical Society. (c) Schematic of the asymmetric topological pump in an array of surface plasmon polariton waveguides. (d) Experimental data for the near-field measurements of the fields when the source is injected from the left boundary (left panel) and right boundary (right panel), respectively. The figures in (c, d) are reproduced from reference[536]. Copyright 2022 Springer Nature.

It is also possible to implement dynamical edge-to-edge pumping by changing the medium properties in time. Notably, Cheng et al. demonstrated this effect using an acoustic waveguide sandwiched between two layers of a 1D chain of tubes[549]. Here, the pumping is achieved thanks to an adiabatic glide of the upper layer during sonic propagation in the waveguide, **Fig. 10(b).** Besides, temporal pumping has been evidenced in artificial elastic media using piezo-electric patches on a beam[552,553] or magneto-mechanical MMs[550], **Fig. 10(c).** In these cases, 2D topological effects are achieved in 1D using time as a synthetic dimension. One can also directly map the adiabatic evolution of the medium's parameters in a second spatial dimension which permits more accessible experimental characterization[551,554,555], as shown in **Fig. 10(d)** for the case of an elastic medium with a stiffness modulation. In this context, new phenomena have been studied, such as selective acoustic topological pumping[556], dynamical temporal pumping and amplification in time-modulated mechanical beams[557] and non-adiabatic topological pumps[558], which allow a robust transfer of topological edge states without entering the bulk, hence hindering the impact of losses during the process.

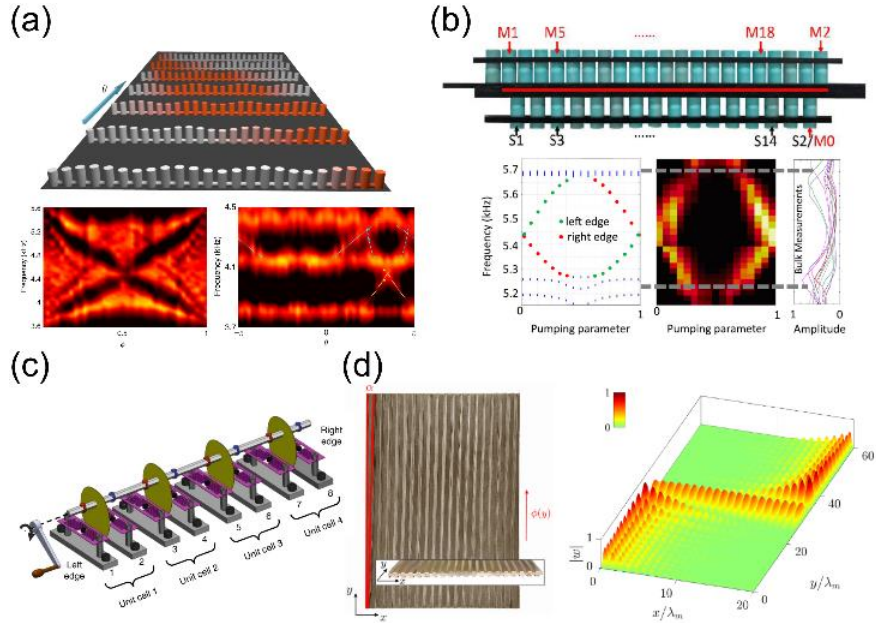

**Figure 10. Adiabatic topological pumping in phononics.** (a) Topological pumping and Hofstadter butterfly in a tight-binding lattice of reconfigurable 3D printed acoustic cavities. Reproduced with permission from[543]. Copyright 2019 Springer Nature. (b) Acoustic topological pumping in a bilayer waveguide structure mediated by the glide of the top layer. Reproduced with permission from[549]. Copyright 2021 American Physical Society. (c) Temporal topological pumping in a chain of magneto-mechanical MMs. Reproduced with permission from[550]. Copyright 2020 Springer Nature. (d) Mechanical topological pumping obtained by mapping the adiabatic evolution of the medium's stiffness in another spatial dimension. Reproduced with permission from[551]. Copyright 2020 American Physical Society.

## 4. Dirac cones in 2D MMs

This section focuses on implementing conical Dirac degeneracies and associated exotic wave phenomena in 2D metamaterials.

### 4.1. Dirac cones in photonics

Since graphene's honeycomb lattice can be reasonably described by a tight-binding Hamiltonian with nearest-neighbor hopping[559], Dirac cones are easy to implement using macroscopic MMs. In the context of 2D photonic crystals, the extremal transmission at the Dirac cones of a photonic band structure was proposed[560], the conical diffraction was measured in a honeycomb photonic lattice supporting Dirac cones[376], and a dramatic enhancement of the spontaneous emission coupling efficiency over a large area was achieved in all-dielectric photonic Dirac cones[561]. Dirac cones carry a topological charge, known as the Berry phase, as denoted in Eq. (14). According to the bulk-edge correspondence, wave localizations at the edge of a finite size sample are predicted, depending on its orientation with regard to the lattice symmetries[523,562]. For instance, when the graphene boundary is in a zigzag shape or bearded shape, zero-energy edge states are present at boundaries. Their bands connect the two Dirac cones of the projected bulk bands experimentally verified in photonic graphene, **Fig. 11(a,b)**[27]. Strained photonic graphene can bring Dirac cones together and annihilate them, lifting their conical degeneracy and trivializing topological charges[563,564]. The strain engineering in photonic graphene also enables the pseudo-magnetic field at optical frequency[378]. Dirac cones usually occur in triangular or honeycomb lattices with three-fold and six-fold rotational symmetries. However, the existence of these Dirac points is exclusively ensured by the inversion symmetry $\mathcal{P}$ and TRS $\mathcal{T}$, which allows the exploration of similar conical dispersions in a broader range of systems, e.g., in a square lattice with mirror symmetry[565]. In particular, the accidental degenerate Dirac cones in square lattice were exploited to achieve zero-refractive-index response and cloaking at microwave frequency, **Fig. 11(c)** [566].

Dirac cones in photonics have been used to simulate relativistic fermionic phenomena like Klein tunneling and *Zitterbewegung* effects[377,567,568]. Moreover, multiple degrees of freedom (DoFs) in photonic crystals like the valley, chirality, and pseudo-spin DoFs can be exploited selectively by artificial gauge fields acting on light, resulting in a single Dirac cone of bulk bands at one of the valleys and bandgap opening at the other valley, as shown in **Fig. 11(d)**. These exotic optical responses emerge at the crossover between distinct topological photonic phases: quantum Hall phase (or quantum spin Hall phase) and valley Hall phase. The proposed structures have marked implications on photonic transport, like the spin-valley polarized one-way Klein tunneling: depending on the spin and valley polarization, photons may experience unimpeded penetration through potential barriers; The second consequence of the single Dirac-like cone is, that topological edge states coexist within the Dirac continuum for the opposite valley and pseudospin polarizations, which may enable a new class of spin-valley controlled devices[569]. A single Dirac cone realized with competing synthetic gauge fields in photonics was demonstrated by the microwave experiment, and the one-way refraction occurs between the photonic crystal and empty waveguide, implying the existence of a single-Dirac cone, **Fig. 11(e)**[570]. Remarkably, tilted Dirac cones in the energy bands were discovered in 2D photonic lattice either in the anisotropic form[571] or with non-symmorphic symmetry[572], which might be composed of linear or parabolic dispersion of photonic modes and flat bands of dipole excitons[573]. The tilted Dirac cones may possess high-order topological charges, and their peculiar dispersion was experimentally measured via polariton photoluminescence in photonic orbital graphene, **Fig. 11(f)** [574].

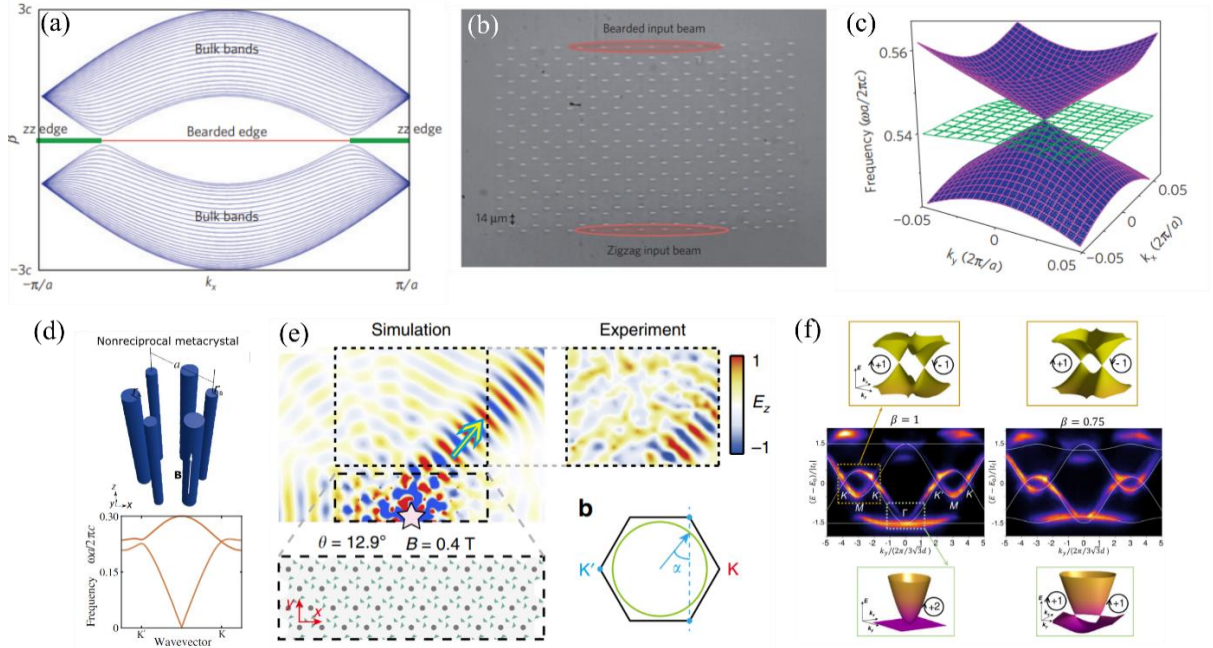

**Figure 11. Dirac cones in photonics.** (a) Projected band structure of photonic graphene ribbons with zigzag or bearded edge at the boundary. (b) Image of photonic graphene made of coupled waveguide arrays[27]. Copyright 2014 Springer Nature. (c) Three-folded Dirac cones due to accidental degeneracy in the photonic crystal[566]. Copyright 2011 Springer Nature. (d) Schematic of photonic crystal realizing the single Dirac cone at one valley and topological band gap at the other valley illustrated in the lower panel[569]. Copyright 2018 Science. (e) Refraction of single Dirac valley into an empty waveguide[570]. Copyright 2020 Springer Nature. (f) The tilt Dirac cones measured by measured polariton photoluminescence intensity show different topological charges[574]. Copyright 2019 American Physical Society.

## 4.2. Dirac cones in phononics

In the phononic domain, a honeycomb lattice of acoustic cavities linked with coupling tubes also mimics graphene's tight-binding Hamiltonian and presents Dirac cones at the corner of the Brillouin zone[170,575]. Similar to graphene, they are protected by both $\mathcal{P}$ and $\mathcal{T}$ symmetries, expanding their existence to numerous systems such as sonic crystals[576–578] as well as locally resonant MMs supporting airborne surface acoustic waves (SAW) [579,580], whose openings to the surrounding space allow for a straightforward measurement of the band structure and direct interaction with sound at the surface of the medium, **Fig. 12(a).** Elastic Dirac points have also been observed, both at the macroscale[581] and microscale[582], **Fig. 12(b).** Moreover, similar to graphene, phononic Dirac cones are linked to remarkable wave phenomena such as the phononic analog of *Zitterbewegung*[576,582], and acoustic Klein tunneling[583] or edge modes stemming from the cones topological charge[578].

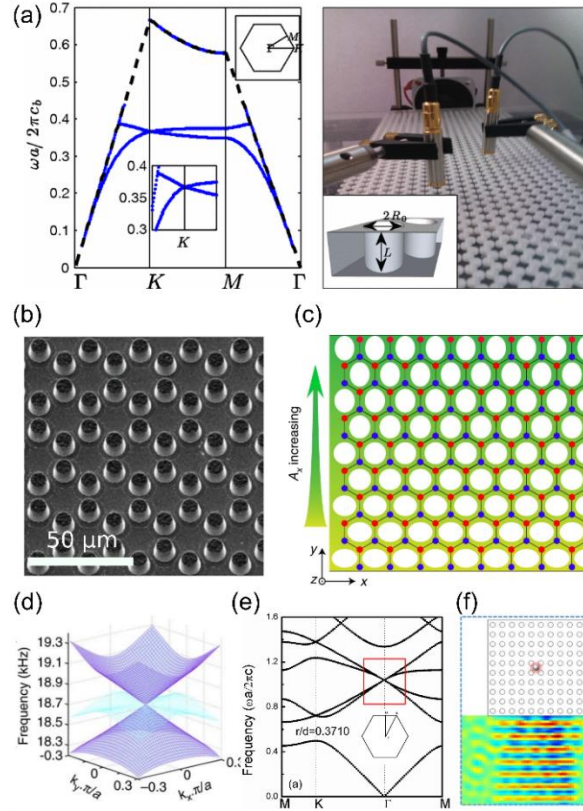

**Figure 12. Dirac cones in phononics**. (a) Acoustic Dirac cone within a honeycomb lattice of holes. Reproduced with permission from[579]. Copyright 2012 American Physical Society. (b) Honeycomb lattice of micro-pillars on a substrate. Reproduced with permission from[582]. Copyright 2016 Springer Nature. (c) Strain pattern applied on a honeycomb sonic crystal to mimic an artificial gauge field. Reproduced with permission from[171]. Copyright 2019 Springer Nature. (d) Triply degenerate sonic Dirac cone in the center of the Brillouin zone. Reproduced with permission from[584]. Copyright 2017 Springer Nature. (e) Double acoustic Dirac cones consisting of a quadruple degeneracy in the center of the Brillouin zone. Reproduced with permission from[585]. Copyright 2014 Springer Nature. (f) Acoustic wave collimation obtained through the zero index properties of the triply degenerate Dirac cone. Reproduced with permission from[584]. Copyright 2017 Springer Nature.

Furthermore, precisely engineered global deformations, such as strain fields, can act as an effective gauge field for phononic waves in the medium, leading to acoustic analogues of Landau levels along with corresponding helical edge states. This has been evidenced for sound in tightly-coupled cavities[170] and sonic crystals[171], **Fig. 12(c)**, but also for mechanical waves[172,586]. Notably, on-chip realizations have been proposed at the nanoscale using a silicon slab pierced with snowflake holes[173] and demonstrated experimentally at the microscale with a phononic crystal made of triangular pillars etched on one side of a silicon substrate[174].

Moreover, these conical degeneracies are not limited to honeycomb symmetries but also exist at the corners of the Brillouin zone in Kagome lattices [587]. Accidental degeneracies in square lattices present triply degenerate Dirac-like cones with an extra flat band in the center of the Brillouin zone $\Gamma$ [584,588], **Fig. 12(d)**, while a honeycomb lattice of steel cylinders in the air with the proper filling ratio also induce accidentally a four-band degeneracy in the shape of two overlapping Dirac cones in $\Gamma$ [585], **Fig. 12(e)**. These other conical degeneracies in $\Gamma$ have different topological properties and enable a zero index medium that can be used for wave collimation, **Fig. 12(f)**. Similar features are found in the self-dual phase of a twisted Kagome lattice[589]. Finally, Dirac cones and their inherent chiral properties have also been demonstrated in soft elastic strips made of a homogeneous silicone elastomer, opening the door to possible analogies with biological tissues[590].

## 5. Topological phases in 2D MMs with broken time-reversal symmetry

This section presents different topological phases of 2D artificial materials with broken T-symmetry. We explain how breaking TRS for classical waves results in the opening of topological bandgaps carrying one-way topological edge modes, classical analogs of the quantum-Hall effect. Then, we present the topological phases based on temporal/spatial modulations achieving analogs of the Floquet TIs or anomalous Floquet TIs in MMs.

### 5.1. Chern insulators in photonics

The most popular 2D topological phase of matter relying on broken TRS is the Quantum-Hall effect[332–334]. The exciting features of chiral edge states, like immunity from backscattering and strong robustness against defects, have attracted much attention in the physics community in the past decades and motivated the transposition of these concepts to classical waves.

The first proposal in 2D topological photonics can be drawn back to Haldane and Raghu in 2008[15]. They proposed the photonic analog of QHEss and disclosed edge states exist on the gyromagnetic photonic crystal. A nonzero Chern number above the photonic bandgap is obtained when the static magnetic field is applied perpendicularly to the photonic crystal. According to the bulk-edge correspondence[7], chiral edge states propagate in a one-way manner at the boundary of the photonic crystal. They are immune to scattering and robust against the disorder in their pathway. To observe the topological phenomena, the gyromagnetic photonic crystal embedded between parallel metal plates was designed and measured in the microwave experiment[480,591]. A strong magneto-optic response breaks the TRS of materials such that the analog of the QHEs in photonics was achieved. The one-way propagating edge states were observed, and their robustness was demonstrated, even when the scatterers were inserted in the propagating pathway of the edge states,

**Fig. 13(a,b)**. More proposals and experiments were performed in a similar fashion[592–596], and the large Chern numbers in higher frequency bandgaps were experimentally confirmed by the Fourier transform of the spatial spectrum[28,29].

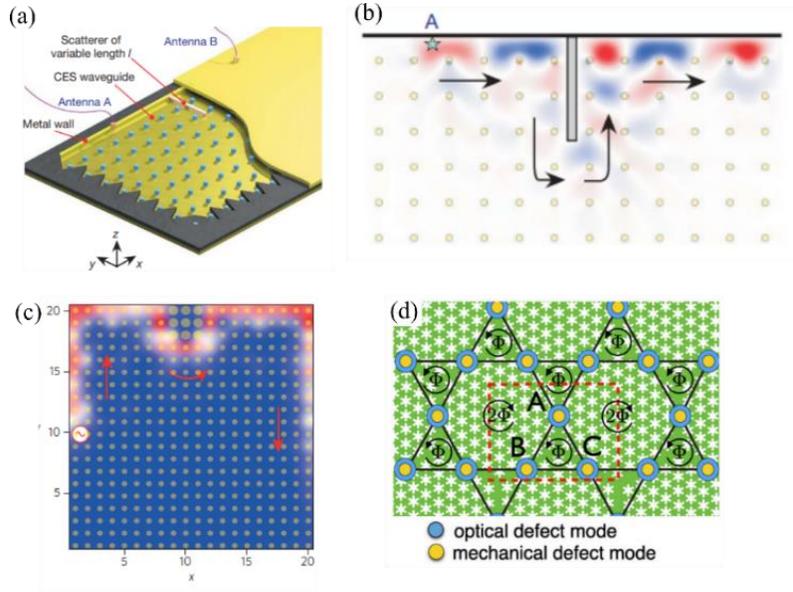

**Figure 13**. **Chern insulators in photonics.** (a) Schematic of magneto-optic lattice embedded between two metal mirrors. (b) Microwave experiment for the robust demonstration of topological edge propagation[17]. Copyright 2009 Springer Nature. (c) Photonic one-way edge mode in a dynamically modulated resonator lattice[20]. Copyright 2012 Springer Nature. (d) A kagome optomechanical array supporting the topological phase of both sound and light[37]. Copyright 2015 American Physical Society.

If the magneto-optic photonic crystal is scaled down to the on-chip design in the optical frequency range, the Ohmic loss of the magneto-optical material is substantial. Moreover, the magneto-optic effect is weak at optical frequencies. Thus, the associated bandgap is negligible compared to its operating frequency. These drawbacks significantly limit the application of magnetized photonic Chern insulators on the integration platform. Researchers have searched for alternative methods to implement non-reciprocal propagation of the edge states. Many photonic systems support the active design whose refractive index can be modulated with electro-optic modulators (EOMs) or acousto-optic interaction. Therefore, it is possible to dynamically modulate the couplings in the resonators and create an artificial gauge field in photonics, **Fig. 13(c),** providing promising platforms for emulating the QHEs and observing robust chiral edge states[20,21,597]. For example, the optomechanical crystal formed by the 2D array of resonators, which couple the photon and phonon in a coherent way, demonstrate a high degree of controllability, **Fig. 13(d)**. A variety of topological phenomena were theoretically predicted in such a platform[37,598]. Non-reciprocity and corresponding synthetic gauge fields were experimentally realized in a few sites of the optomechanical system[599–601]. Topological edge states with nonzero Chern number have been realized in other forms, including quasi-static electronic waves in a circuit, with negative impedance converters employed[602] and magnetoplasmons in graphene[603,604].

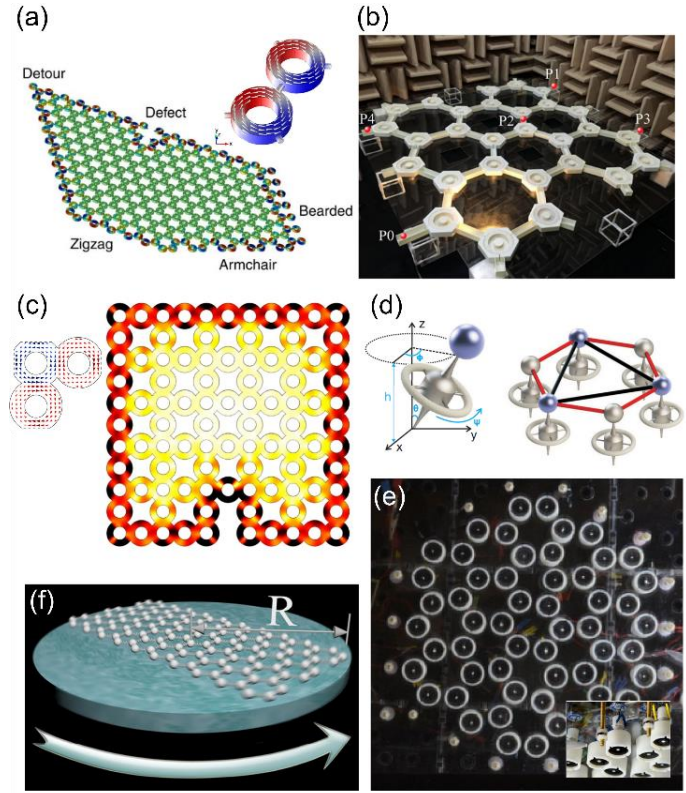

**Figure 14**. **Chern insulators in phononics.** (a) Acoustic Chern insulator achieved based on airflow within a honeycomb lattice. Reproduced with permission from[605]. Copyright 2015 Springer Nature. (b) Experimental realization of the model presented in (a). Reproduced with permission from[606]. Copyright 2019 American Physical Society. (c) Topological one-way edge mode in an acoustic Chern insulator implemented thanks to active particles. Reproduced with permission from[607]. Copyright 2017 Springer Nature. (d) Proposal of a mechanical Chern insulator using gyroscopes. Reproduced with permission from[58]. Copyright 2015 American Physical Society. (e) Experimental realization of a mechanical Chern insulator using such rotating parts. Reproduced with permission from[59]. (f) Concept of inducing the Chern insulator phase in a mechanical spring-mass lattice using the Coriolis force. Reproduced with permission from[608]. Copyright 2015 Institute of Physics.

### 5.2. Chern insulators in phononics

Remarkably, topological phononic modes can exist at the edge of microtubules in biological cells[609]. In that case, weak magnetic fields induce a Lorentz force on ions, breaking TRS. However, this feature cannot be generalized to sound and elastic waves at a macroscopic scale because the matter is magnetically inert.

Hence starting the search for new ideas to transpose Chern insulators in phononic media. Firstly, by considering the simple orbiting electron picture, one can think of mimicking a magnetic bias by introducing a constant motion of the whole medium carrying the waves. Hence, Roux et al. have demonstrated that introducing a vortex within an irrotational fluid dislocates the wave-front of an acoustic wave passing through it, obtaining an analog of the Aharonov-Bohm effect for sound[208]. A few years later, Fleury et al. implemented a 3-ports circular cavity with a rotating airflow that acts as a circulator for sound[610]. Such a non-reciprocal device can then be organized in a honeycomb lattice which lifts the Dirac cones degeneracy of the static medium when the airflow starts rotating[605,611]. The resulting bandgap is described by a non-zero Chern number which accounts for robust one-way guided acoustic modes at the edges of the MM, effectively mimicking a Chern insulator for sound, **Fig. 14(a)**. Similar effects appear in a triangular lattice of rotating scatterers in a viscous fluid[38] and a square lattice with flow-biased cavities having a Dirac-like point at $\Gamma$ [612].

The genuine character of the bias flow approach to breaking $\mathcal{T}$ has enabled the direct transposition of the Chern insulator concepts in acoustics MMs. It has also provoked the study of new phases, such as anti-chiral edge states[613]. Ding et al. successfully implemented an acoustic Chern insulator using high-quality factor cavities[606], **Fig. 14(b)**. However, issues like the high velocity of the flow, inherent loss, and bulky samples have severely hindered the experimental implementation of these concepts. A promising alternative can be found in active liquids, where self-propelling particles can generate a spontaneous flow within a tight-binding lattice or even continuous media[607,614–616], **Fig. 14(c)**. The versatility of active matter shows great potential for acoustic TRS breaking and paves the way toward colloidal, soft-matter, and chemically designed topological acoustic devices.

If the motion of the whole medium carrying waves is tedious in acoustics, this is not the case in mechanical systems where rotating parts such as gyroscopes are commercially available. Honeycomb tight-binding lattices with masses attached to individual gyroscopes have been proven to have bandgaps with a non-zero Chern number which host one-way topologically protected edge modes[58,59], **Fig. 14(d).** Notably, this has been demonstrated experimentally using a lattice of rotators and recording their motion with high-velocity cameras[59], **Fig. 14(e).** This setup was then successfully extended to other lattices[617] and even amorphous media[618]. Another route consists in rotating the entire mechanical sample. The Coriolis force induced by the non-inertial reference frame can be seen as a uniform effective magnetic field opening topological bandgaps in honeycomb mass-spring lattices[608], **Fig. 14(f).** An alternate path directly implements the Haldane model in an active mechanical MM with feedback loops[619].

### 5.3. Floquet TIs in photonics

Another approach to breaking TRS and obtaining the nonreciprocal edge states is temporal modulation of the medium's properties. A time-Floquet system is a typical example where the corresponding Hamiltonian is *periodically* driven in time. This has drastic consequences on the related band structure: the bands of the static system are repeated along the frequency axis with a period equal to the modulation frequency of the periodically driven source[370].

This Floquet approach can be employed in the photonic system producing synthetic gauge fields by periodical modulating the dielectric parameters in time[20,21], achieving cyclotron motion of light without magnetic materials, thus, supporting non-reciprocal edge states. Compared to other approaches, it does not rely on spatial symmetries and the magneto-optic material. However, it requires a modulation frequency larger than the bandwidth of the host frequency and strong modulation strength, making it hard to implement in the photonic experiment. An alternative way to achieve this is utilizing the propagating geometries in the coupled silicon waveguides[23]. To introduce the synthetic gauge field in the system and create chiral topological edge states, the $\mathcal{T}$ symmetry has to be broken by temporal modulation. The spatial modulation by helical rotating the waveguides in $z$ axis can emulate this temporal modulation, **Fig. 15(a)**. The coordinate frame is transformed by such helical design as $x' = x + R\cos(\Omega z)$, $y' = y + R\sin(\Omega z)$, and $z' = z$, where $R$ is the helix radius, and $\Omega$ is the modulation frequency in z, which exceeds the bandwidth of quasi-bands to avoid the mixing with other order Floquet modes. Using the Floquet technique, quasi-frequency bands of the Floquet system are obtained and degeneracy at Dirac points is lifted up due to the temporal modulation. As a result, the topological chiral edge states propagating in one-way direction at the boundaries of the coupled waveguides were observed. In addition to directly tailoring the chiral couplings between waveguides, the onsite potential of the waveguides can be periodically modulated via an intermediate 'chained' waveguide, the couplings between primary sites are mediated by the 'chained' waveguide, thus achieving the rich topological phases[620], **Fig. 15(b)**.

Interestingly, if the driven period is long enough, or the evanescent couplings between waveguides are very strong such that the periodicity of the quasienergy bands $\Omega$ is less than the energy range of the Floquet modes belonging to the same Floquet order, quasienergy bands between different Floquet orders couples and form new bandgap, leading to the anomalous topological edge states in the new bandgap even when Chern number of the Floquet bands is zero[621]. Anomalous Floquet phases and the corresponding topological edge states were theoretically proposed and observed in several photonic platforms, such as the network model of ring resonators[622], the slowly varied coupled waveguides[32,623,624], and designer surface plasmon[625]. For example, a bipartite square lattice in **Fig. 15(c)** was proposed to realize the anomalous FTIs[623,624]. The coupling to the neighboring waveguides occurs in four steps; in each step, hopping takes place only along a certain direction. Consequently, bulk modes have zero Chern number, but integer winding number and the chiral edge states highly robust to distortion were observed. The anomalous Floquet TI was also recently realized on a metal-oxide-semiconductor (CMOS) chip, as shown in **Fig. 15(d)**. It consists of switched-capacitor networks and possesses a topological bandgap order of magnitude larger than any previous demonstration, spanning from DC to GHz frequencies[626]. More importantly, such a topological device was exploited in the 5G communication and showed promising functions like multi-antenna full-duplex wireless operation and true-time-delay-based broadband beamforming.

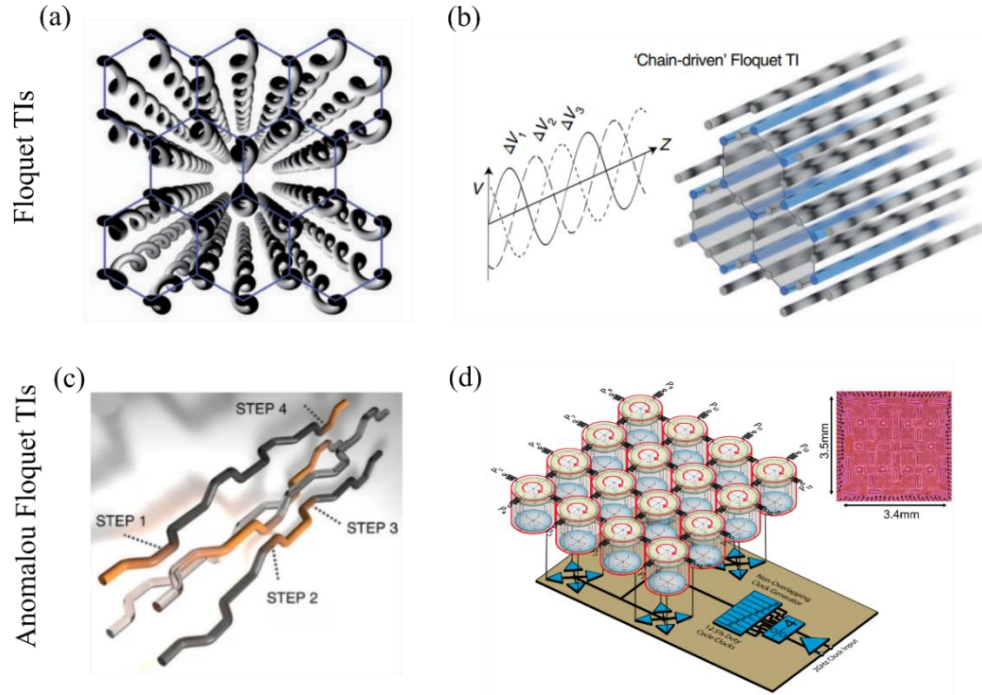

**Figure 15. Photonic TIs achieved by Floquet modulation.** (a) Schematic of spiral propagating waveguides emulating the Floquet TIs[23]. Copyright 2013 Springer Nature. (b) Schematic of 'on-chip' driven photonic Floquet TI, where the onsite potentials of the lattice are modulated cyclically along z direction[620]. Copyright 2022 Springer Nature. (c) Schematic of the coupled waveguides with slow modulation in spacing[623]. Copyright 2017 Springer Nature. (d) Schematic anomalous Floquet TI on a CMOS chip made of temporally switched capacitor networks[626]. Copyright 2022 Springer Nature.

### 5.4. Floquet TIs in phononics

Temporal modulation of the system's parameters has also been proposed in acoustic MMs[57]. In this work, Fleury et al. consider a tight-binding honeycomb lattice of meta-atoms that consist of three tightly coupled cavities whose bulk modulus, or volume, periodically changes over time. Adding a local rotating phase pattern in the modulation at the scale of the trimer results in breaking the Dirac cone of the medium. The open bandgap is topologically non-trivial and exhibits one-way edge modes that are robust against spatial disorder and phase disorder of the modulation, **Fig. 16(a)**. Like the Chern insulator described above, the time-modulation of cavities' acoustic properties comes with severe practical issues such as loss and an intricate synchronization of many devices. Experimental demonstration of non-reciprocal transmission in a dimer has been achieved by mechanically changing the volume of the cavities[627]. An alternate option is to implement carefully designed electric feedback circuits, which have been used in the context of Willis coupling[628], non-linear rainbow trapping [629], or PT-symmetry[630]. This Floquet topological phase is more straightforward in mechanical samples thanks to the wide availability of piezoelectric actuators. Hence, Darabi et al. designed a thin plate consisting of a honeycomb lattice of trimers[631] similar to[57]. Each cavity's Young modulus can be controlled with a piezoelectric patch, **Fig. 16(b)**. The authors experimentally demonstrated a mechanical Floquet TI with reconfigurable robust edge states[631].

As mentioned above, time modulation can be challenging in acoustic experiments. Fortunately, a direct mapping between Floquet Hamiltonians and scattering network systems[622,632] allows for more practical implementations in passive media. Hence, a so-called anomalous Floquet TI has been demonstrated in a square lattice of circular acoustic cavities whose counter-rotating modes are linked through adequate inter-cell forward couplings[633–636], **Fig. 16(c)**. Besides, mapping the time dimension to a third spatial dimension makes it possible to mimic anomalous Floquet topological phases in passive media. The couplings' chirality along the third dimension only effectively breaks $\mathcal{T}$ [637], **Fig. 16(d)**.

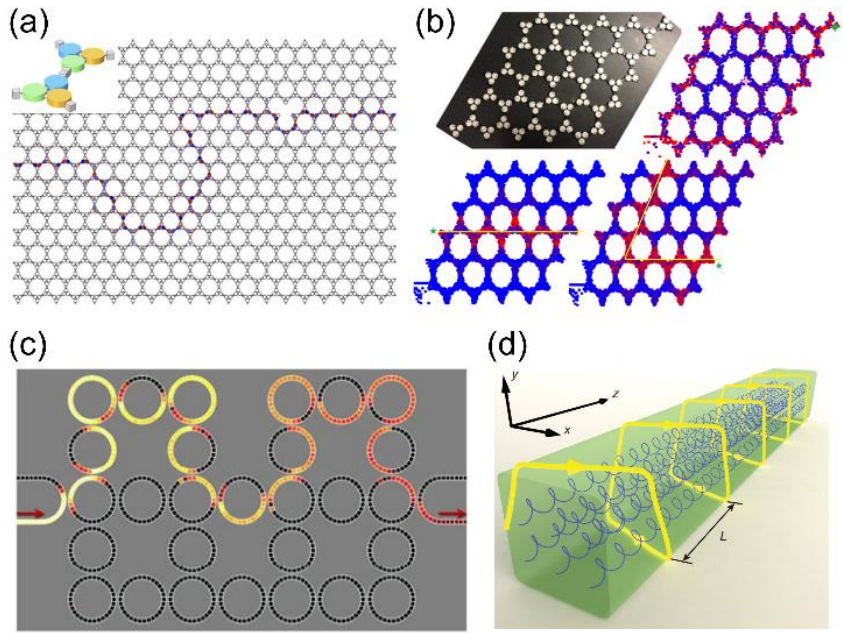

**Figure 16**. **Floquet TIs in phononics.** (a) Floquet TI for sound. Reproduced with permission from[57]. Copyright 2016 Springer Nature. (b) Experimental realization of a reconfigurable Floquet TI for mechanical vibrations. Reproduced with permission from[631]. Copyright 2020 American Association for the Advancement of Science. (c) Acoustic anomalous Floquet TI. Reproduced with permission from[636]. Copyright 2016 Springer Nature. (d) Concept of a topological Floquet phase by mapping the time modulation in a higher spatial dimension. Reproduced with permission from[637]. Copyright 2019 American Physical Society.

# 6. Topological phases in 2D MMs with T-symmetry

This section focuses on topological phases of a 2D system in which TRS is conserved. We present classical analogues of the quantum spin-Hall effect, explain how to design relevant pseudo-spins degree of freedom, and introduce the pseudo-spin orbital couplings for classical waves. We further review the different examples of classical analogs of the valley-Hall effect.

## 6.1. Photonic spin-Hall insulator

The conservation of $\mathcal{T}$ alone in a classical system cannot guarantee the Kramers degeneracy and does not provide the condition of simulating the classic analog of QSHE. However, we can exploit duality in electromagnetic waves or introduce external symmetries to mimic the role of the fermionic TR operator in a bosonic system[22,30]. For example, when $\hat{\epsilon} = \hat{\mu}$, the Maxwell equations turn into the two copies of second order differential equations[22],

$$\hat{\epsilon}^{-1}\nabla \times \{\hat{\epsilon}^{-1}\nabla \times \boldsymbol{E}(\boldsymbol{r})\} = k^2\boldsymbol{E}(\boldsymbol{r}) \tag{17}$$

$$\hat{\epsilon}^{-1}\nabla \times \{\hat{\epsilon}^{-1}\nabla \times \boldsymbol{H}(\boldsymbol{r})\} = k^2\boldsymbol{H}(\boldsymbol{r}), \tag{18}$$

Consequently, double degeneracy of the photonic band structures occurs everywhere in the Brillouin zone, providing DoFs to mimic the pseudo spin degree of freedom. Furthermore, QSHE requires including SOI in the model to break the band degeneracy and induce topological transition. However, no direct analog spin-orbit coupling exists in photonics because of the neutral charge of photons. The first example of photonic TI with conserved $\mathcal{T}$ symmetry exploits synthetic spin-orbit coupling by introducing magneto-electric couplings (bianisotropy) in the constitutional relations of the material

$$\boldsymbol{D} = \hat{\epsilon}\boldsymbol{E} + \hat{\chi}\boldsymbol{H}, \tag{19}$$

$$\boldsymbol{B} = \hat{\mu}\boldsymbol{H} + \hat{\chi}^{\dagger}\boldsymbol{E}. \tag{20}$$

To preserve TRS, the bianisotropic tensor $\hat{\chi}$ needs to be complex-valued such that the Maxwell equations are separated into two decoupled Helmholtz equations

$$\mathcal{L}_0\psi^{\pm} = \pm\mathcal{L}_1\psi^{\pm}, \tag{21}$$

where $\psi^{\pm}$ represent the pseudo-spin states, $\mathcal{L}_0$ acts as the unperturbed Hamiltonian, and $\mathcal{L}_1$ introduces synthetic gauge potential in pseudo spin/down states with opposite signs, emulating pseudo-spin orbit interactions and opening the bulk bandgap, **Fig. 17(a)**. The effective Hamiltonian near Dirac cone with k.p approximation reads

$$\hat{\mathcal{H}} = v_0\hat{s}_0(\delta k_x\hat{\tau}_z\hat{\sigma}_x + \delta k_y\hat{\sigma}_y) + m\hat{\tau}_z\hat{s}_z\hat{\sigma}_z, \tag{22}$$

where $\hat{\sigma}_i$, $\hat{\tau}_i$ and $\hat{s}_i$ are Pauli matrices, $v_0$ is the effective velocity near the Dirac cone, and the mass term $m$ is proportional to bianisotropic strength, directly leading to the formation of the topological bandgap. **Eq.(22)** is equivalent to the electronic Hamiltonian of the Kane-Mele model, and it depicts the picture of two copies of the QHEs displaying their respective topological phases but connected via $\mathcal{T}$ symmetry.

Consequently, such bi-anisotropic photonic crystal with duality supports the helical edge states at the boundaries. It displays the spin-locking propagation of electromagnetic waves with unprecedented robustness to defects and disorder, drastically altering our view of the scattering of electromagnetic waves.

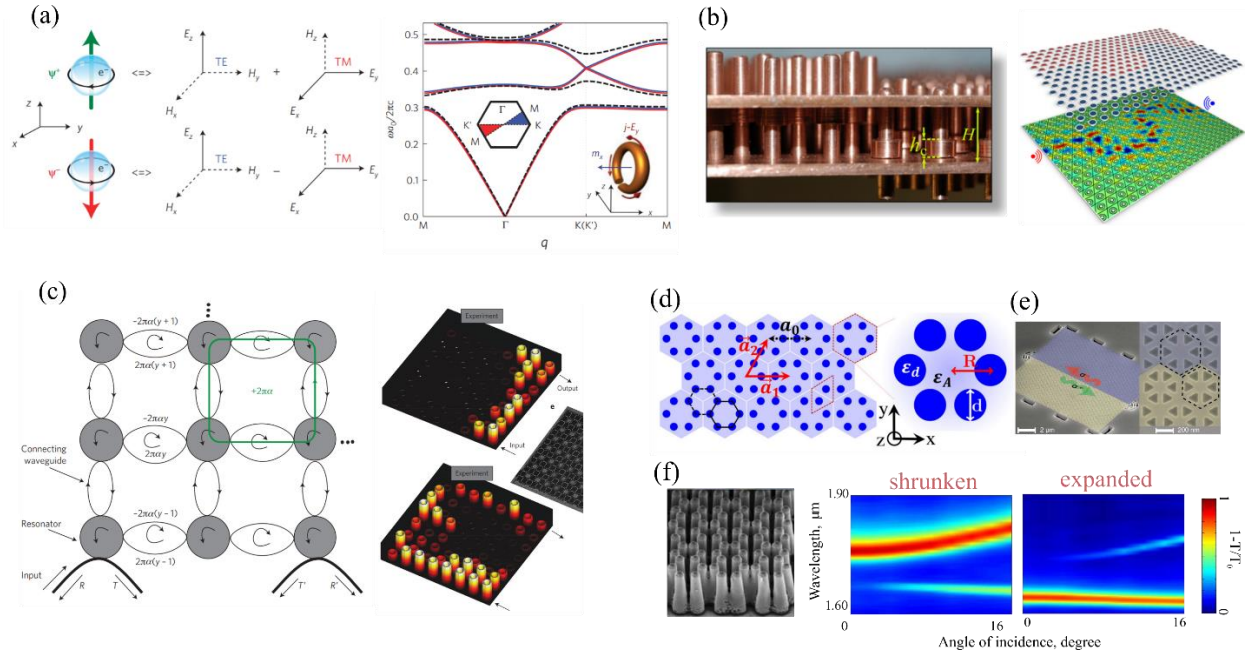

**Figure 17**. **Photonic spin-Hall insulator.** (a) Construction of photonic pseudo-spin states and the topological photonic band structures with duality in electromagnetic wave and bianisotropic response[22]. Copyright 2013 Springer Nature. (b) Reconfigurable MM design achieving the QSHE and robust electromagnetic pathway for microwave experiments[31]. Copyright 2016 Springer Nature. (c) Schematic of CROW lattice achieving analog QSHE and propagation of topological edge states[19,25]. Copyright 2011 Springer Nature and 2013 Springer Nature. (d)All-dielectric photonic crystals with crystalline symmetries realize the pseudo-spin states and QSHE[30]. Copyright 2015 American Physical Society. (e) Scanning electron microscope (SEM) image of the fabricated device made of equilateral triangular air holes supporting two topological helical edge modes[49]. Copyright 2018 Science. (f) SEM image of the topological photonic crystal made of silicon pillars (in the left panel) and angle-resolved experimental spectra from far-field probing revealing the band-inversion between the shrunken lattice and expanded lattice[653]. Copyright 2018 Springer Nature.

The topological edge states were observed on bi-anisotropic MMs with careful MM design to match the effective permittivity and permeability[638–640]. In parallel, a more straightforward meta-waveguide design was theoretically proposed and experimentally implemented[31,641,642], **Fig. 17(b).** The meta-waveguide triangular photonic crystal consists of an array of copper pillars sandwiched between parallel copper plates. Their geometries are carefully chosen such that the accidental degeneracy of Dirac cones between quasi-TE and quasi-TM modes are spectrally overlapped. The copper collars are introduced at each pillar, leading to an effective bi-anisotropic response between the electric and magnetic fields when they are not in the center of the pillars, thus opening up the bandgap of bulk Dirac cones. Importantly, these collars are shifted upward or downward to the plates to switch the sign of the topological index. Therefore, a reconfigurable and arbitrary topological electromagnetic pathway can be created by manipulating the collars, and the topological edge states were experimentally demonstrated to survive from the arbitrary pathway and exhibit ballistic transport through the disorder region. In addition, the topological switch was proposed and experimentally demonstrated. A similar idea was also implemented in an electromagnetic dual MS, and the robust transport of line waves was demonstrated between complementary surface impedances[643,644].

Due to the fabrication complexity, achieving a photonic QSH insulator on a chip via the bi-anisotropic design faces a tremendous challenge. Alternatively, researchers have developed a different approach to realize a QSH insulator in a 2D array of coupled-resonator optical waveguides (CROWs), **Fig. 17(c)**. The waveguides support single-mode (TE) propagation at the telecom wavelength, and the synthetic gauge fields for creating the topological order are realized by the vertical shift of link-resonators coupling site resonators. As a result, the topological edge states were imaged, and their robust transport against disorder was demosntrated[19,25]. A scheme for measuring topological invariants was proposed and tested by inserting artificial magnetic flux at the edge of the CROWs array[645,646]. The fine-tune requirement of gauge fields was relaxed in a similar CROWs network model, and the topological phase transition was changed by the variation of coupling strengths between resonators[647]. The idea of using synthetic gauge fields to obtain the topological phase was similarly employed in the circuit paradigm. The phase gradients in the circuit lattice can be easily realized by various circuit elements without breaking $\mathcal{T}$. Thus, the topological propagation of quasi-static waves was achieved in electrical circuits[648–650].

CROWs based on silicon-on-insulators (SOIs) may offer promising applications for robust light steering[651]. However, they cannot be scaled very small to the visible spectral range. To achieve the topological optical devices in the visible light regime, all-dielectric designs with crystalline symmetries, which do not require a magnetic field and are constructed from deformed photonic crystals (shrunken/expanded lattices), were proposed[30], **Fig. 17(d)**. The basic idea is to describe a honeycomb lattice as a triangular lattice of hexagons, resulting in an artificial folding of the band structure. In this way, the original Dirac cones move from the corner of the Brillouin zone and overlap in its center. Then, shrinking or expanding six meta-molecules (referred to as hexamer) toward the center of the unit cell enables the band-folding of the energy bands, consequently opening a bandgap between two pairs of degenerate bands and forming band inversion between shrunken and expanded lattices. The current fabrication technique can implement this scheme, and the relevant photonic QSH insulator experiments have been carried out from microwave to near-infrared and visible spectral regimes[652–655]. It was used as the quantum optic interface[49] coupled with the quantum emitter, **Fig. 17(e)**, and the properties of spin-momentum locking edge states were utilized in generating vortex laser beams [656]. Interestingly, the open nature of the topological photonics allows to directly observe the topological transition accompanied by the inversion of bright and dark modes, **Fig. 17(f)**, and retrieve the topological properties from the measured far-field scattering characteristics[653].

## 6.2. Phononic spin-Hall insulator

At first sight, it also seems complicated to transpose these concepts to sound or mechanical waves. Indeed, similarly to photons and contrary to electrons, they behave as bosons which means that their time-reversal operator squares to $T^2 = 1$ preventing the existence of Kramers pairs. Nevertheless, several works have successfully demonstrated the possibility of associating extra DoFs, namely pseudo-spins, to phononic waves. A first example was proposed in the tight-binding lattice of pendula dimers coupled with multilayer spring arrangements designed to mimic spin-orbit interaction[43,657], **Fig. 18(a)**. Here, the polarization of each dimer act as a pseudo-spin DoF, and robust helical edge wave propagation has been evidenced experimentally[43,658]. Another proposal is to add a second layer with interlayer couplings whose chirality reproduces the spin-orbit interaction[658,659], **Fig. 18(b)**. Interestingly, Deng et al. showed that cutting the phononic crystal in the middle of the unit cell flips the corresponding pseudo-spin while keeping the topological protection and designed a so-called spin-flipper[659], **Fig. 18(c)**.

Furthermore, a careful design of the medium's spatial symmetries $\xi$ can create pseudo-spins and the related pseudo-time reversal operator $\xi T$ such as $(\xi T)^2 = -1$. Then, manipulating the lattice structure can lead to an equivalent of pseudo-spin orbital coupling for the particular symmetry $\xi$ causing a bandgap opening with a band inversion. A famous example is a triangular lattice of pillars in the air whose band structure has two overlapping Dirac cones at the center of the Brillouin zone[660]. Changing the scatterers' radius lifts the four-band degeneracy. It results in two pairs of bands separated by a bandgap, whose corresponding dipolar and quadrupolar symmetries are associated with pseudo-spins. Tuning the radius size generates a band inversion linked to two distinguished topological behaviors. An interface between these two topologically distinct crystals carries helical waves whose direction of propagation is locked to the sense of rotation of local acoustic vortices along the interface. Nevertheless, their peculiar features permit to design of topological beam splitters ruled by pseudo-spin conservation[660], **Fig. 18(d).** Opening non-trivial topological bandgap from two overlapping Dirac cones in the center of BZ can also be applied within triangular lattices consisting of six "meta-atoms" that can be either shrunken or expanded[30]. The simplicity of this protocol has permitted to expand of these concepts to numerous platforms, ranging from soda can MMs[661], macroscopic phononic crystals[662,663,672–675,664–671] to on-chip systems[676–680], whose reconfigurability opens the door to practical devices[668,669]. Cha et al. measured robust edge transport using free-standing silicon nitride nano-membranes, paving the way for highly compact topological nano-electromechanical MMs[676], **Fig. 18(e).**

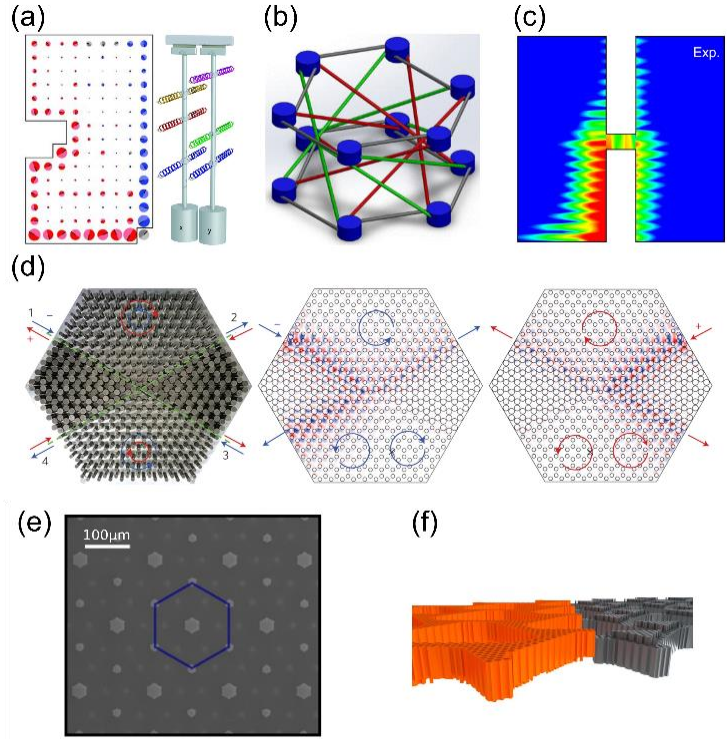

**Figure 18**. **Phononic spin-Hall insulator.** (a) Phononic spin-Hall insulator based on a pendula lattice. Reproduced with permission from[43]. Copyright 2015 American Association for the Advancement of Science. (b) Pseudo-spin orbit coupling induced by chiral inter-layer couplings. Reproduced with permission from[658]. Copyright 2016 American Institute of Physics Publishing. (c) Topological spin-flipper for sound. Reproduced with permission from[659]. Copyright 2020 Springer Nature. (d) Spin-locked sound propagation at a topological intersection. Reproduced with permission from[660]. Copyright 2016 Springer Nature. (e) On-chip spin TI based on the folding technique. Reproduced with permission from[676]. Copyright 2018 Springer Nature. (f) Interface between two carefully designed thin plates whose pseudo-spin topological features rely on symmetry breaking out-of-plane. Reproduced with permission from[681]. Copyright 2015 Springer Nature.

Contrary to scalar acoustic waves, Lamb waves in thin plates have orthogonal polarizations, which can also be used to implement pseudo-spins. Mousavi et al. have shown that the band structure of a properly designed plate with a triangular lattice of triangular holes presents two overlapping Dirac cones in the corner of the Brillouin zone[681], each related to one Lamb wave polarization. Breaking the out-of-plane mirror symmetry creates two pairs of doubly degenerate bands separated by a bandgap with a non-zero spin Chern number, **Fig. 18(f)**. A few years later, Miniaci et al. demonstrated this concept in an experiment with a slightly modified sample with additional holes to facilitate the design[682]. Besides, Zheng et al. have shown that exploiting the rotational DoF of beads within granular media can also lead to this topological phase and be a promising platform for implementing nonlinear effects[683].

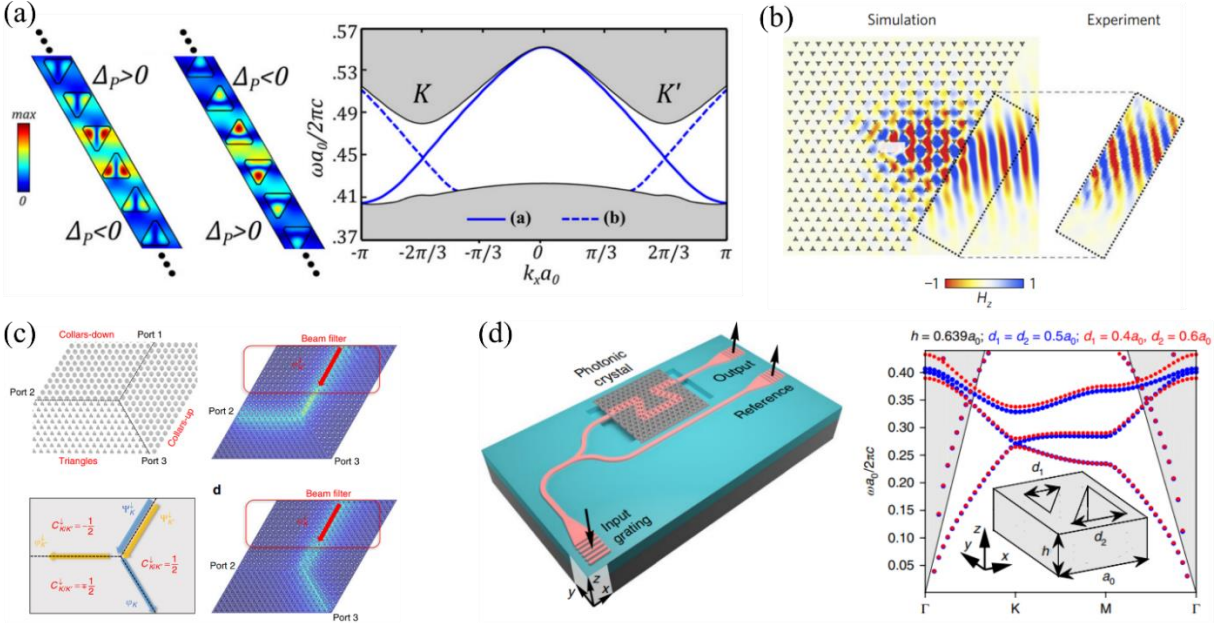

**Figure 19. Photonic valley-Hall insulator.** (a) The photonic design for valley-Hall effect and band structures of supercell supporting valley edge states along zigzag interfaces between two domains with distinct valley index[684]. Copyright 2016 IOP Publishing. (b) Topologically protected fraction of kink states into free space demonstrated by the microwave experiment[688]. Copyright 2018 Springer Nature. (c) Topological Y-junction formed by three regions with different topological indexes, the Y-junction operates as a valley filter with valley-splitting wave transport, in which energy is injected via ports 1 and 2 and 3 only support K′ and K-polarized waves, respectively[696]. Copyright 2018 Springer Nature. (d) Schematic of photonic valley-Hall insulator on a chip (the left panel) and the corresponding band structures exhibiting valley-Hall effect (the right panel)[692]. Copyright 2019 Springer Nature.

### 6.3. Photonic valley-Hall insulator

Since breaking inversion symmetry is straightforward, e.g., reshaping circular Si rods into triangular rods would break inversion symmetry of the photonic crystal, **Fig. 19(a)**, topological valley-Hall insulators have been studied and implemented extensively in various photonic systems[684–686], such as metacrystal embedded between parallel metal plates[687–689], evanescently coupled waveguide arrays[690], surface plasmon crystals[691], on-chip silicon photonics[692693], photonic crystal waveguides made of semiconductor[694]. In particular, When the valley pseudospin is conserved, the valley edge (kink) states exhibit robust refraction into empty waveguide region[688], **Fig. 19(b)**. Interestingly, the valley-Hall and pseudo-spin-Hall effect can coexist in a photonic crystal. As such, due to the valley-Hall effect, the local Berry phase competes with the global Berry phase from the pseudo-spin-Hall effect. At the critical point, a single Dirac cone for bulk modes might be produced, which provides an excellent platform for observing interesting phenomena like valley-polarized transport, and pseudo-spin and valley polarized Klein tunneling effects[569,685]. The heterogeneous interfaces between the spin-Hall insulator and valley-Hall insulator were proposed and experimentally implemented in the microwave spectral range, the heterogeneous interfaces support the propagation of pseudo-spin-valley polarized edge states, and novel functional devices like the pseudospin filtering junctions were experimentally demonstrated on such heterojunction device[695,696]. For example, consider the Y-junction consisting of three different topological regions shown in **Fig. 19(c).** According to the bulk-edge correspondence principle, it supports valley polarized waves from port 1 to port 2/port 3, opening up the possibility of using the valley DoF to manipulate the light flow. The easy fabrication of a photonic valley-Hall insulator enables the possibility of robust light propagation on a chip. **Fig. 19(d)** shows that valley-Hall insulator can be realized in a much small footprint, compatibility with complementary metal–oxide–semiconductor fabrication technology[692,693], and thus allows for optical operation based on valley-Hall insulator at telecommunications wavelengths[697] and even electrically pumped topological lasing with valley edge states[698]. Moreover, a valley plasmonic crystal for the metagate-graphene structure was proposed, which offers a new way for nonmagnetic and dynamically reconfigurable topological nanophotonic devices[699].

## 6.4. Phononic valley-Hall insulator

The simplicity of the inversion symmetry breaking has also enabled phononic analogs of valley-Hall physics. For instance, topological valley edge modes propagate along the interface between two honeycomb sonic crystals whose triangular scatterers have opposite orientations, taking sharp turns with very little backscattering[700–702], **Fig. 20(a)**. Many platforms for sound[703,704,713,705–712], water waves, and mechanical waves[714–723], **Fig. 20(b),** even at the scale of a phononic chip[724–727], have been investigated. The robustness of the edge modes is tightly related to the symmetries of the $\mathcal{P}$-broken lattice, hence providing immunity against defects that do not mix the valleys, allowing for directional antennas[728,729] **(Fig. 20(c)),** topological beam splitters[724] **(Fig. 20(d))** and acoustic delay lines[730] which benefits from the excellent tunability offered by phononic platforms[731]. Valley vortices can also enhance micro-particle manipulation, which could find application in biology or chemistry interfaced with microfluidic systems[732]. Moreover, as phononic analogs of quantum spin and valley-Hall insulators both rely on structural modulations of the medium, the interplay of these two effects results in enhanced wave control, such as valley splitting of helical edge states in patterned elastic plates[733]. Finally, the study of bilayer phononic crystals has shown that adding an extra dimension generates an even richer topological phase diagram, allowing the implementation of layer- and valley-polarization in the same medium[734,735].

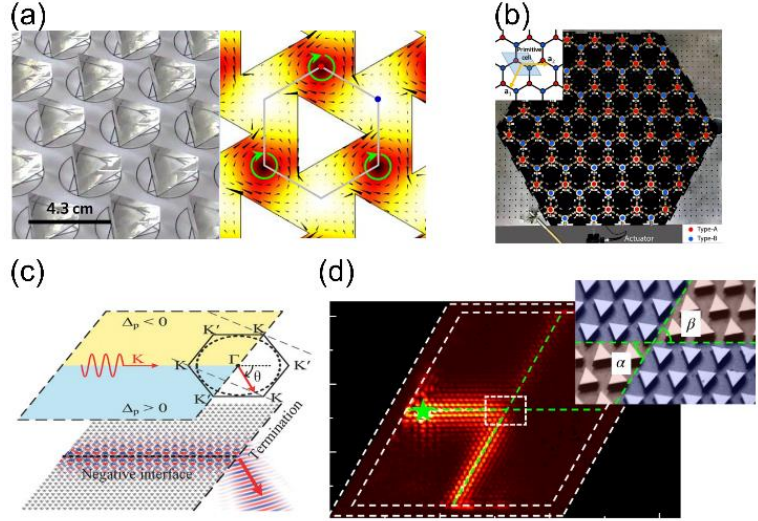

**Figure 20**. **Phononic valley-Hall insulator.** (a) Acoustic valley-Hall insulator made of a triangular lattice of rotated triangular scatterers (left), creating local sonic vortices (right). Reproduced with permission from[701]. Copyright 2017 Springer Nature. (b) Mechanical valley-Hall insulator made of spinners. Reproduced with permission from[722]. Copyright 2018 American Physical Society. (c) Directional acoustic antenna based on sonic valley pseudo-spin. Reproduced with permission from[728]. Copyright 2018 Wiley Online Library. (d) On-chip elastic wave steering at a topological intersection. Reproduced with permission from[724]. Copyright 2018 Springer Nature.

## 7. Topological phases in 3D MMs

In this section, we review 3D topological phases in artificial media. Their features have attracted much attention in the last few years. They have been experimentally realized notably because of their straightforward experimental implementation allowed by modern 3D additive manufacturing techniques. The first part focuses on 3D semimetals in classical MMs englobing media with point, linear and surface degeneracies and their related topological features. The second part presents classical 3D TIs with non-trivial topological surface states.

### 7.1. 3D topological semimetals in photonics

The first photonic Weyl system was realized in double-gyroid structures with inversion symmetry breaking[736,737]. Weyl points belonging to type-I degeneracy were demonstrated in an inversion-breaking double-gyroid photonic crystal at microwave frequency (**Fig. 21(a)**), and optical transmission measurements[738,739], and their extremely broadband topological surface states were also revealed[740]. Multiple Weyl points with topological charges of 2 and 3 were also realized on a printed circuit board (PCB), allowing planar fabrication technology[741]. A theoretical work predicts the simple woodpile photonic crystals also support charge-2 photonic Weyl point[742], verified by the near-infrared experiment[743], **Fig.**

**21(b)**. Type-II Weyl points and the associated Fermi arc-like surface states were observed at optical frequencies in photonic waveguide arrays[744]. Different from the above photonic realization of Weyl degeneracies arising from the spatial symmetries, type-II Weyl points can also be formed by the degeneracies between intrinsic electromagnetic modes of chiral-structured MMs[745,746], and their isofrequency contours were mapped in a microwave experiment[429]. Ideal Weyl points, located at the same energy and separate from other bands, have been realized in a metacrystal with non-centrosymmetric D2d point group symmetry and broken P symmetry, **Fig. 21(c)**. Helicoidal surface states connecting the topologically distinct Weyl points were observed[747]. Moreover, chiral zero-energy modes were observed in inhomogeneous Weyl MM in which a gauge field is generated[748]. Extremely broadband topological surface modes were also observed with double Weyl points in a double-helix photonic structure[740]. The alternative approach to realizing photonic Weyl degeneracies is to break the T symmetry of the system via the magnetized plasma or gyromagnetic response[749]. Photonic Weyl points have also been proposed in other platforms, including 3D network models[750], and even 1D spatial photonic crystal+2D synthetic dimension[751] or 2D spatial structure+1D synthetic dimension[752], and electrical circuits [753]. Interestingly, exotic electromagnetic scattering in Weyl semimetals were disclosed, showing the ability to tailor the strength of wave–matter interactions at arbitrary wavelengths,[754,755] and the helical phase in the scattering matrix was observed in the angle-resolved measurement of a photonic Weyl MM[756].

Dirac point degeneracies may be found in 3D photonic crystals and MMs in which their stability is guaranteed by point group symmetries or duality symmetry[757,758], as experimentally demonstrated in the microwave region[759]. Besides 0D topological degeneracy in 3D momentum space, 1D degeneracies in the band structures may also possess topological properties, and their forms have various configurations, such as nodal rings, chains, links, and knots[433,434,760]. Reconfigurable and flexible MMs have been demonstrated to support the 1D topological degeneracies, and their experimental realization in photonics so far includes the nodal line, nodal chain, and hourglass nodal line[761–763], and the non-Abelian version of nodal link[764]. For example, in **Fig. 21(d)**, the nodal chains[765], having linear band-touching rings chain in momentum space, were mapped in a metallic-mesh photonic crystal, and their drumhead surface states were observed in angle-resolved transmission[762]. Interestingly, the nodal chain and nodal link may exhibit non-Abelian features and have the properties of quaternions[766], and their dispersions were experimentally mapped in bi-anisotropic MMs[767]. Furthermore, 2D topological degeneracies in photonics and MMs were also theoretically investigated[768,769].

### 7.2. 3D topological semimetals in phononics

As mentioned in the previous sections, in acoustic and mechanical artificial media, breaking $\mathcal{P}$ is more accessible than breaking $\mathcal{T}$. Hence, the first proposal for achieving acoustic Weyl points consists of stacking tight-binding honeycomb lattices of cavities with carefully designed inter-layer hoppings to break $\mathcal{P}$ and induce chirality along the out-of-plane axis[770], **Fig. 22(a)**. Another strategy is to use a screw symmetry along the out-of-plane axis, enabling a straightforward sample design[425,771,772]. Hence, Li et al. successfully demonstrated Weyl points in the band structure of a 3D printed chiral 3D MM, measuring the angular transmission of the sample as a function of the angle of incidence. They also directly measured the robust propagation of topological surface states acting as acoustic versions of Fermi arcs[771], **Fig. 22(b)**. Remarkably, He et al. took advantage of the acoustic Fermi arcs' open contours to obtain topological negative refraction[773]. Indeed, carefully designing a *woodpile* sonic crystal, they fully transmitted negatively refracted waves without any reflection at the interface, **Fig. 22(c)**.

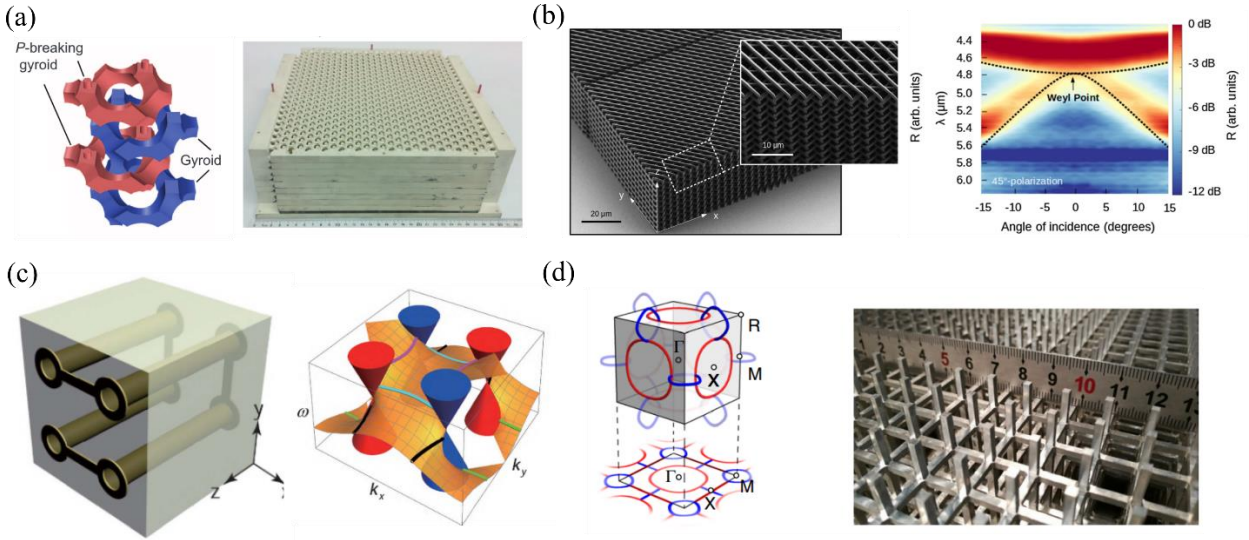

**Figure 21. 3D topological semimetals in photonics.** (a) Weyl points realized on Gyroid photonic crystal with inversion-breaking. The unit cell design, fabrication sample and the angle-resolved transmittance spectrum reveal Weyl points shown from left to right [738]. Copyright 2015 Science. (b) Image of fabricated chiral woodpile photonic crystal and angle-resolved FTIR reflection spectra[743]. Copyright 2020 American Physical Society. (c) Unit cell design of Idea Weyl points, the fabricated sample, and the corresponding helicoidal structure of topological surface states are shown from left to right [747]. Copyright 2018 Science. (d) The structure of nodal chains in the Brillouin zone and the image of the fabricated sample made of Al alloy supporting the nodal chains[762]. Copyright 2018 Springer Nature.

Due to their robustness against small perturbations, some structural anisotropy can tilt the Weyl dispersion without changing its topological charge. The resulting type-II Weyl points become a transition point between ellipsoid and hyperboloid iso-energy surfaces, known as the *Lifshitz* transition. It had been proposed in an acoustic tight-binding lattice of stacked dimerized chains[777] and realized experimentally in 3D printed phononic crystals[774,778], **Fig. 22(d)**. Interestingly, when critically tilted, the Weyl MM behaves as a zero-index medium[779]. Besides, quadratic Weyl points hosting a larger topological charge have also been experimentally characterized in a chirally stacked metacrystal[780].

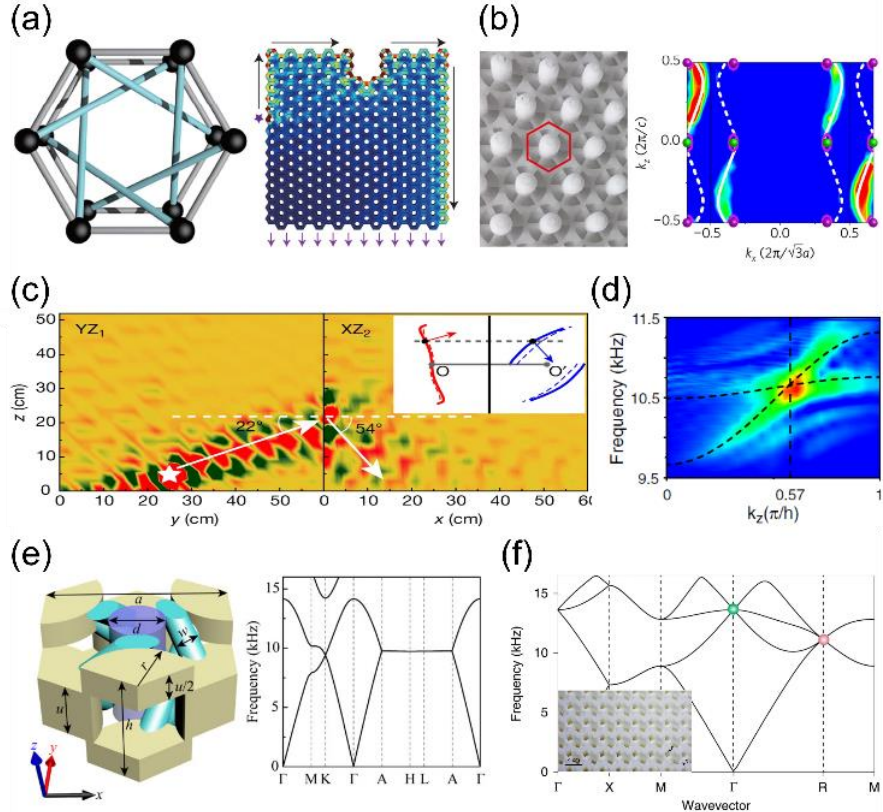

**Figure 22**. **3D topological semimetals in phononics.** (a) Tight-binding Proposal for an acoustic Weyl semimetal with chiral interlayer couplings (left) generating topologically protected surface states (right). Reproduced with permission from[770]. Copyright 2015 Springer Nature. (b) Experimental realization of an acoustic Weyl semimetal with screw symmetry (left) creating acoustic versions of Fermi arcs in the reciprocal space (right). Reproduced with permission from[771]. Copyright 2018 Springer Nature. (c) Topological negative refraction at an acoustic Weyl semimetal surface using the one-way robust surface waves. Reproduced with permission from[773]. Copyright 2018 Springer Nature. (d) Measurement of a tilted type-II Weyl point in an artificial sonic medium. Reproduced with permission from[774]. Copyright 2020 American Physical Society. (e) Carefully engineered glide couplings in an acoustic crystal (left) lead to a nodal surface in the band structure (right). Reproduced with permission from[775]. Copyright 2020 American Association for the Advancement of Science. (f) Triply-degenerate point in a non-symmorphic 3D phononic crystal. Reproduced with permission from[776]. Copyright 2019 Springer Nature.

Another intriguing feature of Weyl modes linked to their chirality occurs in the presence of a magnetic field. Indeed, the resulting $0^{\text{th}}$-order Landau level is not flat but has a linear dispersion determined by the chirality of the Weyl point and the direction of the magnetic field. Starting from a Weyl sonic crystal and then carefully designing a sub-lattice potential along the chiral and $P$-broken direction, Peri et al. induced an effective axial field which resulted in sonic chiral Landau levels[781].

Multiple point-degeneracies also exist in 3D phononic media, which can carry larger topological charges related to multiple surface states[155,776,782], **Fig. 22(e,f)**. On the contrary, 3D Dirac cones are not related to one-way Fermi arcs but can generate helicoidal surface states characterized by a pseudo-spin Chern number[783,784]. They can also be engineered to obtain a 3D zero-index material[785] and serve as a starting point to induce Weyl features[786]. In addition to point degeneracies, 3D media allows a large family of curve-touching features. Hence, nodal lines, rings, or chains have been demonstrated in sonic systems[787–791]. They carry bulk topological features resulting in so-called drumhead surface states with a flat dispersion allowing strong field localization and good frequency stability against perturbation. In the case of a straight line, it transforms into a strongly anisotropic waterslide surface state[790]. Finally, acoustic nodal surfaces obtained through non-symmorphic symmetries can also carry a topological charge and 1-way surface arcs[775,792]. Although it is more challenging to obtain similar results in mechanical systems because of uncontrolled mode conversion, elastic versions of Weyl physics have been evidenced[793–795]. Nodal lines have also been identified in mechanical metacrystals[796] and building alternatives such as self-assembly protocols have been proposed[797].

### 7.3. Photonic 3D TIs

Since 2D TIs can be further extended to 3D TI, in which bulk bands are fully gapped in the whole 3D BZ with gapless topological surface states populated, tremendous efforts in the MM communities have been investigated to implement 3D topological phases by breaking T or introducing pseudo-spins in 3D artificial media by controlling their spatial structure.

The first 3D photonic topological phase with broken TRS was built in the form of a nonsymmorphic photonic crystal, with the cubic unit cell containing four gyroelectric pillars oriented along different lattice vectors, **Fig. 23(a)**. The magnetic fields were applied to the rods with alternative signs. As a result, Dirac cone degeneracies in BZ were lifted, **Fig. 23(b)**. The resulting band structures host a single surface Dirac cone protected by glide-reflection symmetries[798]. The quantum Hall phase is characterized by a 3D Chern vector. Recent work shows that the Chern vector's value and direction are tunable by changing magnetic fields applied to the 3D cubic photonic crystal[799].

Since the proposal of achieving topological surface modes with broken TRS requires an alternative magnetic basis on different pillars, which causes a tremendous challenge for the experimental implementation of it, researchers are looking for the other approach with TRS and achieving analog QSHE in 3D photonics. The protocol for finding 3D topological photonics is similar to the 2D case. First, the material and geometry parameters of the dielectric disk are optimized to overlap the two 3D Dirac cones in a frequency range, and no other modes for any momentum vector in the Brillouin zone are present in this frequency range. Second, the bi-anisotropic response in the material is introduced by the partial removal of the disk in such a way that the mirror symmetry in the axial direction $\sigma_z$ is broken. Subsequently, the modes at Dirac cones are mixed and open up a complete bandgap in the 3D Brillouin zone. An effective Hamiltonian near Dirac points is found based on electromagnetic perturbation theory and first-principle simulation[800]

$$\hat{\mathcal{H}} = v_\perp \hat{s}_0 \big( \delta k_x \hat{\sigma}_x + \delta k_y \hat{\sigma}_y \big) + v_{||} \hat{s}_y \hat{\sigma}_z \delta k_z + m \hat{s}_z \hat{\sigma}_z, \tag{23}$$

where $\hat{s}_i$ and $\hat{\sigma}_i$ are Pauli matrices operating in the subspaces of pseudo spin and orbital angular momentum of photonic modes, respectively, $m$ is the effective mass term induced by the bianisotropy, and $v_{||(\perp)}$ is the out-of-plane (in-plane) Dirac velocity. The effective photonic Hamiltonian is equivalent to a family of condensed matter Hamiltonians describing 3D TI[801]. One of the most important consequences of nontrivial topological invariance is the emergence of gapless surface states. The large-scale full-vector numerical simulations were carried out on the topological domain wall, representing an interface between two topological crystals inverted with respect to each other along the z-direction (left panel of **Fig. 23(c)**), which effectively results in the reversal of bianisotropy. The 3D band diagram of the topological structure shown in the right panel of **Fig. 23(c)** exposes one of the Dirac cones corresponding to the surface states near the K-point. Following this theoretical work, the experimental realization of the weak type 3D photonic TI was achieved[792]. The split-ring resonators are used to induce the strong magneto-electric coupling, and the conical Dirac-like dispersion of surface modes was observed from near-field measurement, **Fig. 23(d)**. A similar work demonstrated that the 3D tetragonal photonic crystal with anisotropic response supports surface modes with quadratic band degeneracy at high symmetry points in BZ[802]. Very recently, a 3D photonic crystal without spin-orbit coupling was proposed and experimentally probed. It emulates the model of the crystalline TI proposed by Fu[803], and the emerging topological surface states are self-guided on its surface[804].

## 7.4. Phononic 3D TI

A few examples of phononic 3D TIs have been demonstrated. For example, a stacked double-layer honeycomb lattice with tailored geometric parameters exhibits a bandgap populated with pseudo-spin-valley coupled saddle surface states[806]. Starting from a 3D hybrid Dirac cone and tuning the inter- and intra-unit cell couplings opens a bandgap linked to a topological band inversion[807]. An interface between two different topological phases hosts a single nearly gap-less conical-like dispersion for acoustic surface states. Finally, another folding band scheme involving Weyl points with opposite chirality allows for generating a similar phase[808]. The corresponding surface states exhibit pseudo-spin locking and can be used in the topological surface-wave splitter.

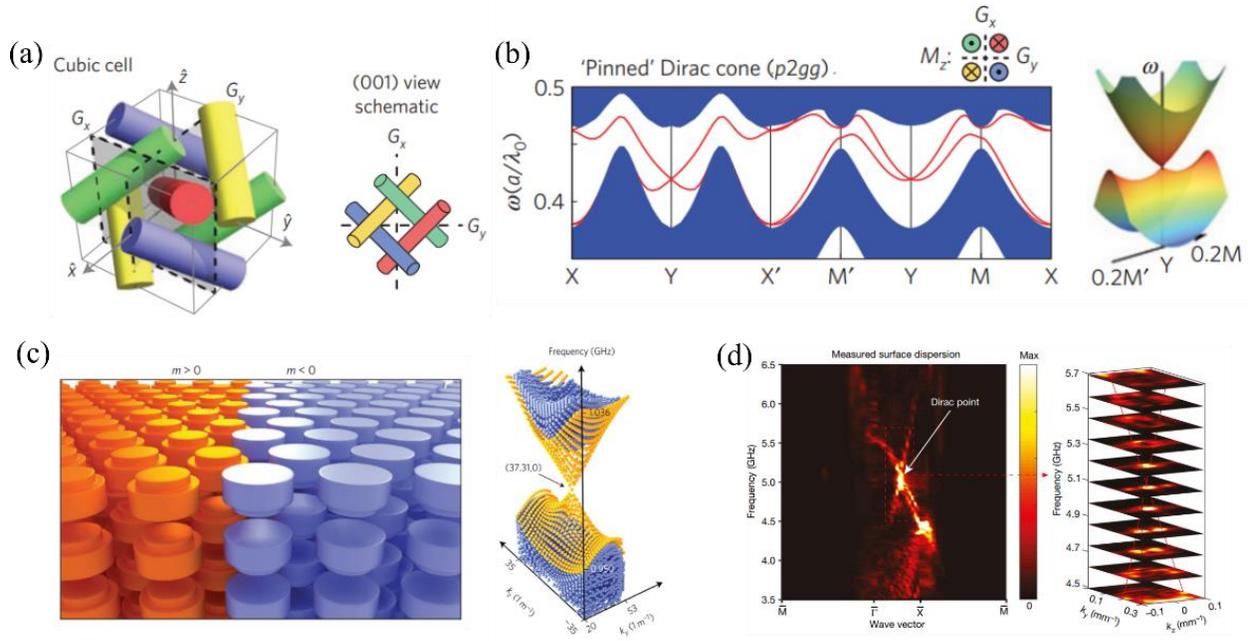

**Figure 23. 3D gapped topological photonics.** (a) Unit cell of nonsymmorphic photonic crystals in which TRS is broken. (b) Bulk and surface band structures of photonic crystals with a meticulous manipulation of the gyroelectric responses of four pillars[798]. Copyright 2016 Springer Nature. (c) Weak type photonic TIs and the according surface Dirac cones[800]. Copyright 2017 Springer Nature. (d) Surface Dirac cone dispersions were mapped by microwave experiment[805]. Copyright 2019 Springer Nature.

## 8. Higher-order topological phases in MMs

As presented in the previous sections, conventional topological phases are media with a gapped bulk and gapless boundary or surface states whose existence is dictated by topological invariants and show robustness against defects preserving relevant symmetries. A new type of topological phase has been introduced in condensed matters named higher-order TIs (HOTIs)[369,448–453,457]. Motivated by the search for higher-order topological phases, the 2D quadrupole topological phases with robust 0D corner states were recently observed using acoustic[461,462], microwave[455,456], and phononic MMs[454,465]. Localized corner 0D states demonstrate high robustness against certain fabrication disorders ubiquitous in MMs. This topological stability and robustness make them very interesting for applications. Namely, directional localization of light in 0D states provides the possibility of topological optical switches, energy dividers, topological lasers, and other devices[34,35,809,810].

This section reviews the classical versions of these unconventional topological phases. In the first part, we focus on 2D examples and corner states. In a 2D higher-order TI, 0D corner states exist in the middle of the boundary modes gap. The experimental realization of second-order TIs is classified into two systems: quantized multipole moment insulators and higher dimensional generalization of the SSH model. We extend it to their 3D and even higher dimensions counterparts in classical platforms in the second part.

## 8.1. Quadrupole TIs in 2D

The first class describes lattices without bulk dipole moment but quantized fractional multiple moments acting as topological invariants. In a 2D system, the topological invariant is called quadrupole moment, and the system is referred to as quadrupole TIs (QTIs)[451369]. To obtain this phase in a realistic system, one has to implement a tight-binding model with positive and negative couplings designed to get a $\pi$-flux per plaquette (area of the unit cell). Although it appears challenging in photonics, researchers realized it in microwave experiments and identified the second-order corner states with spectroscopic measurements[455]. By carefully designing the links' shape between neighboring cells, they realized a microwave circuit with a quantized quadrupole moment $q_{xy} = \frac{1}{2}$, **Fig. 24(a)**. The evidence of topological corner modes is provided by the measured absorptance spectrum in the right panel of **Fig. 24(a)**, they possess certain robustness and are always localized at the corners although the edge lattice is deformed. Photonic QTIs have also been successfully realized in other platforms including CROWs[465], plasmonic MMs[811], and electrical circuits[456,812], in which the hopping phases can be tuned easily. For example, the link rings of CROWs denoted by red color in **Fig. 24(b)** are placed slightly off from the center between the two site rings such that a synthetic flux $\pi$ is introduced in the unit cell. Furthermore, recent works show that the quadrupole topological phase can be realized without flux-threading, e.g., via magneto-optic effect[813] or twisting angle in a lateral heterostructure[814].

In phononics, Serra-Garcia et al. successfully realized the QTIs in a phononic plate made of a square lattice of quadrupolar resonators[454]. Using a laser-vibrometer and a shaker as a source, they could directly measure the field localization at the corners of the sample, **Fig. 25(a)**. For acoustic waves, this topological phase has been induced in tight-binding lattices of cavities using their dipolar resonance mode[815]. Indeed, using straight and bent coupling tubes between cavities, one can respectively mimic positive and negative couplings, which leads to acoustic corner modes with a $\frac{1}{2}$ topological quadrupole charge when implemented in a square lattice, **Fig. 25(b).** Such quantized multipolar moments have also been found in non-symmorphic $p4g$ lattices, where the non-commutative glide symmetries generate the corresponding topological charges[816,817]. Such anomalous topological quadrupole insulators open the door to practical samples beyond tight-binding systems.

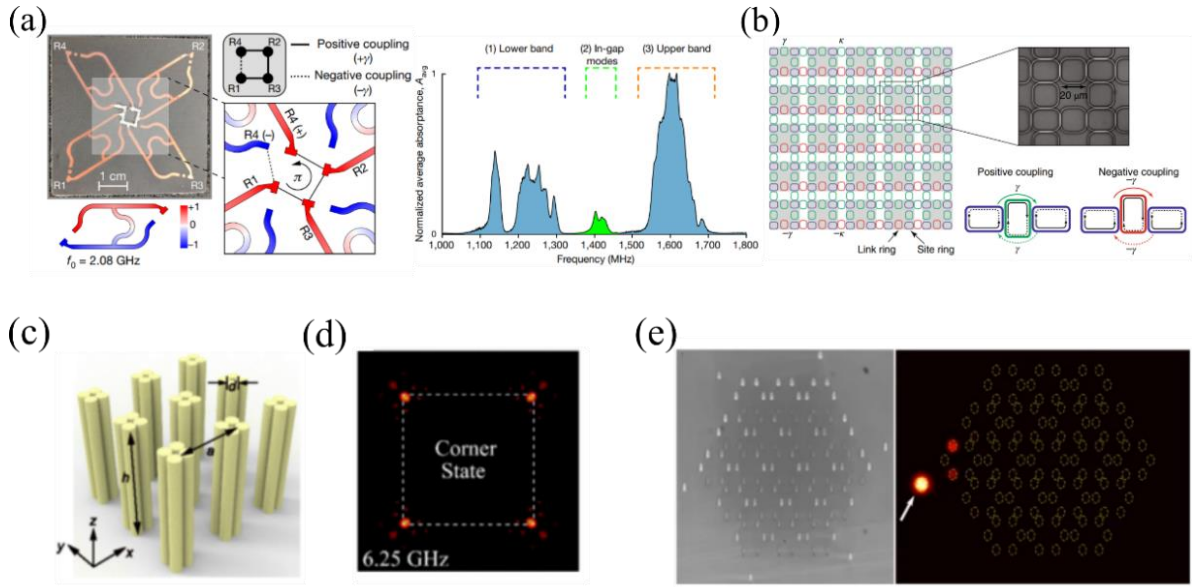

**Figure 24**. **Higher-order topological phases in photonic MMs.** (a) Quadrupole lattice made of a coupled-resonator has a synthetic magnetic flux by introducing a negative coupling, the measured power absorptance spectrum of all the resonators in the array for bulk modes (cyan color) and corner modes (green color) measured from the microwave experiment[455]. Copyright 2018 Springer Nature. (b) Schematic of the 2D lattice of ring resonators in which a gauge flux is introduced via the red link rings[465]. (c) Schematic of square photonic crystal slab representing 2D SSH model[464]. Copyright 2019 American Physical Society. (d)Near field measurement of corner modes in the square lattice at microwave frequency[463]. Copyright 2019 American Physical Society. (e) Microscope image of the waveguide array (left panel) supporting the zero corner mode shown in the right panel[459]. Copyright 2018 Springer Nature.

## 8.2. 2D SSH model

The second class of 2D higher-order TIs relies on the 2D generalization of the SSH model[452]. It differs from the previous phase because it shows no multipole moment but bulk polarizations linked to the system space-group symmetries[462]. In this case, negative and positive couplings are unnecessary as the non-trivial topology is induced through inequivalent inter and intra-unit cell couplings. This allows for a more straightforward design that applies to many photonic systems, including square lattices and Kagome lattices[452,458]. The second-order topological phases with non-zero bulk-polarization have been extensively examined and realized in coupled waveguide arrays[466], photonic crystal slabs[355,463,541,818–820], **Fig. 24(c-d)**, surface wave photonics[821], plasmonic lattices[822], and circuit board[823]. In addition, new type corner modes were disclosed in the far-neighbor interactions of electromagnetic waves[467,824], and their spatial profiles were imaged by near-field measurement at optical frequency[825,826]. Different type topological corner modes with zero bulk polarization and protected by mirror symmetries were revealed in the "hexamer" photonic crystals[459,827], **Fig. 24(e)**, and analog higher-order quantum spin Hall effects were proposed based on this platform[828]. In the active device-oriented direction, the nonlinear effects were investigated in the second-order corner modes[468,829,830], and topological lasing on corner modes was also realized[831]. When the energies of the topological modes do not lie within the bulk bandgap, which might occur for the corner states in the 2D lattice, fractional charge density arising from filled bulk bands is the key indicator to identifying the higher-order topology of the crystalline insulator[832]. The indicator of higher-order topology in terms of mode density was measured in the circuit MMs[833] and was further employed to probe the topology of disclination defects in photonic crystalline insulator[834,835]. Moreover, the embedded features of the corner modes with the bulk continuum enable a new avenue to achieve optical bound states[836].

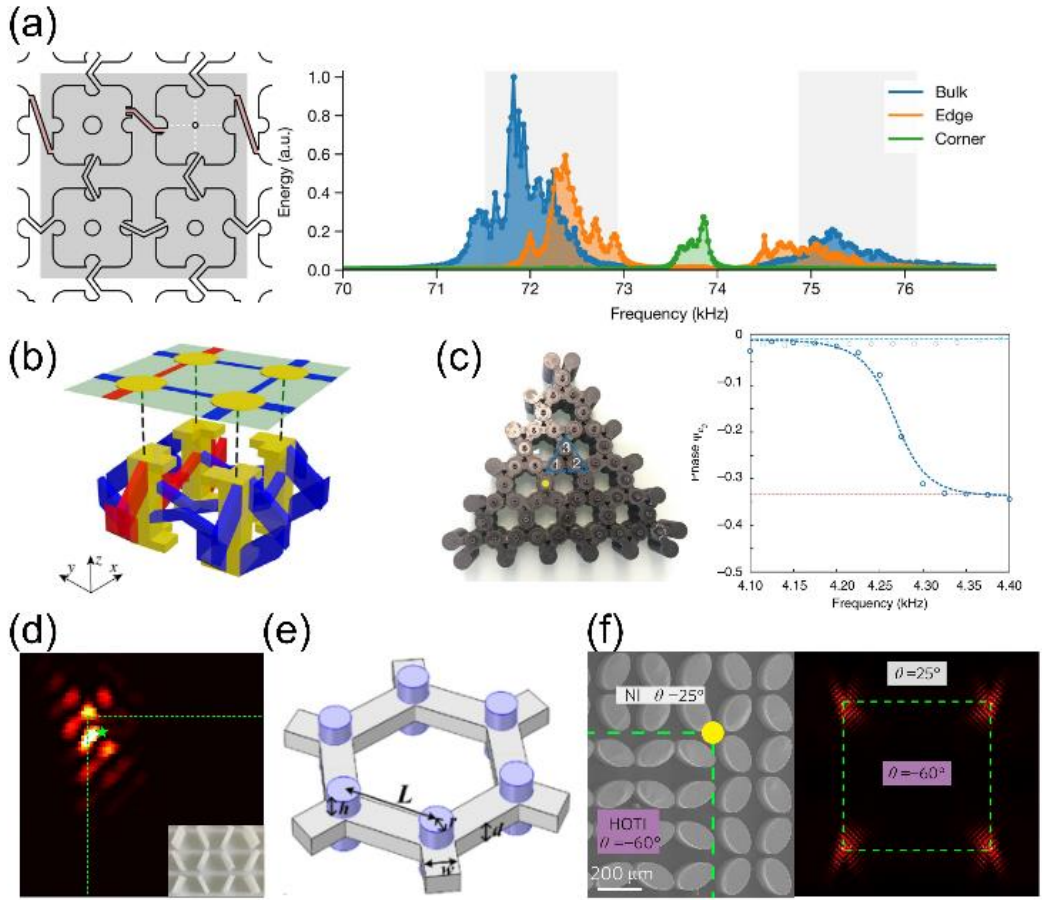

**Figure 25**. **Higher-order topological phases in phononic MMs.** (a) Carefully designed couplings within an elastic plate MM (left) generate a higher-order quadrupole insulator with topological corner modes in its bandgap (right). Reproduced with permission from[454]. Copyright 2018 Springer Nature. (b) Unit cell of an acoustic higher-order quadrupole insulator. Reproduced with permission from[815]. Copyright 2020 American Physical Society. (c) Breathing Kagome tight-binding acoustic lattice as a 2D generalization of the SSH model along with the bulk polarization measurement. Reproduced with permission from[462]. Copyright 2019 Springer Nature. (d) Topological corner mode in an acoustic crystal with rotated blocks. Reproduced with permission from[460]. Copyright 2019 Springer Nature. (e) Mechanical honeycomb lattice as a 3D

generalization of the SSH model. Reproduced with permission from[847]. Copyright 2019 American Physical Society. (f) On-chip proposal of a mechanical crystal based on rotated blocks (left) and corresponding corner modes (right). Reproduced with permission from[853]. Copyright 2021 Elsevier.

Such higher-order phases also emerge in phononics. For example, it was demonstrated in an acoustic breathing Kagome lattice made of tight-bonded cavities[461,462], **Fig. 25(c)**. Increasing the ratio between inter- and intra-cell couplings induces a bulk polarization $(p_1, p_2) = (-\frac{1}{3}, -\frac{1}{3})$, which results in topological corner modes, evidenced thanks to in-situ pressure field measurement in each cavities of the sample. The bulk polarization transition was also experimentally revealed by measuring the quantity related to the rotational symmetry of the lattice, **Fig. 25(c)**. This approach generalizes to other symmetries such as square lattices which have been implemented in tight-binding lattices[837,838] and at the subwavelength scale for surface waves in locally resonant MMs[821,839]. Due to long-range interactions, the corner modes are no longer pinned to the bandgap's center. This chiral symmetry breaking can also happen in coupled cavities media[840]. In extreme configurations, edge and corner modes might fall within the bulk bands, becoming bound states in the continuum[841]. These generalized SSH lattices can be combined with pseudo-spins feature providing an additional DoF to corner modes[842–844].

Besides, these multidimensional topological transitions also occur in sonic crystals whose unit cell comprises four blocks[460,845]. Adjusting their mutual orientation allows reconfiguring the medium from conventional pseudo-spin TIs to higher-order phases, as shown in **Fig. 25(d)**. Topological corner states behavior has also been studied in the context of anomalous Floquet physics[846]. The straightforward implementation of these concepts has been translated to mechanical media made of tight-binding lattices[847,848] (**Fig. 25(e)**) and thin plates decorated with pillars[849–852]. An on-chip implementation with micro-pillars has also been proposed[853], **Fig. 25(f)**.

### 8.3. Higher-order topological phases in 3D

Generally, increasing the dimension of a system gives rise to even richer topological phases. Hence 3D higher-order TIs[369,453] and topological semimetals[854] not only host corner modes (0D) but also hinge states (1D) and surface states (2D). The 3D higher-order topological phase implementations in photonics have been limited in electrical circuits because of the complexity of the sample fabrication[855–858]. In addition, bounded states at partial dislocations of the 3D higher-order TIs were demonstrated in the circuit experiment[859]. The inherent flexibility and tunability of the circuit allow achieving the dimensions suppressing the physical limitation, e.g., the recent circuit realization of a 4D hexadecapole insulator[860].

Phononic higher-order topological phases relying on quantum multipole moments have been demonstrated in 3D tight-binding lattices of cavities in bent tubes to mimic negative couplings[861,862]. The resulting octupole moment is related to gapped surface and hinges modes with eight corner modes measured in *situ*, **Fig. 26(a)**. 3D generalization of the SSH model possessing nontrivial bulk polarization has also been demonstrated for sound. A 3D polyhedral lattice of tightly coupled cavities with an inter-cell intra-cell coupling ratio larger than one also presents the dimensional hierarchy of surface, hinges, and corner modes[863], **Fig. 26(b)**. Corner modes also exist within a rhombohedral-like anisotropic 3D lattice of cavities[864], and the hierarchy of multidimensional modes has been evidenced in cubic lattices[865–867].

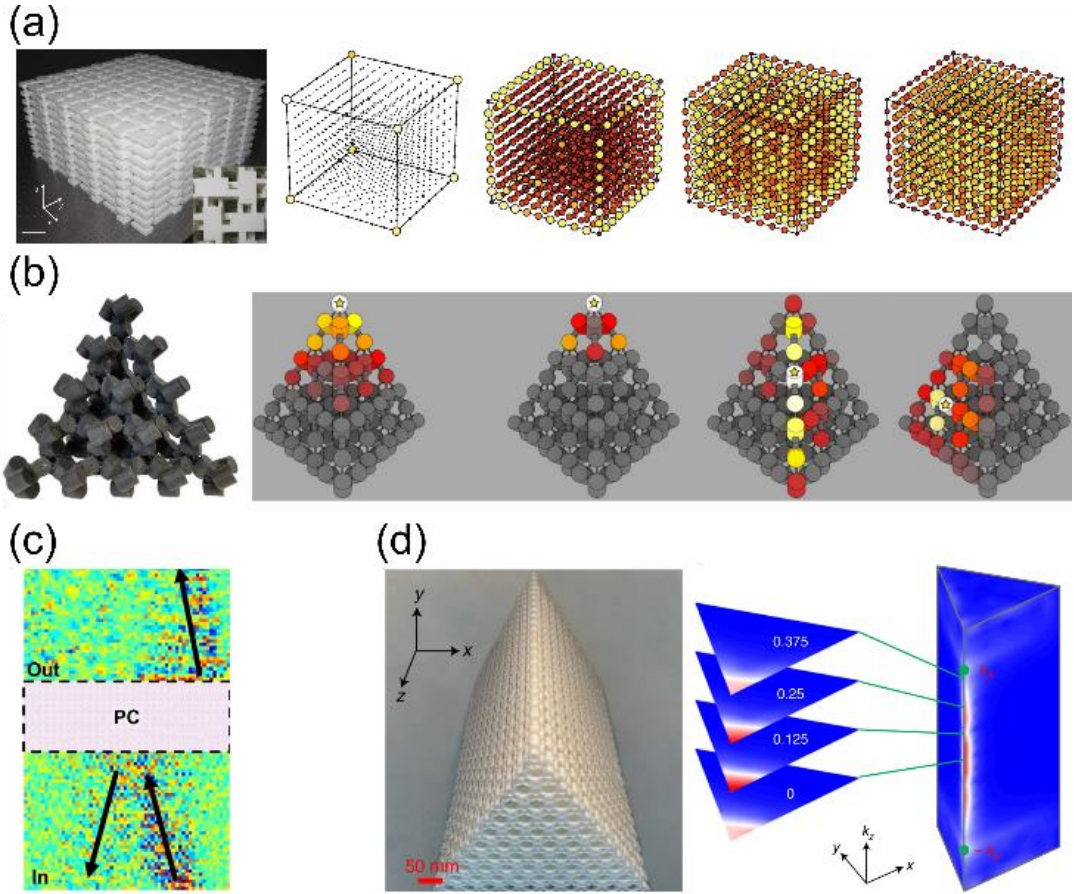

**Figure 26**. **Higher-order topological phases in 3D.** (a) Acoustic quantized octupole higher-order insulator (left) and the corresponding dimensional hierarchy of its lower-dimensional states (right). Reproduced with permission from[862]. Copyright 2020 Springer Nature. (b) 3D generalization of the SSH model in the case of a sonic crystal polyhedron (left) and the corresponding hierarchy of its lower-dimensional states (right). Reproduced with permission from[863]. Copyright 2020 American Association for the Advancement of Science. (c) Negative refraction of sound using the surface mode of a higher-order TI. Reproduced with permission from[868]. Copyright 2021 Elsevier. (d) Acoustic Weyl semimetal (left) presenting lower-dimensional hinge-arc states (right). Reproduced with permission from[869]. Copyright 2021 Springer Nature.

Overall, these peculiar 3D topological phases present great versatility. A unique medium can act as a cavity, a waveguide, or a surface wave medium simply by changing the working frequency. Using the several multidimensional low dimension states of a single 3D sonic crystal with non-symmorphic glide symmetries, Xu et al. demonstrated 2D negative refraction mediated by a surface state and a 3D acoustic interferometer using hinge states[868], **Fig. 26(c).** 3D waveguides using hinge states have also been evidenced experimentally[870].

Quite remarkably, higher-order and conventional topology coexist in the same medium[869,871–874]. Weyl acoustic crystals supporting topologically protected Fermi arcs and higher-order arc hinge states have been evidenced in stacks of Kagome breathing lattices with double helix interlayer[869] couplings, as shown in **Fig. 26(d)**, and chiral sonic crystal with a uniaxial screw symmetry[871]. The challenges linked to mode conversion and 3D printing hinder the transposition of these concepts in mechanical devices for now.

## 9. Other forms of topological MMs

In the above discussions, we have reviewed topological states in Hermitian (lossless) and linear systems with periodicity in space or temporal modulation. However, topological phases can occur in other systems like non-Hermitian materials, synthetic space, nonlinear systems, or non-periodic/quasiperiodic structures. In this section, we list a few but not exclusive topics studying other types of topological states appearing in MMs and briefly review essential works relevant to these topics.

### 9.1. Non-Hermitian topological MMs

The concept of non-Hermiticity has been introduced in TIs and extensively studied in recent years, leading to the richer topological phases beyond the description of Bloch band theory and the generalization of biorthogonal bulk-boundary correspondence of non-Hermitian systems[875–883]. One of the fascinating theoretical questions being explored in the recent non-Hermitian literature is what effect non-Hermiticity, and the PT-symmetric gain/loss, in particular, has on topological phases of matters. To answer this question, the simplest example of a 1D non-Hermitian topological model was studied in an early work, in which the passive onsite potentials are introduced in terms of the decaying of quantum walk[884]. If the imaginary onsite potentials with alternative signs are added in the two sublattices of the SSH model, the PT-symmetric lattice or even the PT-symmetric interface is formed by two PT-symmetric lattices which are PT-symmetric with each other. Such models have been shown to support non-Hermitian variants of the SSH midgap states[497] and "anomalous" edge states that are intrinsically non-Hermitian[885]. Furthermore, suppose the nonreciprocal couplings are introduced in the model. In that case, the so-call non-Hermitian skin effects occur, in which the properties of the bulk eigenstates are extremely sensitive to the boundary conditions, and bulk states might not extend over the lattice but localize at either end of the lattice with open boundaries, thus leading to the breakdown of bulk-boundary correspondence[886]. To faithfully predict the appearance of topological edge modes, non-Bloch band theory or notion of biorthogonal quantum mechanics or auxiliary generalized Brillouin zone was established for the non-Hermitian systems with open boundaries[876,878,880,882].

Distinct from the condensed matter counterparts, photonic systems are intrinsic non-Hermitian since the material absorption and radiation losses are unavoidable. Consequently, the non-Hermitian models can be simply implemented in photonics. For instance, the non-Hermitian behavior was engineered in evanescently coupled photonic waveguides by spatially wiggling the waveguides[498]. Some intriguing connections between PT symmetry and band topology have been discovered in PT-symmetric photonics. For example, the topological interface states at the PT-symmetric interface were experimentally demonstrated in the passive variant version of PT-symmetric coupled waveguide lattices[499] and 1D photonic crystal made of dielectric-resonator supporting the defect modes[486]. Furthermore, the interaction between the nonlinearity and the non-Hermitian physics was also explored in the PT-symmetric SSH model[887]. In parallel, the non-Hermitian skin effects on the topological phases have been investigated in 1D coupled ring resonators[888] and experimentally verified in electrical circuits[889] and via coupled optical fiber loops[890] in which a highly efficient funnel for light was also demonstrated.

In a 2D topological system, whether the topological chiral/helical edge propagation remains robust in the presence of the non-Hermiticity has attracted enormous research efforts from condense matters and photonics communities[875,891,892]. For example, the topological stability of edge states in 2D NH systems has been analyzed[893]. Some works have sought to formulate topological invariants for NH models[894,895], including a definition of bulk topological invariants for NH Chern-like TIs. Non-Hermitian skin effects were also introduced in the quantum Hall insulator, and the non-Bloch framework was established[877]. For 2D non-Hermitian topological photonics, lossless edge states preserving the PT symmetry phase are located at PT symmetric interfaces irrespective of the cut shape. The bulk-interface correspondence principle is still valid as long as the topological gap of the complex bulk spectrum remains open[896]. Interestingly, PT-symmetric photonic waveguide lattices with judiciously designed refractive index landscape and alternating loss were fabricated, and non-Hermitian topological phase transition was demonstrated[897]. Moreover, robust light steering was achieved in reconfigurable non-Hermitian CROWs via a spatial light modulator[651].

Phononic systems also provide a great platform to genuinely control the loss and gain contribution in a system, leading to the new experimental realization of non-Hermitian topological phenomena. Tailoring the loss distribution within acoustic lattices, it is possible to induce topological boundary modes [898,899]. Adding gain in the picture and combining it with the valley-Hall effect generates topological edge states with enhanced amplitude[900]. Remarkably, Hu et al. showed topological "audio lasing" of whispering gallery modes with a specific handedness using the electro-thermoacoustic coupling in carbon nanotube films[901].

Another technique based on circuitry and feedback loops allows controlling the gain within phononic lattices[902–904].

Besides, other intriguing phenomena such as the non-Hermitian skin-effect[903–907] (NHSE) have also been investigated in phononic systems, or non-Hermitian higher-order topological phases are predicted and experimentally verified in the acoustics[899,906,908–914], opening new avenues for topological phononic devices with peculiar non-Hermitian features. The NHSE effect is possible in non-Hermitian 1D or 2D systems and consists of the localization of many modes at system edges and boundaries[876,881,890,915–920]. In non-Hermitian passive systems, the NHSE effect is hidden in the complex frequency plane and can be revealed with gain media or excitations with complex signals. NHSE profoundly impacts band topology as it reflects a novel point-gap topology unique to non-Hermitian systems and leads to a breakdown of the conventional bulk-boundary correspondence[881,886,906,921,922]. Due to its unconventional physical properties, NHSE has also been useful in various applications, including wave funneling[890], enhanced sensing[923,924], and promising for topological lasing[925].

## 9.2. Synthetic TIs

Topological photonics has excellent potential for being realized in microscale devices, i.e., photonic integrated circuits. However, typical topological photonic systems usually require complex or bulky elements, thereby facing tremendous challenges for experimental realization in a miniaturized chip, hindering their applications in integrated photonics. Motivated by these factors, researchers have proposed an alternative approach by adding the synthetic dimensions to low dimensional systems to observe the topological phenomena, thus enabling demonstration on a relatively small scale and access to the higher dimensional topology[96,751,934,935,926–933]. The approach utilizes versatile DoFs of photons or tunable parameters to expand the additional dimensions and construct the gauge fields in *synthetic space*.

The DoFs exploited in the literature include, for instance, longitudinal modes supported by cavities, mode profiles in waveguide arrays, or spin/orbit angular momentums in optical beams. 2D TIs are realized by placing the electric-optic modulators (EOMs) on 1D coupled resonators[928,936] or exploiting the spin, and orbital angular momentum of photons in a 0D cavity[937], or controlling the acoustic pumps and EOMs introduced to an optomechanical ring resonator[938]. Synthetic dimensions also offer the ability to realize 3D Weyl points in 2D cavity arrays[929] and obtain analog 4D QHEs in a 3D lattice composed of coupled resonators[939]. Recent experimental work demonstrated the topological physics of a Hall ladder based on a ring resonator, and it paves a new way for possible implementation on the integrated platform[934].

Alternatively, the synthetic dimensions can be produced using the periodically modulated parameters that act as the synthetic dimension's momentum variables. For example, the 1D topological pump illustrated above is equivalent to realizing the 2D Harper model with the second dimension represented by the phason parameter[527]. The photonic Weyl points were attained in a specially designed four-layer photonic crystal with varying thickness of the layers as the synthetic parameter[751] and in 2D semiconductors in which the magnetic bias plays the role of the third dimension[752]. In addition, the ideal type-II Weyl point was proposed in 1D twisted photonic crystals[940]. The synthetic parameter dimension also allows exploring higher-dimensional topological physics, such as the 4D QHEs[221,386], which cannot be directly implemented due to physical limitations. In contrast, they have been implemented in propagating photonic waveguides and other platforms[931,932].

If harnessing the additional DoFs of phonons to implement synthetic dimensions is still challenging experimentally, this is not the case with parametric synthetic dimensions. Indeed, as discussed above, 1D quasi-periodic lattices acting as topological pumps can already be described as the 2D system with the phason as an extra synthetic dimension. Remarkably, the phason's features can also be extended to higher spatial dimensions allowing for the mapping of the 4D quantum- Hall effect[941,942] and higher-dimensional topological wave transport[943] in phononics. Moreover, phononic 3D topological phases often result in bulky and lossy samples, severely hindering the desired effects. It is possible to circumvent these issues using parametric synthetic dimensions, as evidenced by the demonstration of acoustic Weyl points[944–947] and 3D TIs[948] in synthetic dimensions. Finally, this parametric synthetic dimension principle also allows for obtaining even greater multidimensionality and versatility in higher-order topological phases[949,950].

### 9.3. Nonlinear topological MMs

Nonlinear effects in photonics naturally arise when the optical power intensity is relatively high, and how the topological concepts and nonlinearity in photonics intertwine with each other remains elusive. However, the practical implementations of nonlinear phenomena and applications have been realized in many topological photonic systems[95]. For example, topological solitons and soliton-like edge states were predicted and observed in photonic waveguides[951–954] and polariton lattices[502]. Topological harmonic generations in 1D topological MMs[955,956] and 2D topological photonic MS were experimentally demonstrated[957]. Self-induced topological edge states were disclosed in nonlinear circuit arrays[958]. Nonreciprocal topological edge states were realized in nonlinear photonic crystals through external driving fields[959,960]. Also, nonlinearity-induced topological corner modes were proposed and investigated in various photonic systems[829,830,961]. Interestingly, some topological phenomena in nonlinear optics have no equivalent counterparts in condense matters and have been observed in photonics. For example, a topologically protected traveling-wave parametric amplifier was demonstrated by the squeezing of light[962], and topologically protected four-wave mixing was also demonstrated in a patterned graphene plasmonic MS[963]. In particular, topological lasers with topological protection were achieved in 0D cavity[831], 1D arrays[64,65] and 2D lattices[33–35,698]. In the quantum regime, Laughlin states made of light were created using a gas of strongly interacting, lowest-Landau-level polaritons[964].

In phonics, especially in mechanical waves, non-linear phenomena arise frequently and are prone to alter wave propagation dramatically. In particular, the impact of nonlinearities on the topological modes' robustness has been investigated in 1D lattices[965–967]. On the other hand, it is possible to harness non-linear effects to induce non-trivial topological phases within a phononic media[968,969], hence getting on-demand phononic waveguides or cavities controlled by the intensity of the feeding signal.

## 9.4. Disordered topological phases in MMs

Theoretical works suggest the emergence of protected edge states and quantized topological transport can be induced by the addition of sufficient disorder in the trivial insulator[970–972]. A disordered induced TI is referred to as Anderson TI and was experimentally demonstrated in photonic waveguides by adding the random waveguides[973,974] and in a disorder gyromagnetic photonic crystal[975]. The competition between Anderson localization, topological protection, and non-Hermitian skin effects was explored in the disordered photonic MMs[976] and quantum walks recently[977].

The role of disorder in topological phases has recently been investigated in phononics[978]. For instance, by controlling the level of disorder within a 1D phononic chain, the system can be driven from a topologically trivial phase to a 1D topological Anderson insulator[978,979] and used directly as a filtering device[979]. Topological corner modes can also appear within aperiodic lattices[980]. An amorphous mechanical lattice made of gyroscopes also presents a topological phase linked to dramatic non-reciprocal propagation at the edge of the medium[618]. Finally, phononic topological defects can be used as robust waveguides[981,982] or topological cavities[983–986], and spatial Kekulé distortions can create cavity modes showing strong robustness against spatial perturbations[987–989], expanding the phononic topological phases beyond the periodicity requirements.

## 9.5. Mechanical zero-frequency topological phases

Topological mechanical MMs can be organized into two categories: high frequency and zero-frequency phases[4,39]. Till this point, we have always referred to the former family of systems. In the following, we focus on zero-energy topological modes existing in a gap at 0 frequency. Kane and Lubensky first introduced the topological description of these mechanical media by drawing parallels with particle-hole symmetries in isostatic lattices[41]. Their study of so-called Maxwell frames permits introducing a winding number directly related to the number of localized modes at the medium's edges. Depending on the winding number sign, the latter takes two forms: floppy modes or self-stress states analogs of particles and holes in electronic TIs[990]. While the former are localized free motion of the system, the latter consists of intrinsic stresses occurring without outside forces. The topological nature of these zero energy states protects them against deformations as long as the system remains in the same topological phase, offering a promising way to design mechanical devices with robust properties. Topological floppy modes have been measured in a 1D chain of rotors mimicking the SSH model[991,992]. Notably, using simple LEGO parts and non-linear excitations, Chen et al. managed to propagate these modes through the bulk, transforming the system from an insulator to a conductor. It is also possible to induce self-stress states or floppy mode at the domain wall between topologically different chains, **Fig. 27(a)**.

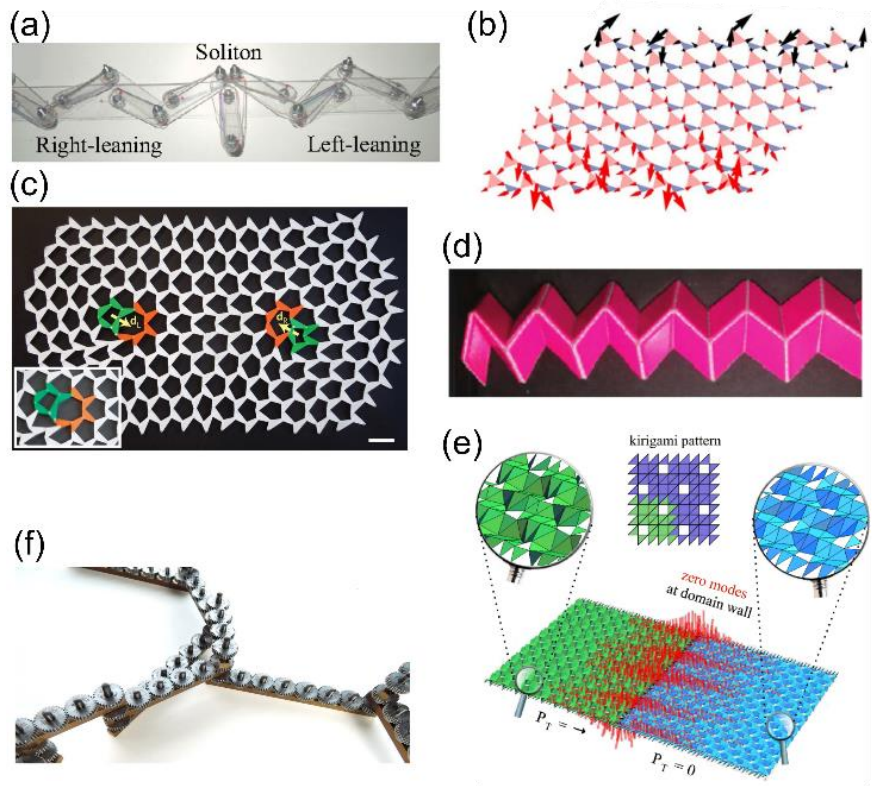

**Figure 27**. **Mechanical zero-frequency topological phases.** (a) 1D SSH model mechanical MM equivalent with a topological zero-energy mode. Reproduced with permission from[991]. Copyright 2014 Academy of Science. (b) Topological floppy modes at the edges of a distorted mechanical Kagome lattice. Reproduced with permission from[993]. Copyright 2017 Springer Nature. (c) Topological defects in a distorted mechanical Kagome lattice linked to states of self-stress. Reproduced with permission from[42]. Copyright 2015 Springer Nature. (d) Origami equivalent of a 1D SSH model with a topological floppy mode on the left side (soft mode). Reproduced with permission from[994]. Copyright 2016 American Physical Society. (e) Interface between two topologically different kirigami mechanical MMs hosts a zero-energy mode. Reproduced with permission from[994]. Copyright 2016 American Physical Society. (f) Geared MM as another platform hosting topological zero-energy modes. Reproduced with permission from[995]. Copyright 2016 American Physical Society.

These concepts have also been transposed to 2D Maxwell lattices. For instance, a deformed Kagome lattice is a gapped system at zero frequency, prone to host 1D floppy modes at its edges[993], **Fig.**

**27(b)**. Progressive changes in the lattice shape can redirect these zero-energy modes on different boundaries giving these MMs a strong tunability[993]. Moreover, topological defects such as dislocations generate self-stress states able to precipitate the buckling of the system when loads are added[42], **Fig. 27(c)**. This selective buckling or failure allows envisioning applications in robotics or sensing with topological mechanical structures. Furthermore, it is possible to design analogs of Weyl fermions in 2D Maxwell lattices[996–998] and Weyl lines in 3D, opening the door to even richer topological phenomena[999,1000].

Another potential application of these specific topological concepts is linked to origami and kirigami MMs[1001]. Indeed, topological floppy modes can be engineered to design robust folding patterns[994,1002–1004]. Notably, a quasi-1D origami strip has been proposed as an equivalent of the SSH model, **Fig. 27(d)**. For the case of 2D media, kirigami designs have been proposed to implement topological zero-modes at domain walls, **Fig. 27(e)**. The great flexibility of such media and their topological engineering make them a promising platform for mechanical MMs.

Topological zero-frequency modes also appear in geared MMs[995] (**Fig. 27(f)**), jammed matter[1005], quasi-crystals[1006], dislocated fiber networks[1007], non-orientable Mobius strips[1008] and have been extended to continuum elastic media[1009]. Their presence in fishbone mechanical systems is also related to the apparition of static non-reciprocity[1010]. We encourage the reader to consult the references provided in this section for a more detailed explanation of these zero-frequency mechanical topological phases. The powerful topological approach to tailoring these floppy modes or self-stressed states associated with specific chemical engineering, notably in the field of polymers, could open the door to chemically designed mechanical MMs with enhanced properties.

## 9.6. Topological properties of scattering anomalies

As an additional interesting opportunity emerging from topology in the context of classical waves, in the last few years there has been a growing interest in so-called anomalies in light scattering and their topological properties. Here we present basic information about scattering anomalies focusing on their topological features, and direct the interested readers to more detailed reviews[1011–1014].

A *bound state in the continuum* (BIC), also known as an embedded eigenstate (EE), is an eigenmode of an optically open system with an unboundedly large radiative Q-factor although lying in the continuum of unbounded states[1011,1015,1024,1016–1023]. BICs can be realized in either symmetry-protected scenarios, e.g., in the case of a hedgehog-like collection of dipole polarisations longitudinally arranged over a sphere[1017,1024] or over a plane[1020], or exploiting the destructive interference of at least two strongly coupled resonant modes coupled to the same radiation channel[1015,1025–1032]. Due to their very high Q-factor and topological features, BICs offer exciting perspectives for applications in sensing[1020,1021,1033], lasing[1014,1034–1037], high-harmonic generation[1038,1039], enforced nonreciprocity[1040,1041], thermal emission[1042,1043], energy transfer and harvesting[1044,1045], polarisation control[1046,1047], and vortex beam generation[1035]. Photonic BICs supported by periodic systems have topological features in the form of a polarization singularity in the space of wave vectors. Their robustness is related to their topological nature, rooted in the topological charge conservation[1027,1030,1048,1049]. The topological nature of the BIC has led to several research studies on BIC charge fusion to create even more unusual resonances[1050,1051]. The topological properties of BICs have been beneficial for polarization control[1052–1054] and circularly polarized states generation[1055–1057]. Recently, the generation of vortex beams through BICs[1058] and efficient topological vortex laser generation[1035] was demonstrated, showing the potential of topological phenomena in radiative and scattering processes. A further connection between new topological phenomena in higher-order angular states and BIC was also recently established in [467], revealing the far-reaching topological consequences that non-radiating states can have on the topological states of electromagnetic structures. The topological features of these states can be studied through the S-matrix formalism, specifically by analyzing the complex reflection coefficient, where phase vortices arise[1052,1059].

Varying the coupling strengths between at least two modes of an active system, their coalescence on the real frequency axis can rise to an *exceptional point* (EP). The EP in a PT-symmetric structure coincides with the spontaneous symmetry-breaking threshold, at which the unbroken PT-symmetric phase abruptly transits to the PT-broken phase. As an illustration of EPs in PT-symmetric systems, we consider a pair of coupled single-mode waveguides, where balanced gain and loss are implemented in the two waveguides or two coupled resonators[1060,1061]. The equation governing these systems reads $\frac{d}{d\xi}\begin{pmatrix} a_1 \\ a_2 \end{pmatrix} = -i\hat{H}\begin{pmatrix} a_1 \\ a_2 \end{pmatrix}$, where $a_1$ and $a_2$ are either waveguiding modes or localized resonator modes, $\xi$ is either the coordinate or time, respectively. The Hamiltonian of this system can be derived from the standard coupled-mode equations [1062,1063,1072,1064–1071]

$$\hat{H} = \begin{pmatrix} \beta_{0,1} - i\gamma_1 & \kappa \\ \kappa^* & \beta_{0,2} - i\gamma_2 \end{pmatrix},$$  (24)

where $\beta_{0,1}$ and $\beta_{0,2}$ are either propagation constants in each waveguide or eigenfrequencies of the localized modes, $\gamma_i$ takes account of the gain/loss magnitude in the waveguide (mode) $i$, and $\kappa$ denotes the reciprocal coupling between the waveguide pair. Eq.(24) has the following eigenvalues

$$\beta_\pm = \beta_{0,\text{ave}} - i\gamma_\text{ave} \pm \sqrt{|\kappa|^2 + (\beta_{0,\text{dif}} + i\gamma_\text{dif})^2} \,,$$  (25)

where $\beta_{0,\text{ave}} = (\beta_{0,1} + \beta_{0,2})/2$ , $\beta_{0,\text{dif}} = (\beta_{0,1} - \beta_{0,2})/2$ , $\gamma_\text{ave} = (\gamma_1 + \gamma_2)/2$ , $\gamma_\text{dif} = (\gamma_1 - \gamma_2)/2$ . In the particular case when $\beta_{0,1} = \beta_{0,2} \equiv \beta_0$ and $\gamma_2 = -\gamma_1 \equiv \gamma$ , the eigenvalues of the two modes become $\beta_\pm = \beta_0 \pm \sqrt{|\kappa|^2 - \gamma^2}$ . Eq. (25) shows that if the coupling is less than a certain critical value ( $\kappa < \kappa_\text{PT} = \gamma$ ), the system possesses two modes, one lossy and one amplifying. The lossy mode decays exponentially in time when excited, whereas the gainy one exhibits exponential growth. When the coupling strength is large enough ($\kappa > \kappa_{PT}$), the system resides in the strong coupling regime where the coherent energy exchange between the elements compensates for the decay and stabilizes the system at the real frequency axis. The critical point $\kappa = \kappa_{PT}$ gives rise to an EP[475,476,1011,1073–1077]. Exactly at the transition threshold where $\gamma_2 = -\gamma_1$, the two modes merge to a single eigenstate with propagation constant $\beta_0$, featuring a non-Hermitian degeneracy. The coalescence of two poles can also happen in the lower complex plane, giving rise to complex EPs (cEPs). Even though such cEPs reside in the complex frequency plane, they can be revealed with either parameter variations[1078] or complex frequency excitations[1079]. In the latter case, cEPs are excited with exponentially decaying signals because cEPs in passive systems lie in the lower half-plane of complex frequency. EPs are of great interest for various applications. EPs enable unidirectional light transport, leading to the observation of anisotropic transmission resonances[473,1080–1082] and directional power flow in microlaser cavities[1083–1085]. The accidental coalescence of eigenmodes offers rich topological features, enabling dynamic chiral-mode conversion [1071,1086,1087], and advanced optical sensing [1088–1090 474,475]. PT-symmetric systems hold a great promise for imaging[1091 630,1092–1095], lasing [1076], cloaking and transformation optics[1096,1097], new mechanisms of wave transport[1098–1100], topology[1101], light scattering engineering[1011,1102,1103], and wireless power transfer [1063,1104,1105].

PT-symmetric systems enable also other interesting topological properties. For example, if one starts from a Hermitian system with a Dirac cone, the introduction of balanced gain and loss can move both cones inside each other, forming an exceptional ring in reciprocal space[1106]. In Weyl systems, recent studies show that dissipation can also turn Weyl points into rings[1107]. A more detailed study reveals that in non-Hermitian systems, the GP becomes complex[147,153,1108–1113]: its real part is modified compared with the corresponding closed system, and its imaginary part gives rise to *geometric dephasing*[1114–1116]. This includes coupling to a dissipative[1114] or noisy[1117,1118] environment.

## 10. Conclusion and perspectives

In this work, we have reviewed the role that topological phases can play in tailoring the response of artificial materials to classical waves in spatial and synthetic dimensions. Although topological MMs have a shorter history than TIs in condensed matter physics, significant exciting research has been developed in this relatively young field. Several intriguing topological phenomena may be unfeasible to be implemented in electronic systems, but they can be readily captured for classical waves using metamaterials. Artificial materials provide excellent tunability and freedom of design, making them excellent platforms for implementing topological concepts in photonics and phononics. For instance, the higher-order TIs with quantized multipole moment supports corner charges, but they are challenging to be observed in quantum solids. However, topological corner states in multiple TIs have been achieved in various classical systems[98].

Significant further research efforts are expected in the future to realize novel topological states in classical systems, not only to understand the underlying physics and topology, but also to provide more opportunities for applying topological physics to promising applications, such as robust signal guiding, wave manipulation and on-chip communications, photonic quantum computation with topological protection[340] and more. Furthermore, classical platforms have new properties beyond the condensed matter counterparts, benefiting the interdisciplinary fields. Notably, long-range coupling, recently unveiled dualities[589,1119], fragile topology[1120], topological quantum chemistry in photonics[1121], or non-Abelian topological effects[589,1122–1126], can all open new routes to exciting topological phenomena for light propagation, sound and mechanical vibrations in chemistry. Researchers urgently question how to harness this relatively new field in photonics, phononics, and interdisciplinary fields, e.g., how to design and fabricate disorder-immune components for high-speed information transfer and processing in the on-chip design integration with optical or acousto-optic elements. Moreover, the versatile and flexible properties of MMs allow for integrating different active topics like non-Hermitian, nonlinearity, and topological phases into the same platform[1127], which may lead to richer topological phases and more fundamental discoveries. On another front of exploration, the time-varying MMs make the implementation of time crystal and spacetime MMs in classical systems accessible[1128], thus enabling researchers to transfer the concept of topological edge states from spatial boundary to temporal interface[1129] and exploring more exotic phenomena exclusively observed in time-varying media[1130], like the amplified and lasing in photonic time crystal[1131].

Furthermore, advances related to the chemical design of materials, such as the fabrication of low-loss or tunable non-Hermitian properties, could further improve such devices. The precise design and realization of active particles can also be of great interest for the practical implementation of topological phases requiring a fluid in motion[354,607,614,615,1132]. Not only can chemistry, biochemistry or physical chemistry provide exciting alternatives to design topological MM media, notably at the nano- or micro-scale. They can also benefit from the inherently robust features of these devices. Hence, topological cavities offer strong field localization robust to fabrication flows that can be used for sensing, particle manipulation, catalysis, and therapy in the lab on a chip[1133–1136]. Topological zero frequency mechanical modes combined with macro-molecular chemistry may lead to the accurate design of multifunctional materials that can find applications in soft robotics[1137–1140]. We envision a bright future at the intersection between photonics, phononics and chemistry grounded into topological concepts.

**Acknowledgements**

Our work on these topics has been supported by the Simons Foundation and the Air Force Office of Scientific Research.

# References

(1) Aharonov, Y.; Bohm, D. Significance of Electromagnetic Potentials in the Quantum Theory. *Phys. Rev.* **1959**, *115* (3), 485–491. https://doi.org/10.1103/PhysRev.115.485.

(2) Hasan, M. Z.; Kane, C. L. Colloquium: Topological Insulators. *Rev. Mod. Phys.* **2010**, *82* (4), 3045–3067. https://doi.org/10.1103/RevModPhys.82.3045.

(3) Liu, Y.; Chen, X.; Xu, Y. Topological Phononics: From Fundamental Models to Real Materials. *Adv. Funct. Mater.* **2020**, *30* (8), 1904784. https://doi.org/10.1002/adfm.201904784.

(4) Huber, S. D. Topological Mechanics. *Nat. Phys.* **2016**, *12* (7), 621–623. https://doi.org/10.1038/nphys3801.

(5) Cooper, N. R.; Dalibard, J.; Spielman, I. B. Topological Bands for Ultracold Atoms. *Rev. Mod. Phys.* **2019**, *91* (1), 015005. https://doi.org/10.1103/RevModPhys.91.015005.

(6) Bradlyn, B.; Elcoro, L.; Cano, J.; Vergniory, M. G.; Wang, Z.; Felser, C.; Aroyo, M. I.; Bernevig, B. A. Topological Quantum Chemistry. *Nature* **2017**, *547* (7663), 298–305. https://doi.org/10.1038/NATURE23268.

(7) Hatsugai, Y. Chern Number and Edge States in the Integer Quantum Hall Effect. *Phys. Rev. Lett.* **1993**, *71* (22), 3697–3700. https://doi.org/10.1103/PhysRevLett.71.3697.

(8) Ozawa, T.; Price, H. M.; Amo, A.; Goldman, N.; Hafezi, M.; Lu, L.; Rechtsman, M. C.; Schuster, D.; Simon, J.; Zilberberg, O.; Carusotto, I. Topological Photonics. *Rev. Mod. Phys.* **2019**, *91* (1), 015006. https://doi.org/10.1103/revmodphys.91.015006.

(9) Haldane, F. D. M. Model for a Quantum Hall Effect without Landau Levels: Condensed-Matter Realization of the "Parity Anomaly." *Phys. Rev. Lett.* **1988**, *61* (18), 2015–2018. https://doi.org/10.1103/PhysRevLett.61.2015.

(10) Kane, C. L.; Mele, E. J. Quantum Spin Hall Effect in Graphene. *Phys. Rev. Lett.* **2005**, *95* (22), 226801. https://doi.org/10.1103/PhysRevLett.95.226801.

(11) Kane, C. L.; Mele, E. J. Z2 Topological Order and the Quantum Spin Hall Effect. *Phys. Rev. Lett.* **2005**, *95* (14), 146802. https://doi.org/10.1103/PhysRevLett.95.146802.

(12) Xiao, D.; Yao, W.; Niu, Q. Valley-Contrasting Physics in Graphene: Magnetic Moment and Topological Transport. *Phys. Rev. Lett.* **2007**, *99* (23), 236809. https://doi.org/10.1103/PhysRevLett.99.236809.

(13) Qi, X.-L.; Zhang, S.-C. Topological Insulators and Superconductors. *Rev. Mod. Phys.* **2011**, *83* (4), 1057–1110. https://doi.org/10.1103/RevModPhys.83.1057.

(14) Bernevig, B. A.; Hughes, T. L. *Topological Insulators and Topological Superconductors*; Princeton University Press, 2013. https://doi.org/10.1515/9781400846733.

(15) Haldane, F. D. M.; Raghu, S. Possible Realization of Directional Optical Waveguides in Photonic Crystals with Broken Time-Reversal Symmetry. *Phys. Rev. Lett.* **2008**, *100* (1), 013904. https://doi.org/10.1103/PhysRevLett.100.013904.

(16) Raghu, S.; Haldane, F. D. M. Analogs of Quantum-Hall-Effect Edge States in Photonic Crystals. *Phys. Rev. A* **2008**, *78* (3), 033834. https://doi.org/10.1103/PhysRevA.78.033834.

(17) Wang, Z.; Chong, Y.; Joannopoulos, J. D.; Soljačić, M. Observation of Unidirectional Backscattering-Immune Topological Electromagnetic States. *Nature* **2009**, *461* (7265), 772–775.

https://doi.org/10.1038/nature08293.

(18)   Feng, L.; Ayache, M.; Huang, J.; Xu, Y. L.; Lu, M. H.; Chen, Y. F.; Fainman, Y.; Scherer, A.; Sze, S. M. Nonreciprocal Light Propagation in a Silicon Photonic Circuit. *Science (80-. ).* **2011**, *333* (6043), 729–733. https://doi.org/10.1126/science.1206038.

(19)   Hafezi, M.; Demler, E. A.; Lukin, M. D.; Taylor, J. M. Robust Optical Delay Lines with Topological Protection. *Nat. Phys.* **2011**, *7* (11), 907–912. https://doi.org/10.1038/nphys2063.

(20)   Fang, K.; Yu, Z.; Fan, S. Realizing Effective Magnetic Field for Photons by Controlling the Phase of Dynamic Modulation. *Nat. Photonics* **2012**, *6* (11), 782–787. https://doi.org/10.1038/NPHOTON.2012.236.

(21)   Fang, K.; Fan, S. Controlling the Flow of Light Using the Inhomogeneous Effective Gauge Field That Emerges from Dynamic Modulation. *Phys. Rev. Lett.* **2013**, *111* (20), 203901. https://doi.org/10.1103/PhysRevLett.111.203901.

(22)   Khanikaev, A. B.; Hossein Mousavi, S.; Tse, W. K.; Kargarian, M.; MacDonald, A. H.; Shvets, G. Photonic Topological Insulators. *Nat. Mater.* **2013**, *12* (3), 233–239. https://doi.org/10.1038/nmat3520.

(23)   Rechtsman, M. C.; Zeuner, J. M.; Plotnik, Y.; Lumer, Y.; Podolsky, D.; Dreisow, F.; Nolte, S.; Segev, M.; Szameit, A. Photonic Floquet Topological Insulators. *Nature* **2013**, *496* (7444), 196–200. https://doi.org/10.1038/nature12066.

(24)   Khanikaev, A. B.; Shvets, G. Two-Dimensional Topological Photonics. *Nat. Photonics* **2017**, *11* (12), 763–773. https://doi.org/10.1038/s41566-017-0048-5.

(25)   Hafezi, M.; Mittal, S.; Fan, J.; Migdall, A.; Taylor, J. M. Imaging Topological Edge States in Silicon Photonics. *Nat. Photonics* **2013**, *7* (12), 1001–1005. https://doi.org/10.1038/nphoton.2013.274.

(26)   Lu, L.; Joannopoulos, J. D.; Soljačić, M. Topological Photonics. *Nat. Photonics* **2014**, *8* (11), 821–829. https://doi.org/10.1038/nphoton.2014.248.

(27)   Plotnik, Y.; Rechtsman, M. C.; Song, D.; Heinrich, M.; Zeuner, J. M.; Nolte, S.; Lumer, Y.; Malkova, N.; Xu, J.; Szameit, A.; Chen, Z.; Segev, M. Observation of Unconventional Edge States in "Photonic Graphene." *Nat. Mater.* **2014**, *13* (1), 57–62. https://doi.org/10.1038/nmat3783.

(28)   Skirlo, S. A.; Lu, L.; Soljačić, M. Multimode One-Way Waveguides of Large Chern Numbers. *Phys. Rev. Lett.* **2014**, *113* (11), 113904. https://doi.org/10.1103/PhysRevLett.113.113904.

(29)   Skirlo, S. A.; Lu, L.; Igarashi, Y.; Yan, Q.; Joannopoulos, J.; Soljačić, M. Experimental Observation of Large Chern Numbers in Photonic Crystals. *Phys. Rev. Lett.* **2015**, *115* (25), 253901. https://doi.org/10.1103/PhysRevLett.115.253901.

(30)   Wu, L.-H.; Hu, X. Scheme for Achieving a Topological Photonic Crystal by Using Dielectric Material. *Phys. Rev. Lett.* **2015**, *114* (22), 223901. https://doi.org/10.1103/PhysRevLett.114.223901.

(31)   Cheng, X.; Jouvaud, C.; Ni, X.; Mousavi, S. H.; Genack, A. Z.; Khanikaev, A. B. Robust Reconfigurable Electromagnetic Pathways within a Photonic Topological Insulator. *Nat. Mater.* **2016**, *15* (5), 542–548. https://doi.org/10.1038/nmat4573.

(32)   Leykam, D.; Rechtsman, M. C.; Chong, Y. D. Anomalous Topological Phases and Unpaired Dirac Cones in Photonic Floquet Topological Insulators. *Phys. Rev. Lett.* **2016**, *117* (1), 013902.

https://doi.org/10.1103/PhysRevLett.117.013902.

(33)    Bahari, B.; Ndao, A.; Vallini, F.; El Amili, A.; Fainman, Y.; Kanté, B. Nonreciprocal Lasing in Topological Cavities of Arbitrary Geometries. *Science (80-. ).* **2017**, *358* (6363), 636–640. https://doi.org/10.1126/SCIENCE.AAO4551.

(34)    Bandres, M. A.; Wittek, S.; Harari, G.; Parto, M.; Ren, J.; Segev, M.; Christodoulides, D. N.; Khajavikhan, M. Topological Insulator Laser: Experiments. *Science (80-. ).* **2018**, *359* (6381).

(35)    Harari, G.; Bandres, M. A.; Lumer, Y.; Rechtsman, M. C.; Chong, Y. D.; Khajavikhan, M.; Christodoulides, D. N.; Segev, M. Topological Insulator Laser: Theory. *Science (80-. ).* **2018**, *359* (6381). https://doi.org/10.1126/science.aar4003.

(36)    Hafezi, M.; Taylor, J. M. Topological Physics with Light. *Phys. Today* **2014**, *67* (5), 68–69. https://doi.org/10.1063/PT.3.2394.

(37)    Peano, V.; Brendel, C.; Schmidt, M.; Marquardt, F. Topological Phases of Sound and Light. *Phys. Rev. X* **2015**, *5* (3), 031011. https://doi.org/10.1103/PhysRevX.5.031011.

(38)    Yang, Z.; Gao, F.; Shi, X.; Lin, X.; Gao, Z.; Chong, Y.; Zhang, B. Topological Acoustics. *Phys. Rev. Lett.* **2015**, *114* (11), 114301. https://doi.org/10.1103/PhysRevLett.114.114301.

(39)    Süsstrunk, R.; Huber, S. D. Classification of Topological Phonons in Linear Mechanical Metamaterials. *Proc. Natl. Acad. Sci.* **2016**, *113* (33), E4767–E4775. https://doi.org/10.1073/pnas.1605462113.

(40)    Xiao, M.; Ma, G.; Yang, Z.; Sheng, P.; Zhang, Z. Q.; Chan, C. T. Geometric Phase and Band Inversion in Periodic Acoustic Systems. *Nat. Phys.* **2015**, *11* (3), 240–244. https://doi.org/10.1038/NPHYS3228.

(41)    Kane, C. L.; Lubensky, T. C. Topological Boundary Modes in Isostatic Lattices. *Nat. Phys.* **2014**, *10* (1), 39–45. https://doi.org/10.1038/nphys2835.

(42)    Paulose, J.; Chen, B. G. G.; Vitelli, V. Topological Modes Bound to Dislocations in Mechanical Metamaterials. *Nat. Phys.* **2015**, *11* (2), 153–156. https://doi.org/10.1038/NPHYS3185.

(43)    Süsstrunk, R.; Huber, S. D. Observation of Phononic Helical Edge States in a Mechanical Topological Insulator. *Science (80-. ).* **2015**, *349* (6243), 47–50. https://doi.org/10.1126/science.aab0239.

(44)    Ma, G.; Xiao, M.; Chan, C. T. Topological Phases in Acoustic and Mechanical Systems. *Nat. Rev. Phys.* **2019**, *1* (4), 281–294. https://doi.org/10.1038/S42254-019-0030-X.

(45)    Alù, A. Topological Acoustics. *Acoust. Today* **2021**, *17* (3), 13. https://doi.org/10.1121/AT.2021.17.3.13.

(46)    Zhang, X.; Xiao, M.; Cheng, Y.; Lu, M.-H.; Christensen, J. Topological Sound. *Commun. Phys.* **2018**, *1* (1), 97. https://doi.org/10.1038/s42005-018-0094-4.

(47)    Miniaci, M.; Pal, R. K. Design of Topological Elastic Waveguides. *J. Appl. Phys.* **2021**, *130* (14), 141101. https://doi.org/10.1063/5.0057288.

(48)    Kitaev, A. Y. Unpaired Majorana Fermions in Quantum Wires. *Physics-Uspekhi* **2001**, *44* (10S), 131–136. https://doi.org/10.1070/1063-7869/44/10S/S29.

(49)    Barik, S.; Karasahin, A.; Flower, C.; Cai, T.; Miyake, H.; DeGottardi, W.; Hafezi, M.; Waks, E. A Topological Quantum Optics Interface. *Science (80-. ).* **2018**, *359* (6376), 666–668.

https://doi.org/10.1126/SCIENCE.AAQ0327.

(50)    Dalibard, J.; Gerbier, F.; Juzeliūnas, G.; Öhberg, P. Colloquium : Artificial Gauge Potentials for Neutral Atoms. *Rev. Mod. Phys.* **2011**, *83* (4), 1523–1543. https://doi.org/10.1103/RevModPhys.83.1523.

(51)    Eckardt, A. Colloquium: Atomic Quantum Gases in Periodically Driven Optical Lattices. *Rev. Mod. Phys.* **2017**, *89* (1), 011004.

(52)    Pesin, D.; MacDonald, A. H. Spintronics and Pseudospintronics in Graphene and Topological Insulators. *Nat. Mater.* **2012**, *11* (5), 409–416. https://doi.org/10.1038/NMAT3305.

(53)    Nayak, C.; Simon, S. H.; Stern, A.; Freedman, M.; Das Sarma, S. Non-Abelian Anyons and Topological Quantum Computation. *Rev. Mod. Phys.* **2008**, *80* (3), 1083–1159. https://doi.org/10.1103/REVMODPHYS.80.1083.

(54)    Stanescu, T. D.; Tewari, S. Majorana Fermions in Semiconductor Nanowires: Fundamentals, Modeling, and Experiment. *J. Phys. Condens. Matter* **2013**, *25* (23), 233201. https://doi.org/10.1088/0953-8984/25/23/233201.

(55)    Alicea, J. New Directions in the Pursuit of Majorana Fermions in Solid State Systems. *Reports Prog. Phys.* **2012**, *75* (7), 076501. https://doi.org/10.1088/0034-4885/75/7/076501.

(56)    Beenakker, C. W. J. Search for Majorana Fermions in Superconductors. *Annu. Rev. Condens. Matter Phys.* **2013**, *4* (1), 113–136. https://doi.org/10.1146/annurev-conmatphys-030212-184337.

(57)    Fleury, R.; Khanikaev, A. B.; Alù, A. Floquet Topological Insulators for Sound. *Nat. Commun.* **2016**, *7* (1), 11744. https://doi.org/10.1038/ncomms11744.

(58)    Wang, P.; Lu, L.; Bertoldi, K. Topological Phononic Crystals with One-Way Elastic Edge Waves. *Phys. Rev. Lett.* **2015**, *115* (10), 104302. https://doi.org/10.1103/PhysRevLett.115.104302.

(59)    Nash, L. M.; Kleckner, D.; Read, A.; Vitelli, V.; Turner, A. M.; Irvine, W. T. M. Topological Mechanics of Gyroscopic Metamaterials. *Proc. Natl. Acad. Sci.* **2015**, *112* (47), 14495–14500. https://doi.org/10.1073/pnas.1507413112.

(60)    Pilozzi, L.; Conti, C. Topological Lasing in Resonant Photonic Structures. *Phys. Rev. B* **2016**, *93* (19), 195317. https://doi.org/10.1103/PhysRevB.93.195317.

(61)    Jeong, K.-Y. Recent Progress in Nanolaser Technology. *Adv. Mater.* **2020**, *32* (51), 2001996. https://doi.org/10.1002/adma.202001996.

(62)    Ota, Y.; Katsumi, R.; Watanabe, K.; Iwamoto, S.; Arakawa, Y. Topological Photonic Crystal Nanocavity Laser. *Commun. Phys.* **2018**, *1* (1), 86. https://doi.org/10.1038/s42005-018-0083-7.

(63)    Chen, Z.; Segev, M. Highlighting Photonics: Looking into the next Decade. *eLight* **2021**, *1* (1), 2. https://doi.org/10.1186/s43593-021-00002-y.

(64)    Parto, M.; Wittek, S.; Hodaei, H.; Harari, G.; Bandres, M. A.; Ren, J.; Rechtsman, M. C.; Segev, M.; Christodoulides, D. N.; Khajavikhan, M. Edge-Mode Lasing in 1D Topological Active Arrays. *Phys. Rev. Lett.* **2018**, *120* (11), 113901. https://doi.org/10.1103/PhysRevLett.120.113901.

(65)    St-Jean, P.; Goblot, V.; Galopin, E.; Lemaître, A.; Ozawa, T.; Le Gratiet, L.; Sagnes, I.; Bloch, J.; Amo, A. Lasing in Topological Edge States of a One-Dimensional Lattice. *Nat. Photonics* **2017**, *11* (10), 651–656. https://doi.org/10.1038/S41566-017-0006-2.

(66)    Lodahl, P.; Mahmoodian, S.; Stobbe, S.; Rauschenbeutel, A.; Schneeweiss, P.; Volz, J.; Pichler,

H.; Zoller, P. Chiral Quantum Optics. *Nature*. Nature Publishing Group January 25, 2017, pp 473–480. https://doi.org/10.1038/nature21037.

(67) Carusotto, I.; Ciuti, C. Quantum Fluids of Light. *Rev. Mod. Phys.* **2013**, *85* (1), 299–366. https://doi.org/10.1103/RevModPhys.85.299.

(68) J.B. Pendry; A. J. Holden; D. J. Robbins; W. J. Stewart. Magnetism from Conductors, and Enhanced Non-Linear Phenomena. *IEEE Trans. Microw. Theory Tech.* **1999**, *47* (11), 2075–2084.

(69) Krasnok, A.; Tymchenko, M.; Alù, A. Nonlinear Metasurfaces: A Paradigm Shift in Nonlinear Optics. *Mater. Today* **2018**, *21* (1), 8–21. https://doi.org/10.1016/j.mattod.2017.06.007.

(70) Lepeshov, S.; Krasnok, A. Tunable Phase-Change Metasurfaces. *Nat. Nanotechnol.* **2021**, *16* (6), 615–616. https://doi.org/10.1038/s41565-021-00892-6.

(71) Hu, G.; Krasnok, A.; Mazor, Y.; Qiu, C.; Alù, A. Moiré Hyperbolic Metasurfaces. *Nano Lett.* **2020**, *20* (5), 3217–3224. https://doi.org/10.1021/acs.nanolett.9b05319.

(72) Chen, M. K.; Wu, Y.; Feng, L.; Fan, Q.; Lu, M.; Xu, T.; Tsai, D. P. Principles, Functions, and Applications of Optical Meta-Lens. *Adv. Opt. Mater.* **2021**, *9* (4), 2001414. https://doi.org/10.1002/adom.202001414.

(73) Kuznetsov, A. I.; Miroshnichenko, A. E.; Brongersma, M. L.; Kivshar, Y. S.; Luk'yanchuk, B. Optically Resonant Dielectric Nanostructures. *Science (80-. ).* **2016**, *354* (6314), aag2472. https://doi.org/10.1126/science.aag2472.

(74) Staude, I.; Schilling, J. Metamaterial-Inspired Silicon Nanophotonics. *Nat. Photonics* **2017**, *11* (5), 274–284. https://doi.org/10.1038/nphoton.2017.39.

(75) Yu, N.; Capasso, F. Flat Optics with Designer Metasurfaces. *Nat. Mater.* **2014**, *13* (2), 139–150. https://doi.org/10.1038/nmat3839.

(76) Urbas, A. M.; Jacob, Z.; Negro, L. D.; Engheta, N.; Boardman, A. D.; Egan, P.; Khanikaev, A. B.; Menon, V.; Ferrera, M.; Kinsey, N.; DeVault, C.; Kim, J.; Shalaev, V.; Boltasseva, A.; Valentine, J.; Pfeiffer, C.; Grbic, A.; Narimanov, E.; Zhu, L.; Fan, S.; Alù, A.; Poutrina, E.; Litchinitser, N. M.; Noginov, M. A.; MacDonald, K. F.; Plum, E.; Liu, X.; Nealey, P. F.; Kagan, C. R.; Murray, C. B.; Pawlak, D. A.; Smolyaninov, I. I.; Smolyaninova, V. N.; Chanda, D. Roadmap on Optical Metamaterials. *J. Opt.* **2016**, *18* (9), 093005. https://doi.org/10.1088/2040-8978/18/9/093005.

(77) Iyer, A. K.; Alu, A.; Epstein, A. Metamaterials and Metasurfaces—Historical Context, Recent Advances, and Future Directions. *IEEE Trans. Antennas Propag.* **2020**, *68* (3), 1223–1231. https://doi.org/10.1109/TAP.2020.2969732.

(78) Wu, P. C.; Chen, J.-W.; Yin, C.; Lai, Y.; Chung, T. L.; Liao, C. Y.; Chen, B. H.; Lee, K.; Chuang, C.; Wang, C.; Tsai, D. P. Visible Metasurfaces for On-Chip Polarimetry. *ACS Photonics* **2018**, *5* (7), 2568–2573. https://doi.org/10.1021/acsphotonics.7b01527.

(79) Yulaev, A.; Zhu, W.; Zhang, C.; Westly, D. A.; Lezec, H. J.; Agrawal, A.; Aksyuk, V. Metasurface-Integrated Photonic Platform for Versatile Free-Space Beam Projection with Polarization Control. *ACS Photonics* **2019**, *6* (11), 2902–2909. https://doi.org/10.1021/acsphotonics.9b01000.

(80) Shaltout, A. M.; Shalaev, V. M.; Brongersma, M. L. Spatiotemporal Light Control with Active Metasurfaces. *Science (80-. ).* **2019**, *364* (6441). https://doi.org/10.1126/science.aat3100.

(81) Jin, C.; Afsharnia, M.; Berlich, R.; Fasold, S.; Zou, C.; Arslan, D.; Staude, I.; Pertsch, T.;

Setzpfandt, F. Dielectric Metasurfaces for Distance Measurements and Three-Dimensional Imaging. *Adv. Photonics* **2019**, *1* (03), 1. https://doi.org/10.1117/1.AP.1.3.036001.

(82)  Li, G.; Zhang, S.; Zentgraf, T. Nonlinear Photonic Metasurfaces. *Nat. Rev. Mater.* **2017**, *2* (5), 17010. https://doi.org/10.1038/natrevmats.2017.10.

(83)  Glybovski, S. B.; Tretyakov, S. A.; Belov, P. A.; Kivshar, Y. S.; Simovski, C. R. Metasurfaces: From Microwaves to Visible. *Phys. Rep.* **2016**, *634*, 1–72. https://doi.org/10.1016/j.physrep.2016.04.004.

(84)  Engheta, N.; Ziolkowski, R. W. *Metamaterials*; Engheta, N., Ziolkowski, R. W., Eds.; John Wiley & Sons, Inc.: Hoboken, NJ, USA, 2006. https://doi.org/10.1002/0471784192.

(85)  Joannopoulos, J.; Johnson, S.; Winn, J.; Meade, R. *Photonic Crystals: Molding the Flow of Light*; 2008.

(86)  Askari, M.; Hutchins, D. A.; Thomas, P. J.; Astolfi, L.; Watson, R. L.; Abdi, M.; Ricci, M.; Laureti, S.; Nie, L.; Freear, S.; Wildman, R.; Tuck, C.; Clarke, M.; Woods, E.; Clare, A. T. Additive Manufacturing of Metamaterials: A Review. *Addit. Manuf.* **2020**, *36*, 101562. https://doi.org/10.1016/j.addma.2020.101562.

(87)  Wu, X.; Su, Y.; Shi, J. Perspective of Additive Manufacturing for Metamaterials Development. *Smart Mater. Struct.* **2019**, *28* (9), 093001. https://doi.org/10.1088/1361-665X/ab2eb6.

(88)  Xie, B. Photonics Meets Topology. *Opt. Express* **2018**, *26*, 24531–24550.

(89)  Rider, M. S.; Palmer, S. J.; Pocock, S. R.; Xiao, X.; Arroyo Huidobro, P.; Giannini, V. A Perspective on Topological Nanophotonics: Current Status and Future Challenges. *J. Appl. Phys.* **2019**, *125* (12), 120901. https://doi.org/10.1063/1.5086433.

(90)  Yuan, L. Synthetic Dimension in Photonics. *Optica* **2018**, *5*, 1396–1405.

(91)  Sun, X.-C.; He, C.; Liu, X.-P.; Lu, M.-H.; Zhu, S.-N.; Chen, Y.-F. Two-Dimensional Topological Photonic Systems. *Prog. Quantum Electron.* **2017**, *55*, 52–73. https://doi.org/10.1016/j.pquantelec.2017.07.004.

(92)  Kim, M.; Jacob, Z.; Rho, J. Recent Advances in 2D, 3D and Higher-Order Topological Photonics. *Light Sci. Appl.* **2020**, *9* (1), 130. https://doi.org/10.1038/s41377-020-0331-y.

(93)  Tang, G.; He, X.; Shi, F.; Liu, J.; Chen, X.; Dong, J. Topological Photonic Crystals: Physics, Designs, and Applications. *Laser Photon. Rev.* **2022**, *16* (4), 2100300. https://doi.org/10.1002/lpor.202100300.

(94)  Ota, Y.; Takata, K.; Ozawa, T.; Amo, A.; Jia, Z.; Kante, B.; Notomi, M.; Arakawa, Y.; Iwamoto, S. Active Topological Photonics. *Nanophotonics* **2020**, *9* (3), 547–567. https://doi.org/10.1515/nanoph-2019-0376.

(95)  Smirnova, D.; Leykam, D.; Chong, Y.; Kivshar, Y. Nonlinear Topological Photonics. *Appl. Phys. Rev.* **2020**, *7* (2), 021306. https://doi.org/10.1063/1.5142397.

(96)  Lustig, E.; Segev, M. Topological Photonics in Synthetic Dimensions. *Adv. Opt. Photonics* **2021**, *13* (2), 426–461. https://doi.org/10.1364/AOP.418074.

(97)  Xue, H.; Yang, Y.; Zhang, B. Topological Valley Photonics: Physics and Device Applications. *Adv. Photonics Res.* **2021**, *2* (8), 2100013. https://doi.org/10.1002/adpr.202100013.

(98)  Xie, B.; Wang, H.-X.; Zhang, X.; Zhan, P.; Jiang, J.-H.; Lu, M.; Chen, Y. Higher-Order Band

Topology. *Nat. Rev. Phys.* **2021**, *3* (7), 520–532. https://doi.org/10.1038/s42254-021-00323-4.

(99)    Jalife, J. *The Tornadoes of Sudden Cardiac Arrest*; Nature Publishing Group, 2018; Vol. 555, pp 597–598. https://doi.org/10.1038/d41586-018-01950-1.

(100)   Tamburini, F.; Mari, E.; Sponselli, A.; Thidé, B.; Bianchini, A.; Romanato, F. Encoding Many Channels on the Same Frequency through Radio Vorticity: First Experimental Test. *New J. Phys.* **2012**, *14* (3), 033001. https://doi.org/10.1088/1367-2630/14/3/033001.

(101)   Thidé, B.; Then, H.; Sjöholm, J.; Palmer, K.; Bergman, J.; Carozzi, T. D.; Istomin, Y. N.; Ibragimov, N. H.; Khamitova, R. Utilization of Photon Orbital Angular Momentum in the Low-Frequency Radio Domain. *Phys. Rev. Lett.* **2007**, *99* (8), 087701. https://doi.org/10.1103/PhysRevLett.99.087701.

(102)   Christoph, J.; Chebbok, M.; Richter, C.; Schröder-Schetelig, J.; Bittihn, P.; Stein, S.; Uzelac, I.; Fenton, F. H.; Hasenfuß, G.; Gilmour, R. F.; Luther, S. Electromechanical Vortex Filaments during Cardiac Fibrillation. *Nature* **2018**, *555* (7698), 667–672. https://doi.org/10.1038/nature26001.

(103)   Berry, M. Quantal Phase Factors Accompanying Adiabatic Changes. *Proc. R. Soc. London. A. Math. Phys. Sci.* **1984**, *392* (1802), 45–57. https://doi.org/10.1098/rspa.1984.0023.

(104)   Simon, B. Holonomy, the Quantum Adiabatic Theorem, and Berry's Phase. *Phys. Rev. Lett.* **1983**, *51* (24), 2167–2170. https://doi.org/10.1103/PhysRevLett.51.2167.

(105)   Berry, M. V.; Jeffrey, M. R. Chapter 2 Conical Diffraction: Hamilton's Diabolical Point at the Heart of Crystal Optics. In *Progress in Optics*; Elsevier, 2007; Vol. 50, pp 13–50. https://doi.org/10.1016/S0079-6638(07)50002-8.

(106)   Pancharatnam, S. Generalized Theory of Interference, and Its Applications. *Proc. Indian Acad. Sci. - Sect. A* **1956**, *44* (5), 247–262. https://doi.org/10.1007/BF03046050.

(107)   Hamilton, W. Third Supplement to an Essay on the Theory of Systems of Rays. *Trans. R. Ir. Acad.* **1837**, *17*, 1–144.

(108)   Lloyd, H. On the Phenomena Presented by Light in Its Passage along the Axes of Biaxial Crystals. *Phil. Mag.* **1833**, *1*, 112–120.

(109)   Bortolotti, E. Memories and Notes Presented by Fellows. *Rend. R. Acc. Naz. Linc.* **1926**, *4*, 552.

(110)   Rytov, S. On Transition from Wave to Geometrical Optics. *Dokl. Akad. Nauk SSSR* **1938**, *18*, 263–266.

(111)   Vladimirskii, V. The Rotation of a Polarization Plane for Curved Light Ray. *Dokl. Akad. Nauk. SSSR* **1941**, *31*, 222–224.

(112)   KG Budden, M. S. Phase Memory and Additional Memory in W. K. B. Solutions for Wave Propagation in Stratified Media. *Proc. R. Soc. London. A. Math. Phys. Sci.* **1976**, *350* (1660), 27–46. https://doi.org/10.1098/rspa.1976.0093.

(113)   Boulanger, J.; Le Bihan, N.; Catheline, S.; Rossetto, V. Observation of a Non-Adiabatic Geometric Phase for Elastic Waves. *Ann. Phys. (N. Y).* **2012**, *327* (3), 952–958. https://doi.org/10.1016/j.aop.2011.11.014.

(114)   Wang, S.; Ma, G.; Chan, C. T. Topological Transport of Sound Mediated by Spin-Redirection Geometric Phase. *Sci. Adv.* **2018**, *4* (2). https://doi.org/10.1126/sciadv.aaq1475.

(115) Wilczek, F.; Shapere, A. *Geometric Phases in Physics*; World Scientific, 1989.

(116) Markovski, B.; Vinitsky, S. I. *Topological Phases in Quantum Theory*; World Scientific, 1989.

(117) Zwanziger, J. W.; Koenig, M.; Pines, A. Berry's Phase. *Annu. Rev. Phys. Chem.* **1990**, *41* (1), 601–646. https://doi.org/10.1146/annurev.pc.41.100190.003125.

(118) Anandan, J. The Geometric Phase. *Nature* **1992**, *360* (6402), 307–313. https://doi.org/10.1038/360307a0.

(119) Li, H. Z. . *Global Properties of Simple Quantum Systems — Berry's Phase and Others*; Shanghai Scientific and Technical Publishers, 1998.

(120) Bohm, A.; Mostafazadeh, A.; Koizumi, H.; Niu, Q.; Zwanziger, J. *The Geometric Phase in Quantum Systems*; Springer Berlin Heidelberg: Berlin, Heidelberg, 2003. https://doi.org/10.1007/978-3-662-10333-3.

(121) Chruściński, D.; Jamiołkowski, A. *Geometric Phases in Classical and Quantum Mechanics*; Birkhäuser Boston: Boston, MA, 2004. https://doi.org/10.1007/978-0-8176-8176-0.

(122) Xiao, D.; Chang, M.-C.; Niu, Q. Berry Phase Effects on Electronic Properties. *Rev. Mod. Phys.* **2010**, *82* (3), 1959–2007. https://doi.org/10.1103/RevModPhys.82.1959.

(123) Berry, M. Anticipations of the Geometric Phase. *Phys. Today* **1990**, *43*, 34–40.

(124) Berry, M. Geometric Phase Memories. *Nat. Phys.* **2010**, *6* (3), 148–150. https://doi.org/10.1038/nphys1608.

(125) Montgomery, R. Isoholonomic Problems and Some Applications. *Commun. Math. Phys.* **1990**, *128* (3), 565–592. https://doi.org/10.1007/BF02096874.

(126) von Bergmann, J.; von Bergmann, H. Foucault Pendulum through Basic Geometry. *Am. J. Phys.* **2007**, *75* (10), 888–892. https://doi.org/10.1119/1.2757623.

(127) Meir, Y.; Gefen, Y.; Entin-Wohlman, O. Universal Effects of Spin-Orbit Scattering in Mesoscopic Systems. *Phys. Rev. Lett.* **1989**, *63* (7), 798–800. https://doi.org/10.1103/PhysRevLett.63.798.

(128) Loss, D.; Goldbart, P.; Balatsky, A. V. Berry's Phase and Persistent Charge and Spin Currents in Textured Mesoscopic Rings. *Phys. Rev. Lett.* **1990**, *65* (13), 1655–1658. https://doi.org/10.1103/PhysRevLett.65.1655.

(129) Nagasawa, F.; Frustaglia, D.; Saarikoski, H.; Richter, K.; Nitta, J. Control of the Spin Geometric Phase in Semiconductor Quantum Rings. *Nat. Commun.* **2013**, *4* (1), 2526. https://doi.org/10.1038/ncomms3526.

(130) Hannay, J. H. Angle Variable Holonomy in Adiabatic Excursion of an Integrable Hamiltonian. *J. Phys. A. Math. Gen.* **1985**, *18* (2), 221–230. https://doi.org/10.1088/0305-4470/18/2/011.

(131) Aharonov, Y.; Anandan, J. Phase Change during a Cyclic Quantum Evolution. *Phys. Rev. Lett.* **1987**, *58* (16), 1593–1596. https://doi.org/10.1103/PhysRevLett.58.1593.

(132) Zak, J. Berry's Phase for Energy Bands in Solids. *Phys. Rev. Lett.* **1989**, *62* (23), 2747–2750. https://doi.org/10.1103/PhysRevLett.62.2747.

(133) HC Longuet-Higgins, U Öpik, MHL Pryce, R. S. Studies of the Jahn-Teller Effect .II. The Dynamical Problem. *Proc. R. Soc. London. Ser. A. Math. Phys. Sci.* **1958**, *244* (1236), 1–16. https://doi.org/10.1098/rspa.1958.0022.

(134) Mead, C. A.; Truhlar, D. G. On the Determination of Born–Oppenheimer Nuclear Motion Wave Functions Including Complications Due to Conical Intersections and Identical Nuclei. *J. Chem. Phys.* **1979**, *70* (5), 2284–2296. https://doi.org/10.1063/1.437734.

(135) Wilson, K. G. Confinement of Quarks. *Phys. Rev. D* **1974**, *10* (8), 2445. https://doi.org/10.1103/PhysRevD.10.2445.

(136) Sonoda, H. Berry's Phase in Chiral Gauge Theories. *Nucl. Phys. B* **1986**, *266* (2), 410–422. https://doi.org/10.1016/0550-3213(86)90097-0.

(137) Dowker, J. S. A Gravitational Aharonov-Bohm Effect. *Nuovo Cim. B 1967 521* **2015**, *52* (1), 129–135. https://doi.org/10.1007/BF02710657.

(138) Houston, A. J. H.; Gradhand, M.; Dennis, M. R.; Guvendi, A.; Hassanabadi, H. A Gravitational Analogue of the Aharonov-Bohm Effect. *J. Phys. A. Math. Gen.* **1981**, *14* (9), 2353. https://doi.org/10.1088/0305-4470/14/9/030.

(139) Datta, D. P. Geometric Phase in Vacuum Instability: Applications in Quantum Cosmology. *Phys. Rev. D* **1993**, *48* (12), 5746. https://doi.org/10.1103/PhysRevD.48.5746.

(140) Berry, M. V.; Chambers, R. G.; Large, M. D.; Upstill, C.; Walmsley, J. C. *Wavefront Dislocations in the Aharonov-Bohm Effect and Its Water Wave Analogue*; IOP Publishing, 1980; Vol. 1, pp 154–162. https://doi.org/10.1088/0143-0807/1/3/008.

(141) Son, D. T.; Yamamoto, N. Berry Curvature, Triangle Anomalies, and the Chiral Magnetic Effect in Fermi Liquids. *Phys. Rev. Lett.* **2012**, *109* (18), 181602. https://doi.org/10.1103/PhysRevLett.109.181602.

(142) Yarkony, D. R. Diabolical Conical Intersections. *Rev. Mod. Phys.* **1996**, *68* (4), 985–1013. https://doi.org/10.1103/REVMODPHYS.68.985.

(143) Mead, C. A. The Geometric Phase in Molecular Systems. *Rev. Mod. Phys.* **1992**, *64* (1), 51–85. https://doi.org/10.1103/RevModPhys.64.51.

(144) Yuan, D.; Guan, Y.; Chen, W.; Zhao, H.; Yu, S.; Luo, C.; Tan, Y.; Xie, T.; Wang, X.; Sun, Z.; Zhang, D. H.; Yang, X. Observation of the Geometric Phase Effect in the h + HD → H2 + d Reaction. *Science (80-. ).* **2018**, *362* (6420), 1289–1293. https://doi.org/10.1126/SCIENCE.AAV1356.

(145) Kuppermann, A.; Wu, Y. S. M. The Geometric Phase Effect Shows up in Chemical Reactions. *Chem. Phys. Lett.* **1993**, *205* (6), 577–586.

(146) Kendrick, B. K.; Hazra, J.; Balakrishnan, N. The Geometric Phase Controls Ultracold Chemistry. *Nat. Commun.* **2015**, *6* (1), 7918. https://doi.org/10.1038/ncomms8918.

(147) Carollo, A.; Fuentes-Guridi, I.; Santos, M. F.; Vedral, V. Geometric Phase in Open Systems. *Phys. Rev. Lett.* **2003**, *90* (16). https://doi.org/10.1103/PHYSREVLETT.90.160402.

(148) Nazir, A.; Spiller, T. P.; Munro, W. J. Decoherence of Geometric Phase Gates. *Phys. Rev. A* **2002**, *65* (4), 042303. https://doi.org/10.1103/PhysRevA.65.042303.

(149) Yi, X. X.; Wang, L. C.; Wang, W. Geometric Phase in Dephasing Systems. *Phys. Rev. A* **2005**, *71* (4), 044101. https://doi.org/10.1103/PhysRevA.71.044101.

(150) Harney, H. L.; Heiss, W. D. Time Reversal and Exceptional Points. *Eur. Phys. J. D* **2004**, *29* (3), 429–432. https://doi.org/10.1140/EPJD/E2004-00049-7.

(151) Heiss, W. D. Exceptional Points – Their Universal Occurrence and Their Physical Significance. *Czechoslov. J. Phys.* **2004**, *54* (10), 1091–1099. https://doi.org/10.1023/B:CJOP.0000044009.17264.dc.

(152) Berry, M. V.; Dennis, M. R. The Optical Singularities of Birefringent Dichroic Chiral Crystals. *Proc. R. Soc. London. Ser. A Math. Phys. Eng. Sci.* **2003**, *459* (2033), 1261–1292. https://doi.org/10.1098/rspa.2003.1155.

(153) Garrison, J. C.; Wright, E. M. Complex Geometrical Phases for Dissipative Systems. *Phys. Lett. A* **1988**, *128* (3–4), 177–181. https://doi.org/10.1016/0375-9601(88)90905-X.

(154) Massar, S. Applications of the Complex Geometric Phase for Metastable Systems. *Phys. Rev. A* **1996**, *54* (6), 4770–4774. https://doi.org/10.1103/PhysRevA.54.4770.

(155) Deng, W.; Huang, X.; Lu, J.; Li, F.; Ma, J.; Chen, S.; Liu, Z. Acoustic Spin-1 Weyl Semimetal. *Sci. China Physics, Mech. Astron.* **2020**, *63* (8), 287032. https://doi.org/10.1007/s11433-020-1558-8.

(156) Heiss, W. D. Phases of Wave Functions and Level Repulsion. *Eur. Phys. J. D* **1999**, *7* (1), 1–4. https://doi.org/10.1007/S100530050339.

(157) Heiss, W. D. Repulsion of Resonance States and Exceptional Points. *Phys. Rev. E* **2000**, *61* (1), 929–932. https://doi.org/10.1103/PhysRevE.61.929.

(158) Lee, S.-Y.; Ryu, J.-W.; Kim, S. W.; Chung, Y. Geometric Phase around Multiple Exceptional Points. *Phys. Rev. A* **2012**, *85* (6), 064103. https://doi.org/10.1103/PhysRevA.85.064103.

(159) Mailybaev, A. A.; Kirillov, O. N.; Seyranian, A. P. Geometric Phase around Exceptional Points. *Phys. Rev. A* **2005**, *72* (1), 014104. https://doi.org/10.1103/PhysRevA.72.014104.

(160) Cohen, E.; Larocque, H.; Bouchard, F.; Nejadsattari, F.; Gefen, Y.; Karimi, E. Geometric Phase from Aharonov–Bohm to Pancharatnam–Berry and Beyond. *Nat. Rev. Phys.* **2019**, *1* (7), 437–449. https://doi.org/10.1038/S42254-019-0071-1.

(161) Jisha, C. P.; Nolte, S.; Alberucci, A. Geometric Phase in Optics: From Wavefront Manipulation to Waveguiding. *Laser Photonics Rev.* **2021**, *15* (10), 2100003. https://doi.org/10.1002/lpor.202100003.

(162) Wu, T. T.; Yang, C. N. Concept of Nonintegrable Phase Factors and Global Formulation of Gauge Fields. *Phys. Rev. D* **1975**, *12* (12), 3845–3857. https://doi.org/10.1103/PHYSREVD.12.3845.

(163) Nakahara, M. *Geometry, Topology, and Physics*; Graduate Student Series in Physics, 2003.

(164) Aidelsburger, M.; Nascimbene, S.; Goldman, N. Artificial Gauge Fields in Materials and Engineered Systems. *Comptes Rendus Phys.* **2018**, *19* (6), 394–432. https://doi.org/10.1016/j.crhy.2018.03.002.

(165) N Goldman, J. B. P. Z. Topological Quantum Matter with Ultracold Gases in Optical Lattices. *Nat. Phys.* **2016**, *12* (7), 639–645. https://doi.org/10.1038/nphys3803.

(166) Goldman, N.; Juzeliūnas, G.; Öhberg, P.; Spielman, I. B. Light-Induced Gauge Fields for Ultracold Atoms. *Reports Prog. Phys.* **2014**, *77* (12), 126401. https://doi.org/10.1088/0034-4885/77/12/126401.

(167) HAFEZI, M. SYNTHETIC GAUGE FIELDS WITH PHOTONS. *Int. J. Mod. Phys. B* **2014**, *28* (02), 1441002. https://doi.org/10.1142/S0217979214410021.

(168) Vozmediano, M. A. H.; Katsnelson, M. I.; Guinea, F. Gauge Fields in Graphene. *Phys. Rep.* **2010**, *496* (4–5), 109–148. https://doi.org/10.1016/j.physrep.2010.07.003.

(169) Ren, Y.; Qiao, Z.; Niu, Q. Topological Phases in Two-Dimensional Materials: A Review. *Reports Prog. Phys.* **2016**, *79* (6), 066501. https://doi.org/10.1088/0034-4885/79/6/066501.

(170) Yang, Z.; Gao, F.; Yang, Y.; Zhang, B. Strain-Induced Gauge Field and Landau Levels in Acoustic Structures. *Phys. Rev. Lett.* **2017**, *118* (19). https://doi.org/10.1103/PHYSREVLETT.118.194301.

(171) Wen, X.; Qiu, C.; Qi, Y.; Ye, L.; Ke, M.; Zhang, F.; Liu, Z. Acoustic Landau Quantization and Quantum-Hall-like Edge States. *Nat. Phys.* **2019**, *15* (4), 352–356. https://doi.org/10.1038/S41567-019-0446-3.

(172) Abbaszadeh, H.; Souslov, A.; Paulose, J.; Schomerus, H.; Vitelli, V. Sonic Landau Levels and Synthetic Gauge Fields in Mechanical Metamaterials. *Phys. Rev. Lett.* **2017**, *119* (19). https://doi.org/10.1103/PhysRevLett.119.195502.

(173) Brendel, C.; Peano, V.; Painter, O. J.; Marquardt, F. Pseudomagnetic Fields for Sound at the Nanoscale. *Proc. Natl. Acad. Sci.* **2017**, *114* (17), E3390–E3395. https://doi.org/10.1073/pnas.1615503114.

(174) Yan, M.; Deng, W.; Huang, X.; Wu, Y.; Yang, Y.; Lu, J.; Li, F.; Liu, Z. Pseudomagnetic Fields Enabled Manipulation of On-Chip Elastic Waves. *Phys. Rev. Lett.* **2021**, *127* (13), 136401. https://doi.org/10.1103/PHYSREVLETT.127.136401/FIGURES/4/MEDIUM.

(175) Yang, Y.; Ge, Y.; Li, R.; Lin, X.; Jia, D.; Guan, Y.; Yuan, S.; Sun, H.; Chong, Y.; Zhang, B. Demonstration of Negative Refraction Induced by Synthetic Gauge Fields. *Sci. Adv.* **2021**, *7* (50). https://doi.org/10.1126/sciadv.abj2062.

(176) Gao, T.; Estrecho, E.; Bliokh, K. Y.; Liew, T. C. H.; Fraser, M. D.; Brodbeck, S.; Kamp, M.; Schneider, C.; Höfling, S.; Yamamoto, Y.; Nori, F.; Kivshar, Y. S.; Truscott, A. G.; Dall, R. G.; Ostrovskaya, E. A. Observation of Non-Hermitian Degeneracies in a Chaotic Exciton-Polariton Billiard. *Nature* **2015**, *526* (7574), 554–558. https://doi.org/10.1038/nature15522.

(177) Estrecho, E.; Gao, T.; Brodbeck, S.; Kamp, M.; Schneider, C.; Höfling, S.; Truscott, A. G.; Ostrovskaya, E. A. Visualising Berry Phase and Diabolical Points in a Quantum Exciton-Polariton Billiard. *Sci. Rep.* **2016**, *6* (1), 37653. https://doi.org/10.1038/srep37653.

(178) Lim, H.-T.; Togan, E.; Kroner, M.; Miguel-Sanchez, J.; Imamoğlu, A. Electrically Tunable Artificial Gauge Potential for Polaritons. *Nat. Commun.* **2017**, *8* (1), 14540. https://doi.org/10.1038/ncomms14540.

(179) Polimeno, L.; Lerario, G.; De Giorgi, M.; De Marco, L.; Dominici, L.; Todisco, F.; Coriolano, A.; Ardizzone, V.; Pugliese, M.; Prontera, C. T.; Maiorano, V.; Moliterni, A.; Giannini, C.; Olieric, V.; Gigli, G.; Ballarini, D.; Xiong, Q.; Fieramosca, A.; Solnyshkov, D. D.; Malpuech, G.; Sanvitto, D. Tuning of the Berry Curvature in 2D Perovskite Polaritons. *Nat. Nanotechnol.* **2021**, *16* (12), 1349–1354. https://doi.org/10.1038/s41565-021-00977-2.

(180) Karzig, T.; Bardyn, C. E.; Lindner, N. H.; Refael, G. Topological Polaritons. *Phys. Rev. X* **2015**, *5* (3), 031001. https://doi.org/10.1103/PhysRevX.5.031001.

(181) L. D. Landau and E. M. Lifshitz. *The Classical Theory of Fields*; Pergamon Press Ltd.: Oxford, 1971.

(182) Žutić, I.; Fabian, J.; Das Sarma, S. Spintronics: Fundamentals and Applications. *Rev. Mod. Phys.*

**2004**, *76* (2), 323–410. https://doi.org/10.1103/RevModPhys.76.323.

(183) Nagaosa, N.; Sinova, J.; Onoda, S.; MacDonald, A. H.; Ong, N. P. Anomalous Hall Effect. *Rev. Mod. Phys.* **2010**, *82* (2), 1539–1592. https://doi.org/10.1103/RevModPhys.82.1539.

(184) von Klitzing, K. The Quantized Hall Effect. *Rev. Mod. Phys.* **1986**, *58* (3), 519–531. https://doi.org/10.1103/RevModPhys.58.519.

(185) Oka, T.; Aoki, H. Photovoltaic Hall Effect in Graphene. *Phys. Rev. B* **2009**, *79* (8), 081406. https://doi.org/10.1103/PhysRevB.79.081406.

(186) Lindner, N. H.; Refael, G.; Galitski, V. Floquet Topological Insulator in Semiconductor Quantum Wells. *Nat. Phys.* **2011**, *7* (6), 490–495. https://doi.org/10.1038/nphys1926.

(187) Cayssol, J.; Dóra, B.; Simon, F.; Moessner, R. Floquet Topological Insulators. *Phys. status solidi - Rapid Res. Lett.* **2013**, *7* (1–2), 101–108. https://doi.org/10.1002/pssr.201206451.

(188) Jotzu, G.; Messer, M.; Desbuquois, R.; Lebrat, M.; Uehlinger, T.; Greif, D.; Esslinger, T. Experimental Realization of the Topological Haldane Model with Ultracold Fermions. *Nature* **2014**, *515* (7526), 237–240. https://doi.org/10.1038/nature13915.

(189) Mulder, R. A. Gauge-Underdetermination and Shades of Locality in the Aharonov–Bohm Effect. *Found. Phys.* **2021**, *51* (2), 48. https://doi.org/10.1007/s10701-021-00446-9.

(190) Wittig, C. Geometric Phase and Gauge Connection in Polyatomic Molecules. *Phys. Chem. Chem. Phys.* **2012**, *14* (18), 6409. https://doi.org/10.1039/c2cp22974a.

(191) Dittrich, W.; Reuter, M. *Classical and Quantum Dynamics*; Springer Berlin Heidelberg: Berlin, Heidelberg, 1992. https://doi.org/10.1007/978-3-642-97921-7.

(192) Peshkin, M. The Aharonov-Bohm Effect: Why It Cannot Be Eliminated from Quantum Mechanics. *Phys. Rep.* **1981**, *80* (6), 375–386. https://doi.org/10.1016/0370-1573(81)90133-2.

(193) Olariu, S.; Popescu, I. I. The Quantum Effects of Electromagnetic Fluxes. *Rev. Mod. Phys.* **1985**, *57* (2), 339–436. https://doi.org/10.1103/RevModPhys.57.339.

(194) Tollaksen, J.; Khrenniko, A. Y.; Jaeger, G.; Schlosshauer, M.; Weihs, G. Dynamical Quantum Non-Locality. In *AIP Conference Proceedings*; 2011; Vol. 1327, pp 269–288. https://doi.org/10.1063/1.3578708.

(195) Y Aharonov, E Cohen, D. R. Nonlocality of the Aharonov–Bohm Effect. *Phys. Rev. A* **2016**, *93*, 042110. https://doi.org/http://www.jstor.org/stable/188368.

(196) Tonomura, A.; Osakabe, N.; Matsuda, T.; Kawasaki, T.; Endo, J.; Yano, S.; Yamada, H. Evidence for Aharonov-Bohm Effect with Magnetic Field Completely Shielded from Electron Wave. *Phys. Rev. Lett.* **1986**, *56* (8), 792–795. https://doi.org/10.1103/PhysRevLett.56.792.

(197) Chambers, R. G. Shift of an Electron Interference Pattern by Enclosed Magnetic Flux. *Phys. Rev. Lett.* **1960**, *5* (1), 3–5. https://doi.org/10.1103/PhysRevLett.5.3.

(198) Noguchi, A.; Shikano, Y.; Toyoda, K.; Urabe, S. Aharonov–Bohm Effect in the Tunnelling of a Quantum Rotor in a Linear Paul Trap. *Nat. Commun.* **2014**, *5* (1), 3868. https://doi.org/10.1038/ncomms4868.

(199) Webb, R. A.; Washburn, S.; Umbach, C. P.; Laibowitz, R. B. Observation of h/e Aharonov-Bohm Oscillations in Normal-Metal Rings. *Phys. Rev. Lett.* **1985**, *54* (25), 2696–2699. https://doi.org/10.1103/PhysRevLett.54.2696.

(200) Yacoby, A.; Heiblum, M.; Mahalu, D.; Shtrikman, H. Coherence and Phase Sensitive Measurements in a Quantum Dot. *Phys. Rev. Lett.* **1995**, *74* (20), 4047–4050. https://doi.org/10.1103/PhysRevLett.74.4047.

(201) Bachtold, A. Aharonov–Bohm Oscillations in Carbon Nanotubes. *Nature* **1999**, *397*, 673–675.

(202) Ji, Y.; Chung, Y.; Sprinzak, D.; Heiblum, M.; Mahalu, D.; Shtrikman, H. An Electronic Mach–Zehnder Interferometer. *Nature* **2003**, *422* (6930), 415–418. https://doi.org/10.1038/nature01503.

(203) Aidelsburger, M.; Atala, M.; Nascimbène, S.; Trotzky, S.; Chen, Y.-A.; Bloch, I. Experimental Realization of Strong Effective Magnetic Fields in an Optical Lattice. *Phys. Rev. Lett.* **2011**, *107* (25), 255301. https://doi.org/10.1103/PhysRevLett.107.255301.

(204) Duca, L. An Aharonov–Bohm Interferometer for Determining Bloch Band Topology. *Science (80-. ).* **2015**, *347*, 288–292.

(205) Peng, H.; Lai, K.; Kong, D.; Meister, S.; Chen, Y.; Qi, X. L.; Zhang, S. C.; Shen, Z. X.; Cui, Y. Aharonov-Bohm Interference in Topological Insulator Nanoribbons. *Nat. Mater.* **2010**, *9* (3), 225–229. https://doi.org/10.1038/nmat2609.

(206) Bardarson, J. H.; Brouwer, P. W.; Moore, J. E. Aharonov-Bohm Oscillations in Disordered Topological Insulator Nanowires. *Phys. Rev. Lett.* **2010**, *105* (15), 156803. https://doi.org/10.1103/PhysRevLett.105.156803.

(207) Zhang, Y.; Vishwanath, A. Anomalous Aharonov-Bohm Conductance Oscillations from Topological Insulator Surface States. *Phys. Rev. Lett.* **2010**, *105* (20), 206601. https://doi.org/10.1103/PhysRevLett.105.206601.

(208) Roux, P.; de Rosny, J.; Tanter, M.; Fink, M. The Aharonov-Bohm Effect Revisited by an Acoustic Time-Reversal Mirror. *Phys. Rev. Lett.* **1997**, *79* (17), 3170–3173. https://doi.org/10.1103/PhysRevLett.79.3170.

(209) Aharonov, Y.; Casher, A. Topological Quantum Effects for Neutral Particles. *Phys. Rev. Lett.* **1984**, *53* (4), 319–321. https://doi.org/10.1103/PhysRevLett.53.319.

(210) Cimmino, A.; Opat, G. I.; Klein, A. G.; Kaiser, H.; Werner, S. A.; Arif, M.; Clothier, R. Observation of the Topological Aharonov-Casher Phase Shift by Neutron Interferometry. *Phys. Rev. Lett.* **1989**, *63* (4), 380–383. https://doi.org/10.1103/PhysRevLett.63.380.

(211) Elion, W. J.; Wachters, J. J.; Sohn, L. L.; Mooij, J. E. The Aharonov-Casher Effect for Vortices in Josephson-Junction Arrays. *Phys. B Condens. Matter* **1994**, *203* (3–4), 497–503. https://doi.org/10.1016/0921-4526(94)90102-3.

(212) König, M.; Tschetschetkin, A.; Hankiewicz, E. M.; Sinova, J.; Hock, V.; Daumer, V.; Schäfer, M.; Becker, C. R.; Buhmann, H.; Molenkamp, L. W. Direct Observation of the Aharonov-Casher Phase. *Phys. Rev. Lett.* **2006**, *96* (7), 076804. https://doi.org/10.1103/PhysRevLett.96.076804.

(213) Vaidman, L. Role of Potentials in the Aharonov-Bohm Effect. *Phys. Rev. A* **2012**, *86* (4), 040101. https://doi.org/10.1103/PhysRevA.86.040101.

(214) Aharonov, Y.; Cohen, E.; Rohrlich, D. Comment on "Role of Potentials in the Aharonov-Bohm Effect." *Phys. Rev. A* **2015**, *92* (2), 026101. https://doi.org/10.1103/PhysRevA.92.026101.

(215) Pearle, P.; Rizzi, A. Quantum-Mechanical Inclusion of the Source in the Aharonov-Bohm Effects. *Phys. Rev. A* **2017**, *95* (5), 052123. https://doi.org/10.1103/PhysRevA.95.052123.

(216) Pearle, P.; Rizzi, A. Quantized Vector Potential and Alternative Views of the Magnetic Aharonov-

Bohm Phase Shift. *Phys. Rev. A* **2017**, *95* (5), 052124. https://doi.org/10.1103/PhysRevA.95.052124.

(217) Maudlin, T. Ontological Clarity via Canonical Presentation: Electromagnetism and the Aharonov-Bohm Effect. *Entropy* **2018**, *20* (6), 465. https://doi.org/10.3390/e20060465.

(218) Li, B.; Hewak, D. W.; Wang, Q. J. The Transition from Quantum Field Theory to One-Particle Quantum Mechanics and a Proposed Interpretation of Aharonov–Bohm Effect. *Found. Phys.* **2018**, *48* (7), 837–852. https://doi.org/10.1007/s10701-018-0191-y.

(219) Born, M.; Fock, V. Beweis Des Adiabatensatzes. *Zeitschrift für Phys.* **1928**, *51* (3–4), 165–180. https://doi.org/10.1007/BF01343193.

(220) Haldane, F. D. M. Berry Curvature on the Fermi Surface: Anomalous Hall Effect as a Topological Fermi-Liquid Property. *Phys. Rev. Lett.* **2004**, *93* (20), 206602. https://doi.org/10.1103/PhysRevLett.93.206602.

(221) Zhang, S.-C.; Hu, J. A Four-Dimensional Generalization of the Quantum Hall Effect. *Science (80-. ).* **2001**, *294* (5543), 823–828. https://doi.org/10.1126/science.294.5543.823.

(222) Berry, M. V. The Adiabatic Phase and Pancharatnam's Phase for Polarized Light. *J. Mod. Opt.* **1987**, *34* (11), 1401–1407. https://doi.org/10.1080/09500348714551321.

(223) Berry, M. Pancharatnam, Virtuoso of the Poincaré Sphere: An Appreciation. In *A Half-Century of Physical Asymptotics and Other Diversions*; WORLD SCIENTIFIC, 2017; pp 548–551. https://doi.org/10.1142/9789813221215_0044.

(224) Simon, R.; Kimble, H. J.; Sudarshan, E. C. G. Evolving Geometric Phase and Its Dynamical Manifestation as a Frequency Shift: An Optical Experiment. *Phys. Rev. Lett.* **1988**, *61* (1), 19–22. https://doi.org/10.1103/PHYSREVLETT.61.19.

(225) Bhandari, R. Polarization of Light and Topological Phases. *Phys. Rep.* **1997**, *281* (1), 1–64. https://doi.org/10.1016/S0370-1573(96)00029-4.

(226) Berry, M. V.; Klein, S. Geometric Phases from Stacks of Crystal Plates. *J. Mod. Opt.* **1996**, *43* (1), 165–180. https://doi.org/10.1080/09500349608232731.

(227) Bhandari, R.; Samuel, J. Observation of Topological Phase by Use of a Laser Interferometer. *Phys. Rev. Lett.* **1988**, *60* (13), 1211–1213. https://doi.org/10.1103/PHYSREVLETT.60.1211.

(228) Chyba, T. H.; Simon, R.; Wang, L. J.; Mandel, L. Measurement of the Pancharatnam Phase for a Light Beam. *Opt. Lett.* **1988**, *13* (7), 562. https://doi.org/10.1364/OL.13.000562.

(229) Onoda, M.; Murakami, S.; Nagaosa, N. Hall Effect of Light. *Phys. Rev. Lett.* **2004**, *93* (8), 083901. https://doi.org/10.1103/PhysRevLett.93.083901.

(230) Petersen, J.; Volz, J.; Rauschenbeutel, A. Chiral Nanophotonic Waveguide Interface Based on Spin-Orbit Interaction of Light. *Science (80-. ).* **2014**, *346* (6205), 67–71. https://doi.org/10.1126/science.1257671.

(231) Lin, J.; Mueller, J. P. B.; Wang, Q.; Yuan, G.; Antoniou, N.; Yuan, X.; Capasso, F. Polarization-Controlled Tunable Directional Coupling of Surface Plasmon Polaritons. *Science (80-. ).* **2013**, *340* (6130), 331–334. https://doi.org/10.1126/science.1233746.

(232) Rodríguez-Fortuño, F. J.; Marino, G.; Ginzburg, P.; O'Connor, D.; Martínez, A.; Wurtz, G. A.; Zayats, A. V. Near-Field Interference for the Unidirectional Excitation of Electromagnetic Guided Modes. *Science (80-. ).* **2013**, *340* (6130), 328–330. https://doi.org/10.1126/science.1233739.

(233) Lee, S.-Y.; Lee, I.-M.; Park, J.; Oh, S.; Lee, W.; Kim, K.-Y.; Lee, B. Role of Magnetic Induction Currents in Nanoslit Excitation of Surface Plasmon Polaritons. *Phys. Rev. Lett.* **2012**, *108* (21), 213907. https://doi.org/10.1103/PhysRevLett.108.213907.

(234) Huang, L.; Chen, X.; Bai, B.; Tan, Q.; Jin, G.; Zentgraf, T.; Zhang, S. Helicity Dependent Directional Surface Plasmon Polariton Excitation Using a Metasurface with Interfacial Phase Discontinuity. *Light Sci. Appl.* **2013**, *2* (3), e70–e70. https://doi.org/10.1038/lsa.2013.26.

(235) Zhang, Q.; Hu, G.; Ma, W.; Li, P.; Krasnok, A.; Hillenbrand, R.; Alù, A.; Qiu, C.-W. Interface Nano-Optics with van Der Waals Polaritons. *Nature* **2021**, *597* (7875), 187–195. https://doi.org/10.1038/s41586-021-03581-5.

(236) Bliokh, K. Y.; Smirnova, D.; Nori, F. Quantum Spin Hall Effect of Light. *Science (80-. ).* **2015**, *348* (6242), 1448–1451. https://doi.org/10.1126/science.aaa9519.

(237) Mansuripur, M.; Zakharian, A. R.; Wright, E. M. Spin and Orbital Angular Momenta of Light Reflected from a Cone. *Phys. Rev. A.* **2011**, *84* (3), 033813. https://doi.org/10.1103/physreva.84.033813.

(238) Bouchard, F.; Mand, H.; Mirhosseini, M.; Karimi, E.; Boyd, R. W. Achromatic Orbital Angular Momentum Generator. *N. J. Phys.* **2014**, *16*, 123006. https://doi.org/10.1088/1367-2630/16/12/123006.

(239) Radwell, N.; Hawley, R. D.; Götte, J. B.; Franke-Arnold, S. Achromatic Vector Vortex Beams from a Glass Cone. *Nat. Commun. 2016 71* **2016**, *7* (1), 1–6. https://doi.org/10.1038/ncomms10564.

(240) Devlin, R. C.; Ambrosio, A.; Rubin, N. A.; Balthasar Mueller, J. P.; Capasso, F. Arbitrary Spin-to–Orbital Angular Momentum Conversion of Light. *Science (80-. ).* **2017**, *358* (6365), 896–901. https://doi.org/10.1126/SCIENCE.AAO5392.

(241) Yao, A. M.; Padgett, M. J. Orbital Angular Momentum: Origins, Behavior and Applications. *Adv. Opt. Photonics* **2011**, *3* (2), 161. https://doi.org/10.1364/aop.3.000161.

(242) Allen, L.; Beijersbergen, M. W.; Spreeuw, R. J. C.; Woerdman, J. P. Orbital Angular Momentum of Light and the Transformation of Laguerre-Gaussian Laser Modes. *Phys. Rev. A* **1992**, *45* (11), 8185–8189. https://doi.org/10.1103/PhysRevA.45.8185.

(243) Bliokh, K. Y.; Rodríguez-Fortuño, F. J.; Nori, F.; Zayats, A. V. Spin–Orbit Interactions of Light. *Nat. Photonics* **2015**, *9* (12), 796–808. https://doi.org/10.1038/nphoton.2015.201.

(244) Bliokh, K. Y.; Bliokh, Y. P. Conservation of Angular Momentum, Transverse Shift, and Spin Hall Effect in Reflection and Refraction of an Electromagnetic Wave Packet. *Phys. Rev. Lett.* **2006**, *96* (7), 073903. https://doi.org/10.1103/PhysRevLett.96.073903.

(245) Bliokh, K. Y.; Nori, F. Transverse and Longitudinal Angular Momenta of Light. *Phys. Rep.* **2015**, *592*, 1–38. https://doi.org/10.1016/j.physrep.2015.06.003.

(246) Bliokh, K. Y.; Niv, A.; Kleiner, V.; Hasman, E. Geometrodynamics of Spinning Light. *Nat. Photonics* **2008**, *2* (12), 748–753. https://doi.org/10.1038/nphoton.2008.229.

(247) Beijersbergen, M. W.; Coerwinkel, R. P. C.; Kristensen, M.; Woerdman, J. P. Helical-Wavefront Laser Beams Produced with a Spiral Phaseplate. *Opt. Commun.* **1994**, *112* (5–6), 321–327. https://doi.org/10.1016/0030-4018(94)90638-6.

(248) Bazhenov, V. Y.; Soskin, M. S.; Vasnetsov, M. V. Screw Dislocations in Light Wavefronts. *J.*

*Mod. Opt.* **1992**, *39* (5), 985. https://doi.org/10.1080/09500349214551011.

(249) Marrucci, L.; Manzo, C.; Paparo, D. Optical Spin-to-Orbital Angular Momentum Conversion in Inhomogeneous Anisotropic Media. *Phys. Rev. Lett.* **2006**, *96* (16), 163905. https://doi.org/10.1103/PhysRevLett.96.163905.

(250) Liberman, V. S.; Zel'dovich, B. Y. Spin-Orbit Interaction of a Photon in an Inhomogeneous Medium. *Phys. Rev. A* **1992**, *46* (8), 5199–5207. https://doi.org/10.1103/PhysRevA.46.5199.

(251) Brasselet, E.; Murazawa, N.; Misawa, H.; Juodkazis, S. Optical Vortices from Liquid Crystal Droplets. *Phys. Rev. Lett.* **2009**, *103* (10), 103903. https://doi.org/10.1103/PhysRevLett.103.103903.

(252) Cardano, F.; Karimi, E.; Slussarenko, S.; Marrucci, L.; Lisio, C. De; Santamato, E. Polarization Pattern of Vector Vortex Beams Generated by Q-Plates with Different Topological Charges. *Appl. Opt.* **2012**, *51* (10), C1. https://doi.org/10.1364/ao.51.0000c1.

(253) Beresna, M.; Gecevičius, M.; Kazansky, P. G.; Gertus, T. Radially Polarized Optical Vortex Converter Created by Femtosecond Laser Nanostructuring of Glass. *Appl. Phys. Lett.* **2011**, *98* (20), 201101. https://doi.org/10.1063/1.3590716.

(254) Quabis, S.; Dorn, R.; Leuchs, G. Generation of a Radially Polarized Doughnut Mode of High Quality. *Appl. Phys. B* **2005**, *81* (5), 597–600. https://doi.org/10.1007/s00340-005-1887-1.

(255) Cardano, F.; Karimi, E.; Marrucci, L.; de Lisio, C.; Santamato, E. Generation and Dynamics of Optical Beams with Polarization Singularities. *Opt. Express* **2013**, *21* (7), 8815. https://doi.org/10.1364/OE.21.008815.

(256) Bauer, T.; Banzer, P.; Karimi, E.; Orlov, S.; Rubano, A.; Marrucci, L.; Santamato, E.; Boyd, R. W.; Leuchs, G. Observation of Optical Polarization Möbius Strips. *Science (80-. ).* **2015**, *347* (6225), 964–966. https://doi.org/10.1126/SCIENCE.1260635.

(257) Karimi, E.; Schulz, S. A.; De Leon, I.; Qassim, H.; Upham, J.; Boyd, R. W. Generating Optical Orbital Angular Momentum at Visible Wavelengths Using a Plasmonic Metasurface. *Light Sci. Appl.* **2014**, *3* (5), e167–e167. https://doi.org/10.1038/lsa.2014.48.

(258) Bouchard, F.; De Leon, I.; Schulz, S. A.; Upham, J.; Karimi, E.; Boyd, R. W. Optical Spin-to-Orbital Angular Momentum Conversion in Ultra-Thin Metasurfaces with Arbitrary Topological Charges. *Appl. Phys. Lett.* **2014**, *105* (10), 101905. https://doi.org/10.1063/1.4895620.

(259) POST, E. J. Sagnac Effect. *Rev. Mod. Phys.* **1967**, *39* (2), 475–493. https://doi.org/10.1103/RevModPhys.39.475.

(260) Gustavson, T. L.; Bouyer, P.; Kasevich, M. A. Precision Rotation Measurements with an Atom Interferometer Gyroscope. *Phys. Rev. Lett.* **1997**, *78* (11), 2046–2049. https://doi.org/10.1103/PHYSREVLETT.78.2046.

(261) Lenef, A.; Hammond, T. D.; Smith, E. T.; Chapman, M. S.; Rubenstein, R. A.; Pritchard, D. E. Rotation Sensing with an Atom Interferometer. *Phys. Rev. Lett.* **1997**, *78* (5), 760–763. https://doi.org/10.1103/PHYSREVLETT.78.760.

(262) Jing, H.; Lü, H.; Özdemir, S. K.; Carmon, T.; Nori, F. Nanoparticle Sensing with a Spinning Resonator. *Optica* **2018**, *5* (11), 1424. https://doi.org/10.1364/optica.5.001424.

(263) Yang, Y.; Peng, C.; Zhu, D.; Buljan, H.; Joannopoulos, J. D.; Zhen, B.; Soljačić, M. Synthesis and Observation of Non-Abelian Gauge Fields in Real Space. *Science (80-. ).* **2019**, *365* (6457), 1021–

1025. https://doi.org/10.1126/SCIENCE.AAY3183.

(264) Matos, G. C.; Souza, R. de M. e; Neto, P. A. M.; Impens, F. Quantum Vacuum Sagnac Effect. *Phys. Rev. Lett.* **2021**, *127* (27), 270401. https://doi.org/10.1103/PhysRevLett.127.270401.

(265) Fu, Y.; Shen, C.; Zhu, X.; Li, J.; Liu, Y.; Cummer, S. A.; Xu, Y. Sound Vortex Diffraction via Topological Charge in Phase Gradient Metagratings. *Sci. Adv.* **2020**, *6* (40). https://doi.org/10.1126/sciadv.aba9876.

(266) Liu, B.; Wei, Q.; Su, Z.; Wang, Y.; Huang, L. Multifunctional Acoustic Holography Based on Compact Acoustic Geometric-Phase Meta-Array. *J. Appl. Phys.* **2022**, *131* (18), 185108. https://doi.org/10.1063/5.0085562.

(267) Liu, B.; Su, Z.; Zeng, Y.; Wang, Y.; Huang, L.; Zhang, S. Acoustic Geometric-Phase Meta-Array. *New J. Phys.* **2021**, *23* (11), 113026. https://doi.org/10.1088/1367-2630/ac33f2.

(268) Hou, Z.; Ding, H.; Wang, N.; Fang, X.; Li, Y. Acoustic Vortices via Nonlocal Metagratings. *Phys. Rev. Appl.* **2021**, *16* (1), 014002. https://doi.org/10.1103/PhysRevApplied.16.014002.

(269) Zou, Z.; Lirette, R.; Zhang, L. Orbital Angular Momentum Reversal and Asymmetry in Acoustic Vortex Beam Reflection. *Phys. Rev. Lett.* **2020**, *125* (7), 074301. https://doi.org/10.1103/PhysRevLett.125.074301.

(270) Baresch, D.; Thomas, J.-L.; Marchiano, R. Orbital Angular Momentum Transfer to Stably Trapped Elastic Particles in Acoustical Vortex Beams. *Phys. Rev. Lett.* **2018**, *121* (7), 074301. https://doi.org/10.1103/PhysRevLett.121.074301.

(271) Marzo, A.; Caleap, M.; Drinkwater, B. W. Acoustic Virtual Vortices with Tunable Orbital Angular Momentum for Trapping of Mie Particles. *Phys. Rev. Lett.* **2018**, *120* (4), 044301. https://doi.org/10.1103/PhysRevLett.120.044301.

(272) Li, J.; Crivoi, A.; Peng, X.; Shen, L.; Pu, Y.; Fan, Z.; Cummer, S. A. Three Dimensional Acoustic Tweezers with Vortex Streaming. *Commun. Phys.* **2021**, *4* (1), 113. https://doi.org/10.1038/s42005-021-00617-0.

(273) Burns, L.; Bliokh, K. Y.; Nori, F.; Dressel, J. Acoustic versus Electromagnetic Field Theory: Scalar, Vector, Spinor Representations and the Emergence of Acoustic Spin. *New J. Phys.* **2020**, *22* (5), 053050. https://doi.org/10.1088/1367-2630/ab7f91.

(274) Bliokh, K. Y.; Nori, F. Spin and Orbital Angular Momenta of Acoustic Beams. *Phys. Rev. B* **2019**, *99* (17), 174310. https://doi.org/10.1103/PhysRevB.99.174310.

(275) Shi, C.; Zhao, R.; Long, Y.; Yang, S.; Wang, Y.; Chen, H.; Ren, J.; Zhang, X. Observation of Acoustic Spin. *Natl. Sci. Rev.* **2019**, *6* (4), 707–712. https://doi.org/10.1093/nsr/nwz059.

(276) Toftul, I. D.; Bliokh, K. Y.; Petrov, M. I.; Nori, F. Acoustic Radiation Force and Torque on Small Particles as Measures of the Canonical Momentum and Spin Densities. *Phys. Rev. Lett.* **2019**, *123* (18), 183901. https://doi.org/10.1103/PhysRevLett.123.183901.

(277) Ge, H.; Xu, X.-Y.; Liu, L.; Xu, R.; Lin, Z.-K.; Yu, S.-Y.; Bao, M.; Jiang, J.-H.; Lu, M.-H.; Chen, Y.-F. Observation of Acoustic Skyrmions. *Phys. Rev. Lett.* **2021**, *127* (14), 144502. https://doi.org/10.1103/PhysRevLett.127.144502.

(278) Wang, S.; Zhang, G.; Wang, X.; Tong, Q.; Li, J.; Ma, G. Spin-Orbit Interactions of Transverse Sound. *Nat. Commun.* **2021**, *12* (1), 6125. https://doi.org/10.1038/s41467-021-26375-9.

(279) Bliokh, K. Y.; Nori, F. Transverse Spin and Surface Waves in Acoustic Metamaterials. *Phys. Rev.*

*B* **2019**, *99* (2), 020301. https://doi.org/10.1103/PhysRevB.99.020301.

(280) Bliokh, K. Y.; Nori, F. Klein-Gordon Representation of Acoustic Waves and Topological Origin of Surface Acoustic Modes. *Phys. Rev. Lett.* **2019**, *123* (5), 054301. https://doi.org/10.1103/PhysRevLett.123.054301.

(281) Long, Y.; Zhang, D.; Yang, C.; Ge, J.; Chen, H.; Ren, J. Realization of Acoustic Spin Transport in Metasurface Waveguides. *Nat. Commun.* **2020**, *11* (1), 4716. https://doi.org/10.1038/s41467-020-18599-y.

(282) Long, Y.; Ren, J.; Chen, H. Intrinsic Spin of Elastic Waves. *Proc. Natl. Acad. Sci.* **2018**, *115* (40), 9951–9955. https://doi.org/10.1073/pnas.1808534115.

(283) Chaplain, G. J.; De Ponti, J. M.; Craster, R. V. Elastic Orbital Angular Momentum. *Phys. Rev. Lett.* **2022**, *128* (6), 064301. https://doi.org/10.1103/PhysRevLett.128.064301.

(284) Yuan, W.; Yang, C.; Zhang, D.; Long, Y.; Pan, Y.; Zhong, Z.; Chen, H.; Zhao, J.; Ren, J. Observation of Elastic Spin with Chiral Meta-Sources. *Nat. Commun.* **2021**, *12* (1), 6954. https://doi.org/10.1038/s41467-021-27254-z.

(285) Bliokh, K. Y.; Alonso, M. A.; Sugic, D.; Perrin, M.; Nori, F.; Brasselet, E. Polarization Singularities and Möbius Strips in Sound and Water-Surface Waves. *Phys. Fluids* **2021**, *33* (7), 077122. https://doi.org/10.1063/5.0056333.

(286) Bliokh, K. Y.; Punzmann, H.; Xia, H.; Nori, F.; Shats, M. Field Theory Spin and Momentum in Water Waves. *Sci. Adv.* **2022**, *8* (3). https://doi.org/10.1126/sciadv.abm1295.

(287) Westermayr, J.; Marquetand, P. Machine Learning for Electronically Excited States of Molecules. *Chem. Rev.* **2021**, *121* (16), 9873–9926. https://doi.org/10.1021/ACS.CHEMREV.0C00749.

(288) Jayachander Rao, B.; Padmanaban, R.; Mahapatra, S. Nonadiabatic Quantum Wave Packet Dynamics of H + H2 (HD) Reactions. *Chem. Phys.* **2007**, *333* (2–3), 135–147. https://doi.org/10.1016/J.CHEMPHYS.2007.01.012.

(289) Wu, Y. S. M.; Kuppermann, A. Prediction of the Effect of the Geometric Phase on Product Rotational State Distributions and Integral Cross Sections. *Chem. Phys. Lett.* **1993**, *201* (1–4), 178–186.

(290) Fábri, C.; Halász, G. J.; Cederbaum, L. S.; Vibók, Á. Born-Oppenheimer Approximation in Optical Cavities: From Success to Breakdown. *Chem. Sci.* **2021**, *12* (4), 1251–1258. https://doi.org/10.1039/d0sc05164k.

(291) Serrano-Andrés, L.; Merchán, M. Quantum Chemistry of the Excited State: 2005 Overview. *J. Mol. Struct. THEOCHEM* **2005**, *729* (1–2), 99–108. https://doi.org/10.1016/j.theochem.2005.03.020.

(292) Domcke, W.; Yarkony, D. R.; Köppel, H. *Conical Intersections*; Advanced Series in Physical Chemistry; WORLD SCIENTIFIC, 2004; Vol. 15. https://doi.org/10.1142/5406.

(293) Devine, J. A.; Weichman, M. L.; Zhou, X.; Ma, J.; Jiang, B.; Guo, H.; Neumark, D. M. Non-Adiabatic Effects on Excited States of Vinylidene Observed with Slow Photoelectron Velocity-Map Imaging. *J. Am. Chem. Soc.* **2016**, *138* (50), 16417–16425. https://doi.org/10.1021/JACS.6B10233/SUPPL_FILE/JA6B10233_SI_001.PDF.

(294) Xie, Y.; Zhao, H.; Wang, Y.; Huang, Y.; Wang, T.; Xu, X.; Xiao, C.; Sun, Z.; Zhang, D. H.; Yang, X. Quantum Interference in H + HD → H 2 + D between Direct Abstraction and Roaming

Insertion Pathways. *Science (80-. ).* **2020**, *368* (6492), 767–771. https://doi.org/10.1126/science.abb1564.

(295) Jankunas, J.; Sneha, M.; Zare, R. N.; Bouakline, F.; Althorpe, S. C.; Herráez-Aguilar, D.; Aoiz, F. J. Is the Simplest Chemical Reaction Really so Simple? *Proc. Natl. Acad. Sci.* **2014**, *111* (1), 15–20. https://doi.org/10.1073/pnas.1315725111.

(296) Polli, D.; Altoè, P.; Weingart, O.; Spillane, K. M.; Manzoni, C.; Brida, D.; Tomasello, G.; Orlandi, G.; Kukura, P.; Mathies, R. A.; Garavelli, M.; Cerullo, G. Conical Intersection Dynamics of the Primary Photoisomerization Event in Vision. *Nature* **2010**, *467* (7314), 440–443. https://doi.org/10.1038/NATURE09346.

(297) Schultz, T.; Samoylova, E.; Radloff, W.; Hertel, I. V.; Sobolewski, A. L.; Domcke, W. Efficient Deactivation of a Model Base Pair via Excited-State Hydrogen Transfer. *Science (80-. ).* **2004**, *306* (5702), 1765–1768. https://doi.org/10.1126/science.1104038.

(298) Zhang, Y.; Yuan, K.; Yu, S.; Yang, X. Highly Rotationally Excited CH 3 from Methane Photodissociation through Conical Intersection Pathway. *J. Phys. Chem. Lett.* **2010**, *1* (2), 475–479. https://doi.org/10.1021/jz900303e.

(299) Herzberg, G.; Longuet-Higgins, H. C. Intersection of Potential Energy Surfaces in Polyatomic Molecules. *Discuss. Faraday Soc.* **1963**, *35*, 77. https://doi.org/10.1039/df9633500077.

(300) The Intersection of Potential Energy Surfaces in Polyatomic Molecules. *Proc. R. Soc. London. A. Math. Phys. Sci.* **1975**, *344* (1637), 147–156. https://doi.org/10.1098/rspa.1975.0095.

(301) Lepetit, B.; Kuppermann, A. Numerical Study of the Geometric Phase in the H+H2 Reaction. *Chem. Phys. Lett.* **1990**, *166* (5–6), 581–588. https://doi.org/10.1016/0009-2614(90)87154-J.

(302) Kendrick, B. K.; Hazra, J.; Balakrishnan, N. Geometric Phase Effects in the Ultracold D + HD $\rightarrow$ D + HD and D + HD $\leftrightarrow$ H + D 2 Reactions. *New J. Phys.* **2016**, *18* (12), 123020. https://doi.org/10.1088/1367-2630/aa4fd2.

(303) Kendrick, B. K. Geometric Phase Effects in the H+D2→HD+D Reaction. *J. Chem. Phys.* **2000**, *112* (13), 5679–5704. https://doi.org/10.1063/1.481143.

(304) Kendrick, B. K. Geometric Phase Effects in Chemical Reaction Dynamics and Molecular Spectra. *J. Phys. Chem. A* **2003**, *107* (35), 6739–6756. https://doi.org/10.1021/jp021865x.

(305) Bañares, L.; Aoiz, F. J.; Herrero, V. J. Latest Findings on the Dynamics of the Simplest Chemical Reaction. *Phys. Scr.* **2006**, *73* (1), C6–C13. https://doi.org/10.1088/0031-8949/73/1/N02.

(306) Fernández-Alonso, F.; Zare, R. N. Scattering Resonances in the Simplest Chemical Reaction. *Annu. Rev. Phys. Chem.* **2002**, *53* (1), 67–99. https://doi.org/10.1146/annurev.physchem.53.091001.094554.

(307) Clary, D. C. Theoretical Studies on Bimolecular Reaction Dynamics. *Proc. Natl. Acad. Sci.* **2008**, *105* (35), 12649–12653. https://doi.org/10.1073/pnas.0800088105.

(308) Skodje, R. T. Resonances in Bimolecular Chemical Reactions. In *Adv Quantum Chem*; 2012; Vol. 63, pp 119–163. https://doi.org/10.1016/B978-0-12-397009-1.00003-5.

(309) Dalibard, J. Introduction to the Physics of Artificial Gauge Fields. *Proc. Int. Sch. Phys. "Enrico Fermi"* **2016**, *191*, 1–61. https://doi.org/10.3254/978-1-61499-694-1-1.

(310) Gherib, R.; Ryabinkin, I. G.; Izmaylov, A. F. Why Do Mixed Quantum-Classical Methods Describe Short-Time Dynamics through Conical Intersections So Well? Analysis of Geometric

Phase Effects. *J. Chem. Theory Comput.* **2015**, *11* (4), 1375–1382. https://doi.org/10.1021/acs.jctc.5b00072.

(311) Kendrick, B. Geometric Phase Effects in the Vibrational Spectrum of Na3(X). *Phys. Rev. Lett.* **1997**, *79* (13), 2431–2434. https://doi.org/10.1103/PHYSREVLETT.79.2431.

(312) Ham, F. S. Berry's Geometrical Phase and the Sequence of States in the Jahn-Teller Effect. *Phys. Rev. Lett.* **1987**, *58* (7), 725. https://doi.org/10.1103/PhysRevLett.58.725.

(313) Kendrick, B.; Pack, R. T. Geometric Phase Effects in H+O 2 Scattering. I. Surface Function Solutions in the Presence of a Conical Intersection. *J. Chem. Phys.* **1996**, *104* (19), 7475–7501. https://doi.org/10.1063/1.471460.

(314) Kendrick, B.; Pack, R. T. Geometric Phase Effects in H+O 2 Scattering. II. Recombination Resonances and State-to-state Transition Probabilities at Thermal Energies. *J. Chem. Phys.* **1996**, *104* (19), 7502–7514. https://doi.org/10.1063/1.471461.

(315) Juanes-Marcos, J. C.; Althorpe, S. C.; Wrede, E. Theoretical Study of Geometric Phase Effects in the Hydrogen-Exchange Reaction. *Science (80-. ).* **2005**, *309* (5738), 1227–1230. https://doi.org/10.1126/science.1114890.

(316) Ryabinkin, I. G.; Izmaylov, A. F. Geometric Phase Effects in Dynamics Near Conical Intersections: Symmetry Breaking and Spatial Localization. *Phys. Rev. Lett.* **2013**, *111* (22), 220406. https://doi.org/10.1103/PhysRevLett.111.220406.

(317) Ryabinkin, I. G.; Joubert-Doriol, L.; Izmaylov, A. F. When Do We Need to Account for the Geometric Phase in Excited State Dynamics? *J. Chem. Phys.* **2014**, *140* (21), 214116. https://doi.org/10.1063/1.4881147.

(318) Kavokin, A.; Malpuech, G.; Glazov, M. Optical Spin Hall Effect. *Phys. Rev. Lett.* **2005**, *95* (13), 136601. https://doi.org/10.1103/PhysRevLett.95.136601.

(319) Bardyn, C. E.; Karzig, T.; Refael, G.; Liew, T. C. H. Topological Polaritons and Excitons in Garden-Variety Systems. *Phys. Rev. B - Condens. Matter Mater. Phys.* **2015**, *91* (16), 161413. https://doi.org/10.1103/PhysRevB.91.161413.

(320) Nalitov, A. V.; Solnyshkov, D. D.; Malpuech, G. Polariton Z Topological Insulator. *Phys. Rev. Lett.* **2015**, *114* (11), 116401. https://doi.org/10.1103/PhysRevLett.114.116401.

(321) Klembt, S.; Harder, T. H.; Egorov, O. A.; Winkler, K.; Ge, R.; Bandres, M. A.; Emmerling, M.; Worschech, L.; Liew, T. C. H.; Segev, M.; Schneider, C.; Höfling, S. Exciton-Polariton Topological Insulator. *Nature* **2018**, *562* (7728), 552–556. https://doi.org/10.1038/s41586-018-0601-5.

(322) Jacqmin, T.; Carusotto, I.; Sagnes, I.; Abbarchi, M.; Solnyshkov, D. D.; Malpuech, G.; Galopin, E.; Lemaître, A.; Bloch, J.; Amo, A. Direct Observation of Dirac Cones and a Flatband in a Honeycomb Lattice for Polaritons. *Phys. Rev. Lett.* **2014**, *112* (11), 116402. https://doi.org/10.1103/PhysRevLett.112.116402.

(323) Real, B.; Jamadi, O.; Milićević, M.; Pernet, N.; St-Jean, P.; Ozawa, T.; Montambaux, G.; Sagnes, I.; Lemaître, A.; Le Gratiet, L.; Harouri, A.; Ravets, S.; Bloch, J.; Amo, A. Semi-Dirac Transport and Anisotropic Localization in Polariton Honeycomb Lattices. *Phys. Rev. Lett.* **2020**, *125* (18), 186601. https://doi.org/10.1103/PhysRevLett.125.186601.

(324) Milićević, M.; Ozawa, T.; Montambaux, G.; Carusotto, I.; Galopin, E.; Lemaître, A.; Le Gratiet, L.; Sagnes, I.; Bloch, J.; Amo, A. Orbital Edge States in a Photonic Honeycomb Lattice. *Phys.*

*Rev. Lett.* **2017**, *118* (10), 107403. https://doi.org/10.1103/PhysRevLett.118.107403.

(325) Liu, W.; Ji, Z.; Wang, Y.; Modi, G.; Hwang, M.; Zheng, B.; Sorger, V. J.; Pan, A.; Agarwal, R. Generation of Helical Topological Exciton-Polaritons. *Science (80-. ).* **2020**, *370* (6516), 600–604. https://doi.org/10.1126/SCIENCE.ABC4975.

(326) Li, M.; Sinev, I.; Benimetskiy, F.; Ivanova, T.; Khestanova, E.; Kiriushechkina, S.; Vakulenko, A.; Guddala, S.; Skolnick, M.; Menon, V. M.; Krizhanovskii, D.; Alù, A.; Samusev, A.; Khanikaev, A. B. Experimental Observation of Topological Z2 Exciton-Polaritons in Transition Metal Dichalcogenide Monolayers. *Nat. Commun.* **2021**, *12* (1), 4425. https://doi.org/10.1038/s41467-021-24728-y.

(327) Krasnok, A. Photonic Rashba Effect. *Nat. Nanotechnol.* **2020**, *15* (11), 893–894. https://doi.org/10.1038/s41565-020-0764-8.

(328) Rong, K.; Wang, B.; Reuven, A.; Maguid, E.; Cohn, B.; Kleiner, V.; Katznelson, S.; Koren, E.; Hasman, E. Photonic Rashba Effect from Quantum Emitters Mediated by a Berry-Phase Defective Photonic Crystal. *Nat. Nanotechnol.* **2020**, *15* (11), 927–933. https://doi.org/10.1038/s41565-020-0758-6.

(329) Sun, L.; Wang, C.-Y.; Krasnok, A.; Choi, J.; Shi, J.; Gomez-Diaz, J. S.; Zepeda, A.; Gwo, S.; Shih, C.-K.; Alù, A.; Li, X. Separation of Valley Excitons in a MoS2 Monolayer Using a Subwavelength Asymmetric Groove Array. *Nat. Photonics* **2019**, *13* (3), 180–184. https://doi.org/10.1038/s41566-019-0348-z.

(330) Scafirimuto, F.; Urbonas, D.; Scherf, U.; Mahrt, R. F.; Stöferle, T. Room-Temperature Exciton-Polariton Condensation in a Tunable Zero-Dimensional Microcavity. *ACS Photonics* **2018**, *5* (1), 85–89. https://doi.org/10.1021/acsphotonics.7b00557.

(331) Su, R.; Ghosh, S.; Wang, J.; Liu, S.; Diederichs, C.; Liew, T. C. H.; Xiong, Q. Observation of Exciton Polariton Condensation in a Perovskite Lattice at Room Temperature. *Nat. Phys.* **2020**, *16* (3), 301–306. https://doi.org/10.1038/s41567-019-0764-5.

(332) Klitzing, K. V.; Dorda, G.; Pepper, M. New Method for High-Accuracy Determination of the Fine-Structure Constant Based on Quantized Hall Resistance. *Phys. Rev. Lett.* **1980**, *45* (6), 494–497. https://doi.org/10.1103/PhysRevLett.45.494.

(333) Thouless, D. J.; Kohmoto, M.; Nightingale, M. P.; Den Nijs, M. Quantized Hall Conductance in a Two-Dimensional Periodic Potential. *Phys. Rev. Lett.* **1982**, *49* (6), 405–408. https://doi.org/10.1103/PhysRevLett.49.405.

(334) Laughlin, R. B. Quantized Hall Conductivity in Two Dimensions. *Phys. Rev. B* **1981**, *23* (10), 5632–5633. https://doi.org/10.1103/PhysRevB.23.5632.

(335) Karplus, R.; Luttinger, J. M. Hall Effect in Ferromagnetics. *Phys. Rev.* **1954**, *95* (5), 1154–1160. https://doi.org/10.1103/PhysRev.95.1154.

(336) Luttinger, J. M. The Effect of a Magnetic Field on Electrons in a Periodic Potential. *Phys. Rev.* **1951**, *84* (4), 814–817. https://doi.org/10.1103/PHYSREV.84.814.

(337) Hofstadter, D. R. Energy Levels and Wave Functions of Bloch Electrons in Rational and Irrational Magnetic Fields. *Phys. Rev. B* **1976**, *14* (6), 2239–2249. https://doi.org/10.1103/PHYSREVB.14.2239.

(338) Bloch, I.; Dalibard, J.; Zwerger, W. Many-Body Physics with Ultracold Gases. *Rev. Mod. Phys.* **2008**, *80* (3), 885–964. https://doi.org/10.1103/RevModPhys.80.885.

(339) Lu, L.; Joannopoulos, J. D.; Soljačić, M. Topological States in Photonic Systems. *Nat. Phys.* **2016**, *12* (7), 626–629. https://doi.org/10.1038/NPHYS3796.

(340) Blanco-Redondo, A. Topological Nanophotonics: Toward Robust Quantum Circuits. *Proc. IEEE* **2020**, *108* (5), 837–849. https://doi.org/10.1109/JPROC.2019.2939987.

(341) Aidelsburger, M.; Lohse, M.; Schweizer, C.; Atala, M.; Barreiro, J. T.; Nascimbène, S.; Cooper, N. R.; Bloch, I.; Goldman, N. Measuring the Chern Number of Hofstadter Bands with Ultracold Bosonic Atoms. *Nat. Phys.* **2015**, *11* (2), 162–166. https://doi.org/10.1038/nphys3171.

(342) Luu, T. T.; Wörner, H. J. Measurement of the Berry Curvature of Solids Using High-Harmonic Spectroscopy. *Nat. Commun.* **2018**, *9* (1), 916. https://doi.org/10.1038/s41467-018-03397-4.

(343) Chang, M. C.; Niu, Q. Berry Phase, Hyperorbits, and the Hofstadter Spectrum. *Phys. Rev. Lett.* **1995**, *75* (7), 1348–1351. https://doi.org/10.1103/physrevlett.75.1348.

(344) Ma, Q.; Grushin, A. G.; Burch, K. S. Topology and Geometry under the Nonlinear Electromagnetic Spotlight. *Nat. Mater.* **2021**, *20* (12), 1601–1614. https://doi.org/10.1038/s41563-021-00992-7.

(345) Du, Z. Z.; Wang, C. M.; Li, S.; Lu, H.-Z.; Xie, X. C. Disorder-Induced Nonlinear Hall Effect with Time-Reversal Symmetry. *Nat. Commun.* **2019**, *10* (1), 3047. https://doi.org/10.1038/s41467-019-10941-3.

(346) Sinitsyn, N. A. Semiclassical Theories of the Anomalous Hall Effect. *J. Phys. Condens. Matter* **2008**, *20* (2), 023201. https://doi.org/10.1088/0953-8984/20/02/023201.

(347) Jungwirth, T.; Niu, Q.; MacDonald, A. H. Anomalous Hall Effect in Ferromagnetic Semiconductors. *Phys. Rev. Lett.* **2002**, *88* (20), 207208. https://doi.org/10.1103/PhysRevLett.88.207208.

(348) Manna, K.; Sun, Y.; Muechler, L.; Kübler, J.; Felser, C. Heusler, Weyl and Berry. *Nat. Rev. Mater.* **2018**, *3* (8), 244–256. https://doi.org/10.1038/s41578-018-0036-5.

(349) Zhao, P.; Ma, Y.; Wang, H.; Huang, B.; Dai, Y. Room-Temperature Quantum Anomalous Hall Effect in Single-Layer CrP 2 S 6. *J. Phys. Chem. C* **2019**, *123* (23), 14707–14711. https://doi.org/10.1021/acs.jpcc.9b04888.

(350) Smit, J. The Spontaneous Hall Effect in Ferromagnetics I. *Physica* **1955**, *21* (6–10), 877–887. https://doi.org/10.1016/s0031-8914(55)92596-9.

(351) Kim, I.; Iwamoto, S.; Arakawa, Y. Topologically Protected Elastic Waves in One-Dimensional Phononic Crystals of Continuous Media. *Appl. Phys. Express* **2018**, *11* (1), 017201. https://doi.org/10.7567/APEX.11.017201.

(352) Yan, B.; Felser, C. Topological Materials: Weyl Semimetals. *Annu. Rev. Condens. Matter Phys.* **2017**, *8* (1), 337–354. https://doi.org/10.1146/annurev-conmatphys-031016-025458.

(353) Otrokov, M. M.; Klimovskikh, I. I.; Bentmann, H.; Estyunin, D.; Zeugner, A.; Aliev, Z. S.; Gaß, S.; Wolter, A. U. B.; Koroleva, A. V.; Shikin, A. M.; Blanco-Rey, M.; Hoffmann, M.; Rusinov, I. P.; Vyazovskaya, A. Y.; Eremeev, S. V.; Koroteev, Y. M.; Kuznetsov, V. M.; Freyse, F.; Sánchez-Barriga, J.; Amiraslanov, I. R.; Babanly, M. B.; Mamedov, N. T.; Abdullayev, N. A.; Zverev, V. N.; Alfonsov, A.; Kataev, V.; Büchner, B.; Schwier, E. F.; Kumar, S.; Kimura, A.; Petaccia, L.; Di Santo, G.; Vidal, R. C.; Schatz, S.; Kißner, K.; Ünzelmann, M.; Min, C. H.; Moser, S.; Peixoto, T. R. F.; Reinert, F.; Ernst, A.; Echenique, P. M.; Isaeva, A.; Chulkov, E. V. Prediction and Observation of an Antiferromagnetic Topological Insulator. *Nature* **2019**, *576* (7787), 416–422.

https://doi.org/10.1038/s41586-019-1840-9.

(354) Haldane, F. D. M. Nobel Lecture: Topological Quantum Matter. *Rev. Mod. Phys.* **2017**, *89* (4), 040502. https://doi.org/10.1103/RevModPhys.89.040502.

(355) Nagaosa, N.; Morimoto, T.; Tokura, Y. Transport, Magnetic and Optical Properties of Weyl Materials. *Nat. Rev. Mater.* **2020**, *5* (8), 621–636. https://doi.org/10.1038/s41578-020-0208-y.

(356) Krasnok, A.; Alù, A. Valley-Selective Response of Nanostructures Coupled to 2D Transition-Metal Dichalcogenides. *Appl. Sci.* **2018**, *8* (7), 1157. https://doi.org/10.3390/app8071157.

(357) Vitale, S. A.; Nezich, D.; Varghese, J. O.; Kim, P.; Gedik, N.; Jarillo-Herrero, P.; Xiao, D.; Rothschild, M. Valleytronics: Opportunities, Challenges, and Paths Forward. *Small* **2018**, *14* (38), 1801483. https://doi.org/10.1002/smll.201801483.

(358) Schaibley, J. R.; Yu, H.; Clark, G.; Rivera, P.; Ross, J. S.; Seyler, K. L.; Yao, W.; Xu, X. Valleytronics in 2D Materials. *Nat. Rev. Mater.* **2016**, *1* (11), 16055. https://doi.org/10.1038/natrevmats.2016.55.

(359) Zeng, H.; Dai, J.; Yao, W.; Xiao, D.; Cui, X. Valley Polarization in MoS 2 Monolayers by Optical Pumping. *Nat. Nanotechnol.* **2012**, *7* (8), 490–493. https://doi.org/10.1038/nnano.2012.95.

(360) Mak, K. F.; Xiao, D.; Shan, J. Light–Valley Interactions in 2D Semiconductors. *Nat. Photonics* **2018**, *12* (8), 451–460. https://doi.org/10.1038/s41566-018-0204-6.

(361) Onga, M.; Zhang, Y.; Ideue, T.; Iwasa, Y. Exciton Hall Effect in Monolayer MoS2. *Nat. Mater.* **2017**, *16* (12), 1193–1197. https://doi.org/10.1038/nmat4996.

(362) Lee, J.; Mak, K. F.; Shan, J. Electrical Control of the Valley Hall Effect in Bilayer MoS2 Transistors. *Nat. Nanotechnol.* **2016**, *11* (5), 421–425. https://doi.org/10.1038/nnano.2015.337.

(363) Cao, T.; Wang, G.; Han, W.; Ye, H.; Zhu, C.; Shi, J.; Niu, Q.; Tan, P.; Wang, E.; Liu, B.; Feng, J. Valley-Selective Circular Dichroism of Monolayer Molybdenum Disulphide. *Nat. Commun.* **2012**, *3* (1), 887. https://doi.org/10.1038/ncomms1882.

(364) Mak, K. F.; He, K.; Shan, J.; Heinz, T. F. Control of Valley Polarization in Monolayer MoS2 by Optical Helicity. *Nat. Nanotechnol.* **2012**, *7* (8), 494–498. https://doi.org/10.1038/nnano.2012.96.

(365) Huang, L.; Krasnok, A.; Alú, A.; Yu, Y.; Neshev, D.; Miroshnichenko, A. E. Enhanced Light–Matter Interaction in Two-Dimensional Transition Metal Dichalcogenides. *Reports Prog. Phys.* **2022**, *85* (4), 046401. https://doi.org/10.1088/1361-6633/ac45f9.

(366) Chen, P.; Lo, T. W.; Fan, Y.; Wang, S.; Huang, H.; Lei, D. Chiral Coupling of Valley Excitons and Light through Photonic Spin–Orbit Interactions. *Adv. Opt. Mater.* **2020**, *8* (5), 1901233. https://doi.org/10.1002/adom.201901233.

(367) Sie, E. J.; McLver, J. W.; Lee, Y. H.; Fu, L.; Kong, J.; Gedik, N. Valley-Selective Optical Stark Effect in Monolayer WS2. *Nat. Mater.* **2015**, *14* (3), 290–294. https://doi.org/10.1038/nmat4156.

(368) Kim, J.; Hong, X.; Jin, C.; Shi, S.-F.; Chang, C.-Y. S.; Chiu, M.-H.; Li, L.-J.; Wang, F. Ultrafast Generation of Pseudo-Magnetic Field for Valley Excitons in WSe 2 Monolayers. *Science (80-. ).* **2014**, *346* (6214), 1205–1208. https://doi.org/10.1126/science.1258122.

(369) Benalcazar, W. A.; Bernevig, B. A.; Hughes, T. L. Quantized Electric Multipole Insulators. *Science (80-. ).* **2017**, *357* (6346), 61–66. https://doi.org/10.1126/science.aah6442.

(370) Kitagawa, T.; Berg, E.; Rudner, M.; Demler, E. Topological Characterization of Periodically

Driven Quantum Systems. *Phys. Rev. B* **2010**, *82* (23), 235114. https://doi.org/10.1103/PhysRevB.82.235114.

(371) Jiménez-García, K.; LeBlanc, L. J.; Williams, R. A.; Beeler, M. C.; Perry, A. R.; Spielman, I. B. Peierls Substitution in an Engineered Lattice Potential. *Phys. Rev. Lett.* **2012**, *108* (22), 225303. https://doi.org/10.1103/PhysRevLett.108.225303.

(372) Kogut, J. B. An Introduction to Lattice Gauge Theory and Spin Systems. *Rev. Mod. Phys.* **1979**, *51* (4), 659. https://doi.org/10.1103/RevModPhys.51.659.

(373) Parameswaran, S. A.; Roy, R.; Sondhi, S. L. Fractional Quantum Hall Physics in Topological Flat Bands. *Comptes Rendus Phys.* **2013**, *14* (9–10), 816–839. https://doi.org/10.1016/J.CRHY.2013.04.003.

(374) Castro Neto, A. H.; Guinea, F.; Peres, N. M. R.; Novoselov, K. S.; Geim, A. K. The Electronic Properties of Graphene. *Rev. Mod. Phys.* **2009**, *81* (1), 109–162. https://doi.org/10.1103/revmodphys.81.109.

(375) Novoselov, K. S.; Geim, A. K.; Morozov, S. V.; Jiang, D.; Katsnelson, M. I.; Grigorieva, I. V.; Dubonos, S. V.; Firsov, A. A. Two-Dimensional Gas of Massless Dirac Fermions in Graphene. *Nature* **2005**, *438* (7065), 197–200. https://doi.org/10.1038/nature04233.

(376) Peleg, O.; Bartal, G.; Freedman, B.; Manela, O.; Segev, M.; Christodoulides, D. N. Conical Diffraction and Gap Solitons in Honeycomb Photonic Lattices. *Phys. Rev. Lett.* **2007**, *98* (10), 103901. https://doi.org/10.1103/PhysRevLett.98.103901.

(377) Bahat-Treidel, O.; Peleg, O.; Grobman, M.; Shapira, N.; Segev, M.; Pereg-Barnea, T. Klein Tunneling in Deformed Honeycomb Lattices. *Phys. Rev. Lett.* **2010**, *104* (6), 063901. https://doi.org/10.1103/PhysRevLett.104.063901.

(378) Rechtsman, M. C.; Zeuner, J. M.; Tünnermann, A.; Nolte, S.; Segev, M.; Szameit, A. Strain-Induced Pseudomagnetic Field and Photonic Landau Levels in Dielectric Structures. *Nat. Photonics* **2013**, *7* (2), 153–158. https://doi.org/10.1038/NPHOTON.2012.302.

(379) Schomerus, H.; Halpern, N. Y. Parity Anomaly and Landau-Level Lasing in Strained Photonic Honeycomb Lattices. *Phys. Rev. Lett.* **2013**, *110* (1), 013903. https://doi.org/10.1103/PhysRevLett.110.013903.

(380) Rechtsman, M. C.; Plotnik, Y.; Zeuner, J. M.; Song, D.; Chen, Z.; Szameit, A.; Segev, M. Topological Creation and Destruction of Edge States in Photonic Graphene. *Phys. Rev. Lett.* **2013**, *111* (10), 103901. https://doi.org/10.1103/PhysRevLett.111.103901.

(381) Sepkhanov, R. A.; Nilsson, J.; Beenakker, C. W. J. Proposed Method for Detection of the Pseudospin-1/2 Berry Phase in a Photonic Crystal with a Dirac Spectrum. *Phys. Rev. B* **2008**, *78* (4), 045122. https://doi.org/10.1103/PhysRevB.78.045122.

(382) Sepkhanov, R. A.; Ossipov, A.; Beenakker, C. W. J. Extinction of Coherent Backscattering by a Disordered Photonic Crystal with a Dirac Spectrum. *EPL (Europhysics Lett.* **2009**, *85* (1), 14005. https://doi.org/10.1209/0295-5075/85/14005.

(383) Zawadzki, W.; Rusin, T. M. Zitterbewegung (Trembling Motion) of Electrons in Semiconductors: A Review. *J. Phys. Condens. Matter* **2011**, *23* (14), 143201. https://doi.org/10.1088/0953-8984/23/14/143201.

(384) Katsnelson, M. I.; Novoselov, K. S.; Geim, A. K. Chiral Tunnelling and the Klein Paradox in Graphene. *Nat. Phys.* **2006**, *2* (9), 620–625. https://doi.org/10.1038/nphys384.

(385) Ochiai, T.; Onoda, M. Photonic Analog of Graphene Model and Its Extension: Dirac Cone, Symmetry, and Edge States. *Phys. Rev. B* **2009**, *80* (15), 155103. https://doi.org/10.1103/PhysRevB.80.155103.

(386) Qi, X.-L.; Hughes, T. L.; Zhang, S.-C. Topological Field Theory of Time-Reversal Invariant Insulators. *Phys. Rev. B* **2008**, *78* (19), 195424. https://doi.org/10.1103/PhysRevB.78.195424.

(387) Bernevig, B. A.; Hughes, T. L.; Zhang, S.-C. Quantum Spin Hall Effect and Topological Phase Transition in HgTe Quantum Wells. *Science (80-. ).* **2006**, *314* (5806), 1757–1761. https://doi.org/10.1126/science.1133734.

(388) Hsieh, D.; Xia, Y.; Qian, D.; Wray, L.; Dil, J. H.; Meier, F.; Osterwalder, J.; Patthey, L.; Checkelsky, J. G.; Ong, N. P.; Fedorov, A. V.; Lin, H.; Bansil, A.; Grauer, D.; Hor, Y. S.; Cava, R. J.; Hasan, M. Z. A Tunable Topological Insulator in the Spin Helical Dirac Transport Regime. *Nature* **2009**, *460* (7259), 1101–1105. https://doi.org/10.1038/nature08234.

(389) Hsieh, D.; Qian, D.; Wray, L.; Xia, Y.; Hor, Y. S.; Cava, R. J.; Hasan, M. Z. A Topological Dirac Insulator in a Quantum Spin Hall Phase. *Nature* **2008**, *452* (7190), 970–974. https://doi.org/10.1038/nature06843.

(390) Chang, C. Z.; Zhang, J.; Feng, X.; Shen, J.; Zhang, Z.; Guo, M.; Li, K.; Ou, Y.; Wei, P.; Wang, L. L.; Ji, Z. Q.; Feng, Y.; Ji, S.; Chen, X.; Jia, J.; Dai, X.; Fang, Z.; Zhang, S. C.; He, K.; Wang, Y.; Lu, L.; Ma, X. C.; Xue, Q. K. Experimental Observation of the Quantum Anomalous Hall Effect in a Magnetic Topological Insulator. *Science (80-. ).* **2013**, *340* (6129), 167–170. https://doi.org/10.1126/science.1234414.

(391) Yu, R.; Zhang, W.; Zhang, H.-J.; Zhang, S.-C.; Dai, X.; Fang, Z. Quantized Anomalous Hall Effect in Magnetic Topological Insulators. *Science (80-. ).* **2010**, *329* (5987), 61–64. https://doi.org/10.1126/science.1187485.

(392) Liu, C.-X.; Zhang, S.-C.; Qi, X.-L. The Quantum Anomalous Hall Effect: Theory and Experiment. *Annu. Rev. Condens. Matter Phys.* **2016**, *7* (1), 301–321. https://doi.org/10.1146/annurev-conmatphys-031115-011417.

(393) König, M.; Wiedmann, S.; Brüne, C.; Roth, A.; Buhmann, H.; Molenkamp, L. W.; Qi, X. L.; Zhang, S. C. Quantum Spin Hall Insulator State in HgTe Quantum Wells. *Science (80-. ).* **2007**, *318* (5851), 766–770. https://doi.org/10.1126/SCIENCE.1148047.

(394) Knez, I.; Rettner, C. T.; Yang, S.-H.; Parkin, S. S. P.; Du, L.; Du, R.-R.; Sullivan, G. Observation of Edge Transport in the Disordered Regime of Topologically Insulating InAsGaSb Quantum Wells. *Phys. Rev. Lett.* **2014**, *112* (2), 026602. https://doi.org/10.1103/PhysRevLett.112.026602.

(395) Liu, Q.; Zhang, X.; Abdalla, L. B.; Zunger, A. Transforming Common III-V and II-VI Semiconductor Compounds into Topological Heterostructures: The Case of CdTe/InSb Superlattices. *Adv. Funct. Mater.* **2016**, *26* (19), 3259–3267. https://doi.org/10.1002/ADFM.201505357.

(396) Roth, A.; Brüne, C.; Buhmann, H.; Molenkamp, L. W.; Maciejko, J.; Qi, X. L.; Zhang, S. C. Nonlocal Transport in the Quantum Spin Hall State. *Science (80-. ).* **2009**, *325* (5938), 294–297. https://doi.org/10.1126/science.1174736.

(397) Gusev, G. M.; Kvon, Z. D.; Shegai, O. A.; Mikhailov, N. N.; Dvoretsky, S. A.; Portal, J. C. Transport in Disordered Two-Dimensional Topological Insulators. *Phys. Rev. B* **2011**, *84* (12), 121302. https://doi.org/10.1103/PhysRevB.84.121302.

(398) Zholudev, M.; Teppe, F.; Orlita, M.; Consejo, C.; Torres, J.; Dyakonova, N.; Czapkiewicz, M.;

Wróbel, J.; Grabecki, G.; Mikhailov, N.; Dvoretskii, S.; Ikonnikov, A.; Spirin, K.; Aleshkin, V.; Gavrilenko, V.; Knap, W. Magnetospectroscopy of Two-Dimensional HgTe-Based Topological Insulators around the Critical Thickness. *Phys. Rev. B* **2012**, *86* (20), 205420. https://doi.org/10.1103/PhysRevB.86.205420.

(399) Ma, E. Y.; Calvo, M. R.; Wang, J.; Lian, B.; Mühlbauer, M.; Brüne, C.; Cui, Y.-T.; Lai, K.; Kundhikanjana, W.; Yang, Y.; Baenninger, M.; König, M.; Ames, C.; Buhmann, H.; Leubner, P.; Molenkamp, L. W.; Zhang, S.-C.; Goldhaber-Gordon, D.; Kelly, M. A.; Shen, Z.-X. Unexpected Edge Conduction in Mercury Telluride Quantum Wells under Broken Time-Reversal Symmetry. *Nat. Commun.* **2015**, *6* (1), 7252. https://doi.org/10.1038/ncomms8252.

(400) Dantscher, K.-M.; Kozlov, D. A.; Scherr, M. T.; Gebert, S.; Bärenfänger, J.; Durnev, M. V.; Tarasenko, S. A.; Bel'kov, V. V.; Mikhailov, N. N.; Dvoretsky, S. A.; Kvon, Z. D.; Ziegler, J.; Weiss, D.; Ganichev, S. D. Photogalvanic Probing of Helical Edge Channels in Two-Dimensional HgTe Topological Insulators. *Phys. Rev. B* **2017**, *95* (20), 201103. https://doi.org/10.1103/PhysRevB.95.201103.

(401) Kadykov, A. M.; Krishtopenko, S. S.; Jouault, B.; Desrat, W.; Knap, W.; Ruffenach, S.; Consejo, C.; Torres, J.; Morozov, S. V.; Mikhailov, N. N.; Dvoretskii, S. A.; Teppe, F. Temperature-Induced Topological Phase Transition in HgTe Quantum Wells. *Phys. Rev. Lett.* **2018**, *120* (8), 086401. https://doi.org/10.1103/PhysRevLett.120.086401.

(402) Min'kov, G. M.; Sherstobitov, A. A.; Germanenko, A. V.; Rut, O. E.; Dvoretskii, S. A.; Mikhailov, N. N. Conductance of a Lateral p—n Junction in Two-Dimensional HgTe Structures with an Inverted Spectrum: The Role of Edge States. *JETP Lett.* **2015**, *101* (7), 469–473. https://doi.org/10.1134/S0021364015070115.

(403) Tikhonov, E. S.; Shovkun, D. V.; Khrapai, V. S.; Kvon, Z. D.; Mikhailov, N. N.; Dvoretsky, S. A. Shot Noise of the Edge Transport in the Inverted Band HgTe Quantum Wells. *JETP Lett.* **2015**, *101* (10), 708–713. https://doi.org/10.1134/S0021364015100148.

(404) Kononov, A.; Egorov, S. V.; Titova, N.; Kvon, Z. D.; Mikhailov, N. N.; Dvoretsky, S. A.; Deviatov, E. V. Conductance Oscillations at the Interface between a Superconductor and the Helical Edge Channel in a Narrow HgTe Quantum Well. *JETP Lett.* **2015**, *101* (1), 41–46. https://doi.org/10.1134/S0021364015010075.

(405) Tanaka, Y.; Furusaki, A.; Matveev, K. A. Conductance of a Helical Edge Liquid Coupled to a Magnetic Impurity. *Phys. Rev. Lett.* **2011**, *106* (23), 236402. https://doi.org/10.1103/PhysRevLett.106.236402.

(406) Lunde, A. M.; Platero, G. Helical Edge States Coupled to a Spin Bath: Current-Induced Magnetization. *Phys. Rev. B* **2012**, *86* (3), 035112. https://doi.org/10.1103/PhysRevB.86.035112.

(407) Altshuler, B. L.; Aleiner, I. L.; Yudson, V. I. Localization at the Edge of a 2D Topological Insulator by Kondo Impurities with Random Anisotropies. *Phys. Rev. Lett.* **2013**, *111* (8), 086401. https://doi.org/10.1103/PhysRevLett.111.086401.

(408) Kurilovich, P. D.; Kurilovich, V. D.; Burmistrov, I. S.; Goldstein, M. Helical Edge Transport in the Presence of a Magnetic Impurity. *JETP Lett.* **2017**, *106* (9), 593–599. https://doi.org/10.1134/S0021364017210020.

(409) Tarasenko, S. A.; Burkard, G. Limitation of Electron Mobility from Hyperfine Interaction in Ultraclean Quantum Wells and Topological Insulators. *Phys. Rev. B* **2016**, *94* (4), 045309. https://doi.org/10.1103/PhysRevB.94.045309.

(410) Lunde, A. M.; Platero, G. Hyperfine Interactions in Two-Dimensional HgTe Topological Insulators. *Phys. Rev. B* **2013**, *88* (11), 115411. https://doi.org/10.1103/PhysRevB.88.115411.

(411) Del Maestro, A.; Hyart, T.; Rosenow, B. Backscattering between Helical Edge States via Dynamic Nuclear Polarization. *Phys. Rev. B* **2013**, *87* (16), 165440. https://doi.org/10.1103/PhysRevB.87.165440.

(412) Roy, R. Z2 Classification of Quantum Spin Hall Systems: An Approach Using Time-Reversal Invariance. *Phys. Rev. B* **2009**, *79*, 195321.

(413) Sheng, D. N.; Weng, Z. Y.; Sheng, L.; Haldane, F. D. M. Quantum Spin-Hall Effect and Topologically Invariant Chern Numbers. *Phys. Rev. Lett.* **2006**, *97* (3), 36808. https://doi.org/10.1103/PhysRevLett.97.036808.

(414) Kim, Y.; Choi, K.; Ihm, J.; Jin, H. Topological Domain Walls and Quantum Valley Hall Effects in Silicene. *Phys. Rev. B* **2014**, *89* (8), 085429. https://doi.org/10.1103/PhysRevB.89.085429.

(415) Wan, X.; Turner, A. M.; Vishwanath, A.; Savrasov, S. Y. Topological Semimetal and Fermi-Arc Surface States in the Electronic Structure of Pyrochlore Iridates. *Phys. Rev. B* **2011**, *83* (20), 205101. https://doi.org/10.1103/PhysRevB.83.205101.

(416) Burkov, A. A.; Balents, L. Weyl Semimetal in a Topological Insulator Multilayer. *Phys. Rev. Lett.* **2011**, *107* (12), 127205. https://doi.org/10.1103/PhysRevLett.107.127205.

(417) Xu, S.-Y.; Belopolski, I.; Alidoust, N.; Neupane, M.; Bian, G.; Zhang, C.; Sankar, R.; Chang, G.; Yuan, Z.; Lee, C.-C.; Huang, S.-M.; Zheng, H.; Ma, J.; Sanchez, D. S.; Wang, B.; Bansil, A.; Chou, F.; Shibayev, P. P.; Lin, H.; Jia, S.; Hasan, M. Z. Discovery of a Weyl Fermion Semimetal and Topological Fermi Arcs. *Science (80-. ).* **2015**, *349* (6248), 613–617. https://doi.org/10.1126/science.aaa9297.

(418) Xu, S.-Y.; Belopolski, I.; Sanchez, D. S.; Zhang, C.; Chang, G.; Guo, C.; Bian, G.; Yuan, Z.; Lu, H.; Chang, T.-R.; Shibayev, P. P.; Prokopovych, M. L.; Alidoust, N.; Zheng, H.; Lee, C.-C.; Huang, S.-M.; Sankar, R.; Chou, F.; Hsu, C.-H.; Jeng, H.-T.; Bansil, A.; Neupert, T.; Strocov, V. N.; Lin, H.; Jia, S.; Hasan, M. Z. Experimental Discovery of a Topological Weyl Semimetal State in TaP. *Sci. Adv.* **2015**, *1* (10), 031013. https://doi.org/10.1126/sciadv.1501092.

(419) Yang, K. Y.; Lu, Y. M.; Ran, Y. Quantum Hall Effects in a Weyl Semimetal: Possible Application in Pyrochlore Iridates. *Phys. Rev. B* **2011**, *84* (7), 075129. https://doi.org/10.1103/PhysRevB.84.075129.

(420) Soluyanov, A. A.; Gresch, D.; Wang, Z.; Wu, Q.; Troyer, M.; Dai, X.; Bernevig, B. A. Type-II Weyl Semimetals. *Nature* **2015**, *527* (7579), 495–498. https://doi.org/10.1038/nature15768.

(421) Shuichi Murakami. Phase Transition between the Quantum Spin Hall and Insulator Phases in 3D: Emergence of a Topological Gapless Phase. *New J. Phys.* **2007**, *9* (9), 356–356. https://doi.org/10.1088/1367-2630/9/9/356.

(422) Burkov, A. A.; Hook, M. D.; Balents, L. Topological Nodal Semimetals. *Phys. Rev. B* **2011**, *84* (23), 235126. https://doi.org/10.1103/PhysRevB.84.235126.

(423) Xu, G.; Weng, H.; Wang, Z.; Dai, X.; Fang, Z. Chern Semimetal and the Quantized Anomalous Hall Effect in HgCr2Se4. *Phys. Rev. Lett.* **2011**, *107* (18), 186806. https://doi.org/10.1103/PhysRevLett.107.186806.

(424) Young, S. M.; Zaheer, S.; Teo, J. C. Y.; Kane, C. L.; Mele, E. J.; Rappe, A. M. Dirac Semimetal in Three Dimensions. *Phys. Rev. Lett.* **2012**, *108* (14), 140405.

https://doi.org/10.1103/PhysRevLett.108.140405.

(425) Armitage, N. P.; Mele, E. J.; Vishwanath, A. Weyl and Dirac Semimetals in Three-Dimensional Solids. *Rev. Mod. Phys.* **2018**, *90* (1), 015001.

(426) H Weng, C. F. Z. F. B. B. X. D. Weyl Semimetal Phase in Noncentrosymmetric Transition-Metal Monophosphides. *Phys. Rev. X* **2015**, *5*, 011029.

(427) Lv, B. Q.; Weng, H. M.; Fu, B. B.; Wang, X. P.; Miao, H.; Ma, J.; Richard, P.; Huang, X. C.; Zhao, L. X.; Chen, G. F.; Fang, Z.; Dai, X.; Qian, T.; Ding, H. Experimental Discovery of Weyl Semimetal TaAs. *Phys. Rev. X* **2015**, *5* (3). https://doi.org/10.1103/PHYSREVX.5.031013.

(428) Potter, A. C.; Kimchi, I.; Vishwanath, A. Quantum Oscillations from Surface Fermi Arcs in Weyl and Dirac Semimetals. *Nat. Commun.* **2014**, *5* (1), 5161. https://doi.org/10.1038/ncomms6161.

(429) Yang, B.; Guo, Q.; Tremain, B.; Barr, L. E.; Gao, W.; Liu, H.; Béri, B.; Xiang, Y.; Fan, Di.; Hibbins, A. P.; Zhang, S. Direct Observation of Topological Surface-State Arcs in Photonic Metamaterials. *Nat. Commun.* **2017**, *8* (1), 97. https://doi.org/10.1038/s41467-017-00134-1.

(430) HB Nielsen, M. N. The Adler-Bell-Jackiw Anomaly and Weyl Fermions in a Crystal. *Phys. Lett. B* **1983**, *130*, 389–396.

(431) Wieder, B. J.; Kim, Y.; Rappe, A. M.; Kane, C. L. Double Dirac Semimetals in Three Dimensions. *Phys. Rev. Lett.* **2016**, *116* (18), 186402. https://doi.org/ARTN 18640210.1103/PhysRevLett.116.186402.

(432) Schoop, L. M.; Ali, M. N.; Straßer, C.; Topp, A.; Varykhalov, A.; Marchenko, D.; Duppel, V.; Parkin, S. S. P.; Lotsch, B. V.; Ast, C. R. Dirac Cone Protected by Non-Symmorphic Symmetry and Three-Dimensional Dirac Line Node in ZrSiS. *Nat. Commun.* **2016**, *7* (1), 11696. https://doi.org/10.1038/ncomms11696.

(433) Bzdušek, T.; Wu, Q. S.; Rüegg, A.; Sigrist, M.; Soluyanov, A. A. Nodal-Chain Metals. *Nature* **2016**, *538* (7623), 75–78. https://doi.org/10.1038/nature19099.

(434) Bi, R.; Yan, Z.; Lu, L.; Wang, Z. Nodal-Knot Semimetals. *Phys. Rev. B* **2017**, *96* (20), 201305.

(435) M Kohmoto, BI Halperin, Y. W. Y.-S. W. Diophantine Equation for the Three-Dimensional Quantum Hall Effect. *Phys. Rev. B* **1992**, *45*, 13488–13493.

(436) Halperin, B. I. Possible States for a Three-Dimensional Electron Gas in a Strong Magnetic Field. *Jpn. J. Appl. Phys.* **1987**, *26* (S3-3), 1913. https://doi.org/10.7567/JJAPS.26S3.1913.

(437) Moore, J. E.; Balents, L. Topological Invariants of Time-Reversal-Invariant Band Structures. *Phys. Rev. B* **2007**, *75* (12), 121306. https://doi.org/10.1103/PhysRevB.75.121306.

(438) Fu, L.; Kane, C. L.; Mele, E. J. Topological Insulators in Three Dimensions. *Phys. Rev. Lett.* **2007**, *98* (10), 106803. https://doi.org/10.1103/PhysRevLett.98.106803.

(439) Blas, A.-Z. E. de; Axel, F. Diophantine Equation for the 3D Transport Coefficients of Bloch Electrons in a Strong Tilted Magnetic Field with Quantum Hall Effect. *J. Phys. Condens. Matter* **2004**, *16* (43), 7673–7708. https://doi.org/10.1088/0953-8984/16/43/010.

(440) Störmer, H. L.; Eisenstein, J. P.; Gossard, A. C.; Wiegmann, W.; Baldwin, K. Quantization of the Hall Effect in an Anisotropic Three-Dimensional Electronic System. *Phys. Rev. Lett.* **1986**, *56* (1), 85–88. https://doi.org/10.1103/PhysRevLett.56.85.

(441) Zhang, H.; Liu, C.-X.; Qi, X.-L.; Dai, X.; Fang, Z.; Zhang, S.-C. Topological Insulators in Bi2Se3,

Bi2Te3 and Sb2Te3 with a Single Dirac Cone on the Surface. *Nat. Phys.* **2009**, *5* (6), 438–442. https://doi.org/10.1038/nphys1270.

(442) Xia, Y.; Qian, D.; Hsieh, D.; Wray, L.; Pal, A.; Lin, H.; Bansil, A.; Grauer, D.; Hor, Y. S.; Cava, R. J.; Hasan, M. Z. Observation of a Large-Gap Topological-Insulator Class with a Single Dirac Cone on the Surface. *Nat. Phys.* **2009**, *5* (6), 398–402. https://doi.org/10.1038/nphys1274.

(443) Fu, L.; Kane, C. L. Topological Insulators with Inversion Symmetry. *Phys. Rev. B* **2007**, *76* (4), 045302. https://doi.org/10.1103/PhysRevB.76.045302.

(444) Brüne, C.; Liu, C. X.; Novik, E. G.; Hankiewicz, E. M.; Buhmann, H.; Chen, Y. L.; Qi, X. L.; Shen, Z. X.; Zhang, S. C.; Molenkamp, L. W. Quantum Hall Effect from the Topological Surface States of Strained Bulk HgTe. *Phys. Rev. Lett.* **2011**, *106* (12). https://doi.org/10.1103/PhysRevLett.106.126803.

(445) Kozlov, D. A.; Kvon, Z. D.; Olshanetsky, E. B.; Mikhailov, N. N.; Dvoretsky, S. A.; Weiss, D. Transport Properties of a 3D Topological Insulator Based on a Strained High-Mobility HgTe Film. *Phys. Rev. Lett.* **2014**, *112* (19), 196801. https://doi.org/10.1103/PhysRevLett.112.196801.

(446) Dantscher, K.-M.; Kozlov, D. A.; Olbrich, P.; Zoth, C.; Faltermeier, P.; Lindner, M.; Budkin, G. V.; Tarasenko, S. A.; Bel'kov, V. V.; Kvon, Z. D.; Mikhailov, N. N.; Dvoretsky, S. A.; Weiss, D.; Jenichen, B.; Ganichev, S. D. Cyclotron-Resonance-Assisted Photocurrents in Surface States of a Three-Dimensional Topological Insulator Based on a Strained High-Mobility HgTe Film. *Phys. Rev. B* **2015**, *92* (16), 165314. https://doi.org/10.1103/PhysRevB.92.165314.

(447) Barfuss, A.; Dudy, L.; Scholz, M. R.; Roth, H.; Höpfner, P.; Blumenstein, C.; Landolt, G.; Dil, J. H.; Plumb, N. C.; Radovic, M.; Bostwick, A.; Rotenberg, E.; Fleszar, A.; Bihlmayer, G.; Wortmann, D.; Li, G.; Hanke, W.; Claessen, R.; Schäfer, J. Elemental Topological Insulator with Tunable Fermi Level: Strained α-Sn on InSb(001). *Phys. Rev. Lett.* **2013**, *111* (15), 157205. https://doi.org/10.1103/PhysRevLett.111.157205.

(448) Khalaf, E. Higher-Order Topological Insulators and Superconductors Protected by Inversion Symmetry. *Phys. Rev. B* **2018**, *97* (20), 205136. https://doi.org/10.1103/PhysRevB.97.205136.

(449) Geier, M.; Trifunovic, L.; Hoskam, M.; Brouwer, P. W. Second-Order Topological Insulators and Superconductors with an Order-Two Crystalline Symmetry. *Phys. Rev. B* **2018**, *97* (20), 205135. https://doi.org/10.1103/PhysRevB.97.205135.

(450) Song, Z.; Fang, Z.; Fang, C. (D-2) -Dimensional Edge States of Rotation Symmetry Protected Topological States. *Phys. Rev. Lett.* **2017**, *119* (24), 246402. https://doi.org/10.1103/PhysRevLett.119.246402.

(451) Benalcazar, W. A.; Bernevig, B. A.; Hughes, T. L. Electric Multipole Moments, Topological Multipole Moment Pumping, and Chiral Hinge States in Crystalline Insulators. *Phys. Rev. B* **2017**, *96* (24), 245115. https://doi.org/10.1103/PhysRevB.96.245115.

(452) Ezawa, M. Higher-Order Topological Insulators and Semimetals on the Breathing Kagome and Pyrochlore Lattices. *Phys. Rev. Lett.* **2018**, *120* (2), 026801. https://doi.org/10.1103/PhysRevLett.120.026801.

(453) Schindler, F.; Cook, A. M.; Vergniory, M. G.; Wang, Z.; Parkin, S. S. P.; Bernevig, B. A.; Neupert, T. Higher-Order Topological Insulators. *Sci. Adv.* **2018**, *4* (6), eaat0346. https://doi.org/10.1126/sciadv.aat0346.

(454) Serra-Garcia, M.; Peri, V.; Süsstrunk, R.; Bilal, O. R.; Larsen, T.; Villanueva, L. G.; Huber, S. D. Observation of a Phononic Quadrupole Topological Insulator. *Nature* **2018**, *555* (7696), 342–345.

https://doi.org/10.1038/nature25156.

(455) Peterson, C. W.; Benalcazar, W. A.; Hughes, T. L.; Bahl, G. A Quantized Microwave Quadrupole Insulator with Topologically Protected Corner States. *Nature* **2018**, *555* (7696), 346–350. https://doi.org/10.1038/nature25777.

(456) Imhof, S.; Berger, C.; Bayer, F.; Brehm, J.; Molenkamp, L. W.; Kiessling, T.; Schindler, F.; Lee, C. H.; Greiter, M.; Neupert, T.; Thomale, R. Topolectrical-Circuit Realization of Topological Corner Modes. *Nat. Phys.* **2018**, *14* (9), 925–929. https://doi.org/10.1038/s41567-018-0246-1.

(457) Langbehn, J.; Peng, Y.; Trifunovic, L.; von Oppen, F.; Brouwer, P. W. Reflection-Symmetric Second-Order Topological Insulators and Superconductors. *Phys. Rev. Lett.* **2017**, *119* (24), 246401. https://doi.org/10.1103/PhysRevLett.119.246401.

(458) Xie, B.-Y.; Wang, H.-F.; Wang, H.-X.; Zhu, X.-Y.; Jiang, J.-H.; Lu, M.-H.; Chen, Y.-F. Second-Order Photonic Topological Insulator with Corner States. *Phys. Rev. B* **2018**, *98* (20), 205147. https://doi.org/10.1103/PhysRevB.98.205147.

(459) Noh, J.; Benalcazar, W. A.; Huang, S.; Collins, M. J.; Chen, K. P.; Hughes, T. L.; Rechtsman, M. C. Topological Protection of Photonic Mid-Gap Defect Modes. *Nat. Photonics* **2018**, *12* (7), 408–415. https://doi.org/10.1038/S41566-018-0179-3.

(460) Zhang, X.; Wang, H. X.; Lin, Z. K.; Tian, Y.; Xie, B.; Lu, M. H.; Chen, Y. F.; Jiang, J. H. Second-Order Topology and Multidimensional Topological Transitions in Sonic Crystals. *Nat. Phys.* **2019**, *15* (6), 582–588. https://doi.org/10.1038/s41567-019-0472-1.

(461) Xue, H.; Yang, Y.; Gao, F.; Chong, Y.; Zhang, B. Acoustic Higher-Order Topological Insulator on a Kagome Lattice. *Nat. Mater.* **2019**, *18* (2), 108–112. https://doi.org/10.1038/s41563-018-0251-x.

(462) Ni, X.; Weiner, M.; Alù, A.; Khanikaev, A. B. Observation of Higher-Order Topological Acoustic States Protected by Generalized Chiral Symmetry. *Nat. Mater.* **2019**, *18* (2), 113–120. https://doi.org/10.1038/s41563-018-0252-9.

(463) Xie, B.-Y.; Su, G.-X.; Wang, H.-F.; Su, H.; Shen, X.-P.; Zhan, P.; Lu, M.-H.; Wang, Z.-L.; Chen, Y.-F. Visualization of Higher-Order Topological Insulating Phases in Two-Dimensional Dielectric Photonic Crystals. *Phys. Rev. Lett.* **2019**, *122* (23), 233903. https://doi.org/10.1103/PhysRevLett.122.233903.

(464) Chen, X.-D.; Deng, W.; Shi, F.; Zhao, F.; Chen, M.; Dong, J. Direct Observation of Corner States in Second-Order Topological Photonic Crystal Slabs. *Phys. Rev. Lett.* **2019**, *122* (23), 233902. https://doi.org/10.1103/PhysRevLett.122.233902.

(465) Mittal, S.; Orre, V. V.; Zhu, G.; Gorlach, M. A.; Poddubny, A.; Hafezi, M. Photonic Quadrupole Topological Phases. *Nat. Photonics* **2019**, *13* (10), 692–696. https://doi.org/10.1038/s41566-019-0452-0.

(466) El Hassan, A.; Kunst, F. K.; Moritz, A.; Andler, G.; Bergholtz, E. J.; Bourennane, M. *Corner States of Light in Photonic Waveguides*; Nature Publishing Group, 2019; Vol. 13, pp 697–700. https://doi.org/10.1038/s41566-019-0519-y.

(467) Li, M.; Zhirihin, D.; Gorlach, M.; Ni, X.; Filonov, D.; Slobozhanyuk, A.; Alù, A.; Khanikaev, A. B. Higher-Order Topological States in Photonic Kagome Crystals with Long-Range Interactions. *Nat. Photonics* **2020**, *14* (2), 89–94. https://doi.org/10.1038/s41566-019-0561-9.

(468) Banerjee, R.; Mandal, S.; Liew, T. C. H. Coupling between Exciton-Polariton Corner Modes through Edge States. *Phys. Rev. Lett.* **2020**, *124* (6), 063901.

https://doi.org/10.1103/PhysRevLett.124.063901.

(469) Heiss, W. D. Exceptional Points of Non-Hermitian Operators. *J. Phys. A. Math. Gen.* **2004**, *37* (6), 2455–2464. https://doi.org/10.1088/0305-4470/37/6/034.

(470) Berry, M. V. Physics of Nonhermitian Degeneracies. *Czechoslov. J. Phys.* **2004**, *54* (10), 1039–1047. https://doi.org/10.1023/B:CJOP.0000044002.05657.04.

(471) Bender, C. M.; Boettcher, S. Real Spectra in Non-Hermitian Hamiltonian PT- Symmetry. *Phys. Rev. Lett.* **1998**, *80* (24), 5243–5246. https://doi.org/10.1103/PhysRevLett.80.5243.

(472) Ashida, Y.; Gong, Z.; Ueda, M. Non-Hermitian Physics. *Adv. Phys.* **2020**, *69* (3), 249–435. https://doi.org/10.1080/00018732.2021.1876991.

(473) Regensburger, A.; Bersch, C.; Miri, M. A.; Onishchukov, G.; Christodoulides, D. N.; Peschel, U. Parity-Time Synthetic Photonic Lattices. *Nature* **2012**, *488* (7410), 167–171. https://doi.org/10.1038/nature11298.

(474) Özdemir, Ş. K.; Rotter, S.; Nori, F.; Yang, L.; Özdemir, K.; Rotter, S.; Nori, F.; Yang, L. Parity–Time Symmetry and Exceptional Points in Photonics. *Nat. Mater.* **2019**, *18* (8), 783–798. https://doi.org/10.1038/s41563-019-0304-9.

(475) Miri, M.-A. A.; Alù, A. Exceptional Points in Optics and Photonics. *Science (80-. ).* **2019**, *363* (6422), 11–19. https://doi.org/10.1126/science.aar7709.

(476) Krasnok, A.; Nefedkin, N.; Alu, A. Parity-Time Symmetry and Exceptional Points [Electromagnetic Perspectives]. *IEEE Antennas Propag. Mag.* **2021**, *63* (6), 110–121. https://doi.org/10.1109/MAP.2021.3115766.

(477) Xu, J.; Du, Y.-X.; Huang, W.; Zhang, D.-W. Detecting Topological Exceptional Points in a Parity-Time Symmetric System with Cold Atoms. *Opt. Express* **2017**, *25* (14), 15786. https://doi.org/10.1364/OE.25.015786.

(478) Midya, B.; Zhao, H.; Feng, L. Non-Hermitian Photonics Promises Exceptional Topology of Light. *Nat. Commun.* **2018**, *9* (1), 2674. https://doi.org/10.1038/s41467-018-05175-8.

(479) Heeger, A. J.; Kivelson, S.; Schrieffer, J. R.; Su, W. P. Solitons in Conducting Polymers. *Rev. Mod. Phys.* **1988**, *60* (3), 781–850. https://doi.org/10.1103/RevModPhys.60.781.

(480) Malkova, N.; Hromada, I.; Wang, X.; Bryant, G.; Chen, Z. Observation of Optical Shockley-like Surface States in Photonic Superlattices. *Opt. Lett.* **2009**, *34* (11), 1633. https://doi.org/10.1364/OL.34.001633.

(481) Xiao, M.; Zhang, Z. Q.; Chan, C. T. Surface Impedance and Bulk Band Geometric Phases in One-Dimensional Systems. *Phys. Rev. X* **2014**, *4* (2), 021017. https://doi.org/10.1103/PhysRevX.4.021017.

(482) Poshakinskiy, A. V.; Poddubny, A. N.; Pilozzi, L.; Ivchenko, E. L. Radiative Topological States in Resonant Photonic Crystals. *Phys. Rev. Lett.* **2014**, *112* (10), 107403. https://doi.org/10.1103/PhysRevLett.112.107403.

(483) Keil, R.; Zeuner, J. M.; Dreisow, F.; Heinrich, M.; Tünnermann, A.; Nolte, S.; Szameit, A. The Random Mass Dirac Model and Long-Range Correlations on an Integrated Optical Platform. *Nat. Commun.* **2013**, *4* (1), 1368. https://doi.org/10.1038/ncomms2384.

(484) Zhu, B.; Zhong, H.; Ke, Y.; Qin, X.; Sukhorukov, A. A.; Kivshar, Y. S.; Lee, C. Topological Floquet Edge States in Periodically Curved Waveguides. *Phys. Rev. A* **2018**, *98* (1), 013855.

https://doi.org/10.1103/PhysRevA.98.013855.

(485) Saei Ghareh Naz, E.; Fulga, I. C.; Ma, L.; Schmidt, O. G.; van den Brink, J. Topological Phase Transition in a Stretchable Photonic Crystal. *Phys. Rev. A* **2018**, *98* (3), 033830. https://doi.org/10.1103/PhysRevA.98.033830.

(486) Poli, C.; Bellec, M.; Kuhl, U.; Mortessagne, F.; Schomerus, H. Selective Enhancement of Topologically Induced Interface States in a Dielectric Resonator Chain. *Nat. Commun.* **2015**, *6* (1), 6710. https://doi.org/10.1038/ncomms7710.

(487) Slobozhanyuk, A. P.; Poddubny, A. N.; Miroshnichenko, A. E.; Belov, P. A.; Kivshar, Y. S. Subwavelength Topological Edge States in Optically Resonant Dielectric Structures. *Phys. Rev. Lett.* **2015**, *114* (12), 123901. https://doi.org/10.1103/PhysRevLett.114.123901.

(488) Slobozhanyuk, A. P.; Poddubny, A. N.; Sinev, I. S.; Samusev, A. K.; Yu, Y. F.; Kuznetsov, A. I.; Miroshnichenko, A. E.; Kivshar, Y. S. Enhanced Photonic Spin Hall Effect with Subwavelength Topological Edge States. *Laser Photonics Rev.* **2016**, *10* (4), 656–664. https://doi.org/10.1002/lpor.201600042.

(489) Kruk, S.; Slobozhanyuk, A.; Denkova, D.; Poddubny, A.; Kravchenko, I.; Miroshnichenko, A.; Neshev, D.; Kivshar, Y. Edge States and Topological Phase Transitions in Chains of Dielectric Nanoparticles. *Small* **2017**, *13* (11), 1603190. https://doi.org/10.1002/smll.201603190.

(490) Goren, T.; Plekhanov, K.; Appas, F.; Le Hur, K. Topological Zak Phase in Strongly Coupled LC Circuits. *Phys. Rev. B* **2018**, *97* (4), 041106. https://doi.org/10.1103/PhysRevB.97.041106.

(491) Rosenthal, E. I.; Ehrlich, N. K.; Rudner, M. S.; Higginbotham, A. P.; Lehnert, K. W. Topological Phase Transition Measured in a Dissipative Metamaterial. *Phys. Rev. B* **2018**, *97* (22), 1–5. https://doi.org/10.1103/PhysRevB.97.220301.

(492) Poddubny, A.; Miroshnichenko, A.; Slobozhanyuk, A.; Kivshar, Y. Topological Majorana States in Zigzag Chains of Plasmonic Nanoparticles. *ACS Photonics* **2014**, *1* (2), 101–105. https://doi.org/10.1021/PH4000949.

(493) Sinev, I. S.; Mukhin, I. S.; Slobozhanyuk, A. P.; Poddubny, A. N.; Miroshnichenko, A. E.; Samusev, A. K.; Kivshar, Y. S. Mapping Plasmonic Topological States at the Nanoscale. *Nanoscale* **2015**, *7* (28), 11904–11908. https://doi.org/10.1039/C5NR00231A.

(494) Ling, C. W.; Xiao, M.; Chan, C. T.; Yu, S. F.; Fung, K. H. Topological Edge Plasmon Modes between Diatomic Chains of Plasmonic Nanoparticles. *Opt. Express* **2015**, *23* (3), 2021. https://doi.org/10.1364/oe.23.002021.

(495) Bleckmann, F.; Cherpakova, Z.; Linden, S.; Alberti, A. Spectral Imaging of Topological Edge States in Plasmonic Waveguide Arrays. *Phys. Rev. B* **2017**, *96* (4), 045417. https://doi.org/10.1103/PhysRevB.96.045417.

(496) Blanco-Redondo, A.; Andonegui, I.; Collins, M. J.; Harari, G.; Lumer, Y.; Rechtsman, M. C.; Eggleton, B. J.; Segev, M. Topological Optical Waveguiding in Silicon and the Transition between Topological and Trivial Defect States. *Phys. Rev. Lett.* **2016**, *116* (16), 163901. https://doi.org/10.1103/PhysRevLett.116.163901.

(497) Schomerus, H. Topologically Protected Midgap States in Complex Photonic Lattices. *Opt. Lett.* **2013**, *38* (11), 1912. https://doi.org/10.1364/ol.38.001912.

(498) Zeuner, J. M.; Rechtsman, M. C.; Plotnik, Y.; Lumer, Y.; Nolte, S.; Rudner, M. S.; Segev, M.; Szameit, A. Observation of a Topological Transition in the Bulk of a Non-Hermitian System.

*Phys. Rev. Lett.* **2015**, *115* (4), 040402. https://doi.org/10.1103/PhysRevLett.115.040402.

(499) Weimann, S.; Kremer, M.; Plotnik, Y.; Lumer, Y.; Nolte, S.; Makris, K. G.; Segev, M.; Rechtsman, M. C.; Szameit, A. Topologically Protected Bound States in Photonic Parity–Time-Symmetric Crystals. *Nat. Mater.* **2017**, *16* (4), 433–438. https://doi.org/10.1038/NMAT4811.

(500) Solnyshkov, D. D.; Nalitov, A. V.; Malpuech, G. Kibble-Zurek Mechanism in Topologically Nontrivial Zigzag Chains of Polariton Micropillars. *Phys. Rev. Lett.* **2016**, *116* (4), 046402. https://doi.org/10.1103/PhysRevLett.116.046402.

(501) Zhao, H.; Miao, P.; Teimourpour, M. H.; Malzard, S.; El-Ganainy, R.; Schomerus, H.; Feng, L. Topological Hybrid Silicon Microlasers. *Nat. Commun.* **2018**, *9* (1), 1–6. https://doi.org/10.1038/s41467-018-03434-2.

(502) Pernet, N.; St-Jean, P.; Solnyshkov, D. D.; Malpuech, G.; Carlon Zambon, N.; Fontaine, Q.; Real, B.; Jamadi, O.; Lemaître, A.; Morassi, M.; Le Gratiet, L.; Baptiste, T.; Harouri, A.; Sagnes, I.; Amo, A.; Ravets, S.; Bloch, J. Gap Solitons in a One-Dimensional Driven-Dissipative Topological Lattice. *Nat. Phys.* **2022**, *18* (6), 678–684. https://doi.org/10.1038/s41567-022-01599-8.

(503) Meng, Y.; Wu, X.; Zhang, R.-Y.; Li, X.; Hu, P.; Ge, L.; Huang, Y.; Xiang, H.; Han, D.; Wang, S.; Wen, W. Designing Topological Interface States in Phononic Crystals Based on the Full Phase Diagrams. *New J. Phys.* **2018**, *20* (7), 073032. https://doi.org/10.1088/1367-2630/aad136.

(504) Xiao, Y.-X.; Ma, G.; Zhang, Z.-Q.; Chan, C. T. Topological Subspace-Induced Bound State in the Continuum. *Phys. Rev. Lett.* **2017**, *118* (16), 166803. https://doi.org/10.1103/PhysRevLett.118.166803.

(505) Fan, L.; Yu, W.; Zhang, S.; Zhang, H.; Ding, J. Zak Phases and Band Properties in Acoustic Metamaterials with Negative Modulus or Negative Density. *Phys. Rev. B* **2016**, *94* (17), 174307. https://doi.org/10.1103/PhysRevB.94.174307.

(506) Chaunsali, R.; Kim, E.; Thakkar, A.; Kevrekidis, P. G.; Yang, J. Demonstrating an In Situ Topological Band Transition in Cylindrical Granular Chains. *Phys. Rev. Lett.* **2017**, *119* (2), 024301. https://doi.org/10.1103/PhysRevLett.119.024301.

(507) Deymier, P.; Runge, K. One-Dimensional Mass-Spring Chains Supporting Elastic Waves with Non-Conventional Topology. *Crystals* **2016**, *6* (4), 44. https://doi.org/10.3390/cryst6040044.

(508) Chen, H.; Nassar, H.; Huang, G. L. A Study of Topological Effects in 1D and 2D Mechanical Lattices. *J. Mech. Phys. Solids* **2018**, *117*, 22–36. https://doi.org/10.1016/j.jmps.2018.04.013.

(509) Tsai, Y. W.; Wang, Y. T.; Luan, P. G.; Yen, T. J. Topological Phase Transition in a One-Dimensional Elastic String System. *Crystals* **2019**, *9* (6). https://doi.org/10.3390/CRYST9060313.

(510) Yin, J.; Ruzzene, M.; Wen, J.; Yu, D.; Cai, L.; Yue, L. Band Transition and Topological Interface Modes in 1D Elastic Phononic Crystals. *Sci. Rep.* **2018**, *8* (1), 6806. https://doi.org/10.1038/s41598-018-24952-5.

(511) Feng, L.; Huang, K.; Chen, J.; Luo, J.; Huang, H.; Huo, S. Magnetically Tunable Topological Interface States for Lamb Waves in One-Dimensional Magnetoelastic Phononic Crystal Slabs. *AIP Adv.* **2019**, *9* (11), 115201. https://doi.org/10.1063/1.5120054.

(512) Zhang, H.; Liu, B.; Zhang, X.; Wu, Q.; Wang, X. Zone Folding Induced Tunable Topological Interface States in One-Dimensional Phononic Crystal Plates. *Phys. Lett. A* **2019**, *383* (23), 2797–2801. https://doi.org/10.1016/j.physleta.2019.05.045.

(513) Lin, S.; Zhang, L.; Tian, T.; Duan, C.-K.; Du, J. Dynamic Observation of Topological Soliton States in a Programmable Nanomechanical Lattice. *Nano Lett.* **2021**, *21* (2), 1025–1031. https://doi.org/10.1021/acs.nanolett.0c04121.

(514) Dai, H.; Liu, L.; Xia, B.; Yu, D. Experimental Realization of Topological On-Chip Acoustic Tweezers. *Phys. Rev. Appl.* **2021**, *15* (6), 064032. https://doi.org/10.1103/PhysRevApplied.15.064032.

(515) Zangeneh-Nejad, F.; Fleury, R. Topological Fano Resonances. *Phys. Rev. Lett.* **2019**, *122* (1), 014301. https://doi.org/10.1103/PhysRevLett.122.014301.

(516) Wang, W.; Jin, Y.; Wang, W.; Bonello, B.; Djafari-Rouhani, B.; Fleury, R. Robust Fano Resonance in a Topological Mechanical Beam. *Phys. Rev. B* **2020**, *101* (2), 24101. https://doi.org/10.1103/PhysRevB.101.024101.

(517) Zangeneh-Nejad, F.; Fleury, R. Topological Analog Signal Processing. *Nat. Commun.* **2019**, *10* (1), 2058. https://doi.org/10.1038/s41467-019-10086-3.

(518) Esmann, M.; Lamberti, F. R.; Lemaître, A.; Lanzillotti-Kimura, N. D. Topological Acoustics in Coupled Nanocavity Arrays. *Phys. Rev. B* **2018**, *98* (16), 161109. https://doi.org/10.1103/PhysRevB.98.161109.

(519) Esmann, M.; Lamberti, F. R.; Senellart, P.; Favero, I.; Krebs, O.; Lanco, L.; Gomez Carbonell, C.; Lemaître, A.; Lanzillotti-Kimura, N. D. Topological Nanophononic States by Band Inversion. *Phys. Rev. B* **2018**, *97* (15), 155422. https://doi.org/10.1103/PhysRevB.97.155422.

(520) Arregui, G.; Ortíz, O.; Esmann, M.; Sotomayor-Torres, C. M.; Gomez-Carbonell, C.; Mauguin, O.; Perrin, B.; Lemaître, A.; García, P. D.; Lanzillotti-Kimura, N. D. Coherent Generation and Detection of Acoustic Phonons in Topological Nanocavities. *APL Photonics* **2019**, *4* (3), 030805. https://doi.org/10.1063/1.5082728.

(521) Ortiz, O.; Priya, P.; Rodriguez, A.; Lemaitre, A.; Esmann, M.; Esmann, M.; Lanzillotti-Kimura, N. D. Topological Optical and Phononic Interface Mode by Simultaneous Band Inversion. *Opt. Vol. 8, Issue 5, pp. 598-605* **2021**, *8* (5), 598–605. https://doi.org/10.1364/OPTICA.411945.

(522) Dai, H.; Xia, B.; Yu, D. Microparticles Separation Using Acoustic Topological Insulators. *Appl. Phys. Lett.* **2021**, *119* (11), 111601. https://doi.org/10.1063/5.0059873.

(523) Delplace, P.; Ullmo, D.; Montambaux, G. Zak Phase and the Existence of Edge States in Graphene. *Phys. Rev. B* **2011**, *84* (19), 195452. https://doi.org/10.1103/PhysRevB.84.195452.

(524) Li, S.; Zhao, D.; Niu, H.; Zhu, X.; Zang, J. Observation of Elastic Topological States in Soft Materials. *Nat. Commun.* **2018**, *9* (1), 1370. https://doi.org/10.1038/s41467-018-03830-8.

(525) Harper, P. G. Single Band Motion of Conduction Electrons in a Uniform Magnetic Field. *Proc. Phys. Soc. Sect. A* **1955**, *68* (10), 874–878. https://doi.org/10.1088/0370-1298/68/10/304.

(526) Rice, M. J.; Mele, E. J. Elementary Excitations of a Linearly Conjugated Diatomic Polymer. *Phys. Rev. Lett.* **1982**, *49* (19), 1455–1459. https://doi.org/10.1103/PhysRevLett.49.1455.

(527) Kraus, Y. E.; Lahini, Y.; Ringel, Z.; Verbin, M.; Zilberberg, O. Topological States and Adiabatic Pumping in Quasicrystals. *Phys. Rev. Lett.* **2012**, *109* (10), 106402. https://doi.org/10.1103/PhysRevLett.109.106402.

(528) Verbin, M.; Zilberberg, O.; Kraus, Y. E.; Lahini, Y.; Silberberg, Y. Observation of Topological Phase Transitions in Photonic Quasicrystals. *Phys. Rev. Lett.* **2013**, *110* (7), 076403.

https://doi.org/10.1103/PhysRevLett.110.076403.

(529) Ke, Y.; Qin, X.; Mei, F.; Zhong, H.; Kivshar, Y. S.; Lee, C. Topological Phase Transitions and Thouless Pumping of Light in Photonic Waveguide Arrays. *Laser Photonics Rev.* **2016**, *10* (6), 995–1001. https://doi.org/10.1002/LPOR.201600119.

(530) Verbin, M.; Zilberberg, O.; Lahini, Y.; Kraus, Y. E.; Silberberg, Y. Topological Pumping over a Photonic Fibonacci Quasicrystal. *Phys. Rev. B* **2015**, *91* (6), 064201. https://doi.org/10.1103/PhysRevB.91.064201.

(531) Kraus, Y. E.; Zilberberg, O. Topological Equivalence between the Fibonacci Quasicrystal and the Harper Model. *Phys. Rev. Lett.* **2012**, *109* (11). https://doi.org/10.1103/PHYSREVLETT.109.116404.

(532) Jürgensen, M.; Mukherjee, S.; Rechtsman, M. C. Quantized Nonlinear Thouless Pumping. *Nature* **2021**, *596* (7870), 63–67. https://doi.org/10.1038/s41586-021-03688-9.

(533) Nakajima, S.; Tomita, T.; Taie, S.; Ichinose, T.; Ozawa, H.; Wang, L.; Troyer, M.; Takahashi, Y. Topological Thouless Pumping of Ultracold Fermions. *Nat. Phys.* **2016**, *12* (4), 296–300. https://doi.org/10.1038/NPHYS3622.

(534) Lohse, M.; Schweizer, C.; Zilberberg, O.; Aidelsburger, M.; Bloch, I. A Thouless Quantum Pump with Ultracold Bosonic Atoms in an Optical Superlattice. *Nat. Phys.* **2016**, *12* (4), 350–354. https://doi.org/10.1038/nphys3584.

(535) Cerjan, A.; Wang, M.; Huang, S.; Chen, K. P.; Rechtsman, M. C. Thouless Pumping in Disordered Photonic Systems. *Light Sci. Appl.* **2020**, *9* (1), 178. https://doi.org/10.1038/s41377-020-00408-2.

(536) Cheng, Q.; Wang, H.; Ke, Y.; Chen, T.; Yu, Y.; Kivshar, Y. S.; Lee, C.; Pan, Y. Asymmetric Topological Pumping in Nonparaxial Photonics. *Nat. Commun.* **2022**, *13* (1), 249. https://doi.org/10.1038/s41467-021-27773-9.

(537) Fedorova, Z.; Qiu, H.; Linden, S.; Kroha, J. Observation of Topological Transport Quantization by Dissipation in Fast Thouless Pumps. *Nat. Commun.* **2020**, *11* (1), 3758. https://doi.org/10.1038/s41467-020-17510-z.

(538) Hu, W.; Pillay, J. C.; Wu, K.; Pasek, M.; Shum, P. P.; Chong, Y. D. Measurement of a Topological Edge Invariant in a Microwave Network. *Phys. Rev. X* **2015**, *5* (1), 011012. https://doi.org/10.1103/PhysRevX.5.011012.

(539) Mei, F.; You, J.-B.; Nie, W.; Fazio, R.; Zhu, S.-L.; Kwek, L. C. Simulation and Detection of Photonic Chern Insulators in a One-Dimensional Circuit-QED Lattice. *Phys. Rev. A* **2015**, *92* (4), 041805. https://doi.org/10.1103/PhysRevA.92.041805.

(540) Wimmer, M.; Price, H. M.; Carusotto, I.; Peschel, U. Experimental Measurement of the Berry Curvature from Anomalous Transport. *Nat. Phys.* **2017**, *13* (6), 545–550. https://doi.org/10.1038/NPHYS4050.

(541) Apigo, D. J.; Cheng, W.; Dobiszewski, K. F.; Prodan, E.; Prodan, C. Observation of Topological Edge Modes in a Quasiperiodic Acoustic Waveguide. *Phys. Rev. Lett.* **2019**, *122* (9), 095501. https://doi.org/10.1103/PhysRevLett.122.095501.

(542) Richoux, O.; Pagneux, V. Acoustic Characterization of the Hofstadter Butterfly with Resonant Scatterers. *Europhys. Lett.* **2002**, *59* (1), 34–40. https://doi.org/10.1209/epl/i2002-00156-5.

(543) Ni, X.; Chen, K.; Weiner, M.; Apigo, D. J.; Prodan, C.; Alù, A.; Prodan, E.; Khanikaev, A. B.

Observation of Hofstadter Butterfly and Topological Edge States in Reconfigurable Quasi-Periodic Acoustic Crystals. *Commun. Phys.* **2019**, *2* (1), 55. https://doi.org/10.1038/s42005-019-0151-7.

(544) Apigo, D. J.; Qian, K.; Prodan, C.; Prodan, E. Topological Edge Modes by Smart Patterning. *Phys. Rev. Mater.* **2018**, *2* (12), 124203. https://doi.org/10.1103/PhysRevMaterials.2.124203.

(545) Pal, R. K.; Rosa, M. I. N.; Ruzzene, M. Topological Bands and Localized Vibration Modes in Quasiperiodic Beams. *New J. Phys.* **2019**, *21* (9), 093017. https://doi.org/10.1088/1367-2630/ab3cd7.

(546) Xia, Y.; Erturk, A.; Ruzzene, M. Topological Edge States in Quasiperiodic Locally Resonant Metastructures. *Phys. Rev. Appl.* **2020**, *13* (1), 014023. https://doi.org/10.1103/PhysRevApplied.13.014023.

(547) Rosa, M. I. N.; Guo, Y.; Ruzzene, M. Exploring Topology of 1D Quasiperiodic Metastructures through Modulated LEGO Resonators. *Appl. Phys. Lett.* **2021**, *118* (13), 131901. https://doi.org/10.1063/5.0042294.

(548) Martí-Sabaté, M.; Torrent, D. Edge Modes for Flexural Waves in Quasi-Periodic Linear Arrays of Scatterers. *APL Mater.* **2021**, *9* (8), 081107. https://doi.org/10.1063/5.0059097.

(549) Cheng, W.; Prodan, E.; Prodan, C. Experimental Demonstration of Dynamic Topological Pumping across Incommensurate Bilayered Acoustic Metamaterials. *Phys. Rev. Lett.* **2020**, *125* (22), 224301. https://doi.org/10.1103/PhysRevLett.125.224301.

(550) Grinberg, I. H.; Lin, M.; Harris, C.; Benalcazar, W. A.; Peterson, C. W.; Hughes, T. L.; Bahl, G. Robust Temporal Pumping in a Magneto-Mechanical Topological Insulator. *Nat. Commun.* **2020**, *11* (1), 974. https://doi.org/10.1038/s41467-020-14804-0.

(551) Riva, E.; Rosa, M. I. N.; Ruzzene, M. Edge States and Topological Pumping in Stiffness-Modulated Elastic Plates. *Phys. Rev. B* **2020**, *101* (9), 094307. https://doi.org/10.1103/PhysRevB.101.094307.

(552) Xia, Y.; Riva, E.; Rosa, M. I. N.; Cazzulani, G.; Erturk, A.; Braghin, F.; Ruzzene, M. Experimental Observation of Temporal Pumping in Electromechanical Waveguides. *Phys. Rev. Lett.* **2021**, *126* (9), 095501. https://doi.org/10.1103/PhysRevLett.126.095501.

(553) Riva, E.; Casieri, V.; Resta, F.; Braghin, F. Adiabatic Pumping via Avoided Crossings in Stiffness-Modulated Quasiperiodic Beams. *Phys. Rev. B* **2020**, *102* (1), 014305. https://doi.org/10.1103/PhysRevB.102.014305.

(554) Rosa, M. I. N.; Pal, R. K.; Arruda, J. R. F.; Ruzzene, M. Edge States and Topological Pumping in Spatially Modulated Elastic Lattices. *Phys. Rev. Lett.* **2019**, *123* (3), 034301. https://doi.org/10.1103/PhysRevLett.123.034301.

(555) Shen, Y.-X.; Peng, Y.-G.; Zhao, D.-G.; Chen, X.-C.; Zhu, J.; Zhu, X.-F. One-Way Localized Adiabatic Passage in an Acoustic System. *Phys. Rev. Lett.* **2019**, *122* (9), 094501. https://doi.org/10.1103/PhysRevLett.122.094501.

(556) Zeng, L.-S.; Shen, Y.-X.; Peng, Y.-G.; Zhao, D.-G.; Zhu, X.-F. Selective Topological Pumping for Robust, Efficient, and Asymmetric Sound Energy Transfer in a Dynamically Coupled Cavity Chain. *Phys. Rev. Appl.* **2021**, *15* (6), 064018. https://doi.org/10.1103/PhysRevApplied.15.064018.

(557) Brouzos, I.; Kiorpelidis, I.; Diakonos, F. K.; Theocharis, G. Fast, Robust, and Amplified Transfer of Topological Edge Modes on a Time-Varying Mechanical Chain. *Phys. Rev. B* **2020**, *102* (17),

174312. https://doi.org/10.1103/PhysRevB.102.174312.

(558) Chen, Z.-G.; Tang, W.; Zhang, R.-Y.; Chen, Z.; Ma, G. Landau-Zener Transition in the Dynamic Transfer of Acoustic Topological States. *Phys. Rev. Lett.* **2021**, *126* (5), 054301. https://doi.org/10.1103/PhysRevLett.126.054301.

(559) Reich, S.; Maultzsch, J.; Thomsen, C.; Ordejón, P. Tight-Binding Description of Graphene. *Phys. Rev. B* **2002**, *66* (3), 035412. https://doi.org/10.1103/PhysRevB.66.035412.

(560) Sepkhanov, R. A.; Bazaliy, Y. B.; Beenakker, C. W. J. Extremal Transmission at the Dirac Point of a Photonic Band Structure. *Phys. Rev. A* **2007**, *75* (6), 063813. https://doi.org/10.1103/PhysRevA.75.063813.

(561) Bravo-Abad, J.; Joannopoulos, J. D.; Soljačić, M. Enabling Single-Mode Behavior over Large Areas with Photonic Dirac Cones. *Proc. Natl. Acad. Sci.* **2012**, *109* (25), 9761–9765. https://doi.org/10.1073/pnas.1207335109.

(562) de Gail, R.; Fuchs, J.-N.; Goerbig, M. O.; Piéchon, F.; Montambaux, G. Manipulation of Dirac Points in Graphene-like Crystals. *Phys. B Condens. Matter* **2012**, *407* (11), 1948–1952. https://doi.org/10.1016/j.physb.2012.01.072.

(563) Montambaux, G.; Piéchon, F.; Fuchs, J.-N.; Goerbig, M. O. Merging of Dirac Points in a Two-Dimensional Crystal. *Phys. Rev. B* **2009**, *80* (15), 153412. https://doi.org/10.1103/PhysRevB.80.153412.

(564) Bellec, M.; Kuhl, U.; Montambaux, G.; Mortessagne, F. Topological Transition of Dirac Points in a Microwave Experiment. *Phys. Rev. Lett.* **2013**, *110* (3), 033902. https://doi.org/10.1103/physrevlett.110.033902.

(565) He, W.-Y.; Chan, C. T. The Emergence of Dirac Points in Photonic Crystals with Mirror Symmetry. *Sci. Rep.* **2015**, *5* (1), 8186. https://doi.org/10.1038/srep08186.

(566) Huang, X.; Lai, Y.; Hang, Z. H.; Zheng, H.; Chan, C. T. Dirac Cones Induced by Accidental Degeneracy in Photonic Crystals and Zero-Refractive-Index Materials. *Nat. Mater.* **2011**, *10* (8), 582–586. https://doi.org/10.1038/nmat3030.

(567) Longhi, S. Photonic Analog of Zitterbewegung in Binary Waveguide Arrays. *Opt. Lett.* **2010**, *35* (2), 235. https://doi.org/10.1364/ol.35.000235.

(568) Dreisow, F.; Heinrich, M.; Keil, R.; Tünnermann, A.; Nolte, S.; Longhi, S.; Szameit, A. Classical Simulation of Relativistic Zitterbewegung in Photonic Lattices. *Phys. Rev. Lett.* **2010**, *105* (14), 143902. https://doi.org/10.1103/PHYSREVLETT.105.143902/FIGURES/4/MEDIUM.

(569) Ni, X.; Purtseladze, D.; Smirnova, D. A.; Slobozhanyuk, A.; Alù, A.; Khanikaev, A. B. Spin- and Valley-Polarized One-Way Klein Tunneling in Photonic Topological Insulators. *Sci. Adv.* **2018**, *4* (5). https://doi.org/10.1126/sciadv.aap8802.

(570) Liu, G.-G.; Zhou, P.; Yang, Y.; Xue, H.; Ren, X.; Lin, X.; Sun, H.; Bi, L.; Chong, Y.; Zhang, B. Observation of an Unpaired Photonic Dirac Point. *Nat. Commun.* **2020**, *11* (1), 1873. https://doi.org/10.1038/s41467-020-15801-z.

(571) Pyrialakos, G. G.; Nye, N. S.; Kantartzis, N. V.; Christodoulides, D. N. Emergence of Type-II Dirac Points in Graphynelike Photonic Lattices. *Phys. Rev. Lett.* **2017**, *119* (11), 113901. https://doi.org/10.1103/PhysRevLett.119.113901.

(572) Lin, J. Y.; Hu, N. C.; Chen, Y. J.; Lee, C. H.; Zhang, X. Line Nodes, Dirac Points, and Lifshitz

Transition in Two-Dimensional Nonsymmorphic Photonic Crystals. *Phys. Rev. B* **2017**, *96* (7), 075438. https://doi.org/10.1103/PhysRevB.96.075438.

(573) Hu, C.; Li, Z.; Tong, R.; Wu, X.; Xia, Z.; Wang, L.; Li, S.; Huang, Y.; Wang, S.; Hou, B.; Chan, C. T.; Wen, W. Type-II Dirac Photons at Metasurfaces. *Phys. Rev. Lett.* **2018**, *121* (2), 024301. https://doi.org/10.1103/PhysRevLett.121.024301.

(574) Milićević, M.; Montambaux, G.; Ozawa, T.; Jamadi, O.; Real, B.; Sagnes, I.; Lemaître, A.; Le Gratiet, L.; Harouri, A.; Bloch, J.; Amo, A. Type-III and Tilted Dirac Cones Emerging from Flat Bands in Photonic Orbital Graphene. *Phys. Rev. X* **2019**, *9* (3), 031010. https://doi.org/10.1103/PhysRevX.9.031010.

(575) Zheng, L. Y.; Achilleos, V.; Chen, Z. G.; Richoux, O.; Theocharis, G.; Wu, Y.; Mei, J.; Felix, S.; Tournat, V.; Pagneux, V. Acoustic Graphene Network Loaded with Helmholtz Resonators: A First-Principle Modeling, Dirac Cones, Edge and Interface Waves. *New J. Phys.* **2020**, *22* (1), 013029. https://doi.org/10.1088/1367-2630/AB60F1.

(576) Zhang, X.; Liu, Z. Extremal Transmission and Beating Effect of Acoustic Waves in Two-Dimensional Sonic Crystals. *Phys. Rev. Lett.* **2008**, *101* (26), 264303. https://doi.org/10.1103/PhysRevLett.101.264303.

(577) Lu, J.; Qiu, C.; Xu, S.; Ye, Y.; Ke, M.; Liu, Z. Dirac Cones in Two-Dimensional Artificial Crystals for Classical Waves. *Phys. Rev. B* **2014**, *89* (13), 134302. https://doi.org/10.1103/PhysRevB.89.134302.

(578) Zhong, W.; Zhang, X. Acoustic Analog of Monolayer Graphene and Edge States. *Phys. Lett. A* **2011**, *375* (40), 3533–3536. https://doi.org/10.1016/j.physleta.2011.08.027.

(579) Torrent, D.; Sánchez-Dehesa, J. Acoustic Analogue of Graphene: Observation of Dirac Cones in Acoustic Surface Waves. *Phys. Rev. Lett.* **2012**, *108* (17), 174301. https://doi.org/10.1103/PhysRevLett.108.174301.

(580) Yves, S.; Lemoult, F.; Fink, M.; Lerosey, G. Crystalline Soda Can Metamaterial Exhibiting Graphene-like Dispersion at Subwavelength Scale. *Sci. Rep.* **2017**, *7* (1), 15359. https://doi.org/10.1038/s41598-017-15335-3.

(581) Torrent, D.; Mayou, D.; Sánchez-Dehesa, J. Elastic Analog of Graphene: Dirac Cones and Edge States for Flexural Waves in Thin Plates. *Phys. Rev. B* **2013**, *87* (11), 115143. https://doi.org/10.1103/PhysRevB.87.115143.

(582) Yu, S.-Y.; Sun, X.-C.; Ni, X.; Wang, Q.; Yan, X.-J.; He, C.; Liu, X.-P.; Feng, L.; Lu, M.-H.; Chen, Y.-F. Surface Phononic Graphene. *Nat. Mater.* **2016**, *15* (12), 1243–1247. https://doi.org/10.1038/nmat4743.

(583) Jiang, X.; Shi, C.; Li, Z.; Wang, S.; Wang, Y.; Yang, S.; Louie, S. G.; Zhang, X. Direct Observation of Klein Tunneling in Phononic Crystals. *Science (80-. ).* **2020**, *370* (6523), 1447–1450. https://doi.org/10.1126/SCIENCE.ABE2011/SUPPL_FILE/ABE2011-JIANG-SM.PDF.

(584) Dubois, M.; Shi, C.; Zhu, X.; Wang, Y.; Zhang, X. Observation of Acoustic Dirac-like Cone and Double Zero Refractive Index. *Nat. Commun.* **2017**, *8* (1), 14871. https://doi.org/10.1038/ncomms14871.

(585) Chen, Z.-G.; Ni, X.; Wu, Y.; He, C.; Sun, X.-C.; Zheng, L.-Y.; Lu, M.-H.; Chen, Y.-F. Accidental Degeneracy of Double Dirac Cones in a Phononic Crystal. *Sci. Rep.* **2015**, *4* (1), 4613. https://doi.org/10.1038/srep04613.

(586) Luo, J. cheng; Feng, L. yang; Huang, H. bo; Chen, J. jiu. Pseudomagnetic Fields and Landau Levels for Out-of-Plane Elastic Waves in Gradient Snowflake-Shaped Crystals. *Phys. Lett. A* **2019**, *383* (33), 125974. https://doi.org/10.1016/J.PHYSLETA.2019.125974.

(587) Ni, X.; Gorlach, M. A.; Alu, A.; Khanikaev, A. B. Topological Edge States in Acoustic Kagome Lattices. *New J. Phys.* **2017**, *19* (5), 055002. https://doi.org/10.1088/1367-2630/aa6996.

(588) Liu, F.; Huang, X.; Chan, C. T. Dirac Cones at K→=0 in Acoustic Crystals and Zero Refractive Index Acoustic Materials. *Appl. Phys. Lett.* **2012**, *100* (7), 071911. https://doi.org/10.1063/1.3686907.

(589) Fruchart, M.; Zhou, Y.; Vitelli, V. Dualities and Non-Abelian Mechanics. *Nature* **2020**, *577* (7792), 636–640. https://doi.org/10.1038/s41586-020-1932-6.

(590) Lanoy, M.; Lemoult, F.; Eddi, A.; Prada, C. Dirac Cones and Chiral Selection of Elastic Waves in a Soft Strip. *Proc. Natl. Acad. Sci. U. S. A.* **2020**, *117* (48), 30186–30190. https://doi.org/10.1073/PNAS.2010812117/SUPPL_FILE/PNAS.2010812117.SM09.M4V.

(591) Wang, Z.; Chong, Y. D.; Joannopoulos, J. D.; Soljačić, M. Reflection-Free One-Way Edge Modes in a Gyromagnetic Photonic Crystal. *Phys. Rev. Lett.* **2008**, *100* (1), 013905. https://doi.org/10.1103/PhysRevLett.100.013905.

(592) Ao, X.; Lin, Z.; Chan, C. T. One-Way Edge Mode in a Magneto-Optical Honeycomb Photonic Crystal. *Phys. Rev. B* **2009**, *80* (3), 033105. https://doi.org/10.1103/PhysRevB.80.033105.

(593) Poo, Y.; Wu, R.; Lin, Z.; Yang, Y.; Chan, C. T. Experimental Realization of Self-Guiding Unidirectional Electromagnetic Edge States. *Phys. Rev. Lett.* **2011**, *106* (9), 093903. https://doi.org/10.1103/PhysRevLett.106.093903.

(594) J-X Fu, J. L. R.-J. L. L. G. Z.-Y. L. Unidirectional Channel-Drop Filter by One-Way Gyromagnetic Photonic Crystal Waveguides. *Appl. Phys. Lett.* **2011**, *98*, 211104.

(595) Liu, K.; Shen, L.; He, S. One-Way Edge Mode in a Gyromagnetic Photonic Crystal Slab. *Opt. Lett.* **2012**, *37* (19), 4110. https://doi.org/10.1364/OL.37.004110.

(596) Yang, Y.; Poo, Y.; Wu, R.; Gu, Y.; Chen, P. Experimental Demonstration of One-Way Slow Wave in Waveguide Involving Gyromagnetic Photonic Crystals. *Appl. Phys. Lett.* **2013**, *102* (23), 231113. https://doi.org/10.1063/1.4809956.

(597) Fang, K.; Yu, Z.; Fan, S. Photonic Aharonov-Bohm Effect Based on Dynamic Modulation. *Phys. Rev. Lett.* **2012**, *108* (15). https://doi.org/10.1103/PhysRevLett.108.153901.

(598) Schmidt, M.; Kessler, S.; Peano, V.; Painter, O.; Marquardt, F. Optomechanical Creation of Magnetic Fields for Photons on a Lattice. *Optica* **2015**, *2* (7), 635. https://doi.org/10.1364/optica.2.000635.

(599) Ruesink, F.; Miri, M.-A.; Alù, A.; Verhagen, E. Nonreciprocity and Magnetic-Free Isolation Based on Optomechanical Interactions. *Nat. Commun.* **2016**, *7* (1), 13662. https://doi.org/10.1038/ncomms13662.

(600) Fang, K.; Luo, J.; Metelmann, A.; Matheny, M. H.; Marquardt, F.; Clerk, A. A.; Painter, O. Generalized Non-Reciprocity in an Optomechanical Circuit via Synthetic Magnetism and Reservoir Engineering. *Nat. Phys.* **2017**, *13* (5), 465–471. https://doi.org/10.1038/nphys4009.

(601) Mathew, J. P.; Pino, J. del; Verhagen, E. Synthetic Gauge Fields for Phonon Transport in a Nano-Optomechanical System. *Nat. Nanotechnol.* **2020**, *15* (3), 198–202.

https://doi.org/10.1038/S41565-019-0630-8.

(602) Hofmann, T.; Helbig, T.; Lee, C. H.; Greiter, M.; Thomale, R. Chiral Voltage Propagation and Calibration in a Topolectrical Chern Circuit. *Phys. Rev. Lett.* **2019**, *122* (24), 247702. https://doi.org/10.1103/PhysRevLett.122.247702.

(603) Jin, D.; Christensen, T.; Soljačić, M.; Fang, N. X.; Lu, L.; Zhang, X. Infrared Topological Plasmons in Graphene. *Phys. Rev. Lett.* **2017**, *118* (24), 245301. https://doi.org/10.1103/PhysRevLett.118.245301.

(604) Jin, D.; Lu, L.; Wang, Z.; Fang, C.; Joannopoulos, J. D.; Soljačić, M.; Fu, L.; Fang, N. X. Topological Magnetoplasmon. *Nat. Commun.* **2016**, *7* (1), 13486. https://doi.org/10.1038/ncomms13486.

(605) Khanikaev, A. B.; Fleury, R.; Mousavi, S. H.; Alù, A. Topologically Robust Sound Propagation in an Angular-Momentum-Biased Graphene-like Resonator Lattice. *Nat. Commun.* **2015**, *6* (1), 8260. https://doi.org/10.1038/ncomms9260.

(606) Ding, Y.; Peng, Y.; Zhu, Y.; Fan, X.; Yang, J.; Liang, B.; Zhu, X.; Wan, X.; Cheng, J. Experimental Demonstration of Acoustic Chern Insulators. *Phys. Rev. Lett.* **2019**, *122* (1), 014302. https://doi.org/10.1103/PhysRevLett.122.014302.

(607) Souslov, A.; Van Zuiden, B. C.; Bartolo, D.; Vitelli, V. Topological Sound in Active-Liquid Metamaterials. *Nat. Phys.* **2017**, *13* (11), 1091–1094. https://doi.org/10.1038/nphys4193.

(608) Wang, Y.-T.; Luan, P.-G.; Zhang, S. Coriolis Force Induced Topological Order for Classical Mechanical Vibrations. *New J. Phys.* **2015**, *17* (7), 73031. https://doi.org/10.1088/1367-2630/17/7/073031.

(609) Prodan, E.; Prodan, C. Topological Phonon Modes and Their Role in Dynamic Instability of Microtubules. *Phys. Rev. Lett.* **2009**, *103* (24), 248101. https://doi.org/10.1103/PhysRevLett.103.248101.

(610) Fleury, R.; Sounas, D. L.; Sieck, C. F.; Haberman, M. R.; Alù, A. Sound Isolation and Giant Linear Nonreciprocity in a Compact Acoustic Circulator. *Science (80-. ).* **2014**, *343* (6170), 516. https://doi.org/10.1126/science.1246957.

(611) Ni, X.; He, C.; Sun, X.-C.; Liu, X.; Lu, M.-H.; Feng, L.; Chen, Y.-F. Topologically Protected One-Way Edge Mode in Networks of Acoustic Resonators with Circulating Air Flow. *New J. Phys.* **2015**, *17* (5), 053016. https://doi.org/10.1088/1367-2630/17/5/053016.

(612) Chen, Z.-G.; Wu, Y. Tunable Topological Phononic Crystals. *Phys. Rev. Appl.* **2016**, *5* (5), 054021. https://doi.org/10.1103/PhysRevApplied.5.054021.

(613) Yu, L.; Xue, H.; Zhang, B. Antichiral Edge States in an Acoustic Resonator Lattice with Staggered Air Flow. *J. Appl. Phys.* **2021**, *129* (23), 235103. https://doi.org/10.1063/5.0050645.

(614) Souslov, A.; Dasbiswas, K.; Fruchart, M.; Vaikuntanathan, S.; Vitelli, V. Topological Waves in Fluids with Odd Viscosity. *Phys. Rev. Lett.* **2019**, *122* (12), 128001. https://doi.org/10.1103/PhysRevLett.122.128001.

(615) Shankar, S.; Bowick, M. J.; Marchetti, M. C. Topological Sound and Flocking on Curved Surfaces. **2017**, *031039*, 1–15. https://doi.org/10.1103/PhysRevX.7.031039.

(616) Sone, K.; Ashida, Y. Anomalous Topological Active Matter. *Phys. Rev. Lett.* **2019**, *123* (20), 205502. https://doi.org/10.1103/PhysRevLett.123.205502.

(617) Mitchell, N. P.; Nash, L. M.; Irvine, W. T. M. Tunable Band Topology in Gyroscopic Lattices. *Phys. Rev. B* **2018**, *98* (17), 174301. https://doi.org/10.1103/PhysRevB.98.174301.

(618) Mitchell, N. P.; Nash, L. M.; Hexner, D.; Turner, A. M.; Irvine, W. T. M. Amorphous Topological Insulators Constructed from Random Point Sets. *Nat. Phys.* **2018**, *14* (4), 380–385. https://doi.org/10.1038/s41567-017-0024-5.

(619) Sirota, L.; Ilan, R.; Shokef, Y.; Lahini, Y. Non-Newtonian Topological Mechanical Metamaterials Using Feedback Control. *Phys. Rev. Lett.* **2020**, *125* (25), 256802. https://doi.org/10.1103/PhysRevLett.125.256802.

(620) Pyrialakos, G. G.; Beck, J.; Heinrich, M.; Maczewsky, L. J.; Kantartzis, N. V.; Khajavikhan, M.; Szameit, A.; Christodoulides, D. N. Bimorphic Floquet Topological Insulators. *Nat. Mater.* **2022**, *21* (6), 634–639. https://doi.org/10.1038/s41563-022-01238-w.

(621) Rudner, M. S.; Lindner, N. H.; Berg, E.; Levin, M. Anomalous Edge States and the Bulk-Edge Correspondence for Periodically Driven Two-Dimensional Systems. *Phys. Rev. X* **2013**, *3* (3), 031005. https://doi.org/10.1103/PhysRevX.3.031005.

(622) Pasek, M.; Chong, Y. D. Network Models of Photonic Floquet Topological Insulators. *Phys. Rev. B* **2014**, *89* (7), 075113. https://doi.org/10.1103/PhysRevB.89.075113.

(623) Maczewsky, L. J.; Zeuner, J. M.; Nolte, S.; Szameit, A. Observation of Photonic Anomalous Floquet Topological Insulators. *Nat. Commun.* **2017**, *8* (1), 13756. https://doi.org/10.1038/ncomms13756.

(624) Mukherjee, S.; Spracklen, A.; Valiente, M.; Andersson, E.; Öhberg, P.; Goldman, N.; Thomson, R. R. Experimental Observation of Anomalous Topological Edge Modes in a Slowly Driven Photonic Lattice. *Nat. Commun.* **2017**, *8* (1), 13918. https://doi.org/10.1038/ncomms13918.

(625) Gao, F.; Gao, Z.; Shi, X.; Yang, Z.; Lin, X.; Xu, H.; Joannopoulos, J. D.; Soljačić, M.; Chen, H.; Lu, L.; Chong, Y.; Zhang, B. Probing Topological Protection Using a Designer Surface Plasmon Structure. *Nat. Commun.* **2016**, *7* (1), 11619. https://doi.org/10.1038/ncomms11619.

(626) Nagulu, A.; Ni, X.; Kord, A.; Tymchenko, M.; Garikapati, S.; Alù, A.; Krishnaswamy, H. Chip-Scale Floquet Topological Insulators for 5G Wireless Systems. *Nat. Electron.* **2022**, *5* (5), 300–309. https://doi.org/10.1038/s41928-022-00751-9.

(627) Shen, C.; Zhu, X.; Li, J.; Cummer, S. A. Nonreciprocal Acoustic Transmission in Space-Time Modulated Coupled Resonators. *Phys. Rev. B* **2019**, *100* (5), 054302. https://doi.org/10.1103/PhysRevB.100.054302.

(628) Wen, X.; Zhu, X.; Wu, H. W.; Li, J. Realizing Spatiotemporal Effective Media for Acoustic Metamaterials. *Phys. Rev. B* **2021**, *104* (6), L060304. https://doi.org/10.1103/PhysRevB.104.L060304.

(629) Rupin, M.; Lerosey, G.; de Rosny, J.; Lemoult, F. Mimicking the Cochlea with an Active Acoustic Metamaterial. *New J. Phys.* **2019**, *21* (9), 93012. https://doi.org/10.1088/1367-2630/ab3d8f.

(630) Fleury, R.; Sounas, D.; Alù, A. An Invisible Acoustic Sensor Based on Parity-Time Symmetry. *Nat. Commun.* **2015**, *6* (1), 5905. https://doi.org/10.1038/ncomms6905.

(631) Darabi, A.; Ni, X.; Leamy, M.; Alù, A. Reconfigurable Floquet Elastodynamic Topological Insulator Based on Synthetic Angular Momentum Bias. *Sci. Adv.* **2020**, *6* (29), eaba8656. https://doi.org/10.1126/sciadv.aba8656.

(632) Delplace, P.; Fruchart, M.; Tauber, C. Phase Rotation Symmetry and the Topology of Oriented Scattering Networks. *Phys. Rev. B* **2017**, *95* (20), 205413. https://doi.org/10.1103/PhysRevB.95.205413.

(633) Wei, Q.; Tian, Y.; Zuo, S.-Y.; Cheng, Y.; Liu, X.-J. Experimental Demonstration of Topologically Protected Efficient Sound Propagation in an Acoustic Waveguide Network. *Phys. Rev. B* **2017**, *95* (9), 094305. https://doi.org/10.1103/PhysRevB.95.094305.

(634) He, C.; Li, Z.; Ni, X.; Sun, X.-C.; Yu, S.-Y.; Lu, M.-H.; Liu, X.-P.; Chen, Y.-F. Topological Phononic States of Underwater Sound Based on Coupled Ring Resonators. *Appl. Phys. Lett.* **2016**, *108* (3), 031904. https://doi.org/10.1063/1.4940403.

(635) Peng, Y.-G.; Shen, Y.-X.; Zhao, D.-G.; Zhu, X.-F. Low-Loss and Broadband Anomalous Floquet Topological Insulator for Airborne Sound. *Appl. Phys. Lett.* **2017**, *110* (17), 173505. https://doi.org/10.1063/1.4982620.

(636) Peng, Y.-G.; Qin, C.-Z.; Zhao, D.-G.; Shen, Y.-X.; Xu, X.-Y.; Bao, M.; Jia, H.; Zhu, X.-F. Experimental Demonstration of Anomalous Floquet Topological Insulator for Sound. *Nat. Commun.* **2016**, *7* (1), 13368. https://doi.org/10.1038/ncomms13368.

(637) Peng, Y. G.; Li, Y.; Shen, Y. X.; Geng, Z. G.; Zhu, J.; Qiu, C. W.; Zhu, X. F. Chirality-Assisted Three-Dimensional Acoustic Floquet Lattices. *Phys. Rev. Res.* **2019**, *1* (3), 33149. https://doi.org/10.1103/PhysRevResearch.1.033149.

(638) Chen, W.-J.; Jiang, S.-J.; Chen, X.-D.; Zhu, B.; Zhou, L.; Dong, J.-W.; Chan, C. T. Experimental Realization of Photonic Topological Insulator in a Uniaxial Metacrystal Waveguide. *Nat. Commun.* **2014**, *5* (1), 5782. https://doi.org/10.1038/ncomms6782.

(639) Slobozhanyuk, A. P.; Khanikaev, A. B.; Filonov, D. S.; Smirnova, D. A.; Miroshnichenko, A. E.; Kivshar, Y. S. Experimental Demonstration of Topological Effects in Bianisotropic Metamaterials. *Sci. Rep.* **2016**, *6*. https://doi.org/10.1038/srep22270.

(640) He, C.; Sun, X. C.; Liu, X. P.; Lu, M. H.; Chen, Y.; Feng, L.; Chen, Y. F. Photonic Topological Insulator with Broken Time-Reversal Symmetry. *Proc. Natl. Acad. Sci. U. S. A.* **2016**, *113* (18), 4924–4928. https://doi.org/10.1073/pnas.1525502113.

(641) Ma, T.; Khanikaev, A. B.; Mousavi, S. H.; Shvets, G. Guiding Electromagnetic Waves around Sharp Corners: Topologically Protected Photonic Transport in Metawaveguides. *Phys. Rev. Lett.* **2015**, *114* (12), 127401. https://doi.org/10.1103/PhysRevLett.114.127401.

(642) Slobozhanyuk, A.; Shchelokova, A. V.; Ni, X.; Hossein Mousavi, S.; Smirnova, D. A.; Belov, P. A.; Alù, A.; Kivshar, Y. S.; Khanikaev, A. B. Near-Field Imaging of Spin-Locked Edge States in All-Dielectric Topological Metasurfaces. *Appl. Phys. Lett.* **2019**, *114* (3), 031103. https://doi.org/10.1063/1.5055601.

(643) Bisharat, D. J.; Sievenpiper, D. F. Electromagnetic-Dual Metasurfaces for Topological States along a 1D Interface. *Laser Photon. Rev.* **2019**, *13* (10), 1900126. https://doi.org/ARTN 190012610.1002/lpor.201900126.

(644) Bisharat, D. J.; Sievenpiper, D. F. Guiding Waves Along an Infinitesimal Line between Impedance Surfaces. *Phys. Rev. Lett.* **2017**, *119* (10), 2–6. https://doi.org/10.1103/PhysRevLett.119.106802.

(645) Hafezi, M. Measuring Topological Invariants in Photonic Systems. *Phys. Rev. Lett.* **2014**, *112* (21), 210405. https://doi.org/10.1103/PhysRevLett.112.210405.

(646) Mittal, S.; Ganeshan, S.; Fan, J.; Vaezi, A.; Hafezi, M. Measurement of Topological Invariants in

a 2D Photonic System. *Nat. Photonics* **2016**, *10* (3), 180–183. https://doi.org/10.1038/NPHOTON.2016.10.

(647) Liang, G. Q.; Chong, Y. D. Optical Resonator Analog of a Two-Dimensional Topological Insulator. *Phys. Rev. Lett.* **2013**, *110* (20), 203904. https://doi.org/10.1103/PhysRevLett.110.203904.

(648) Ningyuan, J.; Owens, C.; Sommer, A.; Schuster, D.; Simon, J. Time- and Site-Resolved Dynamics in a Topological Circuit. *Phys. Rev. X* **2015**, *5* (2), 021031. https://doi.org/10.1103/PhysRevX.5.021031.

(649) Albert, V. V.; Glazman, L. I.; Jiang, L. Topological Properties of Linear Circuit Lattices. *Phys. Rev. Lett.* **2015**, *114* (17), 173902. https://doi.org/10.1103/PhysRevLett.114.173902.

(650) Lee, C. H.; Imhof, S.; Berger, C.; Bayer, F.; Brehm, J.; Molenkamp, L. W.; Kiessling, T.; Thomale, R. Topolectrical Circuits. *Commun. Phys.* **2018**, *1* (1), 39. https://doi.org/10.1038/s42005-018-0035-2.

(651) Zhao, H.; Qiao, X.; Wu, T.; Midya, B.; Longhi, S.; Feng, L. Non-Hermitian Topological Light Steering. *Science (80-. ).* **2019**, *365* (6458), 1163–1166. https://doi.org/10.1126/science.aay1064.

(652) Yang, Y.; Xu, Y. F.; Xu, T.; Wang, H. X.; Jiang, J. H.; Hu, X.; Hang, Z. H. Visualization of a Unidirectional Electromagnetic Waveguide Using Topological Photonic Crystals Made of Dielectric Materials. *Phys. Rev. Lett.* **2018**, *120* (21). https://doi.org/10.1103/PHYSREVLETT.120.217401.

(653) Gorlach, M. A.; Ni, X.; Smirnova, D. A.; Korobkin, D.; Zhirihin, D.; Slobozhanyuk, A. P.; Belov, P. A.; Alù, A.; Khanikaev, A. B. Far-Field Probing of Leaky Topological States in All-Dielectric Metasurfaces. *Nat. Commun.* **2018**, *9* (1), 909. https://doi.org/10.1038/s41467-018-03330-9.

(654) Peng, S.; Schilder, N. J.; Ni, X.; van de Groep, J.; Brongersma, M. L.; Alù, A.; Khanikaev, A. B.; Atwater, H. A.; Polman, A. Probing the Band Structure of Topological Silicon Photonic Lattices in the Visible Spectrum. *Phys. Rev. Lett.* **2019**, *122* (11), 117401. https://doi.org/10.1103/PhysRevLett.122.117401.

(655) Yves, S.; Fleury, R.; Berthelot, T.; Fink, M.; Lemoult, F.; Lerosey, G. Crystalline Metamaterials for Topological Properties at Subwavelength Scales. *Nat. Commun.* **2017**, *8* (1), 16023. https://doi.org/10.1038/ncomms16023.

(656) Yang, Z. Q.; Shao, Z. K.; Chen, H. Z.; Mao, X. R.; Ma, R. M. Spin-Momentum-Locked Edge Mode for Topological Vortex Lasing. *Phys. Rev. Lett.* **2020**, *125* (1), 013903. https://doi.org/10.1103/PhysRevLett.125.013903.

(657) Salerno, G.; Berardo, A.; Ozawa, T.; Price, H. M.; Taxis, L.; Pugno, N. M.; Carusotto, I. Spin–Orbit Coupling in a Hexagonal Ring of Pendula. *New J. Phys.* **2017**, *19* (5), 55001. https://doi.org/10.1088/1367-2630/aa6c03.

(658) Pal, R. K.; Schaeffer, M.; Ruzzene, M. Helical Edge States and Topological Phase Transitions in Phononic Systems Using Bi-Layered Lattices. *J. Appl. Phys.* **2016**, *119* (8), 084305. https://doi.org/10.1063/1.4942357.

(659) Deng, W.; Huang, X.; Lu, J.; Peri, V.; Li, F.; Huber, S. D.; Liu, Z. Acoustic Spin-Chern Insulator Induced by Synthetic Spin–Orbit Coupling with Spin Conservation Breaking. *Nat. Commun.* **2020**, *11* (1), 3227. https://doi.org/10.1038/s41467-020-17039-1.

(660) He, C.; Ni, X.; Ge, H.; Sun, X. C.; Chen, Y. Bin; Lu, M. H.; Liu, X. P.; Chen, Y. F. Acoustic

Topological Insulator and Robust One-Way Sound Transport. *Nat. Phys.* **2016**, *12* (12), 1124–1129. https://doi.org/10.1038/NPHYS3867.

(661) Yves, S.; Fleury, R.; Lemoult, F.; Fink, M.; Lerosey, G. Topological Acoustic Polaritons: Robust Sound Manipulation at the Subwavelength Scale. *New J. Phys.* **2017**, *19* (7), 075003. https://doi.org/10.1088/1367-2630/aa66f8.

(662) Zhang, Z.; Wei, Q.; Cheng, Y.; Zhang, T.; Wu, D.; Liu, X. Topological Creation of Acoustic Pseudospin Multipoles in a Flow-Free Symmetry-Broken Metamaterial Lattice. *Phys. Rev. Lett.* **2017**, *118* (8), 084303. https://doi.org/10.1103/PhysRevLett.118.084303.

(663) Mei, J.; Chen, Z.; Wu, Y. Pseudo-Time-Reversal Symmetry and Topological Edge States in Two-Dimensional Acoustic Crystals. *Sci. Rep.* **2016**, *6* (1), 32752. https://doi.org/10.1038/srep32752.

(664) Deng, Y.; Ge, H.; Tian, Y.; Lu, M.; Jing, Y. Observation of Zone Folding Induced Acoustic Topological Insulators and the Role of Spin-Mixing Defects. *Phys. Rev. B* **2017**, *96* (18), 184305. https://doi.org/10.1103/PhysRevB.96.184305.

(665) Jia, D.; Sun, H.; Xia, J.; Yuan, S.; Liu, X.; Zhang, C. Acoustic Topological Insulator by Honeycomb Sonic Crystals with Direct and Indirect Band Gaps. *New J. Phys.* **2018**, *20* (9), 93027. https://doi.org/10.1088/1367-2630/aae104.

(666) Yu, S.-Y.; He, C.; Wang, Z.; Liu, F.-K.; Sun, X.-C.; Li, Z.; Lu, H.-Z.; Lu, M.-H.; Liu, X.-P.; Chen, Y.-F. Elastic Pseudospin Transport for Integratable Topological Phononic Circuits. *Nat. Commun.* **2018**, *9* (1), 3072. https://doi.org/10.1038/s41467-018-05461-5.

(667) Foehr, A.; Bilal, O. R.; Huber, S. D.; Daraio, C. Spiral-Based Phononic Plates: From Wave Beaming to Topological Insulators. *Phys. Rev. Lett.* **2018**, *120* (20), 205501. https://doi.org/10.1103/PhysRevLett.120.205501.

(668) Xia, J.-P.; Jia, D.; Sun, H.-X.; Yuan, S.-Q.; Ge, Y.; Si, Q.-R.; Liu, X.-J. Programmable Coding Acoustic Topological Insulator. *Adv. Mater.* **2018**, *30* (46), 1805002. https://doi.org/10.1002/adma.201805002.

(669) Pirie, H.; Sadhuka, S.; Wang, J.; Andrei, R.; Hoffman, J. E. Topological Phononic Logic. *Phys. Rev. Lett.* **2022**, *128* (1), 15501. https://doi.org/10.1103/PhysRevLett.128.015501.

(670) Li, G.-H.; Ma, T.-X.; Wang, Y.-Z.; Wang, Y.-S. Active Control on Topological Immunity of Elastic Wave Metamaterials. *Sci. Rep.* **2020**, *10* (1), 9376. https://doi.org/10.1038/s41598-020-66269-2.

(671) Liu, P.; Li, H.; Zhou, Z.; Pei, Y. Topological Acoustic Tweezer and Pseudo-Spin States of Acoustic Topological Insulators. *Appl. Phys. Lett.* **2022**, *120* (22), 222202. https://doi.org/10.1063/5.0091755.

(672) Chaunsali, R.; Chen, C.-W.; Yang, J. Experimental Demonstration of Topological Waveguiding in Elastic Plates with Local Resonators. *New J. Phys.* **2018**, *20* (11), 113036. https://doi.org/10.1088/1367-2630/aaeb61.

(673) Chen, H.; Nassar, H.; Norris, A. N.; Hu, G. K.; Huang, G. L. Elastic Quantum Spin Hall Effect in Kagome Lattices. *Phys. Rev. B* **2018**, *98* (9), 094302. https://doi.org/10.1103/PhysRevB.98.094302.

(674) Chen, Y.; Liu, X.; Hu, G. Topological Phase Transition in Mechanical Honeycomb Lattice. *J. Mech. Phys. Solids* **2019**, *122*, 54–68. https://doi.org/10.1016/j.jmps.2018.08.021.

(675) Li, J.; Wang, J.; Wu, S.; Mei, J. Pseudospins and Topological Edge States in Elastic Shear Waves. *AIP Adv.* **2017**, *7* (12), 125030. https://doi.org/10.1063/1.5010754.

(676) Cha, J.; Kim, K. W.; Daraio, C. Experimental Realization of On-Chip Topological Nanoelectromechanical Metamaterials. *Nature* **2018**, *564* (7735), 229–233. https://doi.org/10.1038/s41586-018-0764-0.

(677) Brendel, C.; Peano, V.; Painter, O.; Marquardt, F. Snowflake Phononic Topological Insulator at the Nanoscale. *Phys. Rev. B* **2018**, *97* (2). https://doi.org/10.1103/PHYSREVB.97.020102.

(678) Ma, J.; Xi, X.; Li, Y.; Sun, X. Nanomechanical Topological Insulators with an Auxiliary Orbital Degree of Freedom. *Nat. Nanotechnol.* **2021**, *16* (5), 576–583. https://doi.org/10.1038/s41565-021-00868-6.

(679) Zhang, Z.-D.; Yu, S.-Y.; Ge, H.; Wang, J.-Q.; Wang, H.-F.; Liu, K.-F.; Wu, T.; He, C.; Lu, M.-H.; Chen, Y.-F. Topological Surface Acoustic Waves. *Phys. Rev. Appl.* **2021**, *16* (4), 44008. https://doi.org/10.1103/PhysRevApplied.16.044008.

(680) Yu, Z.; Ren, Z.; Lee, J. Phononic Topological Insulators Based on Six-Petal Holey Silicon Structures. *Sci. Rep.* **2019**, *9* (1), 1805. https://doi.org/10.1038/s41598-018-38387-5.

(681) Mousavi, S. H.; Khanikaev, A. B.; Wang, Z. Topologically Protected Elastic Waves in Phononic Metamaterials. *Nat. Commun.* **2015**, *6* (1), 8682. https://doi.org/10.1038/ncomms9682.

(682) Miniaci, M.; Pal, R. K.; Morvan, B.; Ruzzene, M. Experimental Observation of Topologically Protected Helical Edge Modes in Patterned Elastic Plates. *Phys. Rev. X* **2018**, *8* (3), 031074. https://doi.org/10.1103/PhysRevX.8.031074.

(683) Zheng, L. Y.; Theocharis, G.; Tournat, V.; Gusev, V. Quasitopological Rotational Waves in Mechanical Granular Graphene. *Phys. Rev. B* **2018**, *97* (6), 060101. https://doi.org/10.1103/PhysRevB.97.060101.

(684) Ma, T.; Shvets, G. All-Si Valley-Hall Photonic Topological Insulator. *New J. Phys.* **2016**, *18* (2), 025012. https://doi.org/10.1088/1367-2630/18/2/025012.

(685) Chen, X. D.; Zhao, F. L.; Chen, M.; Dong, J. W. Valley-Contrasting Physics in All-Dielectric Photonic Crystals: Orbital Angular Momentum and Topological Propagation. *Phys. Rev. B* **2017**, *96* (2), 20202. https://doi.org/ARTN 02020210.1103/PhysRevB.96.020202.

(686) Gao, Z.; Yang, Z.; Gao, F.; Xue, H.; Yang, Y.; Dong, J.; Zhang, B. Valley Surface-Wave Photonic Crystal and Its Bulk/Edge Transport. *Phys. Rev. B* **2017**, *96* (20), 201402. https://doi.org/10.1103/PhysRevB.96.201402.

(687) Ye, L. P.; Yang, Y. T.; Hang, Z. H.; Qiu, C. Y.; Liu, Z. Y. Observation of Valley-Selective Microwave Transport in Photonic Crystals. *Appl. Phys. Lett.* **2017**, *111* (25), 251107. https://doi.org/Artn 25110710.1063/1.5009597.

(688) Gao, F.; Xue, H. R.; Yang, Z. J.; Lai, K. F.; Yu, Y.; Lin, X.; Chong, Y. D.; Shvets, G.; Zhang, B. L. Topologically Protected Refraction of Robust Kink States in Valley Photonic Crystals. *Nat. Phys.* **2018**, *14* (2), 140–144. https://doi.org/10.1038/Nphys4304.

(689) Chen, X.; Shi, F.; Liu, H.; Lu, J.; Deng, W.; Dai, J.; Cheng, Q.; Dong, J.-W. Tunable Electromagnetic Flow Control in Valley Photonic Crystal Waveguides. *Phys. Rev. Appl.* **2018**, *10* (4), 044002. https://doi.org/10.1103/PhysRevApplied.10.044002.

(690) Noh, J.; Huang, S.; Chen, K. P.; Rechtsman, M. C. Observation of Photonic Topological Valley

Hall Edge States. *Phys. Rev. Lett.* **2018**, *120* (6), 063902. https://doi.org/10.1103/PhysRevLett.120.063902.

(691) Wu, X.; Meng, Y.; Tian, J.; Huang, Y.; Xiang, H.; Han, D.; Wen, W. Direct Observation of Valley-Polarized Topological Edge States in Designer Surface Plasmon Crystals. *Nat. Commun.* **2017**, *8* (1), 1304. https://doi.org/10.1038/s41467-017-01515-2.

(692) Shalaev, M. I.; Walasik, W.; Tsukernik, A.; Xu, Y.; Litchinitser, N. M. Robust Topologically Protected Transport in Photonic Crystals at Telecommunication Wavelengths. *Nat. Nanotechnol.* **2019**, *14* (1), 31–34. https://doi.org/10.1038/s41565-018-0297-6.

(693) He, X.-T.; Liang, E.-T.; Yuan, J.-J.; Qiu, H.-Y.; Chen, X.-D.; Zhao, F.-L.; Dong, J.-W. A Silicon-on-Insulator Slab for Topological Valley Transport. *Nat. Commun.* **2019**, *10* (1), 872. https://doi.org/10.1038/s41467-019-08881-z.

(694) Yamaguchi, T.; Ota, Y.; Katsumi, R.; Watanabe, K.; Ishida, S.; Osada, A.; Arakawa, Y.; Iwamoto, S. GaAs Valley Photonic Crystal Waveguide with Light-Emitting InAs Quantum Dots. *Appl. Phys. Express* **2019**, *12* (6), 62005. https://doi.org/10.7567/1882-0786/ab1cc5.

(695) Ma, T.; Shvets, G. Scattering-Free Edge States between Heterogeneous Photonic Topological Insulators. *Phys. Rev. B* **2017**, *95* (16), 165102. https://doi.org/10.1103/PhysRevB.95.165102.

(696) Kang, Y.; Ni, X.; Cheng, X.; Khanikaev, A. B.; Genack, A. Z. Pseudo-Spin–Valley Coupled Edge States in a Photonic Topological Insulator. *Nat. Commun.* **2018**, *9* (1), 3029. https://doi.org/10.1038/s41467-018-05408-w.

(697) Yang, Y.; Yamagami, Y.; Yu, X.; Pitchappa, P.; Webber, J.; Zhang, B.; Fujita, M.; Nagatsuma, T.; Singh, R. Terahertz Topological Photonics for On-Chip Communication. *Nat. Photonics* **2020**, *14* (7), 446–451. https://doi.org/10.1038/s41566-020-0618-9.

(698) Zeng, Y.; Chattopadhyay, U.; Zhu, B.; Qiang, B.; Li, J.; Jin, Y.; Li, L.; Davies, A. G.; Linfield, E. H.; Zhang, B.; Chong, Y.; Wang, Q. J. Electrically Pumped Topological Laser with Valley Edge Modes. *Nature* **2020**, *578* (7794), 246–250. https://doi.org/10.1038/s41586-020-1981-x.

(699) Jung, M.; Fan, Z.; Shvets, G. Midinfrared Plasmonic Valleytronics in Metagate-Tuned Graphene. *Phys. Rev. Lett.* **2018**, *121* (8), 086807. https://doi.org/10.1103/PhysRevLett.121.086807.

(700) Lu, J.; Qiu, C.; Ke, M.; Liu, Z. Valley Vortex States in Sonic Crystals. *Phys. Rev. Lett.* **2016**, *116* (9), 093901. https://doi.org/10.1103/PhysRevLett.116.093901.

(701) Lu, J.; Qiu, C.; Ye, L.; Fan, X.; Ke, M.; Zhang, F.; Liu, Z. Observation of Topological Valley Transport of Sound in Sonic Crystals. *Nat. Phys.* **2017**, *13* (4), 369–374. https://doi.org/10.1038/nphys3999.

(702) Ye, L.; Qiu, C.; Lu, J.; Wen, X.; Shen, Y.; Ke, M.; Zhang, F.; Liu, Z. Observation of Acoustic Valley Vortex States and Valley-Chirality Locked Beam Splitting. *Phys. Rev. B* **2017**, *95* (17), 174106. https://doi.org/10.1103/PhysRevB.95.174106.

(703) Xia, B.-Z.; Zheng, S.-J.; Liu, T.-T.; Jiao, J.-R.; Chen, N.; Dai, H.-Q.; Yu, D.-J.; Liu, J. Observation of Valleylike Edge States of Sound at a Momentum Away from the High-Symmetry Points. *Phys. Rev. B* **2018**, *97* (15), 155124. https://doi.org/10.1103/PhysRevB.97.155124.

(704) Han, X.; Peng, Y.-G.; Li, L.; Hu, Y.; Mei, C.; Zhao, D.-G.; Zhu, X.-F.; Wang, X. Experimental Demonstration of Acoustic Valley Hall Topological Insulators with the Robust Selection of C 3 v -Symmetric Scatterers. *Phys. Rev. Appl.* **2019**, *12* (1), 014046. https://doi.org/10.1103/PhysRevApplied.12.014046.

(705) Zhu, Z.; Huang, X.; Lu, J.; Yan, M.; Li, F.; Deng, W.; Liu, Z. Negative Refraction and Partition in Acoustic Valley Materials of a Square Lattice. *Phys. Rev. Appl.* **2019**, *12* (2), 024007. https://doi.org/10.1103/PhysRevApplied.12.024007.

(706) Yves, S.; Lerosey, G.; Lemoult, F. Structure-Composition Correspondence in Crystalline Metamaterials for Acoustic Valley-Hall Effect and Unidirectional Sound Guiding. *EPL (Europhysics Lett.* **2020**, *129* (4), 44001. https://doi.org/10.1209/0295-5075/129/44001.

(707) Chen, C.; Chen, T.; Wang, Y.; Wu, J.; Zhu, J. Observation of Topological Locally Resonate and Bragg Edge Modes in a Two-Dimensional Slit-Typed Sonic Crystal. *Appl. Phys. Express* **2019**, *12* (9), 97001. https://doi.org/10.7567/1882-0786/ab354b.

(708) Zhang, Z.; Gu, Y.; Long, H.; Cheng, Y.; Liu, X.; Christensen, J. Subwavelength Acoustic Valley-Hall Topological Insulators Using Soda Cans Honeycomb Lattices. *Research* **2019**, *2019*, 1–8. https://doi.org/10.34133/2019/5385763.

(709) Wang, Z.; Yang, Y.; Li, H.; Jia, H.; Luo, J.; Huang, J.; Wang, Z.; Jiang, B.; Yang, N.; Jin, G.; Yang, H. Multichannel Topological Transport in an Acoustic Valley Hall Insulator. *Phys. Rev. Appl.* **2021**, *15* (2), 024019. https://doi.org/10.1103/PhysRevApplied.15.024019.

(710) Shen, Y.; Qiu, C.; Cai, X.; Ye, L.; Lu, J.; Ke, M.; Liu, Z. Valley-Projected Edge Modes Observed in Underwater Sonic Crystals. *Appl. Phys. Lett.* **2019**, *114* (2), 023501. https://doi.org/10.1063/1.5049856.

(711) Huo, S.; Chen, J.; Huang, H.; Huang, G. Simultaneous Multi-Band Valley-Protected Topological Edge States of Shear Vertical Wave in Two-Dimensional Phononic Crystals with Veins. *Sci. Rep.* **2017**, *7* (1), 10335. https://doi.org/10.1038/s41598-017-10857-2.

(712) Yang, Y.; Yang, Z.; Zhang, B. Acoustic Valley Edge States in a Graphene-like Resonator System. *J. Appl. Phys.* **2018**, *123* (9), 091713. https://doi.org/10.1063/1.5009626.

(713) Wang, Z.; Liu, F.-K.; Yu, S.-Y.; Yan, S.-L.; Lu, M.-H.; Jing, Y.; Chen, Y.-F. Guiding Robust Valley-Dependent Edge States by Surface Acoustic Waves. *J. Appl. Phys.* **2019**, *125* (4), 044502. https://doi.org/10.1063/1.5066034.

(714) Pal, R. K.; Ruzzene, M. Edge Waves in Plates with Resonators: An Elastic Analogue of the Quantum Valley Hall Effect. *New J. Phys.* **2017**, *19* (2), 025001. https://doi.org/10.1088/1367-2630/aa56a2.

(715) Vila, J.; Pal, R. K.; Ruzzene, M. Observation of Topological Valley Modes in an Elastic Hexagonal Lattice. *Phys. Rev. B* **2017**, *96* (13), 134307. https://doi.org/10.1103/PhysRevB.96.134307.

(716) Ma, J.; Sun, K.; Gonella, S. Valley Hall In-Plane Edge States as Building Blocks for Elastodynamic Logic Circuits. *Phys. Rev. Appl.* **2019**, *12* (4), 044015. https://doi.org/10.1103/PhysRevApplied.12.044015.

(717) Zhang, Q.; Chen, Y.; Zhang, K.; Hu, G. Dirac Degeneracy and Elastic Topological Valley Modes Induced by Local Resonant States. *Phys. Rev. B* **2020**, *101* (1), 014101. https://doi.org/10.1103/PhysRevB.101.014101.

(718) Liu, T.-W.; Semperlotti, F. Experimental Evidence of Robust Acoustic Valley Hall Edge States in a Nonresonant Topological Elastic Waveguide. *Phys. Rev. Appl.* **2019**, *11* (1), 014040. https://doi.org/10.1103/PhysRevApplied.11.014040.

(719) Mei, J.; Wang, J.; Zhang, X.; Yu, S.; Wang, Z.; Lu, M.-H. Robust and High-Capacity Phononic

Communications through Topological Edge States by Discrete Degree-of-Freedom Multiplexing. *Phys. Rev. Appl.* **2019**, *12* (5), 054041. https://doi.org/10.1103/PhysRevApplied.12.054041.

(720)  Wang, J.; Mei, J. Topological Valley-Chiral Edge States of Lamb Waves in Elastic Thin Plates. *Appl. Phys. Express* **2018**, *11* (5), 057302. https://doi.org/10.7567/APEX.11.057302.

(721)  Zhu, H.; Liu, T.-W.; Semperlotti, F. Design and Experimental Observation of Valley-Hall Edge States in Diatomic-Graphene-like Elastic Waveguides. *Phys. Rev. B* **2018**, *97* (17), 174301. https://doi.org/10.1103/PhysRevB.97.174301.

(722)  Qian, K.; Apigo, D. J.; Prodan, C.; Barlas, Y.; Prodan, E. Topology of the Valley-Chern Effect. *Phys. Rev. B* **2018**, *98* (15), 155138. https://doi.org/10.1103/PhysRevB.98.155138.

(723)  Jin, Y.; Torrent, D.; Djafari-Rouhani, B. Robustness of Conventional and Topologically Protected Edge States in Phononic Crystal Plates. *Phys. Rev. B* **2018**, *98* (5), 054307. https://doi.org/10.1103/PhysRevB.98.054307.

(724)  Yan, M.; Lu, J.; Li, F.; Deng, W.; Huang, X.; Ma, J.; Liu, Z. On-Chip Valley Topological Materials for Elastic Wave Manipulation. *Nat. Mater.* **2018**, *17* (11), 993–998. https://doi.org/10.1038/s41563-018-0191-5.

(725)  Ma, J.; Xi, X.; Sun, X. Experimental Demonstration of Dual-Band Nano-Electromechanical Valley-Hall Topological Metamaterials. *Adv. Mater.* **2021**, *33* (10), 2006521. https://doi.org/10.1002/adma.202006521.

(726)  Zhang, Q.; Lee, D.; Zheng, L.; Ma, X.; Meyer, S. I.; He, L.; Ye, H.; Gong, Z.; Zhen, B.; Lai, K.; Johnson, A. T. C. Gigahertz Topological Valley Hall Effect in Nanoelectromechanical Phononic Crystals. *Nat. Electron.* **2022**, *5* (3), 157–163. https://doi.org/10.1038/s41928-022-00732-y.

(727)  Ren, H.; Shah, T.; Pfeifer, H.; Brendel, C.; Peano, V.; Marquardt, F.; Painter, O. Topological Phonon Transport in an Optomechanical System. *Nat. Commun.* **2022**, *13* (1), 3476. https://doi.org/10.1038/s41467-022-30941-0.

(728)  Zhang, Z.; Tian, Y.; Wang, Y.; Gao, S.; Cheng, Y.; Liu, X.; Christensen, J. Directional Acoustic Antennas Based on Valley-Hall Topological Insulators. *Adv. Mater.* **2018**, *30* (36), 1803229. https://doi.org/10.1002/adma.201803229.

(729)  Jia, D.; Ge, Y.; Xue, H.; Yuan, S.; Sun, H.; Yang, Y.; Liu, X.; Zhang, B. Topological Refraction in Dual-Band Valley Sonic Crystals. *Phys. Rev. B* **2021**, *103* (14), 144309. https://doi.org/10.1103/PhysRevB.103.144309.

(730)  Zhang, Z.; Tian, Y.; Cheng, Y.; Wei, Q.; Liu, X.; Christensen, J. Topological Acoustic Delay Line. *Phys. Rev. Appl.* **2018**, *9* (3), 34032. https://doi.org/10.1103/PhysRevApplied.9.034032.

(731)  Darabi, A.; Collet, M.; Leamy, M. J. Experimental Realization of a Reconfigurable Electroacoustic Topological Insulator. *Proc. Natl. Acad. Sci.* **2020**, *117* (28), 16138–16142. https://doi.org/10.1073/pnas.1920549117.

(732)  Zhu, J.; Chen, T.; Chen, C.; Ding, W. Valley Vortex Assisted and Topological Protected Microparticles Manipulation with Complicated 2D Patterns in a Star-like Sonic Crystal. *Materials (Basel).* **2021**, *14* (17), 4939.

(733)  Miniaci, M.; Pal, R. K.; Manna, R.; Ruzzene, M. Valley-Based Splitting of Topologically Protected Helical Waves in Elastic Plates. *Phys. Rev. B* **2019**, *100* (2), 024304. https://doi.org/10.1103/PhysRevB.100.024304.

(734) Lu, J.; Qiu, C.; Deng, W.; Huang, X.; Li, F.; Zhang, F.; Chen, S.; Liu, Z. Valley Topological Phases in Bilayer Sonic Crystals. *Phys. Rev. Lett.* **2018**, *120* (11), 116802. https://doi.org/10.1103/PhysRevLett.120.116802.

(735) Jiao, J.; Chen, T.; Dai, H.; Yu, D. Observation of Topological Valley Transport of Elastic Waves in Bilayer Phononic Crystal Slabs. *Phys. Lett. A* **2019**, *383* (34), 125988. https://doi.org/10.1016/j.physleta.2019.125988.

(736) Lu, L.; Fu, L.; Joannopoulos, J. D.; Soljačić, M. Weyl Points and Line Nodes in Gyroid Photonic Crystals. *Nat. Photonics* **2013**, *7* (4), 294–299. https://doi.org/10.1038/nphoton.2013.42.

(737) Wang, L. Y.; Jian, S. K.; Yao, H. Topological Photonic Crystal with Equifrequency Weyl Points. *Phys. Rev. A* **2016**, *93* (6), 61801. https://doi.org/ARTN 06180110.1103/PhysRevA.93.061801.

(738) Lu, L.; Wang, Z.; Ye, D.; Ran, L.; Fu, L.; Joannopoulos, J. D.; Soljačić, M. Experimental Observation of Weyl Points. *Science (80-. ).* **2015**, *349* (6248), 622–624. https://doi.org/10.1126/SCIENCE.AAA9273.

(739) Goi, E.; Yue, Z.; Cumming, B. P.; Gu, M. Observation of Type I Photonic Weyl Points in Optical Frequencies. *Laser Photon. Rev.* **2018**, *12* (2), 1700271. https://doi.org/10.1002/lpor.201700271.

(740) Kim, M.; Gao, W.; Lee, D.; Ha, T.; Kim, T.; Zhang, S.; Rho, J. Extremely Broadband Topological Surface States in a Photonic Topological Metamaterial. *Adv. Opt. Mater.* **2019**, *7* (20), 1900900. https://doi.org/10.1002/adom.201900900.

(741) Chen, W.-J.; Xiao, M.; Chan, C. T. Photonic Crystals Possessing Multiple Weyl Points and the Experimental Observation of Robust Surface States. *Nat. Commun.* **2016**, *7* (1), 13038. https://doi.org/10.1038/ncomms13038.

(742) Chang, M.-L.; Xiao, M.; Chen, W.-J.; Chan, C. T. Multiple Weyl Points and the Sign Change of Their Topological Charges in Woodpile Photonic Crystals. *Phys. Rev. B* **2017**, *95* (12), 125136. https://doi.org/10.1103/PhysRevB.95.125136.

(743) Vaidya, S.; Noh, J.; Cerjan, A.; Jorg, C.; von Freymann, G.; Rechtsman, M. C. Observation of a Charge-2 Photonic Weyl Point in the Infrared. *Phys. Rev. Lett.* **2020**, *125* (25), 253902. https://doi.org/ARTN 25390210.1103/PhysRevLett.125.253902.

(744) Noh, J.; Huang, S.; Leykam, D.; Chong, Y. D.; Chen, K. P.; Rechtsman, M. C. Experimental Observation of Optical Weyl Points and Fermi Arc-like Surface States. *Nat. Phys.* **2017**, *13* (6), 611–617. https://doi.org/10.1038/NPHYS4072.

(745) Gao, W.; Lawrence, M.; Yang, B.; Liu, F.; Fang, F.; Béri, B.; Li, J.; Zhang, S. Topological Photonic Phase in Chiral Hyperbolic Metamaterials. *Phys. Rev. Lett.* **2015**, *114* (3), 037402. https://doi.org/10.1103/PhysRevLett.114.037402.

(746) Xiao, M.; Lin, Q.; Fan, S. Hyperbolic Weyl Point in Reciprocal Chiral Metamaterials. *Phys. Rev. Lett.* **2016**, *117* (5), 057401. https://doi.org/10.1103/PhysRevLett.117.057401.

(747) Yang, B.; Guo, Q.; Tremain, B.; Liu, R.; Barr, L. E.; Yan, Q.; Gao, W.; Liu, H.; Xiang, Y.; Chen, J.; Fang, C.; Hibbins, A.; Lu, L.; Zhang, S. Ideal Weyl Points and Helicoid Surface States in Artificial Photonic Crystal Structures. *Science (80-. ).* **2018**, *359* (6379), 1013–1016. https://doi.org/10.1126/SCIENCE.AAQ1221.

(748) Jia, H.; Zhang, R.; Gao, W.; Guo, Q.; Yang, B.; Hu, J.; Bi, Y.; Xiang, Y.; Liu, C.; Zhang, S. Observation of Chiral Zero Mode in Inhomogeneous Three-Dimensional Weyl Metamaterials. *Science (80-. ).* **2019**, *363* (6423), 148–151.

https://doi.org/10.1126/SCIENCE.AAU7707/SUPPL_FILE/AAU7707_JIA_SM.PDF.

(749) Gao, W.; Yang, B.; Lawrence, M.; Fang, F.; Béri, B.; Zhang, S. Photonic Weyl Degeneracies in Magnetized Plasma. *Nat. Commun.* **2016**, *7* (1), 12435. https://doi.org/10.1038/ncomms12435.

(750) Wang, H.; Zhou, L.; Chong, Y. D. Floquet Weyl Phases in a Three-Dimensional Network Model. *Phys. Rev. B* **2016**, *93* (14), 144114. https://doi.org/10.1103/PhysRevB.93.144114.

(751) Wang, Q.; Xiao, M.; Liu, H.; Zhu, S.; Chan, C. T. Optical Interface States Protected by Synthetic Weyl Points. *Phys. Rev. X* **2017**, *7* (3), 031032. https://doi.org/10.1103/PhysRevX.7.031032.

(752) Wang, D.; Yang, B.; Gao, W.; Jia, H.; Yang, Q.; Chen, X.; Wei, M.; Liu, C.; Navarro-Cía, M.; Han, J.; Zhang, W.; Zhang, S. Photonic Weyl Points Due to Broken Time-Reversal Symmetry in Magnetized Semiconductor. *Nat. Phys.* **2019**, *15* (11), 1150–1155. https://doi.org/10.1038/S41567-019-0612-7.

(753) Lu, Y.; Jia, N.; Su, L.; Owens, C.; Juzeliūnas, G.; Schuster, D. I.; Simon, J. Probing the Berry Curvature and Fermi Arcs of a Weyl Circuit. *Phys. Rev. B* **2019**, *99* (2), 020302. https://doi.org/10.1103/PhysRevB.99.020302.

(754) Zhou, M.; Ying, L.; Lu, L.; Shi, L.; Zi, J.; Yu, Z. Electromagnetic Scattering Laws in Weyl Systems. *Nat. Commun.* **2017**, *8* (1), 1388. https://doi.org/10.1038/s41467-017-01533-0.

(755) Yang, Y.; Gao, W.; Xia, L.; Cheng, H.; Jia, H.; Xiang, Y.; Zhang, S. Spontaneous Emission and Resonant Scattering in Transition from Type I to Type II Photonic Weyl Systems. *Phys. Rev. Lett.* **2019**, *123* (3), 033901. https://doi.org/10.1103/PhysRevLett.123.033901.

(756) Cheng, H.; Gao, W.; Bi, Y.; Liu, W.; Li, Z.; Guo, Q.; Yang, Y.; You, O.; Feng, J.; Sun, H.; Tian, J.; Chen, S.; Zhang, S. Vortical Reflection and Spiraling Fermi Arcs with Weyl Metamaterials. *Phys. Rev. Lett.* **2020**, *125* (9), 093904. https://doi.org/10.1103/PhysRevLett.125.093904.

(757) Wang, H. X.; Xu, L.; Chen, H. Y.; Jiang, J. H. Three-Dimensional Photonic Dirac Points Stabilized by Point Group Symmetry. *Phys. Rev. B* **2016**, *93* (23), 235155. https://doi.org/ARTN 23515510.1103/PhysRevB.93.235155.

(758) Guo, Q.; Yang, B.; Xia, L.; Gao, W.; Liu, H.; Chen, J.; Xiang, Y.; Zhang, S. Three Dimensional Photonic Dirac Points in Metamaterials. *Phys. Rev. Lett.* **2017**, *119* (21), 213901. https://doi.org/10.1103/PhysRevLett.119.213901.

(759) Guo, Q.; You, O.; Yang, B.; Sellman, J. B.; Blythe, E.; Liu, H.; Xiang, Y.; Li, J.; Fan, D.; Chen, J.; Chan, C. T.; Zhang, S. Observation of Three-Dimensional Photonic Dirac Points and Spin-Polarized Surface Arcs. *Phys. Rev. Lett.* **2019**, *122* (20). https://doi.org/10.1103/PHYSREVLETT.122.203903.

(760) Kawakami, T.; Hu, X. Symmetry-Guaranteed Nodal-Line Semimetals in an Fcc Lattice. *Phys. Rev. B* **2017**, *96* (23), 235307. https://doi.org/10.1103/PhysRevB.96.235307.

(761) Gao, W.; Yang, B.; Tremain, B.; Liu, H.; Guo, Q.; Xia, L.; Hibbins, A. P.; Zhang, S. Experimental Observation of Photonic Nodal Line Degeneracies in Metacrystals. *Nat. Commun.* **2018**, *9* (1), 950. https://doi.org/10.1038/s41467-018-03407-5.

(762) Yan, Q.; Liu, R.; Yan, Z.; Liu, B.; Chen, H.; Wang, Z.; Lu, L. Experimental Discovery of Nodal Chains. *Nat. Phys.* **2018**, *14* (5), 461–464. https://doi.org/10.1038/s41567-017-0041-4.

(763) Xia, L.; Guo, Q.; Yang, B.; Han, J.; Liu, C.-X.; Zhang, W.; Zhang, S. Observation of Hourglass Nodal Lines in Photonics. *Phys. Rev. Lett.* **2019**, *122* (10), 103903.

https://doi.org/10.1103/PhysRevLett.122.103903.

(764)   Yang, E.; Yang, B.; You, O.; Chan, H. C.; Mao, P.; Guo, Q.; Ma, S.; Xia, L.; Fan, D.; Xiang, Y.; Zhang, S. Observation of Non-Abelian Nodal Links in Photonics. *Phys. Rev. Lett.* **2020**, *125* (3), 033901. https://doi.org/10.1103/PhysRevLett.125.033901.

(765)   Park, H.; Gao, W.; Zhang, X.; Oh, S. S. Nodal Lines in Momentum Space: Topological Invariants and Recent Realizations in Photonic and Other Systems. *Nanophotonics* **2022**, *11* (11), 2779–2801. https://doi.org/10.1515/nanoph-2021-0692.

(766)   Park, H.; Wong, S.; Zhang, X.; Oh, S. S. Non-Abelian Charged Nodal Links in a Dielectric Photonic Crystal. *ACS Photonics* **2021**, *8* (9), 2746–2754. https://doi.org/10.1021/acsphotonics.1c00876.

(767)   Wang, D.; Yang, B.; Guo, Q.; Zhang, R.-Y.; Xia, L.; Su, X.; Chen, W.-J.; Han, J.; Zhang, S.; Chan, C. T. Intrinsic In-Plane Nodal Chain and Generalized Quaternion Charge Protected Nodal Link in Photonics. *Light Sci. Appl.* **2021**, *10* (1), 83. https://doi.org/10.1038/s41377-021-00523-8.

(768)   Zhong, C. Y.; Chen, Y. P.; Xie, Y. E.; Yang, S. Y. A.; Cohen, M. L.; Zhang, S. B. Towards Three-Dimensional Weyl-Surface Semimetals in Graphene Networks. *Nanoscale* **2016**, *8* (13), 7232–7239. https://doi.org/10.1039/c6nr00882h.

(769)   Kim, M.; Lee, D.; Lee, D.; Rho, J. Topologically Nontrivial Photonic Nodal Surface in a Photonic Metamaterial. *Phys. Rev. B* **2019**, *99* (23), 235423. https://doi.org/10.1103/PhysRevB.99.235423.

(770)   Xiao, M.; Chen, W. J.; He, W. Y.; Chan, C. T. Synthetic Gauge Flux and Weyl Points in Acoustic Systems. *Nat. Phys.* **2015**, *11* (11), 920–924. https://doi.org/10.1038/NPHYS3458.

(771)   Li, F.; Huang, X.; Lu, J.; Ma, J.; Liu, Z. Weyl Points and Fermi Arcs in a Chiral Phononic Crystal. *Nat. Phys.* **2018**, *14* (1), 30–34. https://doi.org/10.1038/NPHYS4275.

(772)   Ge, H.; Ni, X.; Tian, Y.; Gupta, S. K.; Lu, M.-H.; Lin, X.; Huang, W.-D.; Chan, C. T.; Chen, Y.-F. Experimental Observation of Acoustic Weyl Points and Topological Surface States. *Phys. Rev. Appl.* **2018**, *10* (1), 14017. https://doi.org/10.1103/PhysRevApplied.10.014017.

(773)   He, H.; Qiu, C.; Ye, L.; Cai, X.; Fan, X.; Ke, M.; Zhang, F.; Liu, Z. Topological Negative Refraction of Surface Acoustic Waves in a Weyl Phononic Crystal. *Nature* **2018**, *560* (7716), 61–64. https://doi.org/10.1038/S41586-018-0367-9.

(774)   Huang, X.; Deng, W.; Li, F.; Lu, J.; Liu, Z. Ideal Type-II Weyl Phase and Topological Transition in Phononic Crystals. *Phys. Rev. Lett.* **2020**, *124* (20), 206802. https://doi.org/10.1103/PhysRevLett.124.206802.

(775)   Xiao, M.; Ye, L.; Qiu, C.; He, H.; Liu, Z.; Fan, S. Experimental Demonstration of Acoustic Semimetal with Topologically Charged Nodal Surface. *Sci. Adv.* **2020**, *6* (8), 1–7. https://doi.org/10.1126/sciadv.aav2360.

(776)   Yang, Y.; Sun, H.; Xia, J.; Xue, H.; Gao, Z.; Ge, Y.; Jia, D.; Yuan, S.; Chong, Y.; Zhang, B. Topological Triply Degenerate Point with Double Fermi Arcs. *Nat. Phys.* **2019**, *15* (7), 645–649. https://doi.org/10.1038/s41567-019-0502-z.

(777)   Yang, Z.; Zhang, B. Acoustic Type-II Weyl Nodes from Stacking Dimerized Chains. *Phys. Rev. Lett.* **2016**, *117* (22), 224301. https://doi.org/10.1103/PhysRevLett.117.224301.

(778)   Xie, B.; Liu, H.; Cheng, H.; Liu, Z.; Chen, S.; Tian, J. Experimental Realization of Type-II Weyl Points and Fermi Arcs in Phononic Crystal. *Phys. Rev. Lett.* **2019**, *122* (10), 104302.

https://doi.org/10.1103/PhysRevLett.122.104302.

(779) Zangeneh-Nejad, F.; Fleury, R. Zero-Index Weyl Metamaterials. *Phys. Rev. Lett.* **2020**, *125* (5), 54301. https://doi.org/10.1103/PhysRevLett.125.054301.

(780) He, H.; Qiu, C.; Cai, X.; Xiao, M.; Ke, M.; Zhang, F.; Liu, Z. Observation of Quadratic Weyl Points and Double-Helicoid Arcs. *Nat. Commun.* **2020**, *11* (1), 1820. https://doi.org/10.1038/s41467-020-15825-5.

(781) Peri, V.; Serra-Garcia, M.; Ilan, R.; Huber, S. D. Axial-Field-Induced Chiral Channels in an Acoustic Weyl System. *Nat. Phys.* **2019**, *15* (4), 357–361. https://doi.org/10.1038/S41567-019-0415-X.

(782) Shi, X.; Yang, J. Spin-1 Weyl Point and Surface Arc State in a Chiral Phononic Crystal. *Phys. Rev. B* **2020**, *101* (21), 214309. https://doi.org/10.1103/PhysRevB.101.214309.

(783) Cai, X.; Ye, L.; Qiu, C.; Xiao, M.; Yu, R.; Ke, M.; Liu, Z. Symmetry-Enforced Three-Dimensional Dirac Phononic Crystals. *Light Sci. Appl.* **2020**, *9* (1), 38. https://doi.org/10.1038/s41377-020-0273-4.

(784) Cheng, H.; Sha, Y.; Liu, R.; Fang, C.; Lu, L. Discovering Topological Surface States of Dirac Points. *Phys. Rev. Lett.* **2020**, *124* (10), 104301. https://doi.org/10.1103/PhysRevLett.124.104301.

(785) Xu, C.; Ma, G.; Chen, Z.-G.; Luo, J.; Shi, J.; Lai, Y.; Wu, Y. Three-Dimensional Acoustic Double-Zero-Index Medium with a Fourfold Degenerate Dirac-like Point. *Phys. Rev. Lett.* **2020**, *124* (7), 074501. https://doi.org/10.1103/PhysRevLett.124.074501.

(786) Xie, B.; Liu, H.; Cheng, H.; Liu, Z.; Tian, J.; Chen, S. Dirac Points and the Transition towards Weyl Points in Three-Dimensional Sonic Crystals. *Light Sci. Appl.* **2020**, *9* (1), 201. https://doi.org/10.1038/s41377-020-00416-2.

(787) Deng, W.; Lu, J.; Li, F.; Huang, X.; Yan, M.; Ma, J.; Liu, Z. Nodal Rings and Drumhead Surface States in Phononic Crystals. *Nat. Commun.* **2019**, *10* (1), 1769. https://doi.org/10.1038/s41467-019-09820-8.

(788) Geng, Z.-G.; Peng, Y.-G.; Shen, Y.-X.; Ma, Z.; Yu, R.; Gao, J.-H.; Zhu, X.-F. Topological Nodal Line States in Three-Dimensional Ball-and-Stick Sonic Crystals. *Phys. Rev. B* **2019**, *100* (22), 224105. https://doi.org/10.1103/PhysRevB.100.224105.

(789) Lu, J.; Huang, X.; Yan, M.; Li, F.; Deng, W.; Liu, Z. Nodal-Chain Semimetal States and Topological Focusing in Phononic Crystals. *Phys. Rev. Appl.* **2020**, *13* (5), 54080. https://doi.org/10.1103/PhysRevApplied.13.054080.

(790) Qiu, H.; Qiu, C.; Yu, R.; Xiao, M.; He, H.; Ye, L.; Ke, M.; Liu, Z. Straight Nodal Lines and Waterslide Surface States Observed in Acoustic Metacrystals. *Phys. Rev. B* **2019**, *100* (4), 41303. https://doi.org/10.1103/PhysRevB.100.041303.

(791) Zheng, L.-Y.; Zhang, X.-J.; Lu, M.-H.; Chen, Y.-F.; Christensen, J. Knitting Topological Bands in Artificial Sonic Semimetals. *Mater. Today Phys.* **2021**, *16*, 100299. https://doi.org/10.1016/j.mtphys.2020.100299.

(792) Yang, Y.; Xia, J.; Sun, H.; Ge, Y.; Jia, D.; Yuan, S.; Yang, S. A.; Chong, Y.; Zhang, B. Observation of a Topological Nodal Surface and Its Surface-State Arcs in an Artificial Acoustic Crystal. *Nat. Commun.* **2019**, *10* (1), 5185. https://doi.org/10.1038/s41467-019-13258-3.

(793) Wang, Y. T.; Tsai, Y. W. Multiple Weyl and Double-Weyl Points in an Elastic Chiral Lattice. *New

*J. Phys.* **2018**, *20* (8), 83031. https://doi.org/10.1088/1367-2630/aada55.

(794)  Shi, X.; Chaunsali, R.; Li, F.; Yang, J. Elastic Weyl Points and Surface Arc States in Three-Dimensional Structures. *Phys. Rev. Appl.* **2019**, *12* (2), 024058. https://doi.org/10.1103/PhysRevApplied.12.024058.

(795)  Ganti, S. S.; Liu, T.-W.; Semperlotti, F. Weyl Points and Topological Surface States in a Three-Dimensional Sandwich-Type Elastic Lattice. *New J. Phys.* **2020**, *22* (8), 083001. https://doi.org/10.1088/1367-2630/ab9e31.

(796)  Xiong, Z.; Wang, H. X.; Ge, H.; Shi, J.; Luo, J.; Lai, Y.; Lu, M. H.; Jiang, J. H. Topological Node Lines in Mechanical Metacrystals. *Phys. Rev. B* **2018**, *97* (18). https://doi.org/10.1103/PHYSREVB.97.180101.

(797)  Fruchart, M.; Jeon, S.-Y.; Hur, K.; Cheianov, V.; Wiesner, U.; Vitelli, V. Soft Self-Assembly of Weyl Materials for Light and Sound. *Proc. Natl. Acad. Sci.* **2018**, *115* (16). https://doi.org/10.1073/pnas.1720828115.

(798)  Lu, L.; Fang, C.; Fu, L.; Johnson, S. G.; Joannopoulos, J. D.; Soljačić, M. Symmetry-Protected Topological Photonic Crystal in Three Dimensions. *Nat. Phys.* **2016**, *12* (4), 337–340. https://doi.org/10.1038/nphys3611.

(799)  Devescovi, C.; García-Díez, M.; Robredo, I.; Blanco de Paz, M.; Lasa-Alonso, J.; Bradlyn, B.; Mañes, J. L.; G. Vergniory, M.; García-Etxarri, A. Cubic 3D Chern Photonic Insulators with Orientable Large Chern Vectors. *Nat. Commun.* **2021**, *12* (1), 7330. https://doi.org/10.1038/s41467-021-27168-w.

(800)  Slobozhanyuk, A.; Mousavi, S. H.; Ni, X.; Smirnova, D.; Kivshar, Y. S.; Khanikaev, A. B. Three-Dimensional All-Dielectric Photonic Topological Insulator. *Nat. Photonics* **2017**, *11* (2), 130–136. https://doi.org/10.1038/nphoton.2016.253.

(801)  Schnyder, A. P.; Ryu, S.; Furusaki, A.; Ludwig, A. W. W. Classification of Topological Insulators and Superconductors in Three Spatial Dimensions. *Phys. Rev. B* **2008**, *78* (19), 195125. https://doi.org/10.1103/PhysRevB.78.195125.

(802)  Ochiai, T. Gapless Surface States Originating from Accidentally Degenerate Quadratic Band Touching in a Three-Dimensional Tetragonal Photonic Crystal. *Phys. Rev. A* **2017**, *96* (4), 043842. https://doi.org/10.1103/PhysRevA.96.043842.

(803)  Fu, L. Topological Crystalline Insulators. *Phys. Rev. Lett.* **2011**, *106* (10), 106802. https://doi.org/10.1103/PhysRevLett.106.106802.

(804)  Kim, M.; Wang, Z.; Yang, Y.; Teo, H. T.; Rho, J.; Zhang, B. Three-Dimensional Photonic Topological Insulator without Spin–Orbit Coupling. *Nat. Commun.* **2022**, *13* (1), 3499. https://doi.org/10.1038/s41467-022-30909-0.

(805)  Yang, Y.; Gao, Z.; Xue, H.; Zhang, L.; He, M.; Yang, Z.; Singh, R.; Chong, Y.; Zhang, B.; Chen, H. Realization of a Three-Dimensional Photonic Topological Insulator. *Nature* **2019**, *565* (7741), 622–626. https://doi.org/10.1038/s41586-018-0829-0.

(806)  He, C.; Yu, S.-Y.; Ge, H.; Wang, H.; Tian, Y.; Zhang, H.; Sun, X.-C.; Chen, Y. B.; Zhou, J.; Lu, M.-H.; Chen, Y.-F. Three-Dimensional Topological Acoustic Crystals with Pseudospin-Valley Coupled Saddle Surface States. *Nat. Commun.* **2018**, *9* (1), 4555. https://doi.org/10.1038/s41467-018-07030-2.

(807)  He, C.; Yu, S.; Wang, H.; Ge, H.; Ruan, J.; Zhang, H. Hybrid Acoustic Topological Insulator in

Three Dimensions. *Phys. Rev. Lett.* **2019**, *123* (19), 195503. https://doi.org/10.1103/PhysRevLett.123.195503.

(808) He, C.; Lai, H.-S.; He, B.; Yu, S.-Y.; Xu, X.; Lu, M.-H.; Chen, Y.-F. Acoustic Analogues of Three-Dimensional Topological Insulators. *Nat. Commun.* **2020**, *11* (1), 2318. https://doi.org/10.1038/s41467-020-16131-w.

(809) Li, F.-F.; Wang, H.-X.; Xiong, Z.; Lou, Q.; Chen, P.; Wu, R.-X.; Poo, Y.; Jiang, J.-H.; John, S. Topological Light-Trapping on a Dislocation. *Nat. Commun.* **2018**, *9* (1), 2462. https://doi.org/10.1038/s41467-018-04861-x.

(810) Ezawa, M. Topological Switch between Second-Order Topological Insulators and Topological Crystalline Insulators. *Phys. Rev. Lett.* **2018**, *121* (11), 116801. https://doi.org/10.1103/PhysRevLett.121.116801.

(811) Chen, Y.; Lin, Z.-K.; Chen, H.; Jiang, J.-H. Plasmon-Polaritonic Quadrupole Topological Insulators. *Phys. Rev. B* **2020**, *101* (4), 041109. https://doi.org/10.1103/PhysRevB.101.041109.

(812) Serra-Garcia, M.; Süsstrunk, R.; Huber, S. D. Observation of Quadrupole Transitions and Edge Mode Topology in an LC Circuit Network. *Phys. Rev. B* **2019**, *99* (2), 20304. https://doi.org/10.1103/PhysRevB.99.020304.

(813) He, L.; Addison, Z.; Mele, E. J.; Zhen, B. Quadrupole Topological Photonic Crystals. *Nat. Commun.* **2020**, *11* (1), 3119. https://doi.org/10.1038/s41467-020-16916-z.

(814) Zhou, X.; Lin, Z.; Lu, W.; Lai, Y.; Hou, B.; Jiang, J. Twisted Quadrupole Topological Photonic Crystals. *Laser Photon. Rev.* **2020**, *14* (8), 2000010. https://doi.org/10.1002/lpor.202000010.

(815) Qi, Y.; Qiu, C.; Xiao, M.; He, H.; Ke, M.; Liu, Z. Acoustic Realization of Quadrupole Topological Insulators. *Phys. Rev. Lett.* **2020**, *124* (20), 206601. https://doi.org/10.1103/PhysRevLett.124.206601.

(816) Lin, Z.-K.; Wang, H.-X.; Xiong, Z.; Lu, M.-H.; Jiang, J.-H. Anomalous Quadrupole Topological Insulators in Two-Dimensional Nonsymmorphic Sonic Crystals. *Phys. Rev. B* **2020**, *102* (3), 35105. https://doi.org/10.1103/PhysRevB.102.035105.

(817) Zhang, X.; Lin, Z.-K.; Wang, H.-X.; Xiong, Z.; Tian, Y.; Lu, M.-H.; Chen, Y.-F.; Jiang, J.-H. Symmetry-Protected Hierarchy of Anomalous Multipole Topological Band Gaps in Nonsymmorphic Metacrystals. *Nat. Commun.* **2020**, *11* (1), 65. https://doi.org/10.1038/s41467-019-13861-4.

(818) Ota, Y.; Liu, F.; Katsumi, R.; Watanabe, K.; Wakabayashi, K.; Arakawa, Y.; Iwamoto, S. Photonic Crystal Nanocavity Based on a Topological Corner State. *Optica* **2019**, *6* (6), 786. https://doi.org/10.1364/optica.6.000786.

(819) Kim, M.; Rho, J. Topological Edge and Corner States in a Two-Dimensional Photonic Su-Schrieffer-Heeger Lattice. *Nanophotonics* **2020**, *9* (10), 3227–3234. https://doi.org/10.1515/nanoph-2019-0451.

(820) Chen, Y.; Lu, X. C.; Chen, H. Y. Effect of Truncation on Photonic Corner States in a Kagome Lattice. *Opt. Lett.* **2019**, *44* (17), 4251–4254. https://doi.org/10.1364/Ol.44.004251.

(821) Zhang, L.; Yang, Y.; Lin, Z.; Qin, P.; Chen, Q.; Gao, F.; Li, E.; Jiang, J.; Zhang, B.; Chen, H. Higher-Order Topological States in Surface-Wave Photonic Crystals. *Adv. Sci.* **2020**, *7* (6), 1902724. https://doi.org/10.1002/advs.201902724.

(822) Proctor, M.; Blanco de Paz, M.; Bercioux, D.; García-Etxarri, A.; Arroyo Huidobro, P. Higher-Order Topology in Plasmonic Kagome Lattices. *Appl. Phys. Lett.* **2021**, *118* (9), 091105. https://doi.org/10.1063/5.0040955.

(823) Yang, H.; Li, Z.-X.; Liu, Y.; Cao, Y.; Yan, P. Observation of Symmetry-Protected Zero Modes in Topolectrical Circuits. *Phys. Rev. Res.* **2020**, *2* (2), 022028. https://doi.org/10.1103/PhysRevResearch.2.022028.

(824) Proctor, M.; Huidobro, P. A.; Bradlyn, B.; de Paz, M. B.; Vergniory, M. G.; Bercioux, D.; García-Etxarri, A. Robustness of Topological Corner Modes in Photonic Crystals. *Phys. Rev. Res.* **2020**, *2* (4), 042038. https://doi.org/10.1103/PhysRevResearch.2.042038.

(825) Vakulenko, A.; Kiriushechkina, S.; Wang, M.; Li, M.; Zhirihin, D.; Ni, X.; Guddala, S.; Korobkin, D.; Alù, A.; Khanikaev, A. B. Near-Field Characterization of Higher-Order Topological Photonic States at Optical Frequencies. *Adv. Mater.* **2021**, *33* (18), 2004376. https://doi.org/10.1002/adma.202004376.

(826) Lu, J.; Wirth, K. G.; Gao, W.; Heßler, A.; Sain, B.; Taubner, T.; Zentgraf, T. Observing 0D Subwavelength-Localized Modes at ~100 THz Protected by Weak Topology. *Sci. Adv.* **2021**, *7* (49). https://doi.org/10.1126/sciadv.abl3903.

(827) Wu, S.; Jiang, B.; Liu, Y.; Jiang, J.-H. All-Dielectric Photonic Crystal with Unconventional Higher-Order Topology. *Photonics Res.* **2021**, *9* (5), 668. https://doi.org/10.1364/prj.418689.

(828) Xie, B.; Su, G.; Wang, H.-F.; Liu, F.; Hu, L.; Yu, S.-Y.; Zhan, P.; Lu, M.-H.; Wang, Z.; Chen, Y.-F. Higher-Order Quantum Spin Hall Effect in a Photonic Crystal. *Nat. Commun.* **2020**, *11* (1), 3768. https://doi.org/10.1038/s41467-020-17593-8.

(829) Zangeneh-Nejad, F.; Fleury, R. Nonlinear Second-Order Topological Insulators. *Phys. Rev. Lett.* **2019**, *123* (5), 053902. https://doi.org/10.1103/PhysRevLett.123.053902.

(830) Kirsch, M. S.; Zhang, Y.; Kremer, M.; Maczewsky, L. J.; Ivanov, S. K.; Kartashov, Y. V.; Torner, L.; Bauer, D.; Szameit, A.; Heinrich, M. Nonlinear Second-Order Photonic Topological Insulators. *Nat. Phys.* **2021**, *17* (9), 995–1000. https://doi.org/10.1038/s41567-021-01275-3.

(831) Kim, H. R.; Hwang, M. S.; Smirnova, D.; Jeong, K. Y.; Kivshar, Y.; Park, H. G. Multipolar Lasing Modes from Topological Corner States. *Nat. Commun.* **2020**, *11* (1). https://doi.org/10.1038/s41467-020-19609-9.

(832) Benalcazar, W. A.; Li, T.; Hughes, T. L. Quantization of Fractional Corner Charge in Cn -Symmetric Higher-Order Topological Crystalline Insulators. *Phys. Rev. B* **2019**, *99* (24), 245151. https://doi.org/10.1103/PhysRevB.99.245151.

(833) Peterson, C. W.; Li, T.; Benalcazar, W. A.; Hughes, T. L.; Bahl, G. A Fractional Corner Anomaly Reveals Higher-Order Topology. *Science (80-. ).* **2020**, *368* (6495), 1114–1118. https://doi.org/10.1126/science.aba7604.

(834) Liu, Y.; Leung, S.; Li, F.-F.; Lin, Z.-K.; Tao, X.; Poo, Y.; Jiang, J.-H. Bulk–Disclination Correspondence in Topological Crystalline Insulators. *Nature* **2021**, *589* (7842), 381–385. https://doi.org/10.1038/s41586-020-03125-3.

(835) Peterson, C. W.; Li, T.; Jiang, W.; Hughes, T. L.; Bahl, G. Trapped Fractional Charges at Bulk Defects in Topological Insulators. *Nature* **2021**, *589* (7842), 376–380. https://doi.org/10.1038/s41586-020-03117-3.

(836) Cerjan, A.; Jürgensen, M.; Benalcazar, W. A.; Mukherjee, S.; Rechtsman, M. C. Observation of a

Higher-Order Topological Bound State in the Continuum. *Phys. Rev. Lett.* **2020**, *125* (21), 213901. https://doi.org/10.1103/PhysRevLett.125.213901.

(837) Coutant, A.; Achilleos, V.; Richoux, O.; Theocharis, G.; Pagneux, V. Robustness of Topological Corner Modes against Disorder with Application to Acoustic Networks. *Phys. Rev. B* **2020**, *102* (21), 214204. https://doi.org/10.1103/PhysRevB.102.214204.

(838) Coutant, A.; Achilleos, V.; Richoux, O.; Theocharis, G.; Pagneux, V. Topological Two-Dimensional Su–Schrieffer–Heeger Analog Acoustic Networks: Total Reflection at Corners and Corner Induced Modes. *J. Appl. Phys.* **2021**, *129* (12), 125108. https://doi.org/10.1063/5.0042406.

(839) Yue, Z.; Liao, D.; Zhang, Z.; Xiong, W.; Cheng, Y.; Liu, X. Experimental Demonstration of a Reconfigurable Acoustic Second-Order Topological Insulator Using Condensed Soda Cans Array. *Appl. Phys. Lett.* **2021**, *118* (20), 203501. https://doi.org/10.1063/5.0049030.

(840) Chen, Z.-G.; Wang, L.; Zhang, G.; Ma, G. Chiral Symmetry Breaking of Tight-Binding Models in Coupled Acoustic-Cavity Systems. *Phys. Rev. Appl.* **2020**, *14* (2), 024023. https://doi.org/10.1103/PhysRevApplied.14.024023.

(841) Chen, Z.-G.; Xu, C.; Al Jahdali, R.; Mei, J.; Wu, Y. Corner States in a Second-Order Acoustic Topological Insulator as Bound States in the Continuum. *Phys. Rev. B* **2019**, *100* (7), 075120. https://doi.org/10.1103/PhysRevB.100.075120.

(842) Zhang, Z.; Hu, B.; Liu, F.; Cheng, Y.; Liu, X.; Christensen, J. Pseudospin Induced Topological Corner State at Intersecting Sonic Lattices. *Phys. Rev. B* **2020**, *101* (22), 220102. https://doi.org/10.1103/PhysRevB.101.220102.

(843) Zhang, X.; Liu, L.; Lu, M.-H.; Chen, Y.-F. Valley-Selective Topological Corner States in Sonic Crystals. *Phys. Rev. Lett.* **2021**, *126* (15), 156401. https://doi.org/10.1103/PhysRevLett.126.156401.

(844) Lin, Z.-K.; Wu, S.-Q.; Wang, H.-X.; Jiang, J.-H. Higher-Order Topological Spin Hall Effect of Sound. *Chinese Phys. Lett.* **2020**, *37* (7), 074302. https://doi.org/10.1088/0256-307X/37/7/074302.

(845) Xiong, Z.; Lin, Z.-K.; Wang, H.-X.; Zhang, X.; Lu, M.-H.; Chen, Y.-F.; Jiang, J.-H. Corner States and Topological Transitions in Two-Dimensional Higher-Order Topological Sonic Crystals with Inversion Symmetry. *Phys. Rev. B* **2020**, *102* (12), 125144. https://doi.org/10.1103/PhysRevB.102.125144.

(846) Zhu, W.; Xue, H.; Gong, J.; Chong, Y.; Zhang, B. Time-Periodic Corner States from Floquet Higher-Order Topology. *Nat. Commun.* **2022**, *13* (1), 11. https://doi.org/10.1038/s41467-021-27552-6.

(847) Fan, H.; Xia, B.; Tong, L.; Zheng, S.; Yu, D. Elastic Higher-Order Topological Insulator with Topologically Protected Corner States. *Phys. Rev. Lett.* **2019**, *122* (20), 204301. https://doi.org/10.1103/PhysRevLett.122.204301.

(848) Wu, Q.; Chen, H.; Li, X.; Huang, G. In-Plane Second-Order Topologically Protected States in Elastic Kagome Lattices. *Phys. Rev. Appl.* **2020**, *14* (1), 14084. https://doi.org/10.1103/PhysRevApplied.14.014084.

(849) Chen, C. W.; Chaunsali, R.; Christensen, J.; Theocharis, G.; Yang, J. Corner States in a Second-Order Mechanical Topological Insulator. *Commun. Mater.* **2021**, *2* (1), 62. https://doi.org/10.1038/s43246-021-00170-x.

(850) Yang, Z.-Z.; Peng, Y.-Y.; Li, X.; Zou, X.-Y.; Cheng, J.-C. Boundary-Dependent Corner States in

Topological Acoustic Resonator Array. *Appl. Phys. Lett.* **2020**, *117* (11), 113501. https://doi.org/10.1063/5.0017503.

(851) Huo, S.; Huang, H.; Feng, L.; Chen, J. Edge States and Corner Modes in Second-Order Topological Phononic Crystal Plates. *Appl. Phys. Express* **2019**, *12* (9), 094003. https://doi.org/10.7567/1882-0786/ab3514.

(852) Wang, Z.; Wei, Q. An Elastic Higher-Order Topological Insulator Based on Kagome Phononic Crystals. *J. Appl. Phys.* **2021**, *129* (3), 035102. https://doi.org/10.1063/5.0031377.

(853) Wu, Y.; Yan, M.; Lin, Z.-K.; Wang, H.-X.; Li, F.; Jiang, J.-H. On-Chip Higher-Order Topological Micromechanical Metamaterials. *Sci. Bull.* **2021**, *66* (19), 1959–1966. https://doi.org/https://doi.org/10.1016/j.scib.2021.06.024.

(854) Ghorashi, S. A. A.; Li, T.; Hughes, T. L.; Wang, H.-X.; Lin, Z.-K.; Jiang, B.; Guo, G.-Y.; Jiang, J.-H. Higher-Order Weyl Semimetals. *Phys. Rev. Lett.* **2020**, *125* (14), 146401. https://doi.org/10.1103/PhysRevLett.125.146401.

(855) Bao, J.; Zou, D.; Zhang, W.; He, W.; Sun, H.; Zhang, X. Topoelectrical Circuit Octupole Insulator with Topologically Protected Corner States. *Phys. Rev. B* **2019**, *100* (20), 201406. https://doi.org/10.1103/PhysRevB.100.201406.

(856) Ni, X.; Xiao, Z.; Khanikaev, A. B.; Alù, A. Robust Multiplexing with Topolectrical Higher-Order Chern Insulators. *Phys. Rev. Appl.* **2020**, *13* (6), 064031. https://doi.org/10.1103/PhysRevApplied.13.064031.

(857) Ni, X.; Alù, A. Higher-Order Topolectrical Semimetal Realized via Synthetic Gauge Fields. *APL Photonics* **2021**, *6* (5), 050802. https://doi.org/10.1063/5.0041458.

(858) Liu, S.; Ma, S.; Zhang, Q.; Zhang, L.; Yang, C.; You, O.; Gao, W.; Xiang, Y.; Cui, T. J.; Zhang, S. Octupole Corner State in a Three-Dimensional Topological Circuit. *Light Sci. Appl.* **2020**, *9* (1), 145. https://doi.org/10.1038/s41377-020-00381-w.

(859) Yamada, S. S.; Li, T.; Lin, M.; Peterson, C. W.; Hughes, T. L.; Bahl, G. Bound States at Partial Dislocation Defects in Multipole Higher-Order Topological Insulators. *Nat. Commun.* **2022**, *13* (1), 2035. https://doi.org/10.1038/s41467-022-29785-5.

(860) Zhang, W.; Zou, D.; Bao, J.; He, W.; Pei, Q.; Sun, H.; Zhang, X. Topolectrical-Circuit Realization of a Four-Dimensional Hexadecapole Insulator. *Phys. Rev. B* **2020**, *102* (10), 100102. https://doi.org/10.1103/PhysRevB.102.100102.

(861) Ni, X.; Li, M.; Weiner, M.; Alù, A.; Khanikaev, A. B. Demonstration of a Quantized Acoustic Octupole Topological Insulator. *Nat. Commun.* **2020**, *11* (1), 2108. https://doi.org/10.1038/s41467-020-15705-y.

(862) Xue, H.; Ge, Y.; Sun, H.-X.; Wang, Q.; Jia, D.; Guan, Y.-J.; Yuan, S.-Q.; Chong, Y.; Zhang, B. Observation of an Acoustic Octupole Topological Insulator. *Nat. Commun.* **2020**, *11* (1), 2442. https://doi.org/10.1038/s41467-020-16350-1.

(863) Weiner, M.; Ni, X.; Li, M.; Alù, A.; Khanikaev, A. B. Demonstration of a Third-Order Hierarchy of Topological States in a Three-Dimensional Acoustic Metamaterial. *Sci. Adv.* **2020**, *6* (13), 1–7. https://doi.org/10.1126/sciadv.aay4166.

(864) Xue, H.; Yang, Y.; Liu, G.; Gao, F.; Chong, Y.; Zhang, B. Realization of an Acoustic Third-Order Topological Insulator. *Phys. Rev. Lett.* **2019**, *122* (24), 244301. https://doi.org/10.1103/PhysRevLett.122.244301.

(865) Zhang, X.; Xie, B.-Y.; Wang, H.-F.; Xu, X.; Tian, Y.; Jiang, J.-H.; Lu, M.-H.; Chen, Y.-F. Dimensional Hierarchy of Higher-Order Topology in Three-Dimensional Sonic Crystals. *Nat. Commun.* **2019**, *10* (1), 5331. https://doi.org/10.1038/s41467-019-13333-9.

(866) Zheng, S.; Xia, B.; Man, X.; Tong, L.; Jiao, J.; Duan, G.; Yu, D. Three-Dimensional Higher-Order Topological Acoustic System with Multidimensional Topological States. *Phys. Rev. B* **2020**, *102* (10), 104113. https://doi.org/10.1103/PhysRevB.102.104113.

(867) Meng, F.; Chen, Y.; Li, W.; Jia, B.; Huang, X. Realization of Multidimensional Sound Propagation in 3D Acoustic Higher-Order Topological Insulator. *Appl. Phys. Lett.* **2020**, *117* (15), 151903. https://doi.org/10.1063/5.0023033.

(868) Xu, C.; Chen, Z. G.; Zhang, G.; Ma, G.; Wu, Y. Multi-Dimensional Wave Steering with Higher-Order Topological Phononic Crystal. *Sci. Bull.* **2021**, *66* (17), 1740–1745. https://doi.org/10.1016/j.scib.2021.05.013.

(869) Wei, Q.; Zhang, X.; Deng, W.; Lu, J.; Huang, X.; Yan, M.; Chen, G.; Liu, Z.; Jia, S. Higher-Order Topological Semimetal in Acoustic Crystals. *Nat. Mater.* **2021**, *20* (6), 812–817. https://doi.org/10.1038/s41563-021-00933-4.

(870) Wei, Q.; Zhang, X.; Deng, W.; Lu, J.; Huang, X.; Yan, M.; Chen, G.; Liu, Z.; Jia, S. 3D Hinge Transport in Acoustic Higher-Order Topological Insulators. *Phys. Rev. Lett.* **2021**, *127* (25), 255501. https://doi.org/10.1103/PhysRevLett.127.255501.

(871) Luo, L.; Wang, H.-X.; Lin, Z.-K.; Jiang, B.; Wu, Y.; Li, F.; Jiang, J.-H. Observation of a Phononic Higher-Order Weyl Semimetal. *Nat. Mater.* **2021**, *20* (6), 794–799. https://doi.org/10.1038/s41563-021-00985-6.

(872) Yang, Y.; Lu, J.; Yan, M.; Huang, X.; Deng, W.; Liu, Z. Hybrid-Order Topological Insulators in a Phononic Crystal. *Phys. Rev. Lett.* **2021**, *126* (15), 156801. https://doi.org/10.1103/PhysRevLett.126.156801.

(873) Zheng, L.-Y.; Christensen, J. Dirac Hierarchy in Acoustic Topological Insulators. *Phys. Rev. Lett.* **2021**, *127* (15), 156401. https://doi.org/10.1103/PhysRevLett.127.156401.

(874) Qiu, H.; Xiao, M.; Zhang, F.; Qiu, C. Higher-Order Dirac Sonic Crystals. *Phys. Rev. Lett.* **2021**, *127* (14), 146601. https://doi.org/10.1103/PhysRevLett.127.146601.

(875) Bergholtz, E. J.; Budich, J. C.; Kunst, F. K. Exceptional Topology of Non-Hermitian Systems. *Rev. Mod. Phys.* **2021**, *93* (1), 015005. https://doi.org/10.1103/RevModPhys.93.015005.

(876) Yao, S.; Wang, Z. Edge States and Topological Invariants of Non-Hermitian Systems. *Phys. Rev. Lett.* **2018**, *121* (8), 086803. https://doi.org/10.1103/PhysRevLett.121.086803.

(877) Yao, S.; Song, F.; Wang, Z. Non-Hermitian Chern Bands. *Phys. Rev. Lett.* **2018**, *121* (13), 136802. https://doi.org/10.1103/PhysRevLett.121.136802.

(878) Kunst, F. K.; Edvardsson, E.; Budich, J. C.; Bergholtz, E. J. Biorthogonal Bulk-Boundary Correspondence in Non-Hermitian Systems. *Phys. Rev. Lett.* **2018**, *121* (2), 026808. https://doi.org/10.1103/PhysRevLett.121.026808.

(879) Kawabata, K.; Shiozaki, K.; Ueda, M.; Sato, M. Symmetry and Topology in Non-Hermitian Physics. *Phys. Rev. X* **2019**, *9* (4), 041015. https://doi.org/10.1103/PHYSREVX.9.041015/FIGURES/7/THUMBNAIL.

(880) Yokomizo, K.; Murakami, S. Non-Bloch Band Theory of Non-Hermitian Systems. *Phys. Rev.*

*Lett.* **2019**, *123* (6), 066404. https://doi.org/10.1103/PhysRevLett.123.066404.

(881)  Borgnia, D. S.; Kruchkov, A. J.; Slager, R.-J. Non-Hermitian Boundary Modes and Topology. *Phys. Rev. Lett.* **2020**, *124* (5), 056802. https://doi.org/10.1103/PhysRevLett.124.056802.

(882)  Yang, Z.; Zhang, K.; Fang, C.; Hu, J. Non-Hermitian Bulk-Boundary Correspondence and Auxiliary Generalized Brillouin Zone Theory. *Phys. Rev. Lett.* **2020**, *125* (22), 226402. https://doi.org/10.1103/PhysRevLett.125.226402.

(883)  Zhang, K.; Yang, Z.; Fang, C. Correspondence between Winding Numbers and Skin Modes in Non-Hermitian Systems. *Phys. Rev. Lett.* **2020**, *125* (12), 126402. https://doi.org/10.1103/PhysRevLett.125.126402.

(884)  Rudner, M. S.; Levitov, L. S. Topological Transition in a Non-Hermitian Quantum Walk. *Phys. Rev. Lett.* **2009**, *102* (6), 065703. https://doi.org/10.1103/PhysRevLett.102.065703.

(885)  Lee, T. E. Anomalous Edge State in a Non-Hermitian Lattice. *Phys. Rev. Lett.* **2016**, *116* (13), 133903. https://doi.org/10.1103/PhysRevLett.116.133903.

(886)  Okuma, N.; Kawabata, K.; Shiozaki, K.; Sato, M. Topological Origin of Non-Hermitian Skin Effects. *Phys. Rev. Lett.* **2020**, *124* (8), 086801. https://doi.org/10.1103/PhysRevLett.124.086801.

(887)  Xia, S.; Kaltsas, D.; Song, D.; Komis, I.; Xu, J.; Szameit, A.; Buljan, H.; Makris, K. G.; Chen, Z. Nonlinear Tuning of PT Symmetry and Non-Hermitian Topological States. *Science (80-. ).* **2021**, *372* (6537), 72–76. https://doi.org/10.1126/science.abf6873.

(888)  Zhu, X.; Wang, H.; Gupta, S. K.; Zhang, H.; Xie, B.; Lu, M.; Chen, Y. Photonic Non-Hermitian Skin Effect and Non-Bloch Bulk-Boundary Correspondence. *Phys. Rev. Res.* **2020**, *2* (1), 013280. https://doi.org/10.1103/PhysRevResearch.2.013280.

(889)  Helbig, T.; Hofmann, T.; Imhof, S.; Abdelghany, M.; Kiessling, T.; Molenkamp, L. W.; Lee, C. H.; Szameit, A.; Greiter, M.; Thomale, R. Generalized Bulk–Boundary Correspondence in Non-Hermitian Topolectrical Circuits. *Nat. Phys.* **2020**, *16* (7), 747–750. https://doi.org/10.1038/s41567-020-0922-9.

(890)  Weidemann, S.; Kremer, M.; Helbig, T.; Hofmann, T.; Stegmaier, A.; Greiter, M.; Thomale, R.; Szameit, A. Topological Funneling of Light. *Science (80-. ).* **2020**, *368* (6488), 311–314. https://doi.org/10.1126/science.aaz8727.

(891)  Wang, H.; Zhang, X.; Hua, J.; Lei, D.; Lu, M.; Chen, Y. Topological Physics of Non-Hermitian Optics and Photonics: A Review. *J. Opt.* **2021**, *23* (12), 123001. https://doi.org/10.1088/2040-8986/ac2e15.

(892)  Parto, M.; Liu, Y. G. N.; Bahari, B.; Khajavikhan, M.; Christodoulides, D. N. Non-Hermitian and Topological Photonics: Optics at an Exceptional Point. *Nanophotonics* **2020**, *10* (1), 403–423. https://doi.org/10.1515/nanoph-2020-0434.

(893)  Esaki, K.; Sato, M.; Hasebe, K.; Kohmoto, M. Edge States and Topological Phases in Non-Hermitian Systems. *Phys. Rev. B* **2011**, *84* (20), 205128. https://doi.org/10.1103/PhysRevB.84.205128.

(894)  Leykam, D.; Bliokh, K. Y.; Huang, C.; Chong, Y. D.; Nori, F. Edge Modes, Degeneracies, and Topological Numbers in Non-Hermitian Systems. *Phys. Rev. Lett.* **2017**, *118* (4), 040401. https://doi.org/10.1103/PhysRevLett.118.040401.

(895)  Shen, H.; Zhen, B.; Fu, L. Topological Band Theory for Non-Hermitian Hamiltonians. *Phys. Rev.*

*Lett.* **2018**, *120* (14), 146402. https://doi.org/10.1103/PhysRevLett.120.146402.

(896) Ni, X.; Smirnova, D.; Poddubny, A.; Leykam, D.; Chong, Y.; Khanikaev, A. B. PT Phase Transitions of Edge States at PT Symmetric Interfaces in Non-Hermitian Topological Insulators. *Phys. Rev. B* **2018**, *98* (16), 165129. https://doi.org/10.1103/PhysRevB.98.165129.

(897) Kremer, M.; Biesenthal, T.; Maczewsky, L. J.; Heinrich, M.; Thomale, R.; Szameit, A. Demonstration of a Two-Dimensional $${\cal P}{\cal T}$$ P T -Symmetric Crystal. *Nat. Commun.* **2019**, *10* (1), 435. https://doi.org/10.1038/s41467-018-08104-x.

(898) Gao, H.; Xue, H.; Wang, Q.; Gu, Z.; Liu, T.; Zhu, J.; Zhang, B. Observation of Topological Edge States Induced Solely by Non-Hermiticity in an Acoustic Crystal. *Phys. Rev. B* **2020**, *101* (18), 180303. https://doi.org/10.1103/PhysRevB.101.180303.

(899) Gao, H.; Xue, H.; Gu, Z.; Liu, T.; Zhu, J.; Zhang, B. Non-Hermitian Route to Higher-Order Topology in an Acoustic Crystal. *Nat. Commun.* **2021**, *12* (1), 1888. https://doi.org/10.1038/s41467-021-22223-y.

(900) Wang, M.; Ye, L.; Christensen, J.; Liu, Z. Valley Physics in Non-Hermitian Artificial Acoustic Boron Nitride. *Phys. Rev. Lett.* **2018**, *120* (24). https://doi.org/10.1103/PHYSREVLETT.120.246601.

(901) Hu, B.; Zhang, Z.; Zhang, H.; Zheng, L.; Xiong, W.; Yue, Z.; Wang, X.; Xu, J.; Cheng, Y.; Liu, X.; Christensen, J. Non-Hermitian Topological Whispering Gallery. *Nature* **2021**, *597* (7878), 655–659. https://doi.org/10.1038/s41586-021-03833-4.

(902) Brandenbourger, M.; Locsin, X.; Lerner, E.; Coulais, C. Non-Reciprocal Robotic Metamaterials. *Nat. Commun.* **2019**, *10* (1), 4608. https://doi.org/10.1038/s41467-019-12599-3.

(903) Ghatak, A.; Brandenbourger, M.; Van Wezel, J.; Coulais, C. Observation of Non-Hermitian Topology and Its Bulk-Edge Correspondence in an Active Mechanical Metamaterial. *Proc. Natl. Acad. Sci. U. S. A.* **2020**, *117* (47), 29561–29568. https://doi.org/10.1073/pnas.2010580117.

(904) Rosa, M. I. N.; Ruzzene, M. Dynamics and Topology of Non-Hermitian Elastic Lattices with Non-Local Feedback Control Interactions. *New J. Phys.* **2020**, *22* (5), 53004. https://doi.org/10.1088/1367-2630/ab81b6.

(905) Scheibner, C.; Irvine, W. T. M.; Vitelli, V. Non-Hermitian Band Topology and Skin Modes in Active Elastic Media. *Phys. Rev. Lett.* **2020**, *125* (11), 118001. https://doi.org/10.1103/PHYSREVLETT.125.118001.

(906) Zhang, X.; Tian, Y.; Jiang, J.-H.; Lu, M.-H.; Chen, Y.-F. Observation of Higher-Order Non-Hermitian Skin Effect. *Nat. Commun.* **2021**, *12* (1), 5377. https://doi.org/10.1038/s41467-021-25716-y.

(907) Gao, P.; Willatzen, M.; Christensen, J. Anomalous Topological Edge States in Non-Hermitian Piezophononic Media. *Phys. Rev. Lett.* **2020**, *125* (20), 206402. https://doi.org/10.1103/PhysRevLett.125.206402.

(908) Zhang, Z.; Rosendo López, M.; Cheng, Y.; Liu, X.; Christensen, J. Non-Hermitian Sonic Second-Order Topological Insulator. *Phys. Rev. Lett.* **2019**, *122* (19), 195501. https://doi.org/10.1103/PhysRevLett.122.195501.

(909) Liu, T.; Zhang, Y.-R.; Ai, Q.; Gong, Z.; Kawabata, K.; Ueda, M.; Nori, F. Second-Order Topological Phases in Non-Hermitian Systems. *Phys. Rev. Lett.* **2019**, *122* (7), 076801. https://doi.org/10.1103/PhysRevLett.122.076801.

(910) Lee, C. H.; Li, L.; Gong, J. Hybrid Higher-Order Skin-Topological Modes in Nonreciprocal Systems. *Phys. Rev. Lett.* **2019**, *123* (1), 016805. https://doi.org/10.1103/PhysRevLett.123.016805.

(911) Ezawa, M. Non-Hermitian Boundary and Interface States in Nonreciprocal Higher-Order Topological Metals and Electrical Circuits. *Phys. Rev. B* **2019**, *99* (12), 121411. https://doi.org/10.1103/PhysRevB.99.121411.

(912) Luo, X.-W.; Zhang, C. Higher-Order Topological Corner States Induced by Gain and Loss. *Phys. Rev. Lett.* **2019**, *123* (7), 073601. https://doi.org/10.1103/PhysRevLett.123.073601.

(913) Edvardsson, E.; Kunst, F. K.; Bergholtz, E. J. Non-Hermitian Extensions of Higher-Order Topological Phases and Their Biorthogonal Bulk-Boundary Correspondence. *Phys. Rev. B* **2019**, *99* (8), 081302. https://doi.org/10.1103/PhysRevB.99.081302.

(914) Kawabata, K.; Sato, M.; Shiozaki, K. Higher-Order Non-Hermitian Skin Effect. *Phys. Rev. B* **2020**, *102* (20), 205118. https://doi.org/10.1103/PhysRevB.102.205118.

(915) Martinez Alvarez, V. M.; Barrios Vargas, J. E.; Foa Torres, L. E. F. Non-Hermitian Robust Edge States in One Dimension: Anomalous Localization and Eigenspace Condensation at Exceptional Points. *Phys. Rev. B* **2018**, *97* (12), 121401. https://doi.org/10.1103/PhysRevB.97.121401.

(916) Gong, Z.; Ashida, Y.; Kawabata, K.; Takasan, K.; Higashikawa, S.; Ueda, M. Topological Phases of Non-Hermitian Systems. *Phys. Rev. X* **2018**, *8* (3), 031079. https://doi.org/10.1103/PhysRevX.8.031079.

(917) Li, L.; Lee, C. H.; Mu, S.; Gong, J. Critical Non-Hermitian Skin Effect. *Nat. Commun.* **2020**, *11* (1), 5491. https://doi.org/10.1038/s41467-020-18917-4.

(918) Longhi, S. Unraveling the Non-Hermitian Skin Effect in Dissipative Systems. *Phys. Rev. B* **2020**, *102* (20), 201103. https://doi.org/10.1103/PhysRevB.102.201103.

(919) Liu, S.; Shao, R.; Ma, S.; Zhang, L.; You, O.; Wu, H.; Xiang, Y. J.; Cui, T. J.; Zhang, S. Non-Hermitian Skin Effect in a Non-Hermitian Electrical Circuit. *Research* **2021**, *2021*, 1–9. https://doi.org/10.34133/2021/5608038.

(920) Roccati, F. Non-Hermitian Skin Effect as an Impurity Problem. *Phys. Rev. A* **2021**, *104* (2), 022215. https://doi.org/10.1103/PhysRevA.104.022215.

(921) Zhang, K.; Yang, Z.; Fang, C. Universal Non-Hermitian Skin Effect in Two and Higher Dimensions. *Nat. Commun.* **2022**, *13* (1), 2496. https://doi.org/10.1038/s41467-022-30161-6.

(922) Longhi, S. Probing Non-Hermitian Skin Effect and Non-Bloch Phase Transitions. *Phys. Rev. Res.* **2019**, *1* (2), 023013. https://doi.org/10.1103/PhysRevResearch.1.023013.

(923) Budich, J. C.; Bergholtz, E. J. Non-Hermitian Topological Sensors. *Phys. Rev. Lett.* **2020**, *125* (18), 180403. https://doi.org/10.1103/PhysRevLett.125.180403.

(924) McDonald, A.; Clerk, A. A. Exponentially-Enhanced Quantum Sensing with Non-Hermitian Lattice Dynamics. *Nat. Commun.* **2020**, *11* (1), 5382. https://doi.org/10.1038/s41467-020-19090-4.

(925) Longhi, S. Non-Hermitian Gauged Topological Laser Arrays. *Ann. Phys.* **2018**, *530* (7), 1800023. https://doi.org/10.1002/andp.201800023.

(926) Kraus, Y. E.; Ringel, Z.; Zilberberg, O. Four-Dimensional Quantum Hall Effect in a Two-Dimensional Quasicrystal. *Phys. Rev. Lett.* **2013**, *111* (22), 226401. https://doi.org/10.1103/PhysRevLett.111.226401.

(927) Celi, A.; Massignan, P.; Ruseckas, J.; Goldman, N.; Spielman, I. B.; Juzeliūnas, G.; Lewenstein, M. Synthetic Gauge Fields in Synthetic Dimensions. *Phys. Rev. Lett.* **2014**, *112* (4), 043001. https://doi.org/10.1103/PhysRevLett.112.043001.

(928) Luo, X. W.; Zhou, X.; Li, C. F.; Xu, J. S.; Guo, G. C.; Zhou, Z. W. Quantum Simulation of 2D Topological Physics in a 1D Array of Optical Cavities. *Nat. Commun.* **2015**, *6*. https://doi.org/10.1038/NCOMMS8704.

(929) Lin, Q.; Xiao, M.; Yuan, L.; Fan, S. Photonic Weyl Point in a Two-Dimensional Resonator Lattice with a Synthetic Frequency Dimension. *Nat. Commun.* **2016**, *7* (1), 13731. https://doi.org/10.1038/ncomms13731.

(930) Zhou, X.-F.; Luo, X.-W.; Wang, S.; Guo, G.-C.; Zhou, X.; Pu, H.; Zhou, Z.-W. Dynamically Manipulating Topological Physics and Edge Modes in a Single Degenerate Optical Cavity. *Phys. Rev. Lett.* **2017**, *118* (8), 083603. https://doi.org/10.1103/PhysRevLett.118.083603.

(931) Lohse, M.; Schweizer, C.; Price, H. M.; Zilberberg, O.; Bloch, I. Exploring 4D Quantum Hall Physics with a 2D Topological Charge Pump. *Nature* **2018**, *553* (7686), 55–58. https://doi.org/10.1038/nature25000.

(932) Zilberberg, O.; Huang, S.; Guglielmon, J.; Wang, M.; Chen, K. P.; Kraus, Y. E.; Rechtsman, M. C. Photonic Topological Boundary Pumping as a Probe of 4D Quantum Hall Physics. *Nature* **2018**, *553* (7686), 59–62. https://doi.org/10.1038/NATURE25011.

(933) Dutt, A.; Minkov, M.; Lin, Q.; Yuan, L.; Miller, D. A. B.; Fan, S. Experimental Band Structure Spectroscopy along a Synthetic Dimension. *Nat. Commun.* **2019**, *10* (1), 3122. https://doi.org/10.1038/s41467-019-11117-9.

(934) Dutt, A.; Lin, Q.; Yuan, L.; Minkov, M.; Xiao, M.; Fan, S. A Single Photonic Cavity with Two Independent Physical Synthetic Dimensions. *Science (80-. ).* **2020**, *367* (6473), 59–64. https://doi.org/10.1126/science.aaz3071.

(935) Lustig, E.; Weimann, S.; Plotnik, Y.; Lumer, Y.; Bandres, M. A.; Szameit, A.; Segev, M. Photonic Topological Insulator in Synthetic Dimensions. *Nature* **2019**, *567* (7748), 356–360. https://doi.org/10.1038/s41586-019-0943-7.

(936) Yuan, L.; Shi, Y.; Fan, S. Photonic Gauge Potential in a System with a Synthetic Frequency Dimension. *Opt. Lett.* **2016**, *41* (4), 741. https://doi.org/10.1364/OL.41.000741.

(937) Yuan, L.; Lin, Q.; Zhang, A.; Xiao, M.; Chen, X.; Fan, S. Photonic Gauge Potential in One Cavity with Synthetic Frequency and Orbital Angular Momentum Dimensions. *Phys. Rev. Lett.* **2019**, *122* (8), 083903. https://doi.org/10.1103/PhysRevLett.122.083903.

(938) Ni, X.; Kim, S.; Alù, A. Topological Insulator in Two Synthetic Dimensions Based on an Optomechanical Resonator. *Optica* **2021**, *8* (8), 1024–1032. https://doi.org/10.1364/OPTICA.430821.

(939) Ozawa, T.; Price, H. M.; Goldman, N.; Zilberberg, O.; Carusotto, I. Synthetic Dimensions in Integrated Photonics: From Optical Isolation to Four-Dimensional Quantum Hall Physics. *Phys. Rev. A* **2016**, *93* (4), 043827. https://doi.org/10.1103/PhysRevA.93.043827.

(940) Chen, Y.; Wang, H.-X.; Bao, Q.; Jiang, J.-H.; Chen, H. Ideal Type-II Weyl Points in Twisted One-Dimensional Dielectric Photonic Crystals. *Opt. Express* **2021**, *29* (24), 40606. https://doi.org/10.1364/OE.444780.

(941) Rosa, M. I. N.; Ruzzene, M.; Prodan, E. Topological Gaps by Twisting. *Commun. Phys.* **2021**, *4*

(1), 130. https://doi.org/10.1038/s42005-021-00630-3.

(942) Cheng, W.; Prodan, E.; Prodan, C. Revealing the Boundary Weyl Physics of the Four-Dimensional Hall Effect via Phason Engineering in Metamaterials. *Phys. Rev. Appl.* **2021**, *16* (4), 044032. https://doi.org/10.1103/PhysRevApplied.16.044032.

(943) Chen, H.; Zhang, H.; Wu, Q.; Huang, Y.; Nguyen, H.; Prodan, E.; Zhou, X.; Huang, G. Creating Synthetic Spaces for Higher-Order Topological Sound Transport. *Nat. Commun.* **2021**, *12* (1), 5028. https://doi.org/10.1038/s41467-021-25305-z.

(944) Zangeneh-Nejad, F.; Fleury, R. Experimental Observation of the Acoustic Z2 Weyl Semimetallic Phase in Synthetic Dimensions. *Phys. Rev. B* **2020**, *102* (6), 064309. https://doi.org/10.1103/PhysRevB.102.064309.

(945) Wang, Z.; Wang, Z.; Li, H.; Luo, J.; Wang, X.; Liu, Z.; Yang, H. Weyl Points and Nodal Lines in Acoustic Synthetic Parameter Space. *Appl. Phys. Express* **2021**, *14* (7), 077002. https://doi.org/10.35848/1882-0786/ac0c8b.

(946) Li, Z.-W.; Liang, B.; Cheng, J.-C. Type-II Weyl Points in a Synthetic Three-Dimensional Acoustic Lattice. *Appl. Phys. Express* **2022**, *15* (3), 037001. https://doi.org/10.35848/1882-0786/ac516c.

(947) Fan, X.; Qiu, C.; Shen, Y.; He, H.; Xiao, M.; Ke, M.; Liu, Z. Probing Weyl Physics with One-Dimensional Sonic Crystals. *Phys. Rev. Lett.* **2019**, *122* (13), 136802. https://doi.org/10.1103/PhysRevLett.122.136802.

(948) Wang, W.; Chen, Z.-G.; Ma, G. Synthetic Three-Dimensional ZxZ(2) Topological Insulator in an Elastic Metacrystal. *Phys. Rev. Lett.* **2021**, *127* (21), 214302. https://doi.org/10.1103/PhysRevLett.127.214302.

(949) Chen, K.; Weiner, M.; Li, M.; Ni, X.; Alù, A.; Khanikaev, A. B. Nonlocal Topological Insulators: Deterministic Aperiodic Arrays Supporting Localized Topological States Protected by Nonlocal Symmetries. *Proc. Natl. Acad. Sci.* **2021**, *118* (34), e2100691118. https://doi.org/10.1073/pnas.2100691118.

(950) Chen, Z. G.; Zhu, W.; Tan, Y.; Wang, L.; Ma, G. Acoustic Realization of a Four-Dimensional Higher-Order Chern Insulator and Boundary-Modes Engineering. *Phys. Rev. X* **2021**, *11* (1), 11016. https://doi.org/10.1103/PhysRevX.11.011016.

(951) Lumer, Y.; Plotnik, Y.; Rechtsman, M. C.; Segev, M. Self-Localized States in Photonic Topological Insulators. *Phys. Rev. Lett.* **2013**, *111* (24), 243905. https://doi.org/10.1103/PhysRevLett.111.243905.

(952) Leykam, D.; Chong, Y. D. Edge Solitons in Nonlinear-Photonic Topological Insulators. *Phys. Rev. Lett.* **2016**, *117* (14), 143901. https://doi.org/10.1103/PhysRevLett.117.143901.

(953) Mukherjee, S.; Rechtsman, M. C. Observation of Floquet Solitons in a Topological Bandgap. *Science (80-. ).* **2020**, *368* (6493), 856–859. https://doi.org/10.1126/science.aba8725.

(954) Mukherjee, S.; Rechtsman, M. C. Observation of Unidirectional Solitonlike Edge States in Nonlinear Floquet Topological Insulators. *Phys. Rev. X* **2021**, *11* (4), 041057. https://doi.org/10.1103/PhysRevX.11.041057.

(955) Kruk, S.; Poddubny, A.; Smirnova, D.; Wang, L.; Slobozhanyuk, A.; Shorokhov, A.; Kravchenko, I.; Luther-Davies, B.; Kivshar, Y. Nonlinear Light Generation in Topological Nanostructures. *Nature Nanotechnology*. Nature Publishing Group February 1, 2019, pp 126–130. https://doi.org/10.1038/s41565-018-0324-7.

(956) Wang, Y.; Lang, L.-J.; Lee, C. H.; Zhang, B.; Chong, Y. D. Topologically Enhanced Harmonic Generation in a Nonlinear Transmission Line Metamaterial. *Nat. Commun.* **2019**, *10* (1), 1102. https://doi.org/10.1038/s41467-019-08966-9.

(957) Smirnova, D.; Kruk, S.; Leykam, D.; Melik-Gaykazyan, E.; Choi, D.-Y.; Kivshar, Y. Third-Harmonic Generation in Photonic Topological Metasurfaces. *Phys. Rev. Lett.* **2019**, *123* (10), 103901. https://doi.org/10.1103/PhysRevLett.123.103901.

(958) Hadad, Y.; Soric, J. C.; Khanikaev, A. B.; Alú, A. Self-Induced Topological Protection in Nonlinear Circuit Arrays. *Nat. Electron.* **2018**, *1* (3), 178–182. https://doi.org/10.1038/s41928-018-0042-z.

(959) He, L.; Addison, Z.; Jin, J.; Mele, E. J.; Johnson, S. G.; Zhen, B. Floquet Chern Insulators of Light. *Nat. Commun.* **2019**, *10* (1), 4194. https://doi.org/10.1038/s41467-019-12231-4.

(960) Duggan, R.; Mann, S. A.; Alù, A. Nonreciprocal Photonic Topological Order Driven by Uniform Optical Pumping. *Phys. Rev. B* **2020**, *102* (10), 100303. https://doi.org/10.1103/PhysRevB.102.100303.

(961) Hu, Z.; Bongiovanni, D.; Jukić, D.; Jajtić, E.; Xia, S.; Song, D.; Xu, J.; Morandotti, R.; Buljan, H.; Chen, Z. Nonlinear Control of Photonic Higher-Order Topological Bound States in the Continuum. *Light Sci. Appl.* **2021**, *10* (1), 164. https://doi.org/10.1038/s41377-021-00607-5.

(962) Peano, V.; Houde, M.; Brendel, C.; Marquardt, F.; Clerk, A. A. Topological Phase Transitions and Chiral Inelastic Transport Induced by the Squeezing of Light. *Nat. Commun.* **2016**, *7* (1), 10779. https://doi.org/10.1038/ncomms10779.

(963) You, J. W.; Lan, Z.; Panoiu, N. C. Four-Wave Mixing of Topological Edge Plasmons in Graphene Metasurfaces. *Sci. Adv.* **2020**, *6* (13). https://doi.org/10.1126/sciadv.aaz3910.

(964) Clark, L. W.; Schine, N.; Baum, C.; Jia, N.; Simon, J. Observation of Laughlin States Made of Light. *Nature* **2020**, *582* (7810), 41–45. https://doi.org/10.1038/s41586-020-2318-5.

(965) Chaunsali, R.; Xu, H.; Yang, J.; Kevrekidis, P. G.; Theocharis, G. Stability of Topological Edge States under Strong Nonlinear Effects. *Phys. Rev. B* **2021**, *103* (2), 24106. https://doi.org/10.1103/PhysRevB.103.024106.

(966) Vila, J.; Paulino, G. H.; Ruzzene, M. Role of Nonlinearities in Topological Protection: Testing Magnetically Coupled Fidget Spinners. *Phys. Rev. B* **2019**, *99* (12), 125116. https://doi.org/10.1103/PhysRevB.99.125116.

(967) Tempelman, J. R.; Matlack, K. H.; Vakakis, A. F. Topological Protection in a Strongly Nonlinear Interface Lattice. *Phys. Rev. B* **2021**, *104* (17), 174306. https://doi.org/10.1103/PhysRevB.104.174306.

(968) Chaunsali, R.; Theocharis, G. Self-Induced Topological Transition in Phononic Crystals by Nonlinearity Management. *Phys. Rev. B* **2019**, *100* (1), 014302. https://doi.org/10.1103/PhysRevB.100.014302.

(969) Darabi, A.; Leamy, M. J. Tunable Nonlinear Topological Insulator for Acoustic Waves. *Phys. Rev. Appl.* **2019**, *12* (4), 44030. https://doi.org/10.1103/PhysRevApplied.12.044030.

(970) Li, J.; Chu, R.-L.; Jain, J. K.; Shen, S.-Q. Topological Anderson Insulator. *Phys. Rev. Lett.* **2009**, *102* (13), 136806. https://doi.org/10.1103/PhysRevLett.102.136806.

(971) Groth, C. W.; Wimmer, M.; Akhmerov, A. R.; Tworzydło, J.; Beenakker, C. W. J. Theory of the

Topological Anderson Insulator. *Phys. Rev. Lett.* **2009**, *103* (19), 196805. https://doi.org/10.1103/PhysRevLett.103.196805.

(972) Guo, H.-M.; Rosenberg, G.; Refael, G.; Franz, M. Topological Anderson Insulator in Three Dimensions. *Phys. Rev. Lett.* **2010**, *105* (21), 216601. https://doi.org/10.1103/PhysRevLett.105.216601.

(973) Stützer, S.; Plotnik, Y.; Lumer, Y.; Titum, P.; Lindner, N. H.; Segev, M.; Rechtsman, M. C.; Szameit, A. Photonic Topological Anderson Insulators. *Nature* **2018**, *560* (7719), 461–465. https://doi.org/10.1038/s41586-018-0418-2.

(974) Meier, E. J.; An, F. A.; Dauphin, A.; Maffei, M.; Massignan, P.; Hughes, T. L.; Gadway, B. Observation of the Topological Anderson Insulator in Disordered Atomic Wires. *Science (80-. ).* **2018**, *362* (6417), 929–933. https://doi.org/10.1126/science.aat3406.

(975) Liu, G.-G.; Yang, Y.; Ren, X.; Xue, H.; Lin, X.; Hu, Y.-H.; Sun, H.; Peng, B.; Zhou, P.; Chong, Y.; Zhang, B. Topological Anderson Insulator in Disordered Photonic Crystals. *Phys. Rev. Lett.* **2020**, *125* (13), 133603. https://doi.org/10.1103/PhysRevLett.125.133603.

(976) Liu, C.; Gao, W.; Yang, B.; Zhang, S. Disorder-Induced Topological State Transition in Photonic Metamaterials. *Phys. Rev. Lett.* **2017**, *119* (18), 183901. https://doi.org/10.1103/PhysRevLett.119.183901.

(977) Lin, Q.; Li, T.; Xiao, L.; Wang, K.; Yi, W.; Xue, P. Observation of Non-Hermitian Topological Anderson Insulator in Quantum Dynamics. *Nat. Commun.* **2022**, *13* (1), 3229. https://doi.org/10.1038/s41467-022-30938-9.

(978) Shi, X.; Kiorpelidis, I.; Chaunsali, R.; Achilleos, V.; Theocharis, G.; Yang, J. Disorder-Induced Topological Phase Transition in a One-Dimensional Mechanical System. *Phys. Rev. Res.* **2021**, *3* (3), 33012. https://doi.org/10.1103/PhysRevResearch.3.033012.

(979) Zangeneh-Nejad, F.; Fleury, R. Disorder-Induced Signal Filtering with Topological Metamaterials. *Adv. Mater.* **2020**, *32* (28), 2001034. https://doi.org/10.1002/adma.202001034.

(980) Wu, X.; Meng, Y.; Hao, Y.; Zhang, R.-Y.; Li, J.; Zhang, X. Topological Corner Modes Induced by Dirac Vortices in Arbitrary Geometry. *Phys. Rev. Lett.* **2021**, *126* (22), 226802. https://doi.org/10.1103/PhysRevLett.126.226802.

(981) Ye, L.; Qiu, C.; Xiao, M.; Li, T.; Du, J.; Ke, M.; Liu, Z. Topological Dislocation Modes in Three-Dimensional Acoustic Topological Insulators. *Nat. Commun.* **2022**, *13* (1), 508. https://doi.org/10.1038/s41467-022-28182-2.

(982) Xia, B.; Zhang, J.; Tong, L.; Zheng, S.; Man, X. Topologically Valley-Polarized Edge States in Elastic Phononic Plates Yielded by Lattice Defects. *Int. J. Solids Struct.* **2022**, *239–240*, 111413. https://doi.org/10.1016/j.ijsolstr.2021.111413.

(983) Wang, Q.; Ge, Y.; Sun, H.; Xue, H.; Jia, D.; Guan, Y.; Yuan, S.; Zhang, B.; Chong, Y. D. Vortex States in an Acoustic Weyl Crystal with a Topological Lattice Defect. *Nat. Commun.* **2021**, *12* (1), 3654. https://doi.org/10.1038/s41467-021-23963-7.

(984) Deng, Y.; Benalcazar, W. A.; Chen, Z.-G.; Oudich, M.; Ma, G.; Jing, Y. Observation of Degenerate Zero-Energy Topological States at Disclinations in an Acoustic Lattice. *Phys. Rev. Lett.* **2022**, *128* (17), 174301. https://doi.org/10.1103/PhysRevLett.128.174301.

(985) Xia, B.; Jiang, Z.; Tong, L.; Zheng, S.; Man, X. Topological Bound States in Elastic Phononic Plates Induced by Disclinations. *Acta Mech. Sin.* **2022**, *38* (2), 521459.

https://doi.org/10.1007/s10409-021-09083-0.

(986) Qi, Y.; He, H.; Xiao, M. Manipulation of Acoustic Vortex with Topological Dislocation States. *Appl. Phys. Lett.* **2022**, *120* (21), 212202. https://doi.org/10.1063/5.0095543.

(987) Chen, C. W.; Lera, N.; Chaunsali, R.; Torrent, D.; Alvarez, J. V.; Yang, J.; San-Jose, P.; Christensen, J. Mechanical Analogue of a Majorana Bound State. *Adv. Mater.* **2019**, *31* (51), 1904386. https://doi.org/10.1002/adma.201904386.

(988) Gao, P.; Christensen, J. Topological Sound Pumping of Zero-Dimensional Bound States. *Adv. Quantum Technol.* **2020**, *3* (9), 2000065. https://doi.org/https://doi.org/10.1002/qute.202000065.

(989) Gao, P.; Torrent, D.; Cervera, F.; San-Jose, P.; Sánchez-Dehesa, J.; Christensen, J. Majorana-like Zero Modes in Kekulé Distorted Sonic Lattices. *Phys. Rev. Lett.* **2019**, *123* (19), 196601. https://doi.org/10.1103/PhysRevLett.123.196601.

(990) Lubensky, T. C.; Kane, C. L.; Mao, X.; Souslov, A.; Sun, K. Phonons and Elasticity in Critically Coordinated Lattices. *Reports Prog. Phys.* **2015**, *78* (7), 073901. https://doi.org/10.1088/0034-4885/78/7/073901.

(991) Chen, B. G. G.; Upadhyaya, N.; Vitelli, V. Nonlinear Conduction via Solitons in a Topological Mechanical Insulator. **2014**, *111* (36), 13004–13009.

(992) Mao, X.; Lubensky, T. C. Maxwell Lattices and Topological Mechanics. *Annu. Rev. Condens. Matter Phys.* **2018**, *9* (1), 413–433. https://doi.org/10.1146/annurev-conmatphys-033117-054235.

(993) Rocklin, D. Z.; Zhou, S.; Sun, K.; Mao, X. Transformable Topological Mechanical Metamaterials. *Nat. Commun.* **2017**, *8* (1), 14201. https://doi.org/10.1038/ncomms14201.

(994) Chen, B. G. G.; Liu, B.; Evans, A. A.; Paulose, J.; Cohen, I.; Vitelli, V.; Santangelo, C. D. Topological Mechanics of Origami and Kirigami. *Phys. Rev. Lett.* **2016**, *116* (13), 135501. https://doi.org/10.1103/PhysRevLett.116.135501.

(995) Meeussen, A. S.; Paulose, J.; Vitelli, V. Geared Topological Metamaterials with Tunable Mechanical Stability. *Phys. Rev. X* **2016**, *6* (4), 41029. https://doi.org/10.1103/PhysRevX.6.041029.

(996) Rocklin, D. Z.; Chen, B. G.; Falk, M.; Vitelli, V.; Lubensky, T. C. Mechanical Weyl Modes in Topological Maxwell Lattices. *Phys. Rev. Lett.* **2016**, *116* (13), 135503. https://doi.org/10.1103/PhysRevLett.116.135503.

(997) Lera, N.; Alvarez, J. V; Sun, K. Topological Mechanical Metamaterial with Nonrectilinear Constraints. *Phys. Rev. B* **2018**, *98* (1), 14101. https://doi.org/10.1103/PhysRevB.98.014101.

(998) Socolar, J. E. S.; Lubensky, T. C.; Kane, C. L. Mechanical Graphene. *New J. Phys.* **2017**, *19* (2), 025003. https://doi.org/10.1088/1367-2630/aa57bb.

(999) Stenull, O.; Kane, C. L.; Lubensky, T. C. Topological Phonons and Weyl Lines in Three Dimensions. *Phys. Rev. Lett.* **2016**, *117* (6), 068001. https://doi.org/10.1103/PhysRevLett.117.068001.

(1000) Baardink, G.; Souslov, A.; Paulose, J.; Vitelli, V. Localizing Softness and Stress along Loops in 3D Topological Metamaterials. *Proc. Natl. Acad. Sci. U. S. A.* **2018**, *115* (3), 489–494. https://doi.org/10.1073/pnas.1713826115.

(1001) Bertoldi, K.; Vitelli, V.; Christensen, J.; Van Hecke, M. *Flexible Mechanical Metamaterials*; Nature Publishing Group, 2017; Vol. 2, p 17066.

(1002) Yan, Z.; Zhang, F.; Wang, J.; Liu, F.; Guo, X.; Nan, K.; Lin, Q.; Gao, M.; Xiao, D.; Shi, Y.; Qiu, Y.; Luan, H.; Kim, J. H.; Wang, Y.; Luo, H.; Han, M.; Huang, Y.; Zhang, Y.; Rogers, J. A. Controlled Mechanical Buckling for Origami-Inspired Construction of 3D Microstructures in Advanced Materials. *Adv. Funct. Mater.* **2016**, *26* (16), 2629–2639. https://doi.org/10.1002/adfm.201504901.

(1003) Chen, S.; Choi, G. P. T.; Mahadevan, L. Deterministic and Stochastic Control of Kirigami Topology. *Proc. Natl. Acad. Sci.* **2020**, *117* (9), 4511–4517. https://doi.org/10.1073/pnas.1909164117.

(1004) Liu, B. Topological Kinematics of Origami Metamaterials. *Nat. Phys.* **2018**, *14* (8), 811–815. https://doi.org/10.1038/s41567-018-0150-8.

(1005) Sussman, D. M.; Stenull, O.; Lubensky, T. C. Topological Boundary Modes in Jammed Matter. *Soft Matter* **2016**, *12* (28), 6079–6087. https://doi.org/10.1039/c6sm00875e.

(1006) Zhou, D.; Zhang, L.; Mao, X. Topological Boundary Floppy Modes in Quasicrystals. *Phys. Rev. X* **2019**, *9* (2), 21054. https://doi.org/10.1103/PhysRevX.9.021054.

(1007) Zhou, D.; Zhang, L.; Mao, X. Topological Edge Floppy Modes in Disordered Fiber Networks. *Phys. Rev. Lett.* **2018**, *120* (6), 68003. https://doi.org/10.1103/PhysRevLett.120.068003.

(1008) Bartolo, D.; Carpentier, D. Topological Elasticity of Nonorientable Ribbons. *Phys. Rev. X* **2019**, *9* (4), 41058. https://doi.org/10.1103/PhysRevX.9.041058.

(1009) Saremi, A.; Rocklin, Z. Topological Elasticity of Flexible Structures. *Phys. Rev. X* **2020**, *10* (1), 11052. https://doi.org/10.1103/PhysRevX.10.011052.

(1010) Coulais, C.; Sounas, D.; Alù, A. Static Non-Reciprocity in Mechanical Metamaterials. *Nature* **2017**, *542* (7642), 461–464. https://doi.org/10.1038/nature21044.

(1011) Krasnok, A.; Baranov, D.; Li, H.; Miri, M.-A.; Monticone, F.; Alú, A. Anomalies in Light Scattering. *Adv. Opt. Photonics* **2019**, *11* (4), 892. https://doi.org/10.1364/aop.11.000892.

(1012) Krasnok, A.; Alu, A. Active Nanophotonics. *Proc. IEEE* **2020**, *108* (5), 628–654. https://doi.org/10.1109/JPROC.2020.2985048.

(1013) Krasnok, A.; Alù, A. Low-Symmetry Nanophotonics. *ACS Photonics* **2022**, *9* (1), 2–24. https://doi.org/10.1021/acsphotonics.1c00968.

(1014) Ni, J.; Huang, C.; Zhou, L.-M.; Gu, M.; Song, Q.; Kivshar, Y.; Qiu, C.-W. Multidimensional Phase Singularities in Nanophotonics. *Science (80-. ).* **2021**, *374* (6566). https://doi.org/10.1126/science.abj0039.

(1015) Hsu, C. W.; Zhen, B.; Stone, A. D.; Joannopoulos, J. D.; Soljačić, M. Bound States in the Continuum. *Nat. Rev. Mater.* **2016**, *1* (9), 16048. https://doi.org/10.1038/natrevmats.2016.48.

(1016) Lannebère, S.; Silveirinha, M. G. Optical Meta-Atom for Localization of Light with Quantized Energy. *Nat. Commun.* **2015**, *6* (1), 8766. https://doi.org/10.1038/ncomms9766.

(1017) Monticone, F.; Alù, A. Embedded Photonic Eigenvalues in 3D Nanostructures. *Phys. Rev. Lett.* **2014**, *112* (21), 213903. https://doi.org/10.1103/PhysRevLett.112.213903.

(1018) Krasnok, A.; Alú, A. Embedded Scattering Eigenstates Using Resonant Metasurfaces. *J. Opt.* **2018**, *20* (6), 064002. https://doi.org/10.1088/2040-8986/aac1d6.

(1019) Kang, M.; Zhang, Z.; Wu, T.; Zhang, X.; Xu, Q.; Krasnok, A.; Han, J.; Alù, A. Coherent Full

Polarization Control Based on Bound States in the Continuum. *Nat. Commun.* **2022**, *13* (1), 4536. https://doi.org/10.1038/s41467-022-31726-1.

(1020)  Jacobsen, R. E.; Krasnok, A.; Arslanagić, S.; Lavrinenko, A. V.; Alú, A. Boundary-Induced Embedded Eigenstate in a Single Resonator for Advanced Sensing. *ACS Photonics* **2022**, *9* (6), 1936–1943. https://doi.org/10.1021/acsphotonics.1c01840.

(1021)  Sakotic, Z.; Krasnok, A.; Alu, A.; Jankovic, N. Topological Scattering Singularities and Embedded Eigenstates for Polarization Control and Sensing Applications. *Photonics Res.* **2021**, *9* (7), 1310–1323. https://doi.org/10.1364/prj.424247.

(1022)  Nefedkin, N.; Alú, A.; Krasnok, A. Quantum Embedded Superstates. *Adv. Quantum Technol.* **2021**, *4* (6), 2000121. https://doi.org/10.1002/qute.202000121.

(1023)  Shi, T.; Deng, Z.-L.; Geng, G.; Zeng, X.; Zeng, Y.; Hu, G.; Overvig, A.; Li, J.; Qiu, C.-W.; Alú, A.; Kivshar, Y. S.; Li, X. Planar Chiral Metasurfaces with Maximal and Tunable Chiroptical Response Driven by Bound States in the Continuum. *Nat. Commun.* **2022**, *13* (1), 4111. https://doi.org/10.1038/s41467-022-31877-1.

(1024)  Monticone, F.; Sounas, D.; Krasnok, A.; Alú, A. Can a Nonradiating Mode Be Externally Excited? Nonscattering States versus Embedded Eigenstates. *ACS Photonics* **2019**, *6* (12), 3108–3114. https://doi.org/10.1021/acsphotonics.9b01104.

(1025)  Friedrich, H.; Wintgen, D. Physical Realization of Bound States in the Continuum. *Phys. Rev. A* **1985**, *31* (6), 3964–3966. https://doi.org/10.1103/PhysRevA.31.3964.

(1026)  Hsu, C. W.; Zhen, B.; Lee, J.; Chua, S.-L.; Johnson, S. G.; Joannopoulos, J. D.; Soljačić, M. Observation of Trapped Light within the Radiation Continuum. *Nature* **2013**, *499* (7457), 188–191. https://doi.org/10.1038/nature12289.

(1027)  Zhen, B.; Hsu, C. W.; Lu, L.; Stone, A. D.; Soljačić, M. Topological Nature of Optical Bound States in the Continuum. *Phys. Rev. Lett.* **2014**, *113* (25), 257401. https://doi.org/10.1103/PhysRevLett.113.257401.

(1028)  Azzam, S. I.; Shalaev, V. M.; Boltasseva, A.; Kildishev, A. V. Formation of Bound States in the Continuum in Hybrid Plasmonic-Photonic Systems. *Phys. Rev. Lett.* **2018**, *121* (25), 253901. https://doi.org/10.1103/PhysRevLett.121.253901.

(1029)  Lee, J.; Zhen, B.; Chua, S. L.; Qiu, W.; Joannopoulos, J. D.; Soljačić, M.; Shapira, O. Observation and Differentiation of Unique High-Q Optical Resonances near Zero Wave Vector in Macroscopic Photonic Crystal Slabs. *Phys. Rev. Lett.* **2012**, *109* (6), 067401. https://doi.org/10.1103/PhysRevLett.109.067401.

(1030)  Doeleman, H. M.; Monticone, F.; Den Hollander, W.; Alú, A.; Koenderink, A. F. Experimental Observation of a Polarization Vortex at an Optical Bound State in the Continuum. *Nat. Photonics* **2018**, *12* (7), 397–401. https://doi.org/10.1038/s41566-018-0177-5.

(1031)  Ovcharenko, A. I.; Blanchard, C.; Hugonin, J. P.; Sauvan, C. Bound States in the Continuum in Symmetric and Asymmetric Photonic Crystal Slabs. *Phys. Rev. B* **2020**, *101* (15), 155303. https://doi.org/10.1103/PhysRevB.101.155303.

(1032)  K Koshelev, S. L. M. L. A. B. Y. K. Asymmetric Metasurfaces with High-Q Resonances Governed by Bound States in the Continuum. *Phys. Rev. Lett.* **2018**, *121*, 193903.

(1033)  Tittl, A.; Leitis, A.; Liu, M.; Yesilkoy, F.; Choi, D.-Y.; Neshev, D. N.; Kivshar, Y. S.; Altug, H. Imaging-Based Molecular Barcoding with Pixelated Dielectric Metasurfaces. *Science (80-. ).*

**2018**, *360* (6393), 1105–1109. https://doi.org/10.1126/science.aas9768.

(1034)  Kodigala, A.; Lepetit, T.; Gu, Q.; Bahari, B.; Fainman, Y.; Kanté, B. Lasing Action from Photonic Bound States in Continuum. *Nature* **2017**, *541* (7636), 196–199. https://doi.org/10.1038/nature20799.

(1035)  Huang, C.; Zhang, C.; Xiao, S.; Wang, Y.; Fan, Y.; Liu, Y.; Zhang, N.; Qu, G.; Ji, H.; Han, J.; Ge, L.; Kivshar, Y.; Song, Q. Ultrafast Control of Vortex Microlasers. *Science (80-. ).* **2020**, *367* (6481), 1018–1021. https://doi.org/10.1126/science.aba4597.

(1036)  Ha, S. T.; Fu, Y. H.; Emani, N. K.; Pan, Z.; Bakker, R. M.; Paniagua-Domínguez, R.; Kuznetsov, A. I. Directional Lasing in Resonant Semiconductor Nanoantenna Arrays. *Nat. Nanotechnol.* **2018**, *13* (11), 1042–1047. https://doi.org/10.1038/s41565-018-0245-5.

(1037)  Hwang, M.-S.; Lee, H.-C.; Kim, K.-H.; Jeong, K.-Y.; Kwon, S.-H.; Koshelev, K.; Kivshar, Y.; Park, H.-G. Ultralow-Threshold Laser Using Super-Bound States in the Continuum. *Nat. Commun.* **2021**, *12* (1), 4135. https://doi.org/10.1038/s41467-021-24502-0.

(1038)  Carletti, L.; Kruk, S. S.; Bogdanov, A. A.; De Angelis, C.; Kivshar, Y. High-Harmonic Generation at the Nanoscale Boosted by Bound States in the Continuum. *Phys. Rev. Res.* **2019**, *1* (2), 023016. https://doi.org/10.1103/PhysRevResearch.1.023016.

(1039)  Koshelev, K.; Kruk, S.; Melik-Gaykazyan, E.; Choi, J.-H.; Bogdanov, A.; Park, H.-G.; Kivshar, Y. Subwavelength Dielectric Resonators for Nonlinear Nanophotonics. *Science (80-. ).* **2020**, *367* (6475), 288 LP – 292. https://doi.org/10.1126/science.aaz3985.

(1040)  Cotrufo, M.; Alù, A. Excitation of Single-Photon Embedded Eigenstates in Coupled Cavity–Atom Systems. *Optica* **2019**, *6* (6), 799. https://doi.org/10.1364/OPTICA.6.000799.

(1041)  Nefedkin, N.; Cotrufo, M.; Krasnok, A.; Alù, A. Dark-State Induced Quantum Nonreciprocity. *Adv. Quantum Technol.* **2022**, *5* (3), 2100112. https://doi.org/10.1002/qute.202100112.

(1042)  Duggan, R.; Ra'di, Y.; Alù, A. Temporally and Spatially Coherent Emission from Thermal Embedded Eigenstates. *ACS Photonics* **2019**, *6* (11), 2949–2956. https://doi.org/10.1021/acsphotonics.9b01131.

(1043)  Sakotic, Z.; Krasnok, A.; Cselyuszka, N.; Jankovic, N.; Alú, A. Berreman Embedded Eigenstates for Narrow-Band Absorption and Thermal Emission. *Phys. Rev. Appl.* **2020**, *13* (6), 064073. https://doi.org/10.1103/PhysRevApplied.13.064073.

(1044)  Ra'di, Y.; Krasnok, A.; Alù, A. Virtual Critical Coupling. *ACS Photonics* **2020**, *7* (6), 1468–1475. https://doi.org/10.1021/acsphotonics.0c00165.

(1045)  Song, M.; Jayathurathnage, P.; Zanganeh, E.; Krasikova, M.; Smirnov, P.; Belov, P.; Kapitanova, P.; Simovski, C.; Tretyakov, S.; Krasnok, A. Wireless Power Transfer Based on Novel Physical Concepts. *Nat. Electron.* **2021**, *4* (10), 707–716. https://doi.org/10.1038/s41928-021-00658-x.

(1046)  Gorkunov, M. V.; Antonov, A. A.; Tuz, V. R.; Kupriianov, A. S.; Kivshar, Y. S. Bound States in the Continuum Underpin Near-Lossless Maximum Chirality in Dielectric Metasurfaces. *Adv. Opt. Mater.* **2021**, *9* (19), 2100797. https://doi.org/10.1002/adom.202100797.

(1047)  Gorkunov, M. V.; Antonov, A. A.; Kivshar, Y. S. Metasurfaces with Maximum Chirality Empowered by Bound States in the Continuum. *Phys. Rev. Lett.* **2020**, *125* (9), 093903. https://doi.org/10.1103/PhysRevLett.125.093903.

(1048)  Bulgakov, E. N.; Maksimov, D. N. Topological Bound States in the Continuum in Arrays of Dielectric Spheres. *Phys. Rev. Lett.* **2017**, *118* (26), 267401.
https://doi.org/10.1103/PhysRevLett.118.267401.

(1049)  Zhang, Y.; Chen, A.; Liu, W.; Hsu, C. W.; Wang, B.; Guan, F.; Liu, X.; Shi, L.; Lu, L.; Zi, J.
Observation of Polarization Vortices in Momentum Space. *Phys. Rev. Lett.* **2018**, *120* (18),
186103. https://doi.org/10.1103/PhysRevLett.120.186103.

(1050)  Jin, J.; Yin, X.; Ni, L.; Soljačić, M.; Zhen, B.; Peng, C. Topologically Enabled Ultrahigh-Q
Guided Resonances Robust to out-of-Plane Scattering. *Nature* **2019**, *574* (7779), 501–504.
https://doi.org/10.1038/s41586-019-1664-7.

(1051)  X Yin, J. J. M. S. C. P. B. Z.; Yin, X.; Jin, J.; Soljačić, M.; Peng, C.; Zhen, B. Observation of
Topologically Enabled Unidirectional Guided Resonances. *Nature* **2020**, *580* (7804), 467–471.
https://doi.org/10.1038/s41586-020-2181-4.

(1052)  Guo, Y.; Xiao, M.; Fan, S. Topologically Protected Complete Polarization Conversion. *Phys.
Rev. Lett.* **2017**, *119* (16), 167401. https://doi.org/10.1103/PhysRevLett.119.167401.

(1053)  Guo, Y.; Xiao, M.; Zhou, Y.; Fan, S. Arbitrary Polarization Conversion with a Photonic Crystal
Slab. *Adv. Opt. Mater.* **2019**, *7* (14), 1801453. https://doi.org/10.1002/adom.201801453.

(1054)  Wu, Y.; Kang, L.; Bao, H.; Werner, D. H. Exploiting Topological Properties of Mie-Resonance-
Based Hybrid Metasurfaces for Ultrafast Switching of Light Polarization. *ACS Photonics* **2020**, *7*
(9), 2362–2373. https://doi.org/10.1021/acsphotonics.0c00858.

(1055)  Liu, W.; Wang, B.; Zhang, Y.; Wang, J.; Zhao, M.; Guan, F.; Liu, X.; Shi, L.; Zi, J. Circularly
Polarized States Spawning from Bound States in the Continuum. *Phys. Rev. Lett.* **2019**, *123* (11),
116104. https://doi.org/10.1103/PhysRevLett.123.116104.

(1056)  Ye, W.; Gao, Y.; Liu, J. Singular Points of Polarizations in the Momentum Space of Photonic
Crystal Slabs. *Phys. Rev. Lett.* **2020**, *124* (15), 153904.
https://doi.org/10.1103/PhysRevLett.124.153904.

(1057)  Yoda, T.; Notomi, M. Generation and Annihilation of Topologically Protected Bound States in
the Continuum and Circularly Polarized States by Symmetry Breaking. *Phys. Rev. Lett.* **2020**, *125*
(5), 053902. https://doi.org/10.1103/PhysRevLett.125.053902.

(1058)  Wang, B.; Liu, W.; Zhao, M.; Wang, J.; Zhang, Y.; Chen, A.; Guan, F.; Liu, X.; Shi, L.; Zi, J.
Generating Optical Vortex Beams by Momentum-Space Polarization Vortices Centred at Bound
States in the Continuum. *Nat. Photonics* **2020**, *14* (10), 623–628. https://doi.org/10.1038/s41566-
020-0658-1.

(1059)  Berkhout, A.; Koenderink, A. F. Perfect Absorption and Phase Singularities in Plasmon Antenna
Array Etalons. *ACS Photonics* **2019**, *6* (11), 2917–2925.
https://doi.org/10.1021/acsphotonics.9b01019.

(1060)  Rüter, C. E.; Makris, K. G.; El-ganainy, R.; Christodoulides, D. N.; Segev, M.; Kip, D.
Observation of Parity – Time Symmetry in Optics. *Nat. Phys.* **2010**, *6* (3), 192–195.
https://doi.org/10.1038/nphys1515.

(1061)  Guo, A.; Salamo, G. J.; Duchesne, D.; Morandotti, R.; Volatier-Ravat, M.; Aimez, V.; Siviloglou,
G. A.; Christodoulides, D. N. Observation of PT-Symmetry Breaking in Complex Optical
Potentials. *Phys. Rev. Lett.* **2009**, *103* (9), 093902.
https://doi.org/10.1103/PhysRevLett.103.093902.

(1062)  Chen, H.-Z.; Liu, T.; Luan, H.-Y.; Liu, R.-J.; Wang, X.-Y.; Zhu, X.-F.; Li, Y.-B.; Gu, Z.-M.;

Liang, S.-J.; Gao, H.; Lu, L.; Ge, L.; Zhang, S.; Zhu, J.; Ma, R.-M. Revealing the Missing Dimension at an Exceptional Point. *Nat. Phys.* **2020**, *16* (5), 571–578. https://doi.org/10.1038/s41567-020-0807-y.

(1063)  Assawaworrarit, S.; Yu, X.; Fan, S. Robust Wireless Power Transfer Using a Nonlinear Parity-Time-Symmetric Circuit. *Nature* **2017**, *546* (7658), 387–390. https://doi.org/10.1038/NATURE22404.

(1064)  Hodaei, H.; Miri, M.-A.; Heinrich, M.; Christodoulides, D. N.; Khajavikhan, M. Parity-Time-Symmetric Microring Lasers. *Science (80-. ).* **2014**, *346* (6212), 975–978.

(1065)  Zhang, H.; Huang, R.; Zhang, S.; Li, Y.; Qiu, C.; Nori, F.; Jing, H. Breaking Anti-PT Symmetry by Spinning a Resonator. *Nano Lett.* **2020**, *20* (10), 7594–7599. https://doi.org/10.1021/acs.nanolett.0c03119.

(1066)  Mortensen, N. A.; Gonçalves, P. A. D.; Khajavikhan, M.; Christodoulides, D. N.; Tserkezis, C.; Wolff, C.; Ortensen, N. A. S. M.; Onçalves, P. A. D. G.; Hajavikhan, M. E. K.; Hristodoulides, D. E. N. C.; Serkezis, C. H. T.; Olff, C. H. W.; Mortensen, N. A.; Gonçalves, P. A. D.; Khajavikhan, M.; Christodoulides, D. N.; Tserkezis, C.; Wolff, C. Fluctuations and Noise-Limited Sensing near the Exceptional Point of Parity-Time-Symmetric Resonator Systems. *Optica* **2018**, *5* (10), 1342. https://doi.org/10.1364/optica.5.001342.

(1067)  Choi, Y.; Hahn, C.; Yoon, J. W.; Song, S. H. Observation of an Anti-PT-Symmetric Exceptional Point and Energy-Difference Conserving Dynamics in Electrical Circuit Resonators. *Nat. Commun.* **2018**, *9* (1), 2182. https://doi.org/10.1038/s41467-018-04690-y.

(1068)  Zhang, J.; Li, L.; Wang, G.; Feng, X.; Guan, B.-O.; Yao, J. Parity-Time Symmetry in Wavelength Space within a Single Spatial Resonator. *Nat. Commun.* **2020**, *11* (1), 3217. https://doi.org/10.1038/s41467-020-16705-8.

(1069)  Domínguez-Rocha, V.; Thevamaran, R.; Ellis, F. M.; Kottos, T. Environmentally Induced Exceptional Points in Elastodynamics. *Phys. Rev. Appl.* **2020**, *13* (1), 014060. https://doi.org/10.1103/PhysRevApplied.13.014060.

(1070)  Wang, Y. X.; Clerk, A. A. Non-Hermitian Dynamics without Dissipation in Quantum Systems. *Phys. Rev. A* **2019**, *99* (6), 6334. https://doi.org/10.1103/PhysRevA.99.063834.

(1071)  Doppler, J.; Mailybaev, A. A.; Böhm, J.; Kuhl, U.; Girschik, A.; Libisch, F.; Milburn, T. J.; Rabl, P.; Moiseyev, N.; Rotter, S. Dynamically Encircling an Exceptional Point for Asymmetric Mode Switching. *Nature* **2016**, *537* (7618), 76–79. https://doi.org/10.1038/nature18605.

(1072)  Bergman, A.; Duggan, R.; Sharma, K.; Tur, M.; Zadok, A.; Alù, A. Observation of Anti-Parity-Time-Symmetry, Phase Transitions and Exceptional Points in an Optical Fibre. *Nat. Commun.* **2021**, *12* (1), 486. https://doi.org/10.1038/s41467-020-20797-7.

(1073)  Makris, K. G.; El-Ganainy, R.; Christodoulides, D. N.; Musslimani, Z. H. Beam Dynamics in PT Symmetric Optical Lattices. *Phys. Rev. Lett.* **2008**, *100* (10), 103904. https://doi.org/10.1103/PhysRevLett.100.103904.

(1074)  Rüter, C. E.; Makris, K. G.; El-Ganainy, R.; Christodoulides, D. N.; Segev, M.; Kip, D. Observation of Parity–Time Symmetry in Optics. *Nat. Phys.* **2010**, *6* (3), 192–195. https://doi.org/10.1038/nphys1515.

(1075)  Liu, Y.; Hao, T.; Li, W.; Capmany, J.; Zhu, N.; Li, M. Observation of Parity-Time Symmetry in Microwave Photonics. *Light Sci. Appl.* **2018**, *7* (1), 38. https://doi.org/10.1038/s41377-018-0035-8.

(1076) Longhi, S. Parity-Time Symmetry Meets Photonics: A New Twist in Non-Hermitian Optics. *EPL (Europhysics Lett.* **2017**, *120* (6), 64001. https://doi.org/10.1209/0295-5075/120/64001.

(1077) Suchkov, S. V.; Sukhorukov, A. A.; Huang, J.; Dmitriev, S. V.; Lee, C.; Kivshar, Y. S. Nonlinear Switching and Solitons in PT-Symmetric Photonic Systems. *Laser Photonics Rev.* **2016**, *10* (2), 177–213. https://doi.org/10.1002/lpor.201500227.

(1078) Bittner, S.; Dietz, B.; Günther, U.; Harney, H. L.; Miski-Oglu, M.; Richter, A.; Schäfer, F. PT-Symmetry and Spontaneous Symmetry Breaking in a Microwave Billiard. *Phys. Rev. Lett.* **2012**, *108* (2), 024101. https://doi.org/10.1103/PhysRevLett.108.024101.

(1079) Li, H.; Mekawy, A.; Krasnok, A.; Alù, A. Virtual Parity-Time Symmetry. *Phys. Rev. Lett.* **2020**, *124* (19), 193901. https://doi.org/10.1103/PhysRevLett.124.193901.

(1080) Lin, Z.; Ramezani, H.; Eichelkraut, T.; Kottos, T.; Cao, H.; Christodoulides, D. N. Unidirectional Invisibility Induced by PT-Symmetric Periodic Structures. *Phys. Rev. Lett.* **2011**, *106* (21), 213901. https://doi.org/10.1103/PhysRevLett.106.213901.

(1081) Feng, L.; Xu, Y. L.; Fegadolli, W. S.; Lu, M. H.; Oliveira, J. E. B.; Almeida, V. R.; Chen, Y. F.; Scherer, A. Experimental Demonstration of a Unidirectional Reflectionless Parity-Time Metamaterial at Optical Frequencies. *Nat. Mater.* **2013**, *12* (2), 108–113. https://doi.org/10.1038/nmat3495.

(1082) Longhi, S. Half-Spectral Unidirectional Invisibility in Non-Hermitian Periodic Optical Structures. *Opt. Lett.* **2015**, *40* (23), 5694. https://doi.org/10.1364/OL.40.005694.

(1083) Peng, B.; Özdemir, Ş. K.; Liertzer, M.; Chen, W.; Kramer, J.; Yılmaz, H.; Wiersig, J.; Rotter, S.; Yang, L. Chiral Modes and Directional Lasing at Exceptional Points. *Proc. Natl. Acad. Sci.* **2016**, *113* (25), 6845–6850. https://doi.org/10.1073/pnas.1603318113.

(1084) Miao, P.; Zhang, Z.; Sun, J.; Walasik, W.; Longhi, S.; Litchinitser, N. M.; Feng, L. Orbital Angular Momentum Microlaser. *Science (80-. ).* **2016**, *353* (6298), 464–467. https://doi.org/10.1126/science.aaf8533.

(1085) Shu, F.-J.; Zou, C.-L.; Zou, X.-B.; Yang, L. Chiral Symmetry Breaking in a Microring Optical Cavity by Engineered Dissipation. *Phys. Rev. A* **2016**, *94* (1), 013848. https://doi.org/10.1103/PhysRevA.94.013848.

(1086) Xu, H.; Mason, D.; Jiang, L.; Harris, J. G. E. Topological Energy Transfer in an Optomechanical System with Exceptional Points. *Nature* **2016**, *537* (7618), 80–83. https://doi.org/10.1038/nature18604.

(1087) Hassan, A. U.; Zhen, B.; Soljačić, M.; Khajavikhan, M.; Christodoulides, D. N. Dynamically Encircling Exceptional Points: Exact Evolution and Polarization State Conversion. *Phys. Rev. Lett.* **2017**, *118* (9), 093002. https://doi.org/10.1103/PhysRevLett.118.093002.

(1088) Wiersig, J. Enhancing the Sensitivity of Frequency and Energy Splitting Detection by Using Exceptional Points: Application to Microcavity Sensors for Single-Particle Detection. *Phys. Rev. Lett.* **2014**, *112* (20), 203901. https://doi.org/10.1103/PhysRevLett.112.203901.

(1089) Chen, W.; Özdemir, Ş. K.; Zhao, G.; Wiersig, J.; Yang, L. Exceptional Points Enhance Sensing in an Optical Microcavity. *Nature* **2017**, *548* (7666), 192–195. https://doi.org/10.1038/NATURE23281.

(1090) Hodaei, H.; Hassan, A. U.; Wittek, S.; Garcia-Gracia, H.; El-Ganainy, R.; Christodoulides, D. N.; Khajavikhan, M. Enhanced Sensitivity at Higher-Order Exceptional Points. *Nature* **2017**, *548*

(7666), 187–191. https://doi.org/10.1038/nature23280.

(1091)  Savoia, S.; Valagiannopoulos, C. A.; Monticone, F.; Castaldi, G.; Galdi, V.; Alù, A. Magnified Imaging Based on Non-Hermitian Nonlocal Cylindrical Metasurfaces. *Phys. Rev. B* **2017**, *95* (11), 115114. https://doi.org/10.1103/PhysRevB.95.115114.

(1092)  Monticone, F.; Valagiannopoulos, C. A.; Alù, A. Parity-Time Symmetric Nonlocal Metasurfaces: All-Angle Negative Refraction and Volumetric Imaging. *Phys. Rev. X* **2016**, *6* (4), 041018. https://doi.org/10.1103/PhysRevX.6.041018.

(1093)  Fleury, R.; Sounas, D. L.; Alu, A. Parity-Time Symmetry in Acoustics: Theory, Devices, and Potential Applications. *IEEE J. Sel. Top. Quantum Electron.* **2016**, *22* (5), 121–129. https://doi.org/10.1109/JSTQE.2016.2549512.

(1094)  Fleury, R.; Sounas, D. L.; Alù, A. Negative Refraction and Planar Focusing Based on Parity-Time Symmetric Metasurfaces. *Phys. Rev. Lett.* **2014**, *113* (2), 023903. https://doi.org/10.1103/PhysRevLett.113.023903.

(1095)  Valagiannopoulos, C. A.; Monticone, F.; Alù, A. PT-Symmetric Planar Devices for Field Transformation and Imaging. *J. Opt.* **2016**, *18* (4), 044028. https://doi.org/10.1088/2040-8978/18/4/044028.

(1096)  Kord, A.; Sounas, D. L.; Alù, A. Active Microwave Cloaking Using Parity-Time-Symmetric Satellites. *Phys. Rev. Appl.* **2018**, *10* (5), 054040. https://doi.org/10.1103/PhysRevApplied.10.054040.

(1097)  Castaldi, G.; Savoia, S.; Galdi, V.; Alù, A.; Engheta, N. PT-Metamaterials via Complex-Coordinate Transformation Optics. *Phys. Rev. Lett.* **2013**, *110* (17), 173901. https://doi.org/10.1103/PhysRevLett.110.173901.

(1098)  Moccia, M.; Castaldi, G.; Alù, A.; Galdi, V. Line Waves in Non-Hermitian Metasurfaces. *ACS Photonics* **2020**, *7* (8), 2064–2072. https://doi.org/10.1021/acsphotonics.0c00465.

(1099)  Savoia, S.; Castaldi, G.; Galdi, V.; Alù, A.; Engheta, N. PT-Symmetry-Induced Wave Confinement and Guiding in Epsilon-near-Zero Metamaterials. *Phys. Rev. B* **2015**, *91* (11), 115114. https://doi.org/10.1103/PhysRevB.91.115114.

(1100)  Savoia, S.; Castaldi, G.; Galdi, V.; Alù, A.; Engheta, N. Tunneling of Obliquely Incident Waves through PT-Symmetric Epsilon-near-Zero Bilayers. *Phys. Rev. B* **2014**, *89* (8), 085105. https://doi.org/10.1103/PhysRevB.89.085105.

(1101)  Li, M.; Ni, X.; Weiner, M.; Alù, A.; Khanikaev, A. B. Topological Phases and Nonreciprocal Edge States in Non-Hermitian Floquet Insulators. *Phys. Rev. B* **2019**, *100* (4), 045423. https://doi.org/10.1103/PhysRevB.100.045423.

(1102)  Zhang, Y. J.; Li, P.; Galdi, V.; Tong, M. S.; Alù, A. Manipulating the Scattering Pattern with Non-Hermitian Particle Arrays. *Opt. Express* **2020**, *28* (13), 19492. https://doi.org/10.1364/oe.395942.

(1103)  Moccia, M.; Castaldi, G.; Alu, A.; Galdi, V. Harnessing Spectral Singularities in Non-Hermitian Cylindrical Structures. *IEEE Trans. Antennas Propag.* **2020**, *68* (3), 1704–1716. https://doi.org/10.1109/TAP.2019.2927861.

(1104)  Assawaworrarit, S.; Fan, S. Robust and Efficient Wireless Power Transfer Using a Switch-Mode Implementation of a Nonlinear Parity--Time Symmetric Circuit. *Nat. Electron.* **2020**, *3* (5), 273–279.

(1105)  Xiao, Z.; Ra'di, Y.; Tretyakov, S.; Alù, A. Microwave Tunneling and Robust Information Transfer Based on Parity-Time-Symmetric Absorber-Emitter Pairs. *Research* **2019**, *2019*, 1–10. https://doi.org/10.34133/2019/7108494.

(1106)  Zhen, B.; Hsu, C. W.; Igarashi, Y.; Lu, L.; Kaminer, I.; Pick, A.; Chua, S. L.; Joannopoulos, J. D.; Soljačić, M. Spawning Rings of Exceptional Points out of Dirac Cones. *Nature* **2015**, *525* (7569), 354–358. https://doi.org/10.1038/NATURE14889.

(1107)  Shastri, K.; Monticone, F. Dissipation-Induced Topological Transitions in Continuous Weyl Materials. *Phys. Rev. Res.* **2020**, *2* (3), 033065. https://doi.org/10.1103/PhysRevResearch.2.033065.

(1108)  Berry, M. Geometric Amplitude Factors in Adiabatic Quantum Transitions. *Proc. R. Soc. London. Ser. A Math. Phys. Sci.* **1990**, *430* (1879), 405–411. https://doi.org/10.1098/rspa.1990.0096.

(1109)  Zwanziger, J. W.; Rucker, S. P.; Chingas, G. C. Measuring the Geometric Component of the Transition Probability in a Two-Level System. *Phys. Rev. A* **1991**, *43* (7), 3232–3240. https://doi.org/10.1103/PhysRevA.43.3232.

(1110)  Kepler, T. B.; Kagan, M. L. Geometric Phase Shifts under Adiabatic Parameter Changes in Classical Dissipative Systems. *Phys. Rev. Lett.* **1991**, *66* (7), 847–849. https://doi.org/10.1103/PhysRevLett.66.847.

(1111)  Ning, C. Z.; Haken, H. Geometrical Phase and Amplitude Accumulations in Dissipative Systems with Cyclic Attractors. *Phys. Rev. Lett.* **1992**, *68* (14), 2109–2112. https://doi.org/10.1103/PhysRevLett.68.2109.

(1112)  Bliokh, K. Y. The Appearance of a Geometric-Type Instability in Dynamic Systems with Adiabatically Varying Parameters. *J. Phys. A. Math. Gen.* **1999**, *32* (13), 2551–2565. https://doi.org/10.1088/0305-4470/32/13/007.

(1113)  Dietz, B.; Harney, H. L.; Kirillov, O. N.; Miski-Oglu, M.; Richter, A.; Schäfer, F. Exceptional Points in a Microwave Billiard with Time-Reversal Invariance Violation. *Phys. Rev. Lett.* **2011**, *106* (15), 150403. https://doi.org/10.1103/PhysRevLett.106.150403.

(1114)  Whitney, R. S.; Gefen, Y. Berry Phase in a Nonisolated System. *Phys. Rev. Lett.* **2003**, *90* (19), 190402. https://doi.org/10.1103/PhysRevLett.90.190402.

(1115)  Whitney, R. S.; Makhlin, Y.; Shnirman, A.; Gefen, Y. Geometric Nature of the Environment-Induced Berry Phase and Geometric Dephasing. *Phys. Rev. Lett.* **2005**, *94* (7), 070407. https://doi.org/10.1103/PhysRevLett.94.070407.

(1116)  Berger, S. Measurement of Geometric Dephasing Using a Superconducting Qubit. *Nat. Commun.* **2015**, *6*.

(1117)  Gaitan, F. Berry's Phase in the Presence of a Stochastically Evolving Environment: A Geometric Mechanism for Energy-Level Broadening. *Phys. Rev. A* **1998**, *58* (3), 1665–1677. https://doi.org/10.1103/PhysRevA.58.1665.

(1118)  De Chiara, G.; Palma, G. M. Berry Phase for a Spin 1/2 Particle in a Classical Fluctuating Field. *Phys. Rev. Lett.* **2003**, *91* (9), 090404. https://doi.org/10.1103/PhysRevLett.91.090404.

(1119)  Lei, Q.-L.; Zheng, W.; Tang, F.; Wan, X.; Ni, R.; Ma, Y. Self-Assembly of Isostatic Self-Dual Colloidal Crystals. *Phys. Rev. Lett.* **2021**, *127* (1), 018001. https://doi.org/10.1103/PhysRevLett.127.018001.

(1120) Peri, V.; Song, Z. Da; Serra-Garcia, M.; Engeler, P.; Queiroz, R.; Huang, X.; Deng, W.; Liu, Z.; Andrei Bernevig, B.; Huber, S. D. Experimental Characterization of Fragile Topology in an Acoustic Metamaterial. *Science (80-. ).* **2020**, *367* (6479), 797–800. https://doi.org/10.1126/SCIENCE.AAZ7654.

(1121) de Paz, M. B.; Vergniory, M. G.; Bercioux, D.; García-Etxarri, A.; Bradlyn, B. Engineering Fragile Topology in Photonic Crystals: Topological Quantum Chemistry of Light. *Phys. Rev. Res.* **2019**, *1* (3), 032005. https://doi.org/10.1103/PhysRevResearch.1.032005.

(1122) Chen, Z. G.; Zhang, R. Y.; Chan, C. T.; Ma, G. Classical Non-Abelian Braiding of Acoustic Modes. *Nat. Phys.* **2022**, *18* (2), 179–184. https://doi.org/10.1038/s41567-021-01431-9.

(1123) Jiang, B.; Bouhon, A.; Lin, Z.-K.; Zhou, X.; Hou, B.; Li, F.; Slager, R.-J.; Jiang, J.-H. Experimental Observation of Non-Abelian Topological Acoustic Semimetals and Their Phase Transitions. *Nat. Phys.* **2021**, *17* (11), 1239–1246. https://doi.org/10.1038/s41567-021-01340-x.

(1124) You, O.; Liang, S.; Xie, B.; Gao, W.; Ye, W.; Zhu, J.; Zhang, S. Observation of Non-Abelian Thouless Pump. *Phys. Rev. Lett.* **2022**, *128* (24), 244302. https://doi.org/10.1103/PhysRevLett.128.244302.

(1125) Zhang, X.-L.; Yu, F.; Chen, Z.-G.; Tian, Z.-N.; Chen, Q.-D.; Sun, H.-B.; Ma, G. Non-Abelian Braiding on Photonic Chips. *Nat. Photonics* **2022**, *16* (5), 390–395. https://doi.org/10.1038/s41566-022-00976-2.

(1126) Wang, M.; Liu, S.; Ma, Q.; Zhang, R.-Y.; Wang, D.; Guo, Q.; Yang, B.; Ke, M.; Liu, Z.; Chan, C. T. Experimental Observation of Non-Abelian Earring Nodal Links in Phononic Crystals. *Phys. Rev. Lett.* **2022**, *128* (24), 246601. https://doi.org/10.1103/PhysRevLett.128.246601.

(1127) Weidemann, S.; Kremer, M.; Longhi, S.; Szameit, A. Topological Triple Phase Transition in Non-Hermitian Floquet Quasicrystals. *Nature* **2022**, *601* (7893), 354–359. https://doi.org/10.1038/s41586-021-04253-0.

(1128) Caloz, C.; Deck-Leger, Z.-L. Spacetime Metamaterials—Part II: Theory and Applications. *IEEE Trans. Antennas Propag.* **2020**, *68* (3), 1583–1598. https://doi.org/10.1109/TAP.2019.2944216.

(1129) Lustig, E.; Sharabi, Y.; Segev, M. Topological Aspects of Photonic Time Crystals. *Optica* **2018**, *5* (11), 1390. https://doi.org/10.1364/OPTICA.5.001390.

(1130) Galiffi, E.; Tirole, R.; Yin, S.; Li, H.; Vezzoli, S.; Huidobro, P. A.; Silveirinha, M. G.; Sapienza, R.; Alù, A.; Pendry, J. B. Photonics of Time-Varying Media. *Adv. Photonics* **2022**, *4* (01). https://doi.org/10.1117/1.AP.4.1.014002.

(1131) Lyubarov, M.; Lumer, Y.; Dikopoltsev, A.; Lustig, E.; Sharabi, Y.; Segev, M. Amplified Emission and Lasing in Photonic Time Crystals. *Science (80-. ).* **2022**. https://doi.org/10.1126/science.abo3324.

(1132) Banerjee, D.; Souslov, A.; Abanov, A. G.; Vitelli, V. Odd Viscosity in Chiral Active Fluids. *Nat. Commun.* **2017**, *8* (1), 1573. https://doi.org/10.1038/s41467-017-01378-7.

(1133) Abgrall, P.; Gué, A.-M. Lab-on-Chip Technologies: Making a Microfluidic Network and Coupling It into a Complete Microsystem—a Review. *J. Micromechanics Microengineering* **2007**, *17* (5), R15–R49. https://doi.org/10.1088/0960-1317/17/5/R01.

(1134) van Reenen, A.; de Jong, A. M.; den Toonder, J. M. J.; Prins, M. W. J. Integrated Lab-on-Chip Biosensing Systems Based on Magnetic Particle Actuation – a Comprehensive Review. *Lab Chip* **2014**, *14* (12), 1966–1986. https://doi.org/10.1039/C3LC51454D.

(1135) Azizipour, N.; Avazpour, R.; Rosenzweig, D. H.; Sawan, M.; Ajji, A. Evolution of Biochip Technology: A Review from Lab-on-a-Chip to Organ-on-a-Chip. *Micromachines* **2020**, *11* (6), 1–15. https://doi.org/10.3390/mi11060599.

(1136) Andersson, H.; van den Berg, A. Microfluidic Devices for Cellomics: A Review. *Sensors Actuators B Chem.* **2003**, *92* (3), 315–325. https://doi.org/10.1016/S0925-4005(03)00266-1.

(1137) Schmitt, F.; Piccin, O.; Barbé, L.; Bayle, B. Soft Robots Manufacturing: A Review. *Front. Robot. AI* **2018**, *5*. https://doi.org/10.3389/frobt.2018.00084.

(1138) Rus, D.; Tolley, M. T. Design, Fabrication and Control of Origami Robots. *Nat. Rev. Mater.* **2018**, *3* (6), 101–112. https://doi.org/10.1038/s41578-018-0009-8.

(1139) Palagi, S.; Fischer, P. Bioinspired Microrobots. *Nat. Rev. Mater.* **2018**, *3* (6), 113–124. https://doi.org/10.1038/s41578-018-0016-9.

(1140) Wallin, T. J.; Pikul, J.; Shepherd, R. F. 3D Printing of Soft Robotic Systems. *Nat. Rev. Mater.* **2018**, *3* (6), 84–100. https://doi.org/10.1038/s41578-018-0002-2.